\documentclass[11pt,reqno]{amsart}

\usepackage[T1]{fontenc}
\usepackage[margin=1in]{geometry}
\usepackage{amsmath,amssymb,amsthm,mathtools}
\usepackage{bm}
\usepackage{booktabs}
\usepackage{enumitem}
\usepackage{microtype}
\usepackage{xcolor}
\newif\ifshowchanges
\showchangesfalse
\definecolor{revisionblue}{RGB}{24,70,155}
\definecolor{additionpurple}{RGB}{115,35,135}
\newcommand{\rev}[1]{{\ifshowchanges\color{revisionblue}\fi #1}}
\newcommand{\added}[1]{{\ifshowchanges\color{additionpurple}\fi #1}}
\newenvironment{revision}{\par\begingroup\ifshowchanges\color{revisionblue}\fi}{\par\endgroup}
\newenvironment{addition}{\par\begingroup\ifshowchanges\color{additionpurple}\fi}{\par\endgroup}
\usepackage{hyperref}
\usepackage{aliascnt}
\usepackage[nameinlink,noabbrev]{cleveref}
\usepackage{tikz}
\usetikzlibrary{arrows.meta,positioning,fit,calc,shapes.geometric}
\usepackage{array}
\usepackage{longtable}
\usepackage{graphicx}
\usepackage[section]{placeins}
\usepackage{needspace}
\usepackage{url}

\hypersetup{
  colorlinks=true,
  linkcolor=blue!45!black,
  citecolor=blue!45!black,
  urlcolor=blue!55!black,
  pdfauthor={Zhe Ju, Jingcheng Liu, Yitong Yin},
  pdftitle={Fine-Grained Complexity Across the Computational Phase Transition}
}

\allowdisplaybreaks
\numberwithin{equation}{section}
\makeatletter
\renewcommand\paragraph{\@startsection{paragraph}{4}%
  \z@{.35\linespacing\@plus.15\linespacing}{-.5em}%
  {\normalfont\bfseries}}
\makeatother

\newtheorem{theorem}{Theorem}[section]
\newaliascnt{proposition}{theorem}
\newtheorem{proposition}[proposition]{Proposition}
\aliascntresetthe{proposition}
\newaliascnt{lemma}{theorem}
\newtheorem{lemma}[lemma]{Lemma}
\aliascntresetthe{lemma}
\newaliascnt{corollary}{theorem}
\newtheorem{corollary}[corollary]{Corollary}
\aliascntresetthe{corollary}
\newaliascnt{conjecture}{theorem}

\aliascntresetthe{conjecture}
\newaliascnt{assumption}{theorem}

\aliascntresetthe{assumption}
\newaliascnt{hypothesis}{theorem}

\aliascntresetthe{hypothesis}
\theoremstyle{definition}
\newaliascnt{definition}{theorem}
\newtheorem{definition}[definition]{Definition}
\aliascntresetthe{definition}
\newaliascnt{construction}{theorem}
\newtheorem{construction}[construction]{Construction}
\aliascntresetthe{construction}
\newaliascnt{problem}{theorem}
\newtheorem{problem}[problem]{Problem}
\aliascntresetthe{problem}
\theoremstyle{remark}
\newaliascnt{remark}{theorem}
\newtheorem{remark}[remark]{Remark}
\aliascntresetthe{remark}

\crefname{assumption}{assumption}{assumptions}
\Crefname{assumption}{Assumption}{Assumptions}
\crefname{hypothesis}{hypothesis}{hypotheses}
\Crefname{hypothesis}{Hypothesis}{Hypotheses}
\crefname{conjecture}{conjecture}{conjectures}
\Crefname{conjecture}{Conjecture}{Conjectures}
\crefname{problem}{problem}{problems}
\Crefname{problem}{Problem}{Problems}
\crefname{construction}{construction}{constructions}
\Crefname{construction}{Construction}{Constructions}

\newcommand{\E}{\mathbb E}
\newcommand{\Pp}{\mathbb P}
\newcommand{\Var}{\operatorname{Var}}
\newcommand{\Cov}{\operatorname{Cov}}
\newcommand{\KL}{D_{\mathrm{KL}}}
\newcommand{\TV}{d_{\mathrm{TV}}}
\newcommand{\Ind}{\mathcal I}
\newcommand{\Cut}{\operatorname{cut}}
\newcommand{\MaxCut}{\operatorname{MAX\text{-}CUT}}
\newcommand{\supp}{\operatorname{supp}}
\newcommand{\diag}{\operatorname{diag}}
\newcommand{\tr}{\operatorname{tr}}
\newcommand{\spec}{\operatorname{spec}}

\newcommand{\poly}{\operatorname{poly}}

\newcommand{\cF}{\mathcal F}

\newcommand{\cC}{\mathcal C}

\newcommand{\one}{\mathbf 1}
\newcommand{\eps}{\varepsilon}
\newcommand{\lambdaC}{\lambda_{\mathrm c}}
\newcommand{\logshift}{\widehat{\delta}}

\newcommand{\dd}{\,\mathrm d}
\newcommand{\HCCount}{\textnormal{\textsc{HC-Count}}}
\newcommand{\HCSample}{\textnormal{\textsc{HC-Sample}}}

\newcommand{\bits}{\operatorname{bits}}

\title[Fine-grained complexity across the computational phase transition]{Fine-Grained Complexity Across the\\Computational Phase Transition}
\author{Zhe Ju}
\author{Jingcheng Liu}
\author{Yitong Yin}
\thanks{All authors are affiliated with the State Key Laboratory for Novel
Software Technology, New Cornerstone Science Laboratory, Nanjing University,
China. %
E-mail addresses:
\href{mailto:juzhe@smail.nju.edu.cn}{\texttt{juzhe@smail.nju.edu.cn}} (Zhe Ju),
\href{mailto:liu@nju.edu.cn}{\texttt{liu@nju.edu.cn}} (Jingcheng Liu), 
and
\href{mailto:yinyt@nju.edu.cn}{\texttt{yinyt@nju.edu.cn}} (Yitong Yin).}
\date{}

\makeatletter
\newcommand{\synlabelalias}[2]{%
  \@ifundefined{r@#2}{}{%
    \protected@write\@auxout{}{\string\newlabel{#1}{\csname r@#2\endcsname}}}%
  \@ifundefined{r@#2@cref}{}{%
    \protected@write\@auxout{}{\string\newlabel{#1@cref}{\csname r@#2@cref\endcsname}}}%
}
\makeatother

\begin{document}
\pagenumbering{roman}
\begin{abstract}
How does a computational problem transition from easy to hard?
Approximate counting and sampling problems exhibit computational phase
transitions. A canonical example is the hard-core model, whose partition
function is the independence polynomial: polynomial-time tractability extends to the tree-uniqueness
threshold, while any constant relative distance above it yields worst-case
intractability unless $\mathsf{NP}=\mathsf{RP}$.

We quantify the transition from polynomial to exponential mixing for classical Glauber dynamics. 
On $n$-vertex graphs of maximum degree $d\ge3$, 
at fugacity $\lambda=(1+\delta)\lambdaC(d)$ with sufficiently small $\delta\ge0$ that may vanish with $n$, 
its worst-case mixing time has the form
\[
 \exp\!\left\{\Theta\!\left(\frac{n\delta^2}{d}\right)\right\}n^{1+\Theta(1)}.
\]
The upper bound is degree-uniform, and the matching lower bound holds for every fixed~$d$.
Annealing gives approximate counting with the same running-time upper bound at fixed accuracy.

For general sampling and counting algorithms, we establish a robust fine-grained reduction from SAT without assuming a lower bound on SAT's running time.
For any fixed $d\ge3$ and $0<\gamma<1$,
a sampling or approximate-counting algorithm running in $\exp\{O((n\delta^2/d)^{1-\gamma})\}\poly(n)$ time along any one suitable inverse-polynomial distance sequence $\delta=\delta(n)$, yields a randomized SAT algorithm running in $\exp\{\widetilde O(s^{1-\gamma})\}$ time on inputs of binary length $s$.
This power-saving transfer identifies $n\delta^2/d$ as the analogue of SAT input length for hard-core sampling and counting at relative distance $\delta$ above criticality:
beating the Glauber exponent by a fixed power through any sampling or counting method would yield a corresponding power improvement for SAT.

At criticality, we establish $O(\sqrt{n/d})$ spectral independence
using a marginal response bound under external-field perturbations.
{The dependence on $n$ is optimal at every fixed degree.}
Together with a subcritical bound, this yields degree-uniform Glauber
mixing in $\smash[b]{\widetilde O\!\left(n(n/d)^{1+1/(d-2)}\right)}$
time, improving the previous $\widetilde O\!\left(n^{4+4/(d-2)}\right)$
bound.
\end{abstract}
\maketitle
{\footnotesize\setlength{\baselineskip}{10pt}\tableofcontents}\par\removelastskip

\clearpage
\pagenumbering{arabic}
\section{Introduction}
{How does the running time of a computational problem grow as it leaves the polynomial-time regime?} The hard-core model is a
central benchmark for studying this question through \emph{computational
phase transitions}: a physical phase transition on an infinite tree
predicts the boundary of worst-case computational tractability on finite
graphs.
Establishing this correspondence has driven a sustained research effort
across statistical physics, probability, and theoretical computer science.

The hard-core model is a distribution over independent sets of a graph. It assigns to each independent
set $I\in\Ind(G)$ of an $n$-vertex graph $G=(V,E)$ the probability
\[
 \mu_{G,\lambda}(I)=\frac{\lambda^{|I|}}{Z_G(\lambda)},
 \qquad Z_G(\lambda)=\sum_{I\in\Ind(G)}\lambda^{|I|}.
\]
We identify each independent set with its indicator vector in
$\{0,1\}^V$. For a random configuration $X$, write $X_v$ for its
occupation indicator and $|X|=\sum_{v\in V}X_v$.

The fugacity $\lambda>0$ controls the preference for occupation. The two
tasks are to sample approximately from $\mu_{G,\lambda}$ and to
approximate its normalizing constant $Z_G(\lambda)$, a weighted count of
independent sets. For graphs of maximum degree $d\ge3$, the governing
threshold is
\[\lambdaC(d)=\frac{(d-1)^{d-1}}{(d-2)^d}.\]
On the infinite $d$-regular tree, the Gibbs measure is unique up to this
threshold~\cite{Kelly1985}; above it, different boundary conditions sustain
distinct phases.

A long sequence of breakthroughs has made this model a meeting point
of algorithm design and hardness of approximation. In the
\emph{subcritical regime} $\lambda<\lambdaC(d)$, correlation decay on
self-avoiding-walk trees yielded deterministic approximate counting
for fixed $d$ and $\lambda$~\cite{Weitz2006}.
Spectral independence and its connection to tree-recursion contraction
established polynomial-time mixing of Glauber dynamics---the local
Markov chain that updates one randomly chosen vertex at a time
\cite{AnariLiuOveisGharan2020,ChenLiuVigoda2020}.
Field dynamics and entropy methods further sharpened the mixing and
sampling bounds~\cite{ChenFengYinZhang2021,ChenLiuVigoda2021,AnariEtAl2022Entropic}.
Subsequent work established $O(n\log n)$ mixing of Glauber dynamics
at any fixed relative distance below the threshold, with constants
independent of $d$~\cite{ChenFengYinZhang2022,ChenEldan2022}.
At \emph{criticality}, $\lambda=\lambdaC(d)$, its mixing time is
polynomial~\cite{CCYZ2026,ChenJiang2025,ChenJuMiaoYinZhang2026}.

In the \emph{supercritical regime}, phase coexistence causes exponentially
slow mixing of local chains on random regular bipartite
graphs~\cite{DyerFriezeJerrum2002,MosselWeitzWormald2009}.
Phase-gadget reductions established computational hardness
\cite{DyerFriezeJerrum2002,Sly2010,SlySun2014,GalanisStefankovicVigoda2016,GSV2016}: for every fixed $d\ge3$ and fixed
positive relative distance above the threshold, neither approximate
counting nor approximate sampling admits a polynomial-time algorithm
unless $\mathsf{NP}=\mathsf{RP}$, even on $d$-regular graphs.

This threshold theory brings a finer question into focus: how does
computational complexity grow from polynomial to exponential time as
the threshold is crossed? We write
$\lambda=(1+\delta)\lambdaC(d)$, where $\delta\ge0$ is the relative
distance above the threshold.
To resolve this transition, we allow $\delta$ to vary with the graph
size and seek bounds on algorithmic running times and computational
hardness that cover both fixed and vanishing distances.

For algorithms, direct comparison with the critical Glauber chain gives
$\exp\{O(n\delta)\}\poly(n)$.\footnote{The likelihood ratio
$\mu_{G,\lambda}/\mu_{G,\lambdaC(d)}$ varies by at most $(1+\delta)^n$.
Variance and Dirichlet-form comparison~\cite[Chapter~13]{LevinPeres2017}
gives this
factor for relaxation time, and hence the stated mixing bound for bounded
$\delta\ge0$.}
At fixed degree, even $\delta=n^{-1/2}$ gives a stretched-exponential
guarantee by this comparison. Capturing the true growth requires
finite-size correlation bounds that remain effective above the threshold
and under conditioning, where subcritical contraction degenerates.

For computational hardness, existing NP-hardness reductions give intractability at every fixed distance $\delta>0$, using gadgets of constant size. 
They do not reach the regime of inverse-polynomially small distances, where Glauber dynamics passes from polynomial mixing to superpolynomial. 
As $\delta\to0$, the extremal Gibbs measures representing the two phases on the infinite $d$-regular tree converge locally to the same critical measure.
The challenge for lower bounds is to quantify how this vanishing phase distinction can still sustain computational hardness on growing finite graphs.

We establish a robust fine-grained reduction from SAT that relates the running times of sampling and counting to that of SAT as the distance from criticality varies with the graph size.
Our bounds identify $n\delta^2/d$ as the exponential scale of worst-case Glauber mixing, up to polynomial factors.
For general sampling and counting algorithms, the reduction transfers improvements on this scale to corresponding improvements
for SAT, including at subexponential running-time scales.  

\subsection{Main results}\label{sec:problems}

For an $n$-vertex graph with maximum degree $d\ge3$, write
$\lambda=(1+\delta)\lambdaC(d)$, where $\delta\ge0$ is the
relative distance from criticality.
The parameter $\delta$ is part of the input and \emph{may depend
on $n$}, including regimes where $\delta\to0$ as $n\to\infty$.
We define the two tasks.

\begin{problem}[Approximate hard-core counting]\label{prob:count}
Given an $n$-vertex simple graph $G$ with $d=\Delta(G)\ge3$, a rational
$\delta\ge0$, and $\eps\in(0,1/4)$, set
$\lambda=(1+\delta)\lambdaC(d)$ and output $\widehat Z$ such that, with
probability at least $3/4$,
\[
  (1-\eps)Z_G(\lambda)\le\widehat Z\le(1+\eps)Z_G(\lambda).
\]
We denote this task by $\HCCount_d(G,\delta,\eps)$.
\end{problem}

\begin{problem}[Approximate hard-core sampling]\label{prob:sample}
Given an $n$-vertex simple graph $G$ with $d=\Delta(G)\ge3$, a rational
$\delta\ge0$, and $\eta\in(0,1/4)$, set
$\lambda=(1+\delta)\lambdaC(d)$ and output a random independent set whose law $\widehat\mu$
satisfies the total-variation bound
\[
  \TV(\widehat\mu,\mu_{G,\lambda})\le\eta.
\]
We denote this task by $\HCSample_d(G,\delta,\eta)$.
\end{problem}

The algorithmic bounds count randomized arithmetic operations,
with binary overhead stated separately; the computational reductions
use the randomized bit model
(see \cref{sec:computational-conventions}).
All logarithms are natural.
In the upper bounds, $\widetilde O$ suppresses only powers of $\log n$;
their exponents and implicit constants are absolute. Dependence on $d$,
$\delta$, $\eta$, and $\eps$ is displayed explicitly.

\paragraph{Glauber dynamics across the computational phase transition.}

\begin{theorem}[Degree-uniform near-critical algorithms]\label{thm:main-upper}
There are absolute constants $C,\delta_{\mathrm{up}}>0$ and two
randomized algorithms such that, for every $n$-vertex simple graph $G$
with maximum degree $d=\Delta(G)\ge3$,
every rational $0\le\delta\le\delta_{\mathrm{up}}$, and every
$\eta,\eps\in(0,1/4)$,
$\HCSample_d(G,\delta,\eta)$ can be solved in time
\[
 O(n+|E|)+\widetilde O\!\left(
 n\left(\frac nd\right)^{1+\frac1{d-2}}\exp\!\left\{\frac{Cn\delta^2}{d}\right\}
 \log\frac1\eta\right).
\]
$\HCCount_d(G,\delta,\eps)$ can be solved in time
\[
 O(n+|E|)+\widetilde O\!\left(
 n\left(\frac nd\right)^{1+\frac1{d-2}}\exp\!\left\{\frac{Cn\delta^2}{d}\right\}
 \eps^{-2}\right).
\]
The bounds hold uniformly in $d$ and $\delta$, including when $d$ grows and $\delta$ tends
to zero with $n$.
All implicit constants and polylogarithmic exponents are absolute.
\end{theorem}

The sampling bound comes from the mixing time of single-site
Glauber dynamics (\cref{thm:glauber-upper}).
At fixed accuracy, our counting algorithm matches this upper
bound up to polylogarithmic factors, avoiding the overhead
of directly applying generic counting-via-sampling reductions
(\cref{sec:proof-outline-counting}).

We call $\frac{n\delta^2}{d}=O(\log n)$ the \emph{polynomial-time window} of the upper bound.
Next, we present a matching exponential lower bound,
making this window exact for rapid mixing at fixed degree.

\begin{theorem}[Mixing-time transition]
\label{thm:main-glauber-lower}
There is an absolute constant $c>0$ such that, for every fixed
$d\ge3$, there are constants $a_d>0$ and
$\delta_{\mathrm{mix}}(d)>0$ with the following property.
For every $n\ge d+1$ and every rational
$0\le\delta\le\delta_{\mathrm{mix}}(d)$, at fugacity
$\lambda=(1+\delta)\lambdaC(d)$,
\[
 \max_G t_{\mathrm{mix}}^G(1/4)
 \ge a_dn^{3/2}\exp\!\left(c\frac{n\delta^2}{d}\right),
\]
where $t_{\mathrm{mix}}^G(1/4)$ is the mixing time of Glauber dynamics
for $\mu_{G,\lambda}$ from the worst initial state (\cref{sec:prelim}),
and the maximum is over simple $n$-vertex graphs of maximum degree $d$.

Together with the mixing-time bound of \cref{thm:glauber-upper}, this implies that,
for any sequence of rationals
$0\le\delta_n\le\delta_{\mathrm{mix}}(d)$, the worst-case Glauber
mixing time at fugacity $(1+\delta_n)\lambdaC(d)$ is polynomial
in $n$ if and only if
\[
 \delta_n=O\!\left(\sqrt{\frac{d\log n}{n}}\right).
\]
\end{theorem}
Thus, for every fixed $d$, the two-sided scale is
$\exp\{\Theta(n\delta^2/d)\}n^{1+\Theta(1)}$.
The upper bound is degree-uniform, whereas the lower bound
holds for every fixed $d$, even on simple $d$-regular
bipartite graphs when $n\ge2d$ is even
(\cref{cor:glauber-superlinear-lower}).
The sequence $(\delta_n)$ in the polynomial mixing
characterization need not be monotone or computable.

Our lower bounds use random symmetric bipartite graphs,
building on their use in hardness reductions and critical
mixing lower bounds~\cite{SlySun2014,CCYZ2026}.
At criticality, our analysis of this construction improves
the $\Omega_d(n^{4/3})$ mixing lower bound of Chen, Chen, Yin, and
Zhang~\cite[Theorem~1.1 and Section~5.2]{CCYZ2026}
to $\Omega_d(n^{3/2})$.

\begin{remark}[Beyond Glauber dynamics]
\label{rem:local-chain-obstruction}
There are absolute constants $b,c>0$ such that, for every fixed
$d\ge3$, every even $n\ge2d$, and every rational
$0<\delta\le\delta_{\mathrm{mix}}(d)$ with $n\delta^2/d\ge b$,
there is a simple $n$-vertex $d$-regular bipartite graph $G$
on which every irreducible, aperiodic Markov chain on $\Ind(G)$
with stationary law $\mu_{G,(1+\delta)\lambdaC(d)}$
and transitions changing at most $\lfloor n\delta^2/(bd)\rfloor$
occupation indicators has mixing time at least
$\exp\{cn\delta^2/d\}$.
Reversibility is not required.
It is proved in \cref{sec:glauber-lower-proof}.
\end{remark}

\paragraph{Robust fine-grained reductions from SAT}

{At fixed $d$, as $\delta$ decreases from a constant to~$n^{-1/2}$, the 
Glauber mixing time passes from exponential, through subexponential,
to polynomial time.
For example, at $\delta\approx n^{-1/3}$ it is
$\exp\{\Theta(n\delta^2/d)\}=\exp\{\Theta(n^{1/3}/d)\}$,
which is subexponential but superpolynomial.
Is the problem still hard, and what would a faster
algorithm imply?}

{Our reduction from SAT answers both questions.
For the distance sequences below and fixed $0<\gamma<1$,
hard-core sampling or counting at fixed accuracy in
$\poly(n)\exp\{O((n\delta^2/d)^{1-\gamma})\}$ time
would yield a randomized SAT algorithm running in
$\exp\{\widetilde O(s^{1-\gamma})\}$ time on inputs of
length~$s$, preserving the power saving in the exponent
up to polylogarithmic factors.
Such a saving would be a major further breakthrough for SAT,
going well beyond the subexponential bounds announced in recent
preprints~\cite{OpenAI2026ETH,ClaudeAnthropic2026SAT}.
In this sense, $n\delta^2/d$ plays the role of the input
length of SAT.
We remark that the announced algorithms, if confirmed, do not improve on Glauber dynamics for the inverse-polynomial distances covered by our power-saving transfer theorems. 
For illustration, consider $\delta\approx n^{-1/3}$.  While computing $Z_G(\lambda)$ exactly takes time
$2^{O(n/\sqrt{\log n})}$ now at fixed degree,\footnote{For each
$0\le k\le n$, independent sets of size $k$ admit a parsimonious
encoding as a $3$-CNF formula with $O(n)$ variables and $O_d(n)$
clauses, using the edge constraints and a linear-size Boolean
circuit for the occupation count, as in the proof
of~\cite[Corollary~7]{ClaudeAnthropic2026SAT}. For formulas with
$O(n)$ variables and $O_d(n)$ clauses, \cite[Theorem~1.1]{OpenAI2026ETH}
and~\cite[Theorem~1]{ClaudeAnthropic2026SAT} count satisfying
assignments in time $2^{O_d(n/\sqrt{\log n})}$. These $n+1$ counts
recover the independence polynomial, which can be evaluated at any
rational fugacity with polynomial binary overhead. Approximate
sampling then follows by self-reduction.}
 Glauber dynamics only needs
$\exp\{O(n^{1/3}/d)\}\poly(n)$ steps to mix.
}

Our transfer assumes no lower bound on SAT's running time.
It follows the spirit of fine-grained
reductions~\cite{VassilevskaWilliamsWilliams2018}, in which a speedup
for one problem gives some speedup for another.
We first state it for fixed power savings;
\cref{thm:robust-sat-transfer} gives the general formulation.

\begin{theorem}[Robust fine-grained reduction from SAT]
\label{thm:main-transfer}
Fix $d\ge3$, $0<\gamma<1$, and
$\gamma/(1+2\gamma)\le a<1/2$.
Let $(\delta_n)$ be a sequence of positive rationals satisfying
$\delta_n=\Theta(n^{-a})$, with $\delta_n$ computable exactly
from $n$ in time polynomial in $n$.

Suppose either $\HCSample_d(G,\delta_n,1/10)$ or
$\HCCount_d(G,\delta_n,1/10)$ admits a randomized algorithm
that, for all sufficiently large $n$, solves the task on every
simple $n$-vertex $d$-regular graph $G$ in worst-case time
\[
 \poly(n)\exp\!\left(
 O\!\left(\left(\frac{n\delta_n^2}{d}\right)^{1-\gamma}\right)\right).
\]
Then SAT on inputs of binary encoding length $s\ge2$ admits
a randomized algorithm with success probability at least $2/3$
and worst-case running time
\[
 \exp\!\left(\widetilde O(s^{1-\gamma})\right).
\]
Here $\widetilde O$ hides a polynomial in $\log s$
of absolute degree.
\end{theorem}

\Cref{thm:main-transfer} does not require uniformity over the distance parameter $\delta$.
Its hypothesis concerns only one prescribed sequence $(\delta_n)$, which need not be monotone.
The assumed algorithm may be tailored to this sequence and need only work at distance $\delta_n$ on $n$-vertex graphs.

The power-saving conclusion of \cref{thm:main-transfer} gives
$n\delta_n^2/d$ a computational interpretation: it plays the
role of SAT input length for hard-core sampling and counting
at relative distance $\delta_n$ above criticality.
Thus, within the theorem's parameter range, beating the
Glauber exponent by a fixed power through any sampling or
counting method would yield the same power saving for SAT,
up to polylogarithmic factors.

The same reduction from SAT preserves every fixed power
saving $0<\gamma<1$ when $1/3\le a<1/2$.
For $0<a<1/3$, it still transfers fixed power savings,
possibly with a loss.
Quasipolynomial running times transfer as well;
\cref{sec:robust-sat-proof} develops these further consequences
and more general running-time bounds.

The next corollary applies this reduction at distances
a polylogarithmic factor above the polynomial mixing window.

\begin{corollary}[Computational consequences near the polynomial window]
\label{cor:near-window-transfer}
Fix $d\ge3$ and $r>1$.
Let $(\delta_n)$ be a sequence of positive rationals satisfying
\[
 \delta_n=\Theta\!\left(\frac{(\log n)^{r/2}}{\sqrt n}\right),
\]
with $\delta_n$ computable exactly from $n$ in time polynomial
in $n$.
Suppose that TV-$1/10$ sampling, or relative-error-$1/10$ counting
with success probability at least $2/3$, is possible in randomized
polynomial time on all sufficiently large simple $n$-vertex
$d$-regular graphs at fugacity $(1+\delta_n)\lambdaC(d)$.
Then SAT on inputs of binary encoding length $s\ge2$ admits
a randomized algorithm with success probability at least $2/3$
and worst-case running time
\[
 \exp\!\left\{\widetilde O(s^{1/r})\right\}.
\]
\end{corollary}
For example, polynomial-time sampling or counting at
$\delta_n=\Theta(\log n/\sqrt n)$ would yield a SAT algorithm
running in $\exp\{\widetilde O(\sqrt s)\}$ time.

{At fixed power distances above $n^{-1/2}$, the reduction
also yields hardness under $\mathsf{NP}\ne\mathsf{RP}$.}
\begin{corollary}[Computational hardness near the window]
\label{cor:main-np-hardness}
Fix $d\ge3$ and $0<\xi<1/2$.
Let $(\delta_n)$ be a sequence of positive rationals satisfying
$\delta_n=\Theta(n^{-1/2+\xi})$, with $\delta_n$ computable exactly
from $n$ in time polynomial in $n$.
Unless $\mathsf{NP}=\mathsf{RP}$, neither TV-$1/10$ sampling nor
relative-error-$1/10$ counting with success probability at least $2/3$
admits a randomized polynomial-time algorithm on all sufficiently large
simple $n$-vertex $d$-regular graphs at fugacity
$(1+\delta_n)\lambdaC(d)$.
\end{corollary}
For rational $\xi$, one may take
$\delta_n=1/\lceil n^{1/2-\xi}\rceil$.
The graphs in this corollary may be disconnected and need not be
bipartite. It identifies the critical power $n^{-1/2}$ for general
algorithms, while their logarithmic tractability boundary and optimal
intermediate running times remain open. The matching exponential
scale in \cref{thm:main-glauber-lower} concerns
Glauber dynamics.

{The power-law assumption can be relaxed to
$c n^{-a}\le\delta_n\le C n^{-a'}$ for fixed
$0<a'\le a<1/2$ and $c,C>0$, without monotonicity;
see \cref{prop:polynomial-schedule-transfer}.}

\begin{remark}[Coarse approximate counting]
\label{rem:coarse-counting}
There is an absolute constant $c_{\mathrm{coarse}}>0$ such that
the counting conclusions of
\cref{thm:main-transfer,cor:near-window-transfer,cor:main-np-hardness}
remain valid if fixed relative error is replaced by the following
guarantee: on every input graph $G$, the algorithm returns
$\widehat Z_G>0$ satisfying
\[
 e^{-c_{\mathrm{coarse}}n\delta_n^2/d}Z_G(\lambda)
 \le\widehat Z_G\le
 e^{c_{\mathrm{coarse}}n\delta_n^2/d}Z_G(\lambda),
 \qquad \lambda=(1+\delta_n)\lambdaC(d),
\]
with probability at least $2/3$.
All other hypotheses remain unchanged.
It is proved in \cref{sec:robust-sat-proof}.
\end{remark}

\paragraph{Spectral independence through criticality.}
Our upper bounds rely on a new bound on \emph{spectral independence}
(SI) that holds through criticality.
This notion was introduced by Anari, Liu, and Oveis
Gharan~\cite{AnariLiuOveisGharan2020}.
For a law $\mu$, the influence matrix $\Psi_\mu$ records changes in
one-coordinate marginals when another coordinate is fixed. We use diagonal entries equal to one on nonconstant coordinates.
See \cref{def:spectral-independence} for the formal definition
and covariance interpretation. A feasible pinning $\tau$ conditions
on an assignment of positive probability, yielding the law $\mu^\tau$.

\begin{theorem}[Spectral independence through criticality]
\label{thm:main-si}
There is an absolute constant $C>0$ such that, for every $d\ge3$,
every $n$-vertex graph $G$ of maximum degree at most $d$, and every
$-1<\delta\le\delta_{\mathrm{up}}$, the hard-core distribution
$\mu=\mu_{G,\lambda}$ at $\lambda=(1+\delta)\lambdaC(d)$ satisfies,
for every feasible pinning $\tau$,
\[
 \lambda_{\max}(\Psi_{\mu^\tau})
 \le C\left(1+\sqrt{\frac nd}+\frac nd\max\{\delta,0\}\right).
\]
In particular, at criticality ($\delta=0$) it has spectral independence
$O(1+\sqrt{n/d})$ under all pinnings.

For every fixed $d\ge3$, there is $c_d>0$ such that, for every
$n\ge d+1$ and every $0\le\delta\le\delta_{\mathrm{mix}}(d)$,
some simple $n$-vertex graph $G$ of maximum degree $d$ satisfies
\[
 \lambda_{\max}(\Psi_{\mu_{G,(1+\delta)\lambdaC(d)}})
 \ge c_d(\sqrt n+n\delta).
\]
Thus the dependence on $n$ and on the nonnegative distance
$\delta$ is optimal at every fixed degree.
\end{theorem}

Our covariance bound follows from marginal response to finite
external-field perturbations (\cref{sec:proof-outline-si}).
Together with a subcritical estimate, it gives degree-uniform
critical Glauber mixing in
$\widetilde O(n(n/d)^{1+1/(d-2)})$
(\cref{thm:glauber-upper}), improving the previous
$\widetilde O(n^{4+4/(d-2)})$
bound~\cite[Theorem~1.1]{ChenJiang2025}.
The critical SI lower bound improves the
$\Omega_d(n^{1/3})$ bound of~\cite[Theorem~5.1]{CCYZ2026}
to $\Omega_d(\sqrt n)$.

\Needspace{4\baselineskip}
\subsection{Related work}\label{sec:related-work}

\paragraph{Complexity between polynomial and exponential time.}
Running time trades off against density in refuting random constraint
satisfaction problems~\cite{RaghavendraRaoSchramm2017} and against
accuracy in mildly exponential approximation~\cite{ManurangsiTrevisan2018}.
Subexponential reductions~\cite{ImpagliazzoPaturiZane2001} and
reductions from approximate counting to decision
\cite{Traxler2016,DellLapinskas2021} compare running times across problems.
Here accuracy is fixed and distance from criticality varies:
our reduction transfers improvements on the scale
$n\delta^2/d$ to SAT.

\paragraph{Algorithms up to uniqueness.}
Early low-fugacity algorithms
\cite{LubyVigoda1999,DyerGreenhill2000,Vigoda2001}
were followed by Weitz's deterministic approximation below
uniqueness at bounded degree~\cite{Weitz2006}, extensions to
antiferromagnetic two-spin systems
\cite{LiLuYin2012,SinclairSrivastavaThurley2014,LiLuYin2013},
and the zero-freeness approach
\cite{PatelRegts2017,PetersRegts2019,LiuSinclairSrivastava2025}.
At any fixed relative distance below uniqueness, spectral
independence established degree-uniform polynomial Glauber
mixing, with an exponent depending only on that distance
\cite{AnariLiuOveisGharan2020,ChenLiuVigoda2020}.
Entropy factorization yielded $O(n\log n)$ mixing at fixed
degree~\cite{ChenLiuVigoda2021}, while field dynamics improved
the degree-uniform bound to $\widetilde O(n^2)$
\cite{ChenFengYinZhang2021}.
Entropic independence also yielded efficient sampling via
modified dynamics
\cite{AnariEtAl2022Entropic,AnariEtAl2022EntropicII}.
Degree-uniform $O(n\log n)$ Glauber mixing at any fixed relative
distance below uniqueness was subsequently established
\cite{ChenFengYinZhang2022,ChenEldan2022}.

At uniqueness, Chen, Chen, Yin, and Zhang proved
degree-uniform polynomial mixing of hard-core Glauber
dynamics~\cite{CCYZ2026}; Chen and Jiang improved the
mixing-time bound using subcritical SI~\cite{ChenJiang2025}.
Edge-tilting field dynamics extends threshold mixing to further
antiferromagnetic two-spin systems under the hypotheses
of~\cite[Theorem~1.3]{ChenJuMiaoYinZhang2026}.
Our critical SI proof uses marginal response to finite field
perturbations on the original graph, avoiding the graph-to-tree
comparison underlying earlier withdrawn square-root claims
(see~\cite[Remark~1.2]{ChenJiang2025}
and~\cite[Remark~1.6]{CCYZ2026}).

Miao and Yin's spectral-stability criterion
\cite[Theorem~1.4]{MiaoYin2026} extends the spectral-gap
approach of~\cite{CCYZ2026,ChenChenChenYinZhang2025}
to a log-Sobolev bound, building on field dynamics
\cite{ChenFengYinZhang2021}.
Applying it to our finite-size and refined subcritical
covariance estimates yields our Glauber mixing bounds
(\cref{sec:proof-outline-sampling}).

\paragraph{Counting from sampling.}
Classical self-reductions link approximate sampling and
counting~\cite{JerrumValiantVazirani1986}; annealing makes
this connection more efficient
\cite{StefankovicVempalaVigoda2009,Huber2015,Kolmogorov2018,HarrisKolmogorovLiuYinZhang2026}.
Following~\cite{FengGuoYinZhang2021},
Guo, Liu, Yang, Yin, and Zhang use nonadaptive schedules and
whole-product variance control to obtain a sampling-to-counting
reduction optimal over a broad range of
parameters~\cite{GuoLiuYangYinZhang2026}.
Estimators based on correlated Markov-chain trajectories have
also been analyzed using relaxation times and trace
variances~\cite{HaddadanZhuangCousinsUpfal2021}.
Our counter uses Huber's paired-product estimator and
whole-product variance control~\cite[Section~3.3]{Huber2015}.
Our bounds on occupation variance and relaxation time along
the fugacity path make counting nearly as fast as sampling
(\cref{sec:proof-outline-counting}).

\paragraph{Hardness beyond criticality.}
Random regular bipartite graphs exhibit phase coexistence and
bottlenecks~\cite{MosselWeitzWormald2009}.
Dyer, Frieze, and Jerrum introduced them as hardness
gadgets~\cite{DyerFriezeJerrum2002}.
Sly proved hard-core hardness just above
uniqueness~\cite{Sly2010}; Galanis, Ge, \v{S}tefankovi\v{c},
Vigoda, and Yang extended the fugacity
range~\cite{GalanisGeStefankovicVigodaYang2011}.
Work of Sly and Sun~\cite{SlySun2014} and
Galanis, \v{S}tefankovi\v{c}, and
Vigoda~\cite{GalanisStefankovicVigoda2016}
established approximate-counting hardness throughout the tree
nonuniqueness region for antiferromagnetic two-spin systems.
Galanis, \v{S}tefankovi\v{c}, and Vigoda~\cite{GSV2016}
subsequently developed an induced-matrix-norm approach,
establishing hardness for broader classes of antiferromagnetic
spin systems, including colorings and the antiferromagnetic
Potts model.

\paragraph{Hardness near criticality.}
For dense Ising models, Galanis, \v{S}tefankovi\v{c}, and
Vigoda establish nearly sharp critical-window hardness:
approximate counting is NP-hard on $N$ spins with
interaction-matrix spectral diameter at most
$1+N^{-1/2+\xi}$, for every fixed $\xi>0$~\cite{GSVdense2026}.
Their proof controls fluctuations across the coupled system
rather than requiring tight concentration within each gadget.
They ask for analogous critical windows in sparse Ising and
hard-core models.
Our hard-core algorithms at
$\delta=O(\sqrt{d\log n/n})$ and fixed-degree hardness at
$\delta_n=\Theta(n^{-1/2+\xi})$ resolve the critical-power
version of their hard-core question.
The quantitative SAT transfer compares intermediate running
times, while we determine the exact polynomial mixing window
for worst-case Glauber dynamics at every fixed degree.

Our near-critical reduction requires uniform estimates as the
phases merge and the terminal boundary grows: gadgets have size
$\Theta(d\delta^{-2})$, and adjacent gadgets are joined by
$\Theta(d^2\delta^{-1})$ edges.
We combine these estimates with constructive PCPs of
quasilinear size~\cite{Dinur2007PCP} and the cubic construction
in \cref{app:cubic-source}, keeping the logarithmic size loss
explicit in \cref{thm:robust-sat-transfer}.
Coupling the same gadgets along a bipartite expander gives
the unconditional dynamics obstruction.

\paragraph{Algorithms beyond uniqueness on special graph families.}
The uniqueness threshold need not delimit tractability on
restricted graph families.
On random regular graphs, Chen, Chen, Chen, Yin, and Zhang
\cite{ChenChenChenYinZhang2025} establish rapid Glauber mixing
up to fugacities of order $d^{-1/2}$, beyond uniqueness for
large $d$.
Polymer methods give high-fugacity counting and sampling
algorithms on bipartite expanders
\cite{JenssenKeevashPerkins2020}.
More recent algorithms cover all fugacities on random
$d$-regular bipartite graphs for sufficiently large $d$
\cite{KocurekOveisGharanTjowasi2026}, and on bipartite graphs
satisfying suitable spectral expansion conditions
\cite{NarangPerkins2026}.
These bipartite algorithms use auxiliary distributions or
polymer representations and do not imply rapid mixing of
the original Glauber chain.
Subexponential Glauber bounds also arise from geometric
and structural restrictions on graphs~\cite{Marin2025}.
Our results instead track distance from criticality on
worst-case bounded-degree inputs.

\paragraph{Critical Ising models.}
For Glauber dynamics of the Curie--Weiss model at inverse
temperature $\beta=1+\delta$, the critical window is
$\delta\asymp n^{-1/2}$, and mixing is polynomial exactly when
$\delta=O(\sqrt{\log n/n})$
\cite{LevinLuczakPeres2010,DingLubetzkyPeres2009}.
For critical Ising on regular graphs, Bencs, Goldberg, Jenssen, Jerrum, Pete, Regts, and Yin
\cite{BencsEtAl2026} obtain $O_d(\sqrt n\log n)$ total influence
via non-backtracking walks; entropy tensorization gives
$\widetilde O_d(n^{2+2/(d-2)})$ mixing for polynomially bounded
external fields.

Prodromidis and Sly obtain a quartic law for critical Ising
magnetization on random regular graphs via small-subgraph
conditioning, and establish an $\Omega_d(n^{3/2})$  
lower bound on the mixing time of Glauber dynamics
\cite{ProdromidisSly2026}.

\paragraph{Recent SAT algorithms.}
Recent preprints announce deterministic subexponential algorithms
for fixed-width SAT and its counting version
\cite{OpenAI2026ETH,ClaudeAnthropic2026SAT}.
If confirmed, these results would refute the exponential time
hypothesis (ETH)~\cite{ImpagliazzoPaturi2001} and also yield
subexponential hard-core counting and approximate sampling at
fixed degree.

Our reduction assumes no lower bound on SAT's running time.
Its conclusion applies to arbitrary SAT inputs, measured by
their binary encoding length $s$: under the hypotheses of
\cref{thm:main-transfer}, a fixed power saving in the hard-core
exponent $n\delta^2/d$ yields SAT time
$\exp\{\widetilde O(s^{1-\gamma})\}$.
On sparse fixed-width formulas, expressing running time in
terms of the number of variables preserves this fixed power
saving, which goes beyond the logarithmic savings announced
in the recent preprints.
The transfer remains valid as SAT algorithms improve.

\subsection{Acknowledgments}
We would like to thank Pasin Manurangsi for motivating discussions that revived this project and for helpful discussions during the early stages of the project.

\section{Proof outline}
\label{sec:overview}\label{sec:proof-outline}

We deduce the main results from a series of key lemmas and theorems stated below,
proved in later sections.

\begin{revision}
The algorithmic and hardness theories of the hard-core model identify the same computational threshold from opposite sides. 
We build on both theories to determine how the running time of Glauber dynamics grows as this threshold is crossed, 
and to relate the running time of every hard-core sampling or counting algorithm to that of a SAT algorithm. This requires quantitative refinements that remain effective as the distance from criticality vanishes with the graph size.

For the upper bounds, we convert marginal response to external
fields into finite-size covariance bounds, establishing spectral
independence through criticality. Together with a refined subcritical
estimate, these bounds yield the Glauber mixing bound
(\cref{sec:proof-outline-si}).
We then bound the variance of total occupation to obtain approximate
counting by annealing at nearly the cost of sampling
(\cref{sec:proof-outline-counting}).

For the lower bounds, we give a robust fine-grained reduction from
SAT (\cref{sec:proof-outline-reduction}).
Our gadgets encode MAX-CUT within the required size budget as the
phases merge; canceling their shared fluctuations preserves
running-time improvements (\cref{sec:overview-phase-weights}).
Coupling these gadgets along an expander gives an exponential mixing
obstruction even for nonreversible local chains; a one-gadget
argument gives the stronger polynomial factor for Glauber dynamics
(\cref{sec:proof-outline-glauber-lower}).
\end{revision}

Both arguments exploit a critical degeneracy: at the critical fixed
point of the tree recursion on the infinite $d$-regular tree, the two-step
recursion has derivative one but still contracts errors at third order. This degeneracy controls the response
to small fields in the upper bound and produces a quartic free-energy
barrier in the lower bound. Tracking these higher-order terms with
uniform errors resolves the scale $n\delta^2/d$.

Throughout, $\lambdaC=\lambdaC(d)$ and
$\delta_{\mathrm{up}}=2^{-18}$. In the algorithmic lemmas, $d\ge3$ may
be any upper bound on the maximum degree, which vertex deletion
preserves; \cref{thm:main-upper} takes $d=\Delta(G)$. Lower-bound asymptotics
fix $d$ before the graph size grows.

\subsection{\rev{Spectral independence and Glauber mixing through criticality}}
\label{sec:overview-si}\label{sec:proof-outline-si}
\label{sec:si-through-criticality}
\label{sec:overview-sampling}\label{sec:proof-outline-sampling}

For a configuration $X\sim\mu$, write $p_v=\E_\mu X_v$. A \emph{feasible pinning} conditions on an assignment
of positive probability. For the hard-core model, delete occupied
pins and their neighbors, then vacant pins: the remaining law has the
same fugacities on an induced graph of maximum degree at most $d$,
whose vertices are called \emph{free}.
For symmetric matrices, $M\preceq N$ means
$z^{\mathsf T}Mz\le z^{\mathsf T}Nz$ for every real vector $z$;
thus a covariance bound controls the variance of every linear statistic.

\begin{definition}[Spectral independence {\cite{AnariLiuOveisGharan2020}}]
\label{def:spectral-independence}
For a Boolean law $\mu$, omit deterministic coordinates and define
\[
 \Psi_\mu(i,j)=\mu(X_j=1\mid X_i=1)-\mu(X_j=1\mid X_i=0)
 =\frac{\Cov(X_i,X_j)}{p_i(1-p_i)}.
\]
We call $\mu$ \emph{$C$-spectrally independent} if
$\lambda_{\max}(\Psi_\mu)\le C$, and use \emph{under all pinnings}
when the same bound holds after every feasible pinning.
The matrix is similar to a symmetric positive semidefinite matrix, so
\begin{equation}\label{eq:si-covariance-equivalence}
 \lambda_{\max}(\Psi_\mu)\le C
 \quad\Longleftrightarrow\quad
 \Cov_\mu(X)\preceq C\diag\bigl(p_v(1-p_v)\bigr).
\end{equation}
The condition is vacuous if no free coordinate remains. Our diagonal
entries are one, shifting the diagonal-zero convention of
\cite{AnariLiuOveisGharan2020} by the identity.
\end{definition}

Below uniqueness, contraction of the tree recursion bounds the sum of
influences by a geometric series
\cite{AnariLiuOveisGharan2020,ChenLiuVigoda2020,ChenJiang2025}.
At the threshold this contraction vanishes. The critical mixing
arguments of \cite{CCYZ2026,ChenJiang2025} avoid needing a critical SI
bound; we obtain one by controlling finite perturbations of the
marginals themselves.

\paragraph{Marginal response from cubic contraction.}
Weitz's self-avoiding-walk (SAW) tree represents a graph marginal
exactly by a root marginal on a finite tree with pinned leaves
\cite{Weitz2006}. Eliminate the fixed vertices by the pinning rule
above; each remaining node inherits the fugacity of its original vertex.
Write $R_w$ for the occupation odds in the subtree rooted at $w$,
and $R_1,\ldots,R_k$ for the corresponding child-subtree odds. The tree
recursion~\cite{Kelly1985,Weitz2006} reads
\begin{equation}\label{eq:glauber-deletion-recursion}
 R_w=\lambda_w\prod_{i=1}^k\frac{1}{1+R_i},
 \qquad p_w=\frac{R_w}{1+R_w}.
\end{equation}
Following the two-step symmetrization of
\cite{LiuLu2015,ChenLiuYin2023}, we compare log odds two generations
apart. At criticality the error contracts by a cubic amount, uniformly
over the starting odds (\cref{lem:glauber-cubic-deficit}).
Rather than truncate the tree, we propagate finite field perturbations
through its full depth, as in zero-freeness arguments
\cite{PetersRegts2019,LiuSinclairSrivastava2025}.

For a positive vector $\boldsymbol\psi$, the \emph{field tilt} is
$(\boldsymbol\psi*\mu)(\sigma)\propto
\mu(\sigma)\prod_{v:\sigma_v=1}\psi_v$; for hard-core distributions it
multiplies each fugacity by $\psi_v$.
Let $\lambda=\lambdaC e^t$, $t_+=\max\{t,0\}$, and
$\psi_v\in[e^{-h},e^h]$. Suppose the log odds of $\mu$ and
$\boldsymbol\psi*\mu$ differ by at most $s\in[0,1]$ at every grandchild
of a nonroot node, and let $s_{\rm new}$ be their difference at the
node. Then
\[
 s_{\rm new}\le s-cs^3+C(t_+s+h).
\]
The cubic term dominates the other terms at
$s=C_0(\sqrt{t_+}+h^{1/3})$ for a sufficiently large absolute $C_0$, so
this error bound propagates from the leaves to every nonroot node,
even when the perturbations have different signs on the original graph.
A final one-step estimate at the root and conversion from log odds
to marginals give the following bound
(\cref{sec:glauber-cubic-response}).

\begin{lemma}[Hard-core marginal response bound]
\label{lem:glauber-field-response}
There is an absolute constant $C>0$ such that the following holds.
Let $G=(V,E)$ have maximum degree at most $d\ge3$, and let
$\mu=\mu_{G,\lambda}$ with $\lambda>0$ and
$t=\log(\lambda/\lambdaC)\le\delta_{\mathrm{up}}$.
For every $0<h\le h_0=2^{-27}$, every field tilt
$\boldsymbol\psi\in[e^{-h},e^h]^V$, and every $v\in V$,
\begin{equation}\label{eq:glauber-field-response}
 \left|(\boldsymbol\psi*\mu)(X_v=1)-\mu(X_v=1)\right|
 \le C\lambda\bigl(\sqrt{t_+}+h^{1/3}\bigr).
\end{equation}
Here $t_+=\max\{t,0\}$.
\end{lemma}

\paragraph{Finite response gives covariance control.}
A H\"older response bound cannot simply be differentiated at $h=0$.
Instead, we integrate the response along a finite tilt to bound an
exponential moment. For a signed sum $S$ of occupation indicators and
an independent copy $S'$, symmetry gives
$\E\cosh(h(S-S'))\ge1+h^2\Var S$.
Choosing the tilt at the scale of the marginals on its support bounds
this variance. Decomposing a real coefficient vector into signed
level sets then controls every linear statistic. \rev{This conversion
uses only the marginal response hypothesis and applies to any Boolean
law. In the hard-core application, its size parameter is the number
of vertices in the original graph. The statement below is proved in
\cref{sec:glauber-response-covariance}; $I_n$ denotes the identity matrix.}
We use real-valued fields, whereas earlier routes use sector stability or
complex zero-freeness~\cite{AlimohammadiEtAl2021,ChenLiuVigoda2021Stability}.

\begin{lemma}[Covariance from marginal response]
\label{thm:si-field-perturbations}
Let $\mu$ be a law on $\{0,1\}^V$ with $n=|V|\ge1$ and no
deterministic coordinates. For $X\sim\mu$, put
$p_v=\mu(X_v=1)$ and $p_{\max}=\max_v p_v$.
Fix $0<\alpha<1$, $L\ge1$, $h_0>0$, and $u\in[0,1]$.
Suppose the following marginal response bound holds for every $0<h\le h_0$, every field tilt
$\boldsymbol\psi\in[e^{-h},e^h]^V$, and every $v\in V$,
\begin{equation}\label{eq:si-field-perturbation-hypothesis}
 \left|(\boldsymbol\psi*\mu)(X_v=1)-\mu(X_v=1)\right|
 \le Lp_{\max}\bigl(\sqrt u+h^\alpha\bigr).
\end{equation}
Then there is a constant $C=C(L,h_0,\alpha)$ such that
\begin{equation}\label{eq:si-field-covariance-bound}
 \Cov_\mu(X)\preceq
 Cp_{\max}\left(
 1+(np_{\max})^{\frac{1-\alpha}{1+\alpha}}+np_{\max}u\right)I_n.
\end{equation}
\end{lemma}

\added{The dependence on $n,p_{\max},u$ in this bound is optimal
up to constants, already for laws with equal marginals
$p_{\max}\le1/2$; see \cref{prop:response-covariance-sharpness}.
In particular, the exponent $(1-\alpha)/(1+\alpha)$ cannot be
reduced using the marginal response hypothesis alone.}

\begin{proof}[Proof of \Cref{thm:main-si}]
Put $t=\log(1+\delta)$, so $t_+\le\max\{\delta,0\}$.
Since $\lambda=O(1/d)$, the tree recursion and conditioning give
\begin{equation}\label{eq:glauber-variance-lower}
 p_v\le\lambda,\qquad
 p_v(1-p_v)\ge\frac{\lambda}{(1+\lambda)^{d+2}}
 \ge c_0\lambda
\end{equation}
for an absolute $c_0>0$. Indeed, the odds lie between
$\lambda/(1+\lambda)^d$ and $\lambda$.
Thus $p_{\max}\asymp\lambda$ and
$p_{\max}/\min_v p_v(1-p_v)=O(1)$.
\Cref{lem:glauber-field-response} verifies the hypothesis of
\cref{thm:si-field-perturbations} with absolute $L$,
$\alpha=1/3$, $u=t_+<1$, and $h_0=2^{-27}$.
Using \eqref{eq:si-covariance-equivalence}, we obtain
\[
 \lambda_{\max}(\Psi_\mu)
 \le C(1+\sqrt{np_{\max}}+np_{\max}t_+)
 \le C\left(1+\sqrt{\frac nd}+\frac nd\max\{\delta,0\}\right).
\]
Every feasible pinning leaves a hard-core distribution on an induced graph
with at most $n$ vertices and the same degree bound and fugacity, so
the same argument proves the bound under all pinnings.

{The matching lower bound follows from
\cref{cor:glauber-superlinear-lower}, whose construction already
attains it without pinning.}
\end{proof}

\begin{revision}
\paragraph{Applying the spectral-stability criterion.}
To obtain Glauber mixing bounds, we apply the criterion of
Miao and Yin~\cite[Theorem~1.4]{MiaoYin2026}.
For the hard-core model, it uses a nondecreasing rate
$\mathcal C(a)\ge1$ that bounds
$\Cov(X)\preceq\mathcal C(a)\diag(p_v)$ on the free coordinates,
where $p_v=\E X_v$, at every $0<a\le\lambda$ and after every
positive pinning, which conditions an independent set to be occupied.
For the graphs and fugacities considered here, the criterion gives
\[
 t_{\rm mix}(1/4)\le\widetilde O(ne^{I(\lambda)}),
 \qquad
 I(\lambda)=\int_0^\lambda
       \frac{\mathcal C(a)-1}{a}\dd a,
\]
provided the integral is finite. The precise formulation and
verification of its additional marginal hypotheses are given in
\cref{sec:glauber-profile}.

We bound this integral by combining two covariance estimates.
Below uniqueness, \cref{lem:glauber-occupation-bound} gives the rate
$[1-(d-1)\theta]^{-1}$, where
$a=\theta/(1-\theta)^d$ and $0<\theta<1/(d-1)$.
This is the same numerical constant as the branch-tree bound
of~\cite[Lemma~3.6]{ChenJiang2025}.
At $a=(1-y)\lambdaC$, it is at most
$2(1+1/(d-2))/y$.
Taking the smaller of this rate and the finite-size bound from
\cref{thm:main-si} cuts off the subcritical divergence at relative
distance $(1+n/d)^{-1/2}$.
Above criticality, the linear dependence of the SI bound on the
distance from the threshold integrates to a quadratic term.
Consequently, for $\lambda=(1+\delta)\lambdaC$ with
$0\le\delta\le\delta_{\mathrm{up}}$,
\[
 I(\lambda)\le
 \left(1+\frac1{d-2}\right)\log(1+n/d)
 +O\!\left(1+\frac{n\delta^2}{d}\right)
\]
with an absolute implicit constant
(\cref{prop:glauber-integrated-profile}).
The logarithmic and quadratic terms give the polynomial and
exponential factors in the mixing bound, respectively.
The sharper polynomial exponent uses the root vacancy factor
in the subcritical covariance estimate; discarding it would give
$1+2/(d-2)$ instead of $1+1/(d-2)$.

Single-site Glauber dynamics updates a uniformly chosen vertex
by resampling its occupation conditional on the other vertices.
Its worst-start mixing time counts these updates.
Keeping the accuracy dependence gives the following theorem,
proved in \cref{sec:glauber-profile}.
\end{revision}

\begin{theorem}[Near-critical Glauber mixing]\label{thm:glauber-upper}
There is an absolute constant $C>0$ such that, for every $n$-vertex
simple graph $G$ with maximum degree $d=\Delta(G)\ge3$,
every $0\le\delta\le\delta_{\mathrm{up}}$, and $0<\eta<1/4$,
the Glauber dynamics for the hard-core distribution $\mu_{G,\lambda}$,
where $\lambda=(1+\delta)\lambdaC(d)$, has mixing time
\[
 t_{\rm mix}(\eta)\le
 Cn\left(\frac nd\right)^{1+\frac1{d-2}}
 \exp\!\left\{\frac{Cn\delta^2}{d}\right\}
 \left(\log^2 n+\log\frac1\eta\right).
\]
\end{theorem}

\subsection{Counting at nearly the cost of sampling}
\label{sec:overview-counting}\label{sec:proof-outline-counting}

Annealing expresses $Z$ as a product of partition-function ratios
along a fugacity path. We use Huber's paired products and, as
in~\cite{Huber2015,FengGuoYinZhang2021,GuoLiuYangYinZhang2026},
control the variance of the whole product.
With the generic bound of $\widetilde O(\eps^{-2}\log Z)$
samples~\cite{Kolmogorov2018,GuoLiuYangYinZhang2026}, each at the
critical sampling cost, we would lose a factor of order $n/d$, since
$\log Z=\Theta(n/d)$ at criticality.
We avoid this loss using cheap sampling far below uniqueness and
few annealing steps near it. The latter rests on the following bound on
the curvature of $\log Z$ in $\log\lambda$, which is the variance of
the total occupation $|X|$ for $X\sim\mu_{G,\lambda}$.

\begin{lemma}[Total-occupation variance]\label{count:lem-aggregate-curvature}
Let $G$ have $n$ vertices and maximum degree at most $d\ge3$.
In the supercritical case, write $\lambda=(1+\delta)\lambdaC(d)$.
There is an absolute constant $C$ such that
\begin{align}
 \Var_{\mu_{G,\lambda}}(|X|)&\le Cn\lambda\log(2+n\lambda),&&0<\lambda\le\lambdaC(d),
 \label{count:eq-aggregate-critical}\\
 \Var_{\mu_{G,\lambda}}(|X|)&\le Cn\lambda\log(2+n\lambda)\exp\{C(1+n/d)\delta^2\},
 &&\lambdaC(d)<\lambda\le(1+\delta_{\mathrm{up}})\lambdaC(d).
 \label{count:eq-aggregate-hot}
\end{align}
The first bound uses only the tree recursion;
the second also uses the SI bound in
\cref{thm:main-si}.
\end{lemma}

SI alone would give $O(n\lambda(1+\sqrt{n\lambda}))$ at criticality.
\added{Our critical SI lower bound is witnessed by the signed
occupation imbalance between the two sides of a bipartite graph,
whereas annealing depends on total occupation.}

At and below criticality, the stronger estimate comes from
differentiating the two-step recursion
in a common field and tracking the sign of its forcing term: large
forcing occurs only where the coefficients of the differentiated
two-step recursion sum to less than
one (\cref{count:lem-killed-forcing}).
\rev{This gives a self-avoiding-walk bound for the squared response.
Reveal each vertex's occupation at an independent exponential time
and condition on the revealed values.
A walk can remain in the free residual graph only while all its
vertices are unrevealed, which suppresses contributions from long walks.
Integrating the resulting bound on the rate of variance loss gives
the logarithmic factor.}
Above criticality, SI controls the variance lost during an initial
revelation interval of length $O(\delta)$, giving the additional
exponential factor. \Cref{count:sec-aggregate-counting} proves both bounds.

We estimate each product factor by averaging along a separate
Glauber trajectory. \added{At criticality, the occupation-variance bound
allows a schedule with only polylogarithmically many steps within
relative distance $O(\sqrt{d/n})$ below the threshold. Farther below
it, the larger number of steps is compensated by shorter relaxation
times. Above the threshold, the additional costs are absorbed into
$\exp\{O(n\delta^2/d)\}$.} Combining the occupation-variance bound with
the relaxation-time estimates along the annealing schedule gives
the following counting theorem
(\cref{count:lem-positive-product-trajectories,sec:counting-main}).

\begin{theorem}[Near-critical approximate counting]
\label{thm:counting-upper}
\label{prop:glauber-algorithms}\label{prop:main-upper-implementation}
There is a randomized algorithm with the following guarantee. Given
an $n$-vertex simple graph $G$, an integer $d$ satisfying
$\max\{3,\Delta(G)\}\le d\le n-1$, and rational parameters
$0\le\delta\le\delta_{\mathrm{up}}$ and  $0<\eps<1/4$,
let $\lambda=(1+\delta)\lambdaC(d)$. The algorithm returns an estimate
$\widehat Z$ satisfying
\[
 \Pr\!\left[(1-\eps)Z_G(\lambda)\le\widehat Z
                    \le(1+\eps)Z_G(\lambda)\right]\ge\frac34.
\]
Its worst-case number of randomized arithmetic operations is at most
\[
 O(n+|E(G)|)
 +Cn\left(1+\frac nd\right)^{1+\frac1{d-2}}
       \exp\!\left\{\frac{Cn\delta^2}{d}\right\}
       \eps^{-2}\log^4 n.
\]
Here $C$ and all implicit constants are absolute. Arithmetic operations
and binary overhead follow
\cref{sec:computational-conventions}.
\end{theorem}

\begin{proof}[Proof of \Cref{thm:main-upper}]
For sampling, combine \cref{thm:glauber-upper} with the implementation
in \cref{count:sec-implementation}. For counting, take $d=\Delta(G)$
in \cref{thm:counting-upper}.
Since $d\le n-1$ and $1+1/(d-2)\le2$,
$(1+n/d)^{1+1/(d-2)}\le4(n/d)^{1+1/(d-2)}$.
\end{proof}

\subsection{\rev{Robust fine-grained reductions from SAT}}
\label{sec:overview-hardness}\label{sec:proof-outline-hardness}
\label{sec:proof-outline-reduction}

\begin{revision}
Fix $d\ge3$. A cubic source graph $H$ on $m$ vertices is replaced by
one phase gadget per vertex, giving $n\delta^2/d=\Theta(m)$.
The computational reduction uses the following promise problem.
Here $\MaxCut(H)$ is the maximum number of edges crossing a
bipartition of $V(H)$.
The constructive SAT source is proved in \cref{app:cubic-source}.
\end{revision}

\begin{problem}[Additive-gap cubic MAX-CUT]
\label{prob:cubic-gap}
For a fixed rational $\nu>0$, the input is a simple cubic graph $H$ on $m$
vertices and an integer threshold $\tau$. Distinguish
\[
  \MaxCut(H)\ge \tau
  \qquad\text{from}\qquad
  \MaxCut(H)\le \tau-\nu m.
\]
\end{problem}

\begin{addition}
Constructive quasilinear PCPs and a linear-size cubic reduction give a
fixed gap $\nu>0$ and source size
$m=\Theta(s\lceil\log_2s\rceil^{C_{\rm src}})$ from SAT encoding
length $s\ge2$ (\cref{thm:sat-cubic-source}). This is an unconditional
instance reduction; all remaining costs will be accounted for explicitly.
Write $\bits(\delta)$ for the binary encoding length of a rational $\delta$.

\begin{theorem}[Robust fine-grained reduction from SAT: general form]\label{thm:robust-sat-transfer}
\rev{There are absolute constants $K\ge1$, $c_{\rm app}>0$,
and an integer $C_{\rm src}\ge1$ with the following property.}
\rev{For every fixed $d\ge3$ there are effective
constants $C_{\rm rep}(d)>0$, $\delta_0(d)>0$, a polynomial $P_d$, and a
randomized reduction that, given a SAT input of encoding length $s\ge2$
and a rational $0<\delta\le\delta_0(d)$, has the following properties.
Let $m$ be the number of vertices of the cubic graph that
\cref{thm:sat-cubic-source} produces from the input, so that
$m=\Theta(s\lceil\log_2s\rceil^{C_{\rm src}})$.}
\begin{revision}
\begin{enumerate}[label=(\roman*),leftmargin=*]
\item \emph{Queries.} It queries only simple $d$-regular graphs at
fugacity $\lambda=(1+\delta)\lambdaC(d)$, all of the same even order
$n_*$, where
\[
 n_*\le\left\lceil\frac{Kdm}{\delta^2}\right\rceil
 \qquad\text{and}\qquad
 \frac{n_*\delta^2}{d}=\Theta(m).
\]
\item \emph{Repetitions.} It makes at most
$Q=\lceil\exp\{C_{\rm rep}(d)\,(1+m\delta)\}\rceil$
oracle calls.
\item \emph{Sampling.} If $T$ bounds the worst-case time of a
TV-$1/10$ sampler on every possible query graph, then the reduction, with
this sampler as its oracle, solves SAT with one-sided error at most $1/3$,
never accepting an unsatisfiable formula, in time
\[
 Q\,[T+P_d(n_*,\bits(\delta))].
\]
\item \emph{Counting.} If instead $T$ bounds the worst-case
time of a counter that, on every possible query graph $G$, returns with
probability at least $2/3$ a positive estimate $\widehat Z_G$ satisfying
\[
 e^{-c_{\rm app}n_*\delta^2/d}Z_G(\lambda)
 \le\widehat Z_G\le
 e^{c_{\rm app}n_*\delta^2/d}Z_G(\lambda),
\]
then the reduction, with this counter as its oracle, solves SAT with
two-sided error at most $1/3$ in time
\[
 Q\,P_d(n_*,\bits(\delta))\,[T+1].
\]
In particular, the same time bound holds for a relative-error-$1/10$
counter with success probability at least $2/3$.
\end{enumerate}
\end{revision}
All constants are independent of the oracle and of any SAT hardness
assumption.
\end{theorem}

\rev{The graph-size relation transfers an
improvement in the near-critical scale $n\delta^2/d$ to the source
scale $m$; the bound on $Q$ determines whether that improvement survives
the cost of finding a useful realization. Both estimates remain valid
as $\delta$ tends to zero. The proof, including the constructive source
and bounded-time oracle implementation, is in
\cref{sec:lower-main,app:hardness-support}.}

\begin{proof}[Proof of \Cref{thm:main-transfer}]
\rev{Apply \cref{thm:robust-sat-transfer}.}
Let $m$ be as in that theorem, so
$m=\Theta(s\lceil\log_2s\rceil^{C_{\rm src}})$.
\rev{Choose a constant $B>0$ so that its query-size bound is at most
$Bdm/\delta^2+1$.}
Scan $n=2,4,8,\ldots$ and take the first size with
$\rev{n\delta_n^2\ge4Bdm}$.
The two-sided bound $\delta_n=\Theta(n^{-a})$ and the first-crossing
condition give
\[
 n=\Theta\!\left(m^{1/(1-2a)}\right),\qquad
 \frac{n\delta_n^2}{d}=\Theta(m).
\]
This scan and the exact \rev{distance} evaluations take polynomial time in $m$.
At \rev{distance} $\delta_n$, every unpadded query has the same even size
\[
 \rev{n_*\le n/4+1\le n/2}
\]
for sufficiently large $m$. By \cref{lem:regular-padding}, pad each
query to exactly $n$ vertices, preserving its fugacity and accuracy.
The oracle is therefore used only at the \rev{given distance} for its
input size. Moreover,
\[
 1+m\delta_n
 =O\!\left(1+m^{(1-3a)/(1-2a)}\right)
 =O(m^{1-\gamma}),
\]
\rev{where the last inequality uses $a\ge\gamma/(1+2\gamma)$.
This restriction ensures that the repetition cost fits within
$\exp\{O(m^{1-\gamma})\}$; the query-size relation gives the same
bound for the oracle.}
The oracle cost, repetitions, and polynomial overhead thus give time
$\poly(m)\exp\{O(m^{1-\gamma})\}
=\exp\{\widetilde O(s^{1-\gamma})\}$ for either task.
The finitely many inputs before these estimates and the \rev{distance} cutoff
hold are solved directly.
\end{proof}

\begin{revision}
The two corollaries use different choices of the target size,
with $m$ again denoting the cubic source size.
For \cref{cor:near-window-transfer}, matching the size to the
given distance gives $\log n=\Theta(m^{1/r})$ and
$m\delta_n=o(1)$.
The repetition count is therefore bounded, and polynomial work
in $n$ gives SAT time $\exp\{\widetilde O(s^{1/r})\}$.

For \cref{cor:main-np-hardness}, choosing a sufficiently large
target size polynomial in $m$ gives $n_*=o(n)$ and
$m\delta_n=o(1)$.
Padding makes every query have exactly $n$ vertices at the
given distance $\delta_n$.
Under the assumed polynomial-time oracle, the total work is
polynomial in the SAT input length $s$.
The full deductions appear in \cref{sec:robust-sat-proof}.
\end{revision}
\end{addition}

\subsection{\rev{Phase gadgets near criticality}}
\label{sec:overview-phase-weights}

\rev{We now state the phase-weight bounds used in both the reduction and
the Glauber obstruction, then explain their proof. Fix a simple cubic
source graph $H$ on $m$ vertices.}
Classical reductions encode a cut by the two phases of bipartite
gadgets~\cite{Sly2010,SlySun2014,GalanisStefankovicVigoda2016,GSV2016}.
We use the symmetric bipartite double cover of $d$ independent uniform
perfect matchings on $N$ base labels~\cite[Section~5.2]{CCYZ2026},
as in~\cite[Section~4.1]{SlySun2014}. It has $g=2N$ vertices.
For occupied base-label sets $S,T$ on its two sides, put
$\rho=(|S|-|T|)/N$; its sign is the phase, with ties assigned $+$.
A phase sector specifies $\sigma\in\{\pm\}^{V(H)}$, and
$\Cut_H(\sigma)$ counts source edges with opposite endpoint signs.
Since $n=mg$, the target $n\delta^2/d=\Theta(m)$ permits only
$g=\Theta(d\delta^{-2})$ vertices per gadget.

For each source edge, designate $k$ matching edges in each endpoint
gadget, delete their crossed lifts, and reconnect the exposed terminals
between like sides of the two gadgets. Within each gadget, the $3k$ designated edges are distinct edges of a single matching, so the rewiring preserves degrees and creates no parallel edges. This degree-preserving
\emph{parallel connection}~\cite[Section~5]{GSV2016} is balanced between
the two sides.

Let $\mathbb P$ be the law of the independent random gadgets and their
terminal choices, conditioned on each base gadget (its union of matchings)
being simple. Restore
the deleted edges and remove the inter-gadget connections to obtain
$G_{\sqcup}:=\bigsqcup_{v\in V(H)}G_v$. Its partition function is
$Z_{G_{\sqcup}}=\prod_v Z_{G_v}>0$.

\rev{The resulting graph $G$ is simple and $d$-regular, with $n=2mN$
vertices; the full construction is in \cref{constr:graph-reduction}.
Let $Z_\sigma$ be its partition function restricted to phase sector
$\sigma$, and put $A=Z_{G_{\sqcup}}$. All partition functions here use
fugacity $(1+\delta)\lambdaC(d)$.}
We suppress fugacity arguments in partition functions below.
\rev{Define $\mathbb Q$ by}
\[
 \rev{\frac{\dd\mathbb Q}{\dd\mathbb P}
 =\frac{A^2}{\E_{\mathbb P}A^2}.}
\]
The resulting $\mathbb Q$ is the
\emph{reference-square law}, obtained by planting two reference
replicas~\cite{AchlioptasCojaOghlan2008,CojaOghlanEtAl2018}.

\begin{addition}
\begin{lemma}[Bounds for phase weights]
\label{lem:overview-phase-weights}
Fix an accuracy $0<\varepsilon\le1$ and a target reward $J_0\ge2$.
There are constants $c_*,C_*>0$, depending only on
$\varepsilon,J_0$, with the following property. For every fixed
$d\ge3$, there are $C_d>0$ and a rational cutoff
$0<\delta_0(d)\le\delta_{\mathrm{up}}$ such that, for every rational
$0<\delta\le\delta_0(d)$, one can choose an even integer $N$ and an
integer $k\ge1$, independently of $H,m$, satisfying
\begin{gather*}
 c_*d\delta^{-2}\le N\le C_*d\delta^{-2},\qquad
 c_*d^2\delta^{-1}\le k\le C_*d^2\delta^{-1},\\
 J:=\frac{48(d-1)}{d^3}k\delta\in[J_0,J_0+4].
\end{gather*}
For every simple cubic $H$ on $m$ vertices, the construction is valid
and, for every phase vector $\sigma$, putting
$w_\sigma=2^{-m}e^{-J(|E(H)|-\Cut_H(\sigma))}$ gives
\begin{align}
 \E_{\mathbb Q}(Z_\sigma/A)&\le w_\sigma e^{\varepsilon m},
 \label{eq:overview-phase-upper}\\
 \mathbb Q\!\left(Z_\sigma/A\ge\tfrac12w_\sigma e^{-\varepsilon m}\right)
 &\ge\tfrac14\exp\{-C_dm/\sqrt N\}.
 \label{eq:overview-phase-lower}
\end{align}
The constants are independent of $H,m,\sigma$, and there is no
restriction comparing $m$ with $N$. For fixed rational
$\varepsilon,J_0$, the choices are effective.
\end{lemma}

\begin{proof}
Apply \cref{thm:cut-weight-realization}, using the rounding rules in
\eqref{eq:degree-explicit-lower-parameters} with $J_*=J_0$ and the
gadget-size coefficient chosen for the present accuracy $\varepsilon$.
Shrink the cutoff as necessary.
\end{proof}
\end{addition}

\paragraph{Near-critical gadgets are soft spins.}
At this size the Gaussian approximation degenerates as the phases merge,
so their fluctuating amplitudes must remain part of the analysis.
The mean imbalance weights have a quartic law, as in critical Curie--Weiss
magnetization~\cite{EllisNewman1978}. With
$\logshift=\log(1+\delta)$, the mean weight at imbalance $\rho$ is
proportional to $e^{g\Phi_\delta(\rho)}$, where
\[
 \Phi_\delta(\rho)=A_d\logshift\,\rho^2-B_d\rho^4,
 \qquad A_d=\Theta(d),\quad B_d=\Theta(d^3)
\]
(\cref{prop:imbalance-law}; the coefficients are in
\eqref{eq:AB-constants}).
The wells satisfy $\rho_*^2=\Theta(\delta/d^2)$, and their barrier is
$g[\Phi_\delta(\rho_*)-\Phi_\delta(0)]=\Theta(g\delta^2/d)$,
with degree-independent comparison constants. The gadget budget thus
allows a large constant barrier, but not a growing one.
\added{Even after the phase is fixed, $|\rho|/\sqrt{\logshift}$
still has nontrivial fluctuations in the limiting mean law at fixed
$g\delta^2/d$.}
The critical estimate of~\cite[Lemma~5.7]{CCYZ2026} controls mean
occupation-count weights up to constant factors on the $N^{2/3}$ scale.
We need relative precision on the $N^{3/4}$ scale of the wells.
Starting from~\cite[Equation~(51)]{CCYZ2026}, we \rev{keep} the quartic
imbalance term and integrate the two stable count directions
$|S\setminus T|+|T\setminus S|$ and $|S\cap T|$
(\cref{prop:quartic-expansion,prop:bin-integral-mean}).

\paragraph{Fine-grained MAX-CUT encoding through gadget connections.}
\rev{The leading interaction of a parallel connection} in the logarithm
of the mean weight is
$-\tfrac{k}{2}(\rho_u+\rho_v)^2$
(\cref{prop:port-hessian,prop:marked-local-coercivity}).
At the wells, equal signs cost $2k\rho_*^2$ and opposite signs cost
nothing. Hence $k=\Theta(d^2/\delta)$ produces a constant cut reward;
a fixed number of terminals would produce a vanishing reward.

This boundary has $k=\Theta_d(\sqrt N)$, so arbitrary occupation-count
shifts could cost $O(k^2/N)=O_d(1)$ per gadget. \rev{At the critical
occupation proportions,} balanced connections have the same
endpoint means as the
crossed edges they replace. This cancellation reduces the normalized
coefficient-comparison error to $O_d(N^{-1/2})$ per gadget
(\cref{lem:normalized-block-replacement}) on \emph{fixed windows}:
the occupation-count difference $|S|-|T|$ is $O(N^{3/4})$ and the
two stable counts are within $O(N^{1/2})$ of their optimum given the imbalance.
Writing the imbalance window as $|\rho_v|\le R\sqrt{\logshift}$
for fixed $R>0$, the mean restricted sector weight is, up to a
common baseline,
\[
 \int_{\substack{\operatorname{sign}(\rho)=\sigma\\
                 |\rho_v|\le R\sqrt{\logshift},\ \forall v}}
 \exp\Bigl\{g\sum_{v\in V(H)}\Phi_\delta(\rho_v)
   -\frac k2\sum_{uv\in E(H)}(\rho_u+\rho_v)^2\Bigr\}\dd\rho,
\]
within a factor $e^{\pm\omega m}$, where $\omega\to0$ as $\delta\to0$
uniformly in $m$ (\cref{eq:compact-normalized-integral}).
Tails are controlled separately (\cref{prop:wired-first-moment-tails}).
Integrating the amplitudes, rather than freezing them at the wells,
gives a common baseline times $e^{J\Cut_H(\sigma)}$ with error
$e^{\pm\varepsilon m}$ (\cref{lem:amplitude-integration}).

Choose a bounded range of rewards $J$ large enough that the cut gap
dominates these errors and the $2^m$ phase assignments.

\paragraph{Canceling the shared fluctuations.}
The soft-spin model describes mean weights, whereas the reduction uses
one random realization.
For a fixed phase sector, let $\rev{Y=Z_\sigma}$
be its full weight after wiring and $X\le Y$ its weight restricted to
the fixed windows defined in
\cref{eq:restricted-phase-weight}. Upper bounds on the sector weights
follow from their means by Markov's inequality; a lower bound for a
sector with a large cut requires a second moment. Short cycles
produce a common fluctuation factor in these weights: the partition
function of one random gadget has normalized second moment near a
constant $\mathfrak c_1>1$, the cycle multiplier of small-subgraph
conditioning~\cite{RobinsonWormald1992,RobinsonWormald1994,Janson1995,Wormald1999}. Bounding this factor
separately for each gadget would lose $(\mathfrak c_1)^m=e^{\Theta(m)}$,
\rev{even when each local error tends to zero. Such a loss would
overwhelm polynomial or stretched-exponential source algorithms.}

\rev{The reference-square law cancels the common cycle factor through
the identities}
\begin{align*}
 \rev{\E_{\mathbb Q}(X/A)}
 &\rev{=\frac{\E_{\mathbb P}(XA)}{\E_{\mathbb P}A^2},}\\
 \rev{\frac{\E_{\mathbb Q}[(X/A)^2]}
      {[\E_{\mathbb Q}(X/A)]^2}}
 &\rev{=\frac{\E_{\mathbb P}X^2\,\E_{\mathbb P}A^2}
        {[\E_{\mathbb P}(XA)]^2}.}
\end{align*}
\rev{The mean identity also holds for $Y$; the general statement is
\cref{lem:sqplant-cancellation}.}

For the parameters in \cref{lem:overview-phase-weights},
\cref{thm:sqplant-full-moments,cor:sqplant-conditioned-moments}
\rev{show that each of the four normalized moments}
\[
 \rev{\frac{\E_{\mathbb P}A^2}{(\E_{\mathbb P}A)^2},\qquad
 \frac{\E_{\mathbb P}X^2}{(\E_{\mathbb P}X)^2},\qquad
 \frac{\E_{\mathbb P}(XA)}{\E_{\mathbb P}X\E_{\mathbb P}A},\qquad
 \frac{\E_{\mathbb P}(YA)}{\E_{\mathbb P}Y\E_{\mathbb P}A}}
\]
\rev{lies in $(\mathfrak c_1)^m e^{\pm e_Nm}$, where
$e_N=O_d(N^{-1/2})$.} These estimates are uniform in $H,m,\sigma$, with no
restriction on the relative sizes of $m$ and $N$.
Thus the logarithm of the relative second moment is at most $4e_Nm$,
even when the moment itself does not approach one.

The moment estimates \rev{track} each replica's critical quartic direction
while controlling their \emph{overlap}, the matrix of joint occupation
states on base labels. Conditional on the individual occupation counts,
this overlap is strictly stable. Its Gaussian fluctuations give the
leading cycle multiplier, preserved by balanced connections; the
simplicity correction yields $\mathfrak c_1$.
A critical two-point hypercontractive inequality
(\cref{lem:critical-HC-new}), tensorized as in~\cite{GSV2016},
\rev{rules out competing overlaps.}

\begin{addition}
\paragraph{The precision needed for robust reductions.}
\rev{The repetition bound in \cref{thm:robust-sat-transfer} requires
$e_N=O_d(N^{-1/2})$ after integrating over occupation counts and
overlaps and conditioning on simplicity.
Exponential-moment bounds control the type-dependent coefficient
errors.
The simplicity probability is estimated to relative error
$O_d(N^{-1/2})$, uniformly after $O_d(\sqrt N)$ exposed pairs
(\cref{lem:uniform-cycle-free-transfer}); side-swap symmetry removes
a potential first-order variation of its multiplier.
Together, these estimates give accumulated logarithmic error
$O_d(m/\sqrt N)$, with no restriction comparing $m$ and $N$.}
\end{addition}

\rev{Together, these estimates yield the phase-weight bounds in
\cref{lem:overview-phase-weights,thm:cut-weight-realization}.
Paley--Zygmund applied to $X/A$ gives a favorable maximum-cut
sector with probability at least $e^{-O_d(1+m/\sqrt N)}$,
since its full weight $Y$ satisfies $Y\ge X$.
Markov's inequality controls the total weight of low-cut sectors,
with failure probability $e^{-\Omega(m)}$.
After shrinking the distance cutoff, this failure probability is
negligible compared with the favorable-sector probability for
sufficiently large $m$.
On realizations satisfying both bounds, constant-TV sampling
recovers a large cut; the reference $A$ cancels in ratios of sector
weights.
Since $N=\Theta_d(\delta^{-2})$,
$\exp\{O_d(1+m\delta)\}$ independent trials suffice.

For counting, the maximum of $\log(Z_G/A)$ over the trials
separates YES and NO instances with high probability, with a
$\Theta(m)$ gap between the corresponding thresholds
(\cref{lem:square-source-transfer}).}
The oracle estimates both $Z_G$ and $A=Z_{G_{\sqcup}}$ from the same
trial: $G$ and $G_{\sqcup}$ are simple $d$-regular graphs of the same
order. The oracle's accuracy guarantee bounds the sum of the two
logarithmic errors by less than half the gap. Amplification ensures
that both estimates satisfy this guarantee in every trial with high
probability.

Together with bounded-time generation
\rev{(\cref{prop:reduction-implementation}), the trial bound and
$n_*=2mN=\Theta(dm\delta^{-2})$ give
\cref{thm:robust-sat-transfer} after applying the source construction
in \cref{thm:sat-cubic-source}.}

\subsection{\rev{Lower bounds for local Markov chains}}
\label{sec:proof-outline-glauber-lower}

\begin{addition}
Although each gadget has only a constant barrier, coupling $m$ gadgets
along an expander produces a barrier of order $m=\Theta(n\delta^2/d)$.
For the dynamics application, take $H$ to be a cubic bipartite expander.
Its bipartition cuts every source edge, and a phase vector
\rev{at an intermediate Hamming distance from this bipartition}
loses linearly many cut edges.
\rev{Applying the phase-weight estimates to the bipartition phase
and to a band of intermediate phase vectors gives one realization
with a heavy peak and an exponentially lighter separating band.}
A global side-swap automorphism produces a disjoint region of equal
weight.
\rev{A transition changing only a small fraction of the gadget phases
cannot leave the first region without starting in the separating band.
Stationarity then bounds the probability of escape, without requiring
reversibility.}
\begin{revision}
For each fixed $d\ge3$ and sufficiently small rational $\delta>0$,
this gives simple $d$-regular bipartite graphs at the core
construction sizes, with $n\delta^2/d=\Theta(m)$.
Every irreducible, aperiodic chain on $\Ind(G)$ with stationary law
$\mu_{G,(1+\delta)\lambdaC(d)}$ whose transitions change at most
a sufficiently small absolute constant times $m$ occupation
indicators has mixing time at least $\exp\{\Omega(m)\}$.
The precise construction sizes and move bound are stated in
\cref{thm:glauber-obstruction}.
Padding in \cref{prop:glauber-all-sizes} gives the all-size
conclusion of \cref{rem:local-chain-obstruction}.
\end{revision}

\rev{For Glauber dynamics, the one-gadget estimates also give a
stronger polynomial factor.}
At criticality, the occupation imbalance has scale $n^{3/4}$.
Following the maxima of the exact rate function as the fugacity
increases, we compare the weight near a maximum with the weight
near zero imbalance. A Rayleigh-quotient argument then gives
\[
 t_{\rm rel},\ t_{\rm mix}(1/4)
 \ge a_dn^{3/2}\exp\!\left(c\frac{n\delta^2}{d}\right).
\]
The same realization has imbalance variance
$\Omega_d(n^{3/2}+n^2\delta)$, which yields the matching SI lower
bound $\Omega_d(\sqrt n+n\delta)$.
These conclusions are proved in
\cref{thm:single-gadget-lower,cor:glauber-superlinear-lower}.

\begin{revision}
\begin{proof}[Proof of \Cref{thm:main-glauber-lower}]
The lower bound follows from
\cref{cor:glauber-superlinear-lower}, after choosing the common
cutoff $\delta_{\mathrm{mix}}(d)$ sufficiently small that
$\delta_{\mathrm{mix}}(d)\le\delta_{\mathrm{up}}$.
The upper bound used in the characterization follows from
\cref{thm:glauber-upper} at $\eta=1/8$, using
$t_{\rm mix}(1/4)\le t_{\rm mix}(1/8)$ and suppressing
the factor $\log^2 n$.
Taking logarithms of these upper and lower bounds, with $d$
and $a_d>0$ fixed, gives the stated characterization of the
polynomial mixing window.
\end{proof}
\end{revision}
\end{addition}

\begin{revision}
\paragraph{Organization of the remaining proofs.}
After the preliminaries, \cref{sec:upper-main} proves the response and
covariance lemmas and the mixing theorem; \cref{sec:counting-main}
proves the occupation-variance bound and the counting theorem.
\Cref{sec:lower-main} proves the robust SAT reduction and the
dynamics lower bounds using phase-weight estimates established in
\cref{sec:gadget-moments,sec:shared-moments}.
The appendices supply upper-bound implementation, the constructive SAT
source and reduction algorithms, and finite-state tools for gadget moments.
\end{revision}

\section{Preliminaries}
\label{app:preliminaries}\label{sec:prelim}

For nonnegative quantities, $X\asymp Y$ means
$cY\le X\le CY$ for positive constants $c,C$.
A subscript $d$ in $\asymp_d$, $O_d$, or $\Theta_d$ means that the
implicit constants may depend on $d$, but are uniform in the other
parameters over the stated range.

For the technical estimates, let $d\ge3$ be an upper bound on the maximum degree and put
\[
  D=d-1,
  \qquad
  \lambdaC=\lambdaC(d)=\frac{D^D}{(D-1)^{D+1}}.
\]
For a graph $G=(V,E)$, put $n=|V|$ and write $N_G[A]$ for the
closed neighborhood of a set $A\subseteq V$. For positive vertex
fugacities $\bm\lambda=(\lambda_v)_{v\in V}$,
\[
  Z_G(\bm\lambda)=\sum_{I\in\Ind(G)}\prod_{v\in I}\lambda_v,
  \qquad
  \mu_{G,\bm\lambda}(I)=\frac{1}{Z_G(\bm\lambda)}\prod_{v\in I}\lambda_v.
\]
The homogeneous model is recovered by setting all fugacities equal.

We use the pinning convention of \cref{sec:proof-outline-si}:
for a feasible pinning, delete occupied pins and their neighbors,
then vacant pins. The remaining free vertices keep their original
fugacities on the induced graph, with the supplied degree bound unchanged.
Thus uniform graph
bounds also apply after pinning.

\paragraph{Marginal ratios and the tree recursion.}
For $G$ with vertex fugacities $\bm\lambda$, the
\emph{marginal ratio} (occupation odds) at $v$ is
$R_G(v)=\mu_{G,\bm\lambda}(X_v=1)/\mu_{G,\bm\lambda}(X_v=0)$.
Weitz's self-avoiding-walk (SAW) tree preserves the marginal at $v$
\cite[Theorem~3.1]{Weitz2006}. Its nodes represent self-avoiding walks
from $v$, with additional cycle-closing leaves pinned according to fixed
neighbor orders. Each copy of a vertex inherits its fugacity; the same
representation applies to inhomogeneous fugacities.
Eliminate these fixed boundary vertices using the pinning rule above
and keep the component containing the root. We use this form of the SAW
tree throughout. It is finite, depends only on the graph and neighbor
orders, and has the same root marginal as $G$ for every fugacity vector.
The root has at most $d$ children and every nonroot node at most $D$.
Each root-to-node path has distinct original labels and maps injectively
to a self-avoiding walk in $G$.

The odds on this tree satisfy \eqref{eq:glauber-deletion-recursion},
with $\lambda_w$ inherited from the original vertex and $p_w$ the
occupation probability in the subtree rooted at $w$. The regular-tree
form appears in \cite[Equation~(2.9)]{Kelly1985}, and the general
form in \cite[Equation~(1)]{Weitz2006}.

On the infinite $(D+1)$-regular tree, every vertex below the root has
$D$ children, and at fugacity $\lambda$ the recursion becomes the map
$x\mapsto\lambda(1+x)^{-D}$. This map has a unique fixed point $\hat x$,
where its derivative has absolute value $D\hat x/(1+\hat x)$; this is
at most one exactly when $\lambda\le\lambdaC$. This tree-uniqueness
threshold goes back to Kelly~\cite[Section~2]{Kelly1985}: above it the
fixed point is unstable; see also
\cite[Proposition~2.5]{Weitz2006}. At $\lambdaC$ the critical fixed
point is $\hat x=1/(D-1)$,
where the occupation probability $\hat x/(1+\hat x)$ of a vertex in its
own subtree equals $1/D$.
Since the map is decreasing, we follow Liu and Lu~\cite{LiuLu2015} and
Chen, Liu, and Yin~\cite{ChenLiuYin2023} and analyze the two-step
recursion, which expresses the ratio of a vertex through those of its
grandchildren and is increasing in them.

For a graph $G$ with homogeneous fugacity $\lambda>0$, put
$\mu=\mu_{G,\lambda}$, $R_v=R_G(v)$, and $p_v=\mu(X_v=1)$.
The tree recursion and conditioning give
\begin{equation}\label{eq:hc-elementary-residual}
 \frac{\lambda}{(1+\lambda)^d}\le R_v\le\lambda,\qquad
 p_v(1-p_v)\ge\frac{\lambda}{(1+\lambda)^{d+2}},\qquad
 \Pr(X_u=X_v=1)\le\lambda^2\quad(u\ne v).
\end{equation}
Indeed, every child ratio is at most $\lambda$, giving the bounds on $R_v$.
The variance bound follows from $p_v(1-p_v)=R_v/(1+R_v)^2$.
For the pair bound, conditioning on an occupied
vertex either forces another vertex vacant or leaves it in a hard-core
residual. The minimum state probability
$\mu_{\min}=\min_{I\in\Ind(G)}\mu_{G,\lambda}(I)$ also satisfies
$\mu_{\min}\ge(\min\{1,\lambda\}/(1+\lambda))^n$, since
$Z_G(\lambda)\le(1+\lambda)^n$.

\subsection{Markov chains and mixing times}

For probability laws $\nu,\pi$ on a finite state space $\Omega$, write
$\TV(\nu,\pi)=\frac12\sum_{x\in\Omega}|\nu(x)-\pi(x)|$ for their
total-variation (TV) distance.
For an irreducible, aperiodic transition matrix $P$ with
stationary law $\pi$ on a state space with at least two states, and
$0<\eta<1$, let
\[
 t_{\rm mix}(\eta)=\min\{t\in\mathbb Z_{\ge0}:\max_{x\in\Omega}
        \TV(P^t(x,\cdot),\pi)\le\eta\}.
\]
Mixing times count transitions; for Glauber dynamics, each
transition is a single-vertex update.

If $P$ is also reversible and positive semidefinite in $L^2(\pi)$,
its eigenvalues on functions orthogonal to constants lie in $[0,1)$.
Writing $\theta_2$ for the largest of these, we use
$\operatorname{gap}(P)=1-\theta_2$ and
$t_{\rm rel}=1/\operatorname{gap}(P)$.
The transition matrix of Glauber dynamics is reversible and positive semidefinite.
A one-state law is deterministic and is handled directly.
These are the standard finite-chain conventions of~\cite{LevinPeres2017}.

\subsection{Computational conventions}\label{sec:computational-conventions}

Graphs are given by adjacency arrays, and numerical input parameters are
rationals encoded in binary.

The upper bounds use the randomized arithmetic model: rational arithmetic,
comparisons, integer rounding, array access, and independent random words
of $O(\log(n+1))$ bits cost one operation.
The computational reductions use the randomized bit model.
Charging random bits individually adds a factor $O(\log(n+1))$
to the upper bounds; binary implementations also incur the cost of
arithmetic at the required operand lengths.

The algorithmic running-time bounds hold in the worst case over inputs
and internal branches. Numerical approximations, operand-length bounds,
and the implementation of these worst-case guarantees are detailed in
\cref{count:sec-implementation}.

\section{Spectral independence and Glauber mixing}
\label{sec:upper-main}

We prove the three analytic inputs used in
\cref{sec:proof-outline-si}:
\cref{thm:glauber-upper,thm:si-field-perturbations,lem:glauber-field-response}.
First, \cref{sec:glauber-profile} derives the mixing theorem by
integrating the spectral-stability rate supplied by
\cref{thm:main-si,lem:glauber-occupation-bound}. We then prove the two inputs from which the SI
bound was deduced: the general response-to-covariance lemma in
\cref{sec:glauber-response-covariance}, and the hard-core response
bound in \cref{sec:glauber-cubic-response}. None of these response
arguments uses the mixing theorem.

\subsection{Mixing from field-dynamics spectral stability}
\label{sec:glauber-profile}

A \emph{positive pinning} fixes an independent set of vertices to be
occupied and leaves the other coordinates unpinned.

\begin{definition}[Field-dynamics spectral stability {\cite[Definition~1.1]{MiaoYin2026}}]
\label{def:field-dynamics-spectral-stability}
Let $G$ be a finite graph and $\lambda>0$.
The law $\mu_{G,\lambda}$ satisfies \emph{field-dynamics spectral stability}
with rate $\mathcal C:(0,\lambda]\to[0,\infty)$ if, for every
$0<a\le\lambda$ and every positive pinning of $\mu_{G,a}$, the
residual configuration $X$ satisfies
\begin{equation}\label{eq:glauber-stability-covariance}
 \Cov(X)\preceq\mathcal C(a)\diag(p_v),\qquad p_v=\E X_v.
\end{equation}
The inequality is on the free coordinates and is vacuous for an empty
residual. We call $\mathcal C$ a \emph{spectral-stability rate}.
\end{definition}

This condition originates in \cite{CCYZ2026}, with the covariance
formulation from \cite{ChenChenChenYinZhang2025,MiaoYin2026}.
The following specialization of Miao and Yin's theorem gives mixing
and the relaxation bound needed for counting.

\begin{lemma}[Mixing and relaxation from field-dynamics spectral stability]
\label{lem:glauber-stability-mixing}
Let $G$ have $n\ge1$ vertices and let $\lambda>0$.
Suppose $\mu_{G,\lambda}$ satisfies field-dynamics spectral stability
(\cref{def:field-dynamics-spectral-stability}) with a nondecreasing
rate $\mathcal C:(0,\lambda]\to[1,\infty)$.
Set
\[
 I=\int_0^\lambda\frac{\mathcal C(a)-1}{a}\dd a,
\]
and assume $I<\infty$. Let $\overline{\lambda}\ge\max\{1,\lambda\}$, and suppose
$\mu_{G,\lambda}$ satisfies the occupied-marginal lower-bound condition
under positive pinning with parameter $b>0$; namely, for every
independent set $S\in\Ind(G)$ and every vertex
$v\in V(G)\setminus N_G[S]$,
\[
 \mu_{G,\lambda}(X_v=1\mid X_w=1\text{ for all }w\in S)\ge b.
\]
Then the Glauber dynamics for $\mu_{G,\lambda}$ has mixing time
\begin{equation}\label{eq:glauber-field-mixing}
 t_{\mathrm{mix}}(\eta)\le C\overline{\lambda}ne^I
 \left[\log\frac{n\overline{\lambda}}{b}\,
             \log\log\frac{e^2}{\mu_{\min}}
             +\log\frac1\eta\right]
 \qquad(0<\eta<1/4),
\end{equation}
where $\mu_{\min}=\min_{J\in\Ind(G)}\mu_{G,\lambda}(J)$ and $C$ is an
absolute constant. Without the occupied-marginal lower bound, the
relaxation time of the same chain satisfies
\begin{equation}\label{eq:glauber-field-relaxation}
 t_{\mathrm{rel}}\le n(1+\overline{\lambda})e^I.
\end{equation}
\end{lemma}

\begin{proof}
We apply \cite[Theorem~1.4]{MiaoYin2026} to $\mu=\mu_{G,\lambda}$,
whose support $\Ind(G)$ is downward closed. For $0\le u<1$, the
scalar-field tilt in that theorem is $\mu_{G,(1-u)\lambda}$.
By \cref{def:field-dynamics-spectral-stability}, the spectral-stability
condition in that theorem holds with nonincreasing rate
$u\mapsto\mathcal C((1-u)\lambda)\ge1$. The substitution
$a=(1-u)\lambda$ identifies its stability integral with $I$:
\[
 \int_0^1\frac{\mathcal C((1-u)\lambda)-1}{1-u}\dd u
 =I<\infty.
\]

For every $J\in\Ind(G)$ and $v\notin N_G[J]$, we have
$\mu(J\cup\{v\})/\mu(J)=\lambda\le\overline{\lambda}$, verifying the
ratio bound required by the theorem. The occupied-marginal lower bound under
positive pinning is assumed. Thus all hypotheses of the theorem hold,
and its log-Sobolev inequality gives \eqref{eq:glauber-field-mixing}.

For relaxation, \cite[Proposition~2.1]{MiaoYin2026} uses only the
spectral-stability condition and the ratio bound verified above, so
it gives the bound in \eqref{eq:glauber-field-relaxation}.
\end{proof}

The next lemma, whose proof is given in \cref{app:glauber-subcritical},
gives a subcritical spectral-stability bound.

\begin{lemma}[Subcritical spectral stability]
\label{lem:glauber-occupation-bound}
Let $d\ge3$ and let $G$ have maximum degree at most $d$.
For each $0<a<\lambdaC(d)$, define
$\theta_d(a)\in(0,1/(d-1))$ by
\[
 a=\frac{\theta_d(a)}{(1-\theta_d(a))^d}.
\]
Then, for every $0<\lambda<\lambdaC(d)$, the hard-core distribution
$\mu_{G,\lambda}$ with fugacity $\lambda$
satisfies field-dynamics spectral stability
(\cref{def:field-dynamics-spectral-stability}) with rate $\mathcal C$, where
\begin{equation}\label{eq:glauber-subcritical-input}
 \mathcal C(a)=\frac1{1-(d-1)\theta_d(a)}
 \le\frac{2(1+1/(d-2))}{1-a/\lambdaC(d)},
 \qquad 0<a\le\lambda.
\end{equation}
\end{lemma}

Here $\theta=\theta_d(\lambda)$ is the occupation probability of a vertex in
its own subtree, at the fixed point of the \rev{tree recursion} on the infinite
$d$-regular tree. Applying the general-tree bound of Chen and Jiang~\cite[Lemma~3.3]{ChenJiang2025}
directly would introduce an extra factor $1+\theta$. Our proof avoids
this loss by keeping the root vacancy factor $1-p_v$ in
\eqref{eq:glauber-subcritical-row-sum}, where
$p_v=\mu_{G,\lambda}(X_v=1)$.

The next proposition combines \cref{lem:glauber-occupation-bound,thm:main-si}
to bound the stability integral in \cref{lem:glauber-stability-mixing}.
The near-critical bound will give the mixing theorem, while the
subcritical bound will give relaxation-time bounds for counting.

\begin{proposition}[Integrated spectral-stability bound]
\label{prop:glauber-integrated-profile}
There are absolute constants $K\ge2$ and $C>0$ with the following
property. Let $G$ have $n\ge1$ vertices and maximum degree at most
$d\ge3$. For every $0<\lambda\le(1+\delta_{\mathrm{up}})\lambdaC$,
the law $\mu_{G,\lambda}$ satisfies field-dynamics spectral stability
(\cref{def:field-dynamics-spectral-stability}) with the nondecreasing
rate $\mathcal C\ge1$ defined by
\begin{equation}\label{eq:glauber-stability-profile}
 \mathcal C(a)=
 \begin{cases}
 \min\{K(1+\sqrt{n/d}),[1-(d-1)\theta_d(a)]^{-1}\},
     &0<a<\lambdaC,\\
 K\bigl(1+\sqrt{n/d}+(n/d)(a/\lambdaC-1)\bigr),
     &\lambdaC\le a\le(1+\delta_{\mathrm{up}})\lambdaC
 \end{cases}.
\end{equation}
Moreover, its stability integral
\[
 I(\lambda)=\int_0^\lambda\frac{\mathcal C(a)-1}{a}\dd a
\]
is finite and satisfies
\begin{equation}\label{eq:glauber-integrated-stability}
 I(\lambda)\le
 \begin{cases}
 2\left(1+\frac1{d-2}\right)\left[\log\mathcal C(\lambda)+1\right],
     &\lambda<\lambdaC,\\[1mm]
 \left(1+\frac1{d-2}\right)\log(1+n/d)+\frac{Cn\delta^2}{d}+C,
     &\lambda\ge\lambdaC,
 \end{cases}
\end{equation}
where $\delta=\lambda/\lambdaC-1$ in the second case.
\end{proposition}

\begin{proof}
\textbf{Verifying spectral stability.}
Choose $K\ge2$ at least as large as the absolute constant $C$ in
\cref{thm:main-si}. By \eqref{eq:si-covariance-equivalence} and
$\diag(p_v(1-p_v))\preceq\diag(p_v)$, that theorem with
$\delta=a/\lambdaC-1$ gives the spectral-stability rate
$K(1+\sqrt{n/d})$ at every $a<\lambdaC$ and the second branch of
\eqref{eq:glauber-stability-profile} at and above $\lambdaC$.
Below $\lambdaC$, taking the minimum with the rate from
\cref{lem:glauber-occupation-bound} gives the first branch.
The resulting rate is nondecreasing and at least one.

\paragraph{Below criticality.}
Write $c_1=K(1+\sqrt{n/d})$ for the constant rate from
\cref{thm:main-si}, and use the subcritical rate
$c_2=[1-(d-1)\theta_d(a)]^{-1}$ as the integration variable.
Then $\mathcal C(a)=\min\{c_1,c_2\}$, and $c_2$ increases
from one to infinity as $a$ increases from zero to $\lambdaC$.
Writing $\theta=\theta_d(a)$
and $\gamma_d=1+1/(d-2)$, differentiation gives
\[
 \frac{\mathrm d a}{a}
 =\frac{1+(d-1)\theta}{1-\theta}\frac{\mathrm d c_2}{c_2(c_2-1)}
 \le2\gamma_d\frac{\mathrm d c_2}{c_2(c_2-1)},
\]
where the inequality uses $0<\theta<1/(d-1)$.
Changing variables and splitting where the two rates agree,
at $c_2=c_1$, gives
\begin{align}
 I(\lambdaC)
 &\le2\gamma_d\int_1^\infty
       \frac{\min\{c_1,c_2\}-1}{c_2(c_2-1)}\dd c_2\notag\\*
 &=2\gamma_d\left[\int_1^{c_1}\frac{\mathrm d c_2}{c_2}
       +(c_1-1)\int_{c_1}^\infty
       \frac{\mathrm d c_2}{c_2(c_2-1)}\right]\notag\\*
 &=2\gamma_d\left[\log c_1+(c_1-1)
       \log\frac{c_1}{c_1-1}\right]\notag\\*
 &\le2\gamma_d(\log c_1+1)
 \le\gamma_d\log(1+n/d)+C.
 \label{eq:glauber-negative-integral}
\end{align}
Here $\log(1+x)\le x$, $1\le\gamma_d\le2$, and
$c_1\le C\sqrt{1+n/d}$.
For $0<\lambda<\lambdaC$, the same substitution gives
$I(\lambda)\le2\gamma_d\log\mathcal C(\lambda)$ when
$\mathcal C(\lambda)<c_1$; otherwise use $I(\lambda)\le I(\lambdaC)$
and \eqref{eq:glauber-negative-integral}.
This proves the first case of \eqref{eq:glauber-integrated-stability}.

\paragraph{Above criticality.}
Put $r=a/\lambdaC-1$, so $\mathrm d a/a=\mathrm d r/(1+r)$. Then
\begin{align*}
 I((1+\delta)\lambdaC)-I(\lambdaC)
 &=\int_0^\delta
   \frac{K(1+\sqrt{n/d}+(n/d)r)-1}{1+r}\dd r\\
 &\le C\left(\delta+\sqrt{n/d}\,\delta+(n/d)\delta^2\right)
 \le C+\frac{Cn\delta^2}{d}.
\end{align*}
The last inequality uses $\delta\le1$ and
$\sqrt{n/d}\,\delta\le1+(n/d)\delta^2$.
Combining this with \eqref{eq:glauber-negative-integral} proves the second case of
\eqref{eq:glauber-integrated-stability}.
\end{proof}

With \cref{prop:glauber-integrated-profile} in hand, we now apply
\cref{lem:glauber-stability-mixing} to prove \cref{thm:glauber-upper}.

\begin{proof}[Proof of \Cref{thm:glauber-upper}]
\Cref{prop:glauber-integrated-profile} verifies the spectral-stability
hypothesis of \cref{lem:glauber-stability-mixing} and gives
\[
 e^I\le C(1+n/d)^{1+1/(d-2)}
       \exp\!\left\{\frac{Cn\delta^2}{d}\right\}.
\]

One can verify that $1/d\le\lambda\le8$ and $d\lambda\le24$, so we may take
$\overline{\lambda}=\max\{1,\lambda\}\le8$.

By \eqref{eq:hc-elementary-residual}, the occupied-marginal lower-bound
condition holds with $b=e^{-24}/(9d)$: after every positive pinning,
each free vertex satisfies
\[
 p_v\ge\frac{\lambda}{(1+\lambda)^{d+1}}
       \ge b.
\]
The same elementary bounds give
\[
 \mu_{\min}\ge
 \left(\frac{\min\{1,\lambda\}}{1+\lambda}\right)^n
       \ge(9d)^{-n}.
\]

Since $3\le d=\Delta(G)\le n-1$, these choices imply
\[
 \log(n\overline{\lambda}/b)\log\log(e^2/\mu_{\min})\le C\log^2 n.
\]
Applying \cref{lem:glauber-stability-mixing} with the bound on $e^I$
proves the theorem, using $1+n/d\le2n/d$ and $1+1/(d-2)\le2$.
\end{proof}

\subsection{From marginal response to covariance bound}
\label{sec:glauber-response-covariance}

We now prove \cref{thm:si-field-perturbations} by bounding variances
of linear combinations of the coordinates.

\begin{proof}[Proof of \Cref{thm:si-field-perturbations}]
Write $p=\max_v\mu(X_v=1)$ and
$\beta=1/(1+\alpha)\in(1/2,1)$.
Recall that $L\ge1$, $h_0>0$, $u\in[0,1]$, and
\eqref{eq:si-field-perturbation-hypothesis} bounds every coordinate's
response to a field in $[e^{-h},e^h]^V$ by
$Lp(\sqrt u+h^\alpha)$ for $0<h\le h_0$.
For any real vector $c$,
\[
 c^{\mathsf T}\Cov_\mu(X)c=\Var(c^{\mathsf T}X).
\]
By the definition of the positive-semidefinite order, to prove
\eqref{eq:si-field-covariance-bound} it suffices to show that every
$c\in\mathbb R^V$ satisfies
\begin{equation}\label{eq:si-field-variance-target}
 \Var(c^{\mathsf T}X)
 \le Cp\bigl(1+(np)^{2\beta-1}+npu\bigr)\|c\|_2^2.
\end{equation}
Here $2\beta-1=(1-\alpha)/(1+\alpha)$.
We first bound the variance for arbitrary real vectors $c$ with
$p|\supp c|\le1$, then for $c\in\{-1,0,1\}^V$ with
$p|\supp c|\ge1$. Finally, we combine these two bounds to prove
\eqref{eq:si-field-variance-target} for arbitrary real vectors.
Set
$s=\min\{h_0,(8L)^{-1}\}$; constants below depend only on
$L,h_0,\alpha$.

\paragraph{Vectors with small support.}
Let $c\in\mathbb R^V$ be supported on $J\subseteq V$ with
$p|J|\le1$. Put
$N_J=\sum_{i\in J}X_i$, $Z_J(z)=\E e^{zN_J}$, and
$p_i(z)=(e^{z\mathbf1_J}*\mu)(X_i=1)$, where $\mathbf1_J$ is the
indicator vector of $J$. For $0\le z\le s$, differentiating gives
\[
 \frac{\mathrm d}{\mathrm dz}\log Z_J(z)
 =\sum_{i\in J}p_i(z)\le3Lp|J|,
\]
where the marginal response bound
\eqref{eq:si-field-perturbation-hypothesis} gives $p_i(z)\le p+2Lp\le3Lp$.
Integrating from $0$ to $s$ and using $Z_J(0)=1$, we obtain
\[
 \log Z_J(s)\le3Lsp|J|\le3/8,
 \qquad Z_J(s)<2.
\]
By the definition of the tilted law,
$p_i(s)=\E[X_ie^{sN_J}]/Z_J(s)$.
Thus, using $N_J\le e^{sN_J}/s$, for $i\in J$ we have
\[
 \E[X_iN_J]\le\frac{\E[X_ie^{sN_J}]}{s}
 =\frac{Z_J(s)p_i(s)}{s}\le\frac{6L}{s}p.
\]
Cauchy--Schwarz over the occupied coordinates gives
$(c^{\mathsf T}X)^2\le N_J\sum_{i\in J}c_i^2X_i$.
Taking expectations yields
\begin{equation}\label{eq:si-field-small-support}
 \Var(c^{\mathsf T}X)\le\E[(c^{\mathsf T}X)^2]
 \le\frac{6L}{s}p\|c\|_2^2.
\end{equation}

\paragraph{Signed sums on larger supports.}
Fix $c\in\{-1,0,1\}^V$ with $q=p|\supp c|\ge1$.
For independent $X,X'\sim\mu$, put $M=c^{\mathsf T}(X-X')$ and
$F(h)=\log\E e^{hM}$. The law of $M$ is symmetric, so
\begin{equation}\label{eq:glauber-symmetric-tilt}
 e^{F(h)}=\E\cosh(hM)\ge1+\frac{h^2}{2}\E M^2
 =1+h^2\Var(c^{\mathsf T}X).
\end{equation}
Differentiating
$F(h)=\log\E e^{hc^{\mathsf T}X}+\log\E e^{-hc^{\mathsf T}X}$
and applying \eqref{eq:si-field-perturbation-hypothesis} to both tilts
gives, for $0<h\le h_0$,
\[
 F'(h)
 =\sum_i c_i\bigl[(e^{hc}*\mu)(X_i=1)
                  -(e^{-hc}*\mu)(X_i=1)\bigr]
 \le2Lq(\sqrt u+h^\alpha).
\]
Integrating from $0$ to $h$ and using $F(0)=0$ gives
\[
 F(h)=\int_0^h F'(z)\dd z
 \le2Lq(\sqrt u\,h+h^{1+\alpha}).
\]
Choose $h$ so that $F(h)$ stays bounded:
\[
 h=\frac{s}{\max\{q^\beta,q\sqrt u\}}.
\]
Since $q\ge1$, we have $0<h\le s\le h_0$.
Moreover, $q\sqrt u\,h\le s$ and
\[
 qh^{1+\alpha}=(q^\beta h)^{1+\alpha}
 \le s^{1+\alpha}\le s,
\]
because $\beta(1+\alpha)=1$ and $s\le1$. Hence
$0\le F(h)\le4Ls\le1/2$, where nonnegativity follows from
\eqref{eq:glauber-symmetric-tilt}.
Using $e^z-1\le2z$ for $0\le z\le1/2$ in
\eqref{eq:glauber-symmetric-tilt} gives
\[
 \Var(c^{\mathsf T}X)\le\frac{e^{F(h)}-1}{h^2}
 \le\frac{8Ls}{h^2}=\frac{8L}{s}\max\{q^{2\beta},uq^2\}
 \le\frac{8L}{s}(q^{2\beta}+uq^2).
\]
Thus, for a sufficiently large constant $C=C(L,h_0)$,
\begin{equation}\label{eq:si-field-signed-variance}
 \Var(c^{\mathsf T}X)\le C(q^{2\beta}+uq^2),\qquad q\ge1.
\end{equation}

\paragraph{Arbitrary real vectors.}
Let $c\in\mathbb R^V$ be arbitrary.
Relabel coordinates so that $a_1=|c_1|\ge\cdots\ge a_n=|c_n|$,
and put $a_{n+1}=0$. For each $1\le k\le n$, define the vector
$b^{(k)}\in\{-1,0,1\}^n$ by $b_i^{(k)}=\operatorname{sgn}(c_i)$
for $i\le k$ and $b_i^{(k)}=0$ for $i>k$. Then
\[
 c=\sum_{k=1}^n(a_k-a_{k+1})b^{(k)}.
\]
Put $m=\min\{n,\lfloor1/p\rfloor\}$ and set
$c_{\rm head}=\sum_{k=1}^m(a_k-a_{k+1})b^{(k)}$.
For $i\le m$, its $i$th coordinate is
$(c_{\rm head})_i=\operatorname{sgn}(c_i)(a_i-a_{m+1})$,
and for $i>m$ it is zero. Hence $c_{\rm head}$ is supported on at
most $m\le1/p$ coordinates and $\|c_{\rm head}\|_2\le\|c\|_2$.
For each $k>m$ with $a_k-a_{k+1}>0$, we have $a_k>0$, so
$b^{(k)}$ has exactly $k$ nonzero coordinates and $pk>1$.

Since $\sqrt{\Var(c^{\mathsf T}X)}
=\|c^{\mathsf T}(X-\E X)\|_{L^2}$, the $L^2$ triangle inequality
applied to the above decomposition gives
\begin{equation}\label{eq:si-field-standard-deviation}
\begin{aligned}
 \sqrt{\Var(c^{\mathsf T}X)}
 &\le\sqrt{\Var(c_{\rm head}^{\mathsf T}X)}
  +\sum_{k>m}(a_k-a_{k+1})
    \sqrt{\Var((b^{(k)})^{\mathsf T}X)}\\
 &\le C\sqrt p\,\|c\|_2
  +Cp^\beta\sum_{k>m}(a_k-a_{k+1})k^\beta
  +Cp\sqrt u\sum_{k>m}(a_k-a_{k+1})k.
\end{aligned}
\end{equation}
The second inequality uses \eqref{eq:si-field-small-support} for
$c_{\rm head}$ and \eqref{eq:si-field-signed-variance} for each
$b^{(k)}$ with $k>m$ and $a_k-a_{k+1}>0$.
Since $a_k-a_{k+1}\ge0$, extending the first sum to all $1\le k\le n$ gives
\[
 \sum_{k>m}(a_k-a_{k+1})k^\beta
 \le\sum_{k=1}^n(a_k-a_{k+1})k^\beta
 =\sum_{k=1}^n a_k\bigl(k^\beta-(k-1)^\beta\bigr)
 \le\sum_{k=1}^n a_k k^{\beta-1}.
\]
The last inequality uses $k^\beta-(k-1)^\beta\le k^{\beta-1}$
for $0<\beta<1$. Also, $2\beta-2\in(-1,0)$, so
\[
 \sum_{k=1}^n k^{2\beta-2}
 \le\int_0^n x^{2\beta-2}\dd x
 =\frac{n^{2\beta-1}}{2\beta-1}.
\]
Cauchy--Schwarz and $a_k=|c_k|$ therefore give
\[
 \sum_{k=1}^n a_k k^{\beta-1}
 \le\|c\|_2\left(\sum_{k=1}^n k^{2\beta-2}\right)^{1/2}
 \le\frac{n^{\beta-1/2}}{\sqrt{2\beta-1}}\|c\|_2.
\]
For the second sum, we have
\[
 \sum_{k>m}(a_k-a_{k+1})k
 \le\sum_{k=1}^n(a_k-a_{k+1})k
 =\sum_{k=1}^n a_k=\|c\|_1\le\sqrt n\|c\|_2.
\]
Substituting these two estimates into \eqref{eq:si-field-standard-deviation}
and absorbing $(2\beta-1)^{-1/2}$ into the constant gives
\[
 \sqrt{\Var(c^{\mathsf T}X)}
 \le C\left(\sqrt p+p^\beta n^{\beta-1/2}
                    +p\sqrt{nu}\right)\|c\|_2.
\]
Squaring and using $(x+y+z)^2\le3(x^2+y^2+z^2)$ yields, for a
sufficiently large constant $C=C(L,h_0,\alpha)$,
\[
 \Var(c^{\mathsf T}X)
 \le C\bigl(p+p^{2\beta}n^{2\beta-1}+p^2nu\bigr)\|c\|_2^2
 =Cp\bigl(1+(np)^{2\beta-1}+npu\bigr)\|c\|_2^2.
\]
This proves \eqref{eq:si-field-variance-target} for every real vector
$c$, and hence the covariance bound \eqref{eq:si-field-covariance-bound}.
\end{proof}

\begin{proposition}[Sharpness of the response-to-covariance bound]
\label{prop:response-covariance-sharpness}
For every $0<\alpha<1$, integer $n\ge1$, $0<p\le1/2$, and
$u\in[0,1]$, there is a law $\mu$ on $\{0,1\}^n$ with full support
and $\mu(X_i=1)=p$ for every $i$, such that
\[
 \left|(\boldsymbol\psi*\mu)(X_i=1)-p\right|
 \le6p(\sqrt u+h^\alpha)
\]
for every $0<h\le1$, every
$\boldsymbol\psi\in[e^{-h},e^h]^n$, and every $i$, while
\[
 \lambda_{\max}(\Cov_\mu(X))
 \ge\frac p8\left(
 1+(np)^{\frac{1-\alpha}{1+\alpha}}+npu\right).
\]
Thus all three terms in
\eqref{eq:si-field-covariance-bound} are necessary, up to constants,
for general Boolean laws.
\end{proposition}

\begin{proof}
Put
\[
 q=np,\qquad
 a=\max\{1,q\}^{-\alpha/(1+\alpha)},\qquad
 s=\frac p2\max\{a,\sqrt u\}.
\]
Choose a uniform sign $\sigma\in\{-1,1\}$ and, conditional on
$\sigma$, draw the coordinates independently with occupation
probability $p_\sigma=p+\sigma s$.
Since $p/2\le p_\sigma\le3/4$, the resulting law has full support.
Its marginals equal $p$, and
\[
 \Cov_\mu(X)
 =\bigl[p(1-p)-s^2\bigr]I_n+s^2\one\one^{\mathsf T}.
\]
The largest eigenvalue is $p(1-p)-s^2+ns^2$, and
$p(1-p)-s^2\ge3p/8$.
If $q\ge1$, then
\[
 ns^2=\frac p4
 \max\left\{q^{\frac{1-\alpha}{1+\alpha}},qu\right\}
 \ge\frac p8
 \left(q^{\frac{1-\alpha}{1+\alpha}}+qu\right),
\]
which gives the claimed covariance lower bound.
If $q\le1$, both terms inside the last parentheses are at most
one, so the lower bound follows already from $3p/8$.

It remains to verify the response bound for arbitrary signed fields.
Tilting changes both the conditional occupation probabilities
and the probabilities of the two signs.
Write $z_i=\log\psi_i$, so $|z_i|\le h\le1$.
Conditional on $\sigma$, the tilted occupation probability is
\[
 f_{z_i}(p_\sigma),\qquad
 f_z(r)=\frac{re^z}{1+r(e^z-1)}.
\]
For $0<r<1$ and $|z|\le1$,
\[
 \partial_z f_z(r)=f_z(r)(1-f_z(r))\le er.
\]
Consequently,
\[
 |f_{z_i}(p_\sigma)-p_\sigma|
 \le e p_\sigma h\le\frac{3e}{2}ph.
\]
The two components are reweighted by their respective normalizing
factors, so the tilted probabilities of $\sigma=+1$ and $\sigma=-1$
have ratio $e^{\ell_{\boldsymbol\psi}}$, where
\[
 \ell_{\boldsymbol\psi}
 =\sum_{j=1}^n
 \log\frac{1+(p+s)(e^{z_j}-1)}
          {1+(p-s)(e^{z_j}-1)}.
\]
For $0\le r\le1$ and $|z|\le1$,
\[
 \left|\partial_r\log(1+r(e^z-1))\right|
 =\frac{|e^z-1|}{1+r(e^z-1)}
 \le e|z|.
\]
Hence $|\ell_{\boldsymbol\psi}|\le2ensh$.
The difference of these two probabilities therefore has absolute value
\[
 \left|\tanh(\ell_{\boldsymbol\psi}/2)\right|
 \le\min\{ensh,1\}.
\]
Averaging the conditional marginals gives
\[
 \left|(\boldsymbol\psi*\mu)(X_i=1)-p\right|
 \le\frac{3e}{2}ph+s\min\{ensh,1\}.
\]

If $\sqrt u\ge a$, then $s=p\sqrt u/2$, and the second term is
at most $p\sqrt u/2$.
Otherwise $s=pa/2$. Using $\min\{x,1\}\le x^\alpha$ for $x\ge0$
and the definition of $a$, we obtain
\[
 \begin{aligned}
 s\min\{ensh,1\}
 &\le e^\alpha n^\alpha s^{1+\alpha}h^\alpha\\
 &=\frac{e^\alpha}{2^{1+\alpha}}\,
   p\left(\frac{q}{\max\{1,q\}}\right)^\alpha h^\alpha\\
 &\le\frac{e^\alpha}{2^{1+\alpha}}ph^\alpha.
 \end{aligned}
\]
Since $h\le h^\alpha$ and
$3e/2+e^\alpha/2^{1+\alpha}<6$, both cases give the stated
response bound.
\end{proof}

\subsection{Marginal response for the hard-core model}
\label{sec:glauber-cubic-response}

By Weitz's SAW-tree representation in \cref{sec:prelim}, proving
\cref{lem:glauber-field-response} reduces to bounding the change in the
root marginal on a finite tree under a field tilt. We use the two-step
tree recursion to bound changes in log odds by induction on height.
The univariate two-step recursion below describes two levels at
fugacity $\lambdaC$, with $D$ children per node and equal grandchild
log odds, as in \cite[Section~2.1]{ChenLiuYin2023}.
The lemma bounds the change when this common input increases by $s$;
its proof is given in \cref{app:glauber-cubic}.

\begin{lemma}[Critical two-step contraction]
\label{lem:glauber-cubic-deficit}
For $D=d-1\ge2$, let $\ell(x)=\log(1+e^x)$ and define
\[
 g(x)=\log\lambdaC-D\ell(\log\lambdaC-D\ell(x)).
\]
Then $0\le g'(x)\le1$, with equality in the upper bound precisely
at $x=-\log(D-1)$. There is an absolute constant $c>0$ such that,
for every $x\in\mathbb R$ and $0\le s\le1$,
\begin{equation}\label{eq:glauber-integrated-deficit}
 g(x+s)-g(x)\le s-cs^3.
\end{equation}
\end{lemma}

\begin{proof}[Proof of \Cref{lem:glauber-field-response}]
Write $\lambda=\lambdaC e^t$, where $t\le\delta_{\mathrm{up}}$,
and $t_+=\max\{t,0\}$. Fix $0<h\le h_0=2^{-27}$,
$v\in V$, and $\boldsymbol\psi\in[e^{-h},e^h]^V$;
the tilted law has vertex fugacities $\lambda\psi_w$.
Let $T$, rooted at $v$, be the common SAW tree for $\mu$ and
$\boldsymbol\psi*\mu$.
For each node $u$ of $T$, write $y_u$ and $\widetilde y_u$ for its
log odds in the subtree rooted at $u$, with fugacities
inherited from $\mu$ and $\boldsymbol\psi*\mu$, respectively.
By preservation of the root marginals, our goal
\eqref{eq:glauber-field-response} is equivalent to
\[
 \left|\frac{e^{\widetilde y_v}}{1+e^{\widetilde y_v}}
       -\frac{e^{y_v}}{1+e^{y_v}}\right|
 \le C\lambda\bigl(\sqrt{t_+}+h^{1/3}\bigr).
\]

Put $D=d-1$ and use $\ell(x)=\log(1+e^x)$ as in
\cref{lem:glauber-cubic-deficit}.
If $u$ is a copy of the original vertex $w$, taking logarithms in
\eqref{eq:glauber-deletion-recursion} gives
\[
 y_u=\log\lambda-\sum_i\ell(y_i),\qquad
 \widetilde y_u=\log\lambda+\log\psi_w-\sum_i\ell(\widetilde y_i),
\]
where $i$ indexes the children of $u$.
For a nonroot node $u$, index its grandchildren by $(i,j)$, with $j$
indexing the children of $i$. Each of these index sets has size at most $D$.
The main step is to prove that, for every $0<s\le1$, the inequalities
$|\widetilde y_{ij}-y_{ij}|\le s$ at all grandchildren imply
\begin{equation}\label{eq:glauber-two-step-discrepancy}
 |\widetilde y_u-y_u|
 \le s-cs^3+C_1t_+s+C_2h,
\end{equation}
where $C_1,C_2$ are absolute constants.
We will choose $s=A(\sqrt{t_+}+h^{1/3})$, with an absolute constant $A$,
so that the right-hand side is at most $s$, and then prove
$|\widetilde y_u-y_u|\le s$ at every
nonroot node by induction on height. The root, which can have $d$
children, is handled separately at the end.

\paragraph{Reduction to the univariate two-step recursion.}
Fix a nonroot node $u$ and suppose
$|\widetilde y_{ij}-y_{ij}|\le s$ for all its grandchildren.
We first bound $\widetilde y_u-y_u$ using
$\widetilde y_{ij}\le y_{ij}+s$.
Set
\[
 S_i=\sum_j\ell(y_{ij}),\qquad
 \Delta_i=\sum_j\bigl(\ell(y_{ij}+s)-\ell(y_{ij})\bigr).
\]
The two-step recursion is increasing in the grandchild log odds
and in the logarithmic tilt at $u$, and decreasing in the logarithmic
tilts at its children. All these tilts lie in $[-h,h]$ under either
law, with zero tilts under $\mu$. Consequently,
\[
 \begin{aligned}
 \widetilde y_u
 &\le\log\lambda+h-\sum_i\ell(\log\lambda-h-S_i-\Delta_i),\\
 y_u
 &\ge\log\lambda-h-\sum_i\ell(\log\lambda+h-S_i).
 \end{aligned}
\]
Subtracting these bounds gives
\begin{equation}\label{eq:glauber-heterogeneous-envelope}
 \widetilde y_u-y_u\le
 2h+\sum_i\left[
 \ell(\log\lambda+h-S_i)
 -\ell(\log\lambda-h-S_i-\Delta_i)\right].
\end{equation}

We now reduce \eqref{eq:glauber-heterogeneous-envelope} to a univariate
bound using Jensen's inequality, as in \cite[Claim~4.5]{LiuLu2015}
and \cite[Section~7.1]{ChenLiuYin2023}.
For each child, pad its list of grandchild odds
with zeros until it has $D$ entries, assigning log odds $-\infty$
to the added entries and interpreting $\ell(-\infty)=0$.
These entries contribute zero to both $S_i$ and $\Delta_i$.
Choose $x_i\in\mathbb R\cup\{-\infty\}$ by
$\ell(x_i)=S_i/D$. The function
\[
 f_s(z)=\log\bigl(e^s-(e^s-1)e^{-z}\bigr),\qquad z\ge0,
\]
satisfies $f_s(0)=0$ and
\[
 f_s''(z)=
 -\frac{e^s(e^s-1)e^{-z}}{(e^s-(e^s-1)e^{-z})^2}\le0.
\]
Since $f_s(\ell(y))=\ell(y+s)-\ell(y)$, Jensen's inequality gives
\begin{equation}\label{eq:glauber-child-jensen}
 \Delta_i=\sum_{j=1}^D f_s(\ell(y_{ij}))
 \le Df_s(S_i/D)=D\bigl(\ell(x_i+s)-\ell(x_i)\bigr).
\end{equation}
If the child has no children, then $S_i=\Delta_i=0$ and $x_i=-\infty$;
the same inequality holds by continuity.

For a two-step recursion with $D$ children per node and common grandchild
log odds $x$, logarithmic tilts $\pm h$ at the current node and $\mp h$
at its children give the univariate two-step recursions
\[
 G_\pm(x)=\log\lambda\pm h
             -D\ell(\log\lambda\mp h-D\ell(x)).
\]
The $i$th summand on the right-hand side of
\eqref{eq:glauber-heterogeneous-envelope}
increases with $\Delta_i$. By \eqref{eq:glauber-child-jensen}, it is
at most $[G_+(x_i+s)-G_-(x_i)-2h]/D$.
These upper bounds are nonnegative, and there are at most $D$ children.
Thus $\widetilde y_u-y_u$ is at most
$\sup_{x\in\mathbb R}[G_+(x+s)-G_-(x)]$.
Interchanging the two laws gives the same bound on
$y_u-\widetilde y_u$, since both laws have logarithmic tilts in $[-h,h]$.
Hence
\begin{equation}\label{eq:glauber-scalar-envelope}
 |\widetilde y_u-y_u|\le
 \sup_{x\in\mathbb R}
 \bigl[G_+(x+s)-G_-(x)\bigr].
\end{equation}

\paragraph{Choosing the induction bound.}
The function $\ell'(z)=e^z/(1+e^z)$ is increasing and satisfies
$\ell'(z+a)\le e^a\ell'(z)$ for $a\ge0$.
Since $\log\lambda-h\le\log\lambdaC+t_+$, differentiation gives
\[
 G_+'(x)=D^2\ell'(x)\ell'(\log\lambda-h-D\ell(x))
 \le e^{t_+}g'(x).
\]
Integrating and applying \cref{lem:glauber-cubic-deficit} gives
\[
 G_+(x+s)-G_+(x)
 \le e^{t_+}(s-cs^3)
 \le s-cs^3+C_1t_+s.
\]
Moreover, by the mean value theorem, using $\ell'(z)\le e^z$ and
$\ell(x)\ge0$,
\[
 \begin{aligned}
 G_+(x)-G_-(x) \le (2+2D\lambda e^h)h\le(2+4D\lambdaC)h\le C_2h,
 \end{aligned}
\]
where we used $\lambda e^h=\lambdaC e^{t+h}\le2\lambdaC$.
Combining the last two estimates yields
\[
 \sup_{x\in\mathbb R}\bigl[G_+(x+s)-G_-(x)\bigr]
 \le s-cs^3+C_1t_+s+C_2h.
\]
Together with \eqref{eq:glauber-scalar-envelope}, this proves
\eqref{eq:glauber-two-step-discrepancy}.
Put $s=A(\sqrt{t_+}+h^{1/3})$, where $A$ is an absolute constant.
As verified in \cref{app:glauber-response-constants}, these constants
can be chosen so that
\[
 s\le\frac12,\qquad
 \frac{C_1}{A^2}+\frac{C_2}{A^3}\le c,\qquad C_2h\le s
\]
throughout the stated ranges of $t$ and $h$.
Since $t_+\le s^2/A^2$ and $h\le s^3/A^3$, we have
\[
 C_1t_+s+C_2h
 \le\left(\frac{C_1}{A^2}+\frac{C_2}{A^3}\right)s^3\le cs^3.
\]
The right-hand side of \eqref{eq:glauber-two-step-discrepancy} is
therefore at most $s$. Thus $|\widetilde y_{ij}-y_{ij}|\le s$ at all
grandchildren implies $|\widetilde y_u-y_u|\le s$.

\paragraph{Induction and the original root.}
For leaves, the difference in log odds is at most $2h\le C_2h\le s$.
At a nonroot node whose children are leaves, the one-step recursion
and $\ell'\le2\lambdaC$ between corresponding child log odds give
the bound $(2+4D\lambdaC)h\le C_2h\le s$.
These are the two base cases. Applying
\eqref{eq:glauber-two-step-discrepancy} now proves
$|\widetilde y_u-y_u|\le s$ at every nonroot node by induction on height.

The original root can have $d$ children, so we use the one-step
recursion there. Each pair of child log odds differs by at most $s$,
and the corresponding odds are at most $\lambda e^h\le2\lambdaC$.
Hence $|\widetilde y_v-y_v|\le2h+2d\lambdaC s\le26s$,
using $h\le s$ and $d\lambdaC\le12$.
To pass from log odds to marginal probabilities, we use that the derivative of
$x\mapsto e^x/(1+e^x)$ is at most $e^x$.
Both root odds are at most $\lambda e^h\le2\lambda$, so the marginal
difference is at most
$2\lambda\cdot26s=52A\lambda(\sqrt{t_+}+h^{1/3})$.
This proves \eqref{eq:glauber-field-response} with $C=52A$.
\end{proof}

\section{Near-critical approximate counting}
\label{sec:counting-main}\label{app:glauber-upper}\label{app:upper-details}

This section proves \cref{count:lem-aggregate-curvature,thm:counting-upper},
stated in \cref{sec:proof-outline-counting}.
\Cref{count:sec-paired-estimation} gives the general estimator and
trajectory bounds; \cref{count:sec-improved-counting} applies them to
hard-core counting, using the occupation-variance bound proved in
\cref{count:sec-aggregate-counting}.
The implementation details, including worst-case work and sampling,
are deferred to \cref{count:sec-implementation}.

\subsection{Paired products and independent trajectories}
\label{count:sec-paired-estimation}

On a finite state space $\Omega$, fix weights $w(\omega)\ge0$, not
all zero, and an observable $H:\Omega\to\mathbb R$. Let $\mu_b$ have
weights $w(\omega)e^{-bH(\omega)}$, partition function $Z(b)$, and
log partition function $f(b)=\log Z(b)$. Then $f''(b)=\Var_{\mu_b}(H)$.
For an interval $[b,b+u]$ with $u>0$, draw independent configurations
$X\sim\mu_b$ and $Y\sim\mu_{b+u}$, and define the paired factors
\[
 W^-=e^{-uH(X)/2},\qquad W^+=e^{uH(Y)/2}.
\]
Huber's paired-product identities~\cite[Section~2]{Huber2015} give
\begin{equation}\label{count:eq-paired-identities}
 \frac{\E W^+}{\E W^-}=\frac{Z(b)}{Z(b+u)},\qquad
 \frac{\E[(W^+)^2]}{(\E W^+)^2}
 =\frac{\E[(W^-)^2]}{(\E W^-)^2}
 =\exp\{f(b)+f(b+u)-2f(b+u/2)\}.
\end{equation}
Huber bounds the relative variance of the whole product through the
exponents $f(b)+f(b+u)-2f(b+u/2)$~\cite[Section~3.3]{Huber2015}; we use
the following form for one interval.
\begin{lemma}[A relative-variance bound for one interval]
\label{count:lem-band-budget}
Fix an absolute constant $c_*>0$. Consider an interval $[a,c]$ of length
$\ell=c-a>0$ and a known bound $V\ge\int_a^c f''(s)\dd s$.
For any partition of $[a,c]$ into subintervals of length at most $h$
with $hV\le c_*$, the paired factors satisfy
\begin{equation}\label{count:eq-paired-band-budget}
 \sum_i\left[
 \frac{\Var(W_i^+)}{(\E W_i^+)^2}
 +\frac{\Var(W_i^-)}{(\E W_i^-)^2}\right]\le ChV,
\end{equation}
where $W_i^+,W_i^-$ are the factors for subinterval $i$ and $C$ depends
only on $c_*$. In particular, choosing
$m=\max\{1,\lceil\ell V/c_*\rceil\}$ equal subintervals gives
$O(1+\ell V)$ subintervals and $O(1)$ total relative variance.
\end{lemma}

\begin{proof}
For the $i$th subinterval $[b,b+u]$, let $z_i=f(b)+f(b+u)-2f(b+u/2)$.
By \eqref{count:eq-paired-identities}, both paired factors have
relative second moment $e^{z_i}$, and
\[
 z_i=\int_b^{b+u}\min\{s-b,b+u-s\}f''(s)\dd s.
\]
Since $0\le\min\{s-b,b+u-s\}\le u/2\le h/2$ and $f''\ge0$,
each $z_i\ge0$, and summing over the subintervals gives
\[
 \sum_i z_i\le\frac h2\int_a^c f''(s)\dd s
             \le\frac{hV}{2}\le\frac{c_*}{2}.
\]
On $[0,c_*/2]$, we have $e^z-1\le Cz$. Thus the sum in
\eqref{count:eq-paired-band-budget} equals
$2\sum_i(e^{z_i}-1)\le ChV$.
For the stated equal partition, $h=\ell/m$ satisfies $hV\le c_*$,
and $m\le1+\ell V/c_*$, proving the remaining claims.
\end{proof}

We estimate each factor by averaging observations along a separate
Markov-chain trajectory. Different trajectories use independent
randomness, while observations within a trajectory may be correlated.
The following lemma combines standard results
from~\cite{LevinPeres2017,Huber2015}.

\begin{lemma}[Positive products from independent trajectories]
\label{count:lem-positive-product-trajectories}
Let $M\ge1$. For each $j\in[M]=\{1,\ldots,M\}$, let $\mu_j$ be a
probability law on a finite configuration space $\Omega_j$, and let
$g_j:\Omega_j\to[0,\infty)$ be a nonnegative function with positive
mean $m_j=\E_{\mu_j}g_j$. Suppose an irreducible reversible
positive-semidefinite Markov chain on $\Omega_j$ with stationary law
$\mu_j$ has mixing time $t_{\mathrm{mix},j}$ and relaxation time
$t_{\mathrm{rel},j}$, with the conventions of \cref{sec:prelim}.
Let $R_j\ge1$ and $S\ge0$ be given bounds satisfying
$t_{\mathrm{rel},j}\le R_j$ and
\[
 \sum_{j=1}^M\frac{\Var_{\mu_j}(g_j)}{m_j^2}\le S.
\]
Partition $[M]$ into disjoint numerator and denominator sets
$\mathcal N,\mathcal D$, allowing either set to be empty.
\rev{For $0<\eps<1/4$ and $0<\zeta<1$, run the chains independently
from arbitrary initial configurations. Burn in chain $j$ for
$t_{\mathrm{mix},j}(\zeta/M)$ updates, then record the next
$L_j=\lceil C_0R_j\max\{1,S\}\eps^{-2}\rceil$ configurations
$X_{j,1},\ldots,X_{j,L_j}$, where $C_0$ is a sufficiently
large absolute constant.} Form the empirical means
\[
 A_j=\frac1{L_j}\sum_{t=1}^{L_j}g_j(X_{j,t}).
\]
Then the estimator
\[
 \frac{\prod_{j\in\mathcal N}A_j}{\prod_{j\in\mathcal D}A_j},
\]
set to $1$ if the denominator is zero, approximates
$\prod_{j\in\mathcal N}m_j/\prod_{j\in\mathcal D}m_j$
to relative error at most $\eps/2$ with probability at least
$1-\zeta-1/16$, using at most
\begin{equation}\label{count:eq-trajectory-work}
 O\!\left(\eps^{-2}\max\{1,S\}\sum_{j=1}^M R_j
       +\sum_{j=1}^M t_{\mathrm{mix},j}(\zeta/M)\right)
\end{equation}
chain updates.
\end{lemma}

\begin{proof}
For stationary trajectories, $\E A_j=m_j$, and
\cite[Lemma~12.22]{LevinPeres2017} gives
\[
 \sum_{j=1}^M\frac{\Var(A_j)}{m_j^2}
 \le \frac{2\eps^2}{C_0\max\{1,S\}}
       \sum_{j=1}^M\frac{\Var_{\mu_j}(g_j)}{m_j^2}
 \le \frac{2\eps^2}{C_0}.
\]
Independence and \cite[Lemma~1.1]{Huber2015} therefore give, for
$\mathcal A\in\{\mathcal N,\mathcal D\}$,
\[
 \frac{\Var(\prod_{j\in\mathcal A}A_j)}
      {(\prod_{j\in\mathcal A}m_j)^2}
 \le e^{2\eps^2/C_0}-1=O(\eps^2/C_0).
\]
For large enough $C_0$, Chebyshev's inequality and a union bound put
both products within relative error $\eps/8$ of their positive means
with probability at least $15/16$. Their ratio then has relative error
at most $\eps/2$, and its denominator is nonzero.

The burn-in coupling argument in the proof of
\cite[Theorem~12.21]{LevinPeres2017}, applied independently across $j$,
couples each entire recorded trajectory to a stationary one with
disagreement probability at most $\zeta/M$. A union bound thus adds
at most $\zeta$ to the failure probability. Summing the burn-in times
and the $L_j$ observation lengths gives \eqref{count:eq-trajectory-work}.
\end{proof}

\subsection{The counting algorithm and its running time}
\label{count:sec-improved-counting}

Fix the target fugacity $\lambda=(1+\delta)\lambdaC(d)$
as in \cref{thm:counting-upper}, and write $\lambda_b=\lambda e^{-b}$.

\paragraph{Relaxation and mixing times.}

\begin{lemma}[Low-fugacity Glauber dynamics]\label{lem:glauber-low-activity}
Let $G$ have $n\ge1$ vertices and let $a>0$. If
$\alpha_{\mathrm D}:=\Delta(G)a/(1+a)<1$, then Glauber dynamics for
$\mu_{G,a}$ satisfies
\[
 t_{\mathrm{mix}}(\eta)\le
 \left\lceil\frac{n\log(n/\eta)}{1-\alpha_{\mathrm D}}\right\rceil
 \quad(0<\eta<1).
\]
\end{lemma}

\begin{proof}
Changing one neighboring spin changes a vertex's conditional occupation
probability by at most $a/(1+a)$, while nonneighbors have no effect.
Thus the sum of these maximal influences at any vertex is at most
$\Delta(G)a/(1+a)=\alpha_{\mathrm D}$.
This verifies the Dobrushin condition, so the claimed mixing bound
follows from \cite[Observation~10 and Corollary~7]{Hayes2006}.
\end{proof}

The integral bounds in \cref{prop:glauber-integrated-profile} give
the following costs along the annealing path.

\begin{lemma}[Fugacity-dependent relaxation and burn-in]
\label{count:lem-temperature-costs}
Let $G$ be an $n$-vertex simple graph and let $d$ be an integer
satisfying $\max\{3,\Delta(G)\}\le d\le n-1$.
Set
$\lambda_b=(1+\delta)\lambdaC(d)e^{-b}$, where
$0\le\delta\le\delta_{\mathrm{up}}$ and $b\ge0$.
Let $I(\lambda_b)$ be the stability integral of
\cref{prop:glauber-integrated-profile}, and define
\[
 R(b)=n\bigl(1+\max\{1,\lambda_b\}\bigr)e^{I(\lambda_b)}\ge n.
\]
The Glauber dynamics for $\mu_{G,\lambda_b}$ has relaxation time at
most $R(b)$. Moreover, there is an absolute constant $C>0$ such that
\begin{equation}\label{count:eq-temperature-costs}
 R(b)\le
 \begin{cases}
 Cn\left(1+\frac nd\right)^{1+\frac1{d-2}}
    \exp\!\left\{C\left(1+\frac nd\right)(\delta-b)^2\right\},
    &0\le b\le\delta,\\[1mm]
 Cn\left[1+\min\left\{
    \left(1+\frac nd\right)^{1+\frac1{d-2}},
    (b-\delta)^{-2-\frac2{d-2}}\right\}\right],&b>\delta.
 \end{cases}
\end{equation}
Its mixing time satisfies
$t_{\mathrm{mix}}(\eta)\le CR(b)[\log^2(n+2)+\log(1/\eta)]$
for $0<\eta<1/4$.
In particular, $R(b)=O(n)$ for $b\ge\delta+1$.
\end{lemma}

\begin{proof}
By \eqref{eq:glauber-field-relaxation} with
$\overline\lambda=\max\{1,\lambda_b\}$, the quantity $R(b)$ bounds the
relaxation time, so it remains to bound $I(\lambda_b)$ by the two
cases of \eqref{eq:glauber-integrated-stability}. Note that
$1+\max\{1,\lambda_b\}\le6$.
For this calculation write $Q=1+n/d$ and $\gamma_d=1+1/(d-2)$.
For $b\le\delta$, the relative distance satisfies
$\lambda_b/\lambdaC-1\le\delta-b$. If $\lambda_b\ge\lambdaC$,
\eqref{eq:glauber-integrated-stability} at $\lambda_b$ therefore gives
$I(\lambda_b)\le\gamma_d\log Q+CQ(\delta-b)^2+C$. If
$\lambda_b<\lambdaC$, then $I(\lambda_b)\le I(\lambdaC)$, because the
stability integral increases with the fugacity, and
\eqref{eq:glauber-integrated-stability} at $\lambdaC$ gives the same
bound. Exponentiating proves the first line.
For $u=b-\delta>0$,
\cref{lem:glauber-occupation-bound} gives
$\mathcal C(\lambda_b)\le2\gamma_d/(1-e^{-u})\le C\max\{1,u^{-1}\}$.
Also, \eqref{eq:glauber-stability-profile} gives
$\mathcal C(\lambda_b)\le K(1+\sqrt{n/d})\le C\sqrt Q$.
Thus the first case of \eqref{eq:glauber-integrated-stability} implies
\[
 I(\lambda_b)\le C+2\gamma_d\log_+\min\{\sqrt Q,1/u\}.
\]
Exponentiating proves the second line.

For mixing, first suppose $\lambda_b\ge1/(2d)$.
Equation~\eqref{eq:hc-elementary-residual} gives
$p_v\ge e^{-24}/(18d)$ and $\mu_{\min}\ge(18d)^{-n}$;
since $d\le n-1$, the mixing criterion,
\cref{lem:glauber-stability-mixing}, has logarithmic cost
$O(\log^2(n+2)+\log(1/\eta))$, and its
prefactor $\max\{1,\lambda_b\}ne^{I(\lambda_b)}$ is at most $R(b)$.
If $\lambda_b<1/(2d)$, then
$\Delta(G)\lambda_b/(1+\lambda_b)<1/2$, so
\cref{lem:glauber-low-activity} gives
$t_{\mathrm{mix}}(\eta)\le\lceil2n\log(n/\eta)\rceil$,
which suffices since $R(b)\ge n$.
\end{proof}

\begin{proof}[Proof of \Cref{thm:counting-upper}: algorithm and update count]
We first bound the number of Glauber updates.
\Cref{count:sec-implementation} implements the schedule and observations
at finite precision and converts this update bound into the claimed
worst-case arithmetic bound, with polynomial binary overhead.

Apply the framework of \cref{count:sec-paired-estimation} to independent
sets with observable $H(\omega)=|\omega|$. Write
$\mu_b=\mu_{G,\lambda_b}$, $f(b)=\log Z_G(\lambda_b)$, and $Q=1+n/d$.
At fugacity $\lambda_b$, the supercritical parameter in
\cref{count:lem-aggregate-curvature} is $\lambda_b/\lambdaC(d)-1$.
Write $u_+=\max\{u,0\}$. Since $e^b\ge1+b$ for $b\ge0$,
\[
 \left(\frac{\lambda_b}{\lambdaC(d)}-1\right)_+
 =\bigl((1+\delta)e^{-b}-1\bigr)_+
 \le\left(\frac{\delta-b}{1+b}\right)_+
 \le(\delta-b)_+.
\]
Using this inequality, $n\lambda_b\le CQ$, and
$f''(b)=\Var_{\mu_b}(|X|)$, the two bounds in
\cref{count:lem-aggregate-curvature} (proved in
\cref{count:sec-aggregate-counting}) give, for every $b\ge0$,
\begin{equation}\label{count:eq-counting-curvature-envelope}
 f''(b)\le CQ\log(2+Q)
                 \exp\{CQ(\delta-b)_+^2\}.
\end{equation}
By \cref{count:lem-temperature-costs}, burn-in to TV error $\eta$ costs
$CR(b)[\log^2 n+\log(1/\eta)]$ updates.
\synlabelalias{count:eq-counting-relaxation-envelope}{count:eq-temperature-costs}

\paragraph{The annealing schedule.}
Choose an integer $k\ge0$ with $4^k\le Q<4^{k+1}$,
and set $w_0=2^{-k}$.
Thus $1\le Qw_0^2<4$.
We partition the annealing path into intervals in $b$, called bands,
which we subdivide into annealing steps below.
The first band is $b\in[0,\delta+w_0]$, of length $\ell_0=\delta+w_0$.
Throughout this interval, the occupation variance and the relaxation
time of Glauber dynamics at fugacity $\lambda_b$ satisfy
\[
 \begin{aligned}
 f''(b)=\Var_{\mu_b}(|X|)&\le CQ\log(2+Q)e^{CQ\delta^2},\\
 t_{\mathrm{rel}}\le R(b)&\le CnQ^{\gamma_d} e^{CQ\delta^2},
 \end{aligned}
\]
where $\gamma_d=1+1/(d-2)\in[1,2]$.
These upper bounds in fact hold for all $b\ge0$ by
\cref{count:lem-temperature-costs} and
\eqref{count:eq-counting-curvature-envelope}.
On each cold band $b\in[\delta+w,\delta+2w]$, where
$w=2^jw_0$ for $j=0,\ldots,k-1$, we have
\[
 f''(b)\le CQ\log(2+Q),\qquad R(b)\le Cnw^{-2\gamma_d}.
\]
The initial and cold bands together cover $[0,\delta+1]$;
when $k=0$, there are no cold bands.
For the late band $b\in[\delta+1,B]$, choose
\[
 B=\delta+1+\left\lceil\log_2\bigl(16\max\{1,n\lambda\}\bigr)\right\rceil.
\]
Throughout this band, $R(b)\le Cn$, and
\[
 \int_{\delta+1}^B f''(b)\dd b
 =\E_{\mu_{\delta+1}}|X|-\E_{\mu_B}|X|\le n\lambda\le CQ.
\]
There are $O(\log(2+Q))$ bands altogether.

For a band $\mathcal B$ of length $\ell_{\mathcal B}$, let
$V_{\mathcal B}$ be an upper bound on $\int_{\mathcal B}f''(b)\dd b$:
take $C\ell_0Q\log(2+Q)e^{CQ\delta^2}$ initially,
$CQw\log(2+Q)$ in a cold band, and $CQ$ in the late band.
Fix a small absolute $c_*>0$. Divide each band $\mathcal B$ into
$m_{\mathcal B}$ annealing steps of equal length $h_{\mathcal B}$ in $b$,
where
\begin{equation}\label{count:eq-counting-band-mesh}
 m_{\mathcal B}=\max\{1,\lceil\ell_{\mathcal B}V_{\mathcal B}/c_*\rceil\},
 \qquad h_{\mathcal B}=\ell_{\mathcal B}/m_{\mathcal B}.
\end{equation}
Thus successive schedule points within this band differ by
$h_{\mathcal B}$. Since
$h_{\mathcal B}V_{\mathcal B}\le c_*$,
\cref{count:lem-band-budget} gives $O(1)$ total relative variance per band.
For the initial band, let $m_0$ denote this number of steps.
Since $d\le n-1$, we have $Q\delta^2\le2n\delta^2/d$ and
$Q\ell_0^2\le2Q\delta^2+8=O(1+n\delta^2/d)$. Hence
\[
 \begin{aligned}
 m_0\le C\bigl(1+\ell_0^2Q\log(2+Q)e^{CQ\delta^2}\bigr)
 \le C(1+n\delta^2/d)\log(2+Q)e^{Cn\delta^2/d}
 \le C\log(2+Q)e^{C'n\delta^2/d}.
 \end{aligned}
\]
Table~\ref{count:tab-bands} collects these bounds and the number of steps
in each band.

\begin{table}[htbp]
\centering
\small
\setlength{\tabcolsep}{3pt}
\renewcommand{\arraystretch}{1.35}
\begin{tabular}{@{}lccc@{}}
\toprule
Band & $\int_{\mathcal B} f''(b)\dd b$
 & Steps $m_{\mathcal B}$ & $R(b)$\\
\midrule
\begin{tabular}[c]{@{}l@{}}Initial\\$0\le b\le\delta+w_0$\end{tabular}
 & $C\ell_0Q\log(2+Q)e^{CQ\delta^2}$
 & $C\log(2+Q)e^{Cn\delta^2/d}$ & $CnQ^{\gamma_d} e^{Cn\delta^2/d}$\\
\begin{tabular}[c]{@{}l@{}}Cold\\$w\le b-\delta\le2w$\end{tabular}
 & $CQw\log(2+Q)$ & $C(1+Qw^2\log(2+Q))$ & $Cnw^{-2\gamma_d}$\\
\begin{tabular}[c]{@{}l@{}}Late\\$\delta+1\le b\le B$\end{tabular}
 & $CQ$ & $CQ\log(2+Q)$ & $Cn$\\
\bottomrule
\end{tabular}
\caption{Annealing bands.
The last three columns give upper bounds, with absolute constants $C$.}
\label{count:tab-bands}
\end{table}

Paired products along this schedule estimate $Z_G(\lambda)/Z_G(\lambda_B)$.
To recover $Z_G(\lambda)$, we divide by an estimate of
$1/Z_G(\lambda_B)$, the empty-set probability under $\mu_B$.
For this purpose, append the observable
$E_B(X)=\one_{\{X=\emptyset\}}$ at the final endpoint; since
$n\lambda_B\le1/16$,
\[
 \E_{\mu_B}E_B=Z_G(\lambda_B)^{-1}\ge e^{-1/16}.
\]
Each annealing step contributes two observables, $W_i^+$ and $W_i^-$,
and the final endpoint contributes $E_B$. List all these observables,
whose means we will estimate, as $g_1,\ldots,g_M$, where
\[
 M=2\sum_{\mathcal B}m_{\mathcal B}+1.
\]
For each $g_j$, let $b_j$ be the endpoint where it is observed and set
$\mu_j=\mu_{G,\lambda_{b_j}}$.

\added{For each observable $g$, let $A_g$ be its empirical mean along
a separate Glauber trajectory, with lengths specified below.}
The estimator is
\begin{equation}\label{count:eq-paired-return-value}
 \widehat Z_{\rm raw}=
 \frac{\prod_i A_{W_i^+}}{A_{E_B}\prod_i A_{W_i^-}}.
\end{equation}
\rev{By \eqref{count:eq-paired-identities}, the exact means telescope
to $Z_G(\lambda)$, including $\E_{\mu_B}E_B=1/Z_G(\lambda_B)$.}
If $A_{E_B}=0$, return $1$.

We now verify the inputs to \cref{count:lem-positive-product-trajectories}.
Each band contributes $O(1)$ to the sum of relative variances.
There are $O(\log(2+Q))$ bands. The terminal indicator $E_B$ has
relative variance at most $e^{1/16}-1$, since its mean is at least
$e^{-1/16}$. Moreover, Table~\ref{count:tab-bands}
gives $M\le CQ\log^2(2+Q)e^{Cn\delta^2/d}$.
Consequently, using $\log(2+Q)=O(\log n)$,
\[
 \sum_{j=1}^M\frac{\Var_{\mu_j}(g_j)}{(\E_{\mu_j}g_j)^2}\le C\log(2+Q),
 \qquad \log(2+M)=O(n\delta^2/d+\log n).
\]
The latter bound controls the burn-in cost: below, each chain is burned
in to TV error $1/(32M)$.

\paragraph{Summing the relaxation-time bounds.}
The trajectory lemma also requires a relaxation-time bound $R_j$ for
the Glauber chain used to estimate each $\E_{\mu_j}g_j$.
If $g_j$ is $W_i^+$ or $W_i^-$, take $R_j$ from the last column of
Table~\ref{count:tab-bands} for the band containing step $i$.
This ensures $t_{\mathrm{rel},j}\le R(b_j)\le R_j$.
For $E_B$, take $R_j=Cn$, since $R(B)\le Cn$.
We now bound $\sum_jR_j$ by summing over bands.
Each of the $m_{\mathcal B}$ steps in a band has two observables,
$W_i^+$ and $W_i^-$. The sum of their $R_j$ is therefore bounded by
$2m_{\mathcal B}$ times the band's bound on $R(b)$ in
Table~\ref{count:tab-bands}.
We absorb this factor of two into $C$ below.

For a cold band of scale $w$, this bounds the sum of its $R_j$ by
\[
 Cn w^{-2\gamma_d}\bigl(1+Qw^2\log(2+Q)\bigr).
\]
Since $w\ge w_0\ge Q^{-1/2}$, $w\le1$, and $1\le\gamma_d\le2$,
both $w^{-2\gamma_d}$ and $Qw^{2-2\gamma_d}$ are at most $Q^{\gamma_d}$.
Hence the sum for one cold band is at most
$CnQ^{\gamma_d}\log(2+Q)$. There are $O(\log(2+Q))$ cold bands,
so the sum over all their observables is at most
$CnQ^{\gamma_d}\log^2(2+Q)$, with an absolute constant independent of $d$.

For the initial band, multiplying the step count by the bound on
$R(b)$ similarly bounds the sum of its $R_j$ by
\[
 C\log(2+Q)e^{Cn\delta^2/d}\cdot nQ^{\gamma_d}e^{CQ\delta^2}
 \le CnQ^{\gamma_d}\log(2+Q)e^{C'n\delta^2/d},
\]
where we used $Q\delta^2\le2n\delta^2/d$.
The late band has $O(Q\log(2+Q))$ steps and $R_j=Cn$, so the sum
of its $R_j$ is at most $CnQ\log(2+Q)$.
The terminal indicator $E_B$ adds one further term $R_j=Cn$.
Combining these bounds, using $Q^{\gamma_d}\ge Q$, and enlarging $C$,
we obtain
\begin{equation}\label{count:eq-counting-sum-costs}
 \sum_{j=1}^M R_j\le CnQ^{\gamma_d}\log^2(2+Q)e^{Cn\delta^2/d}.
\end{equation}
\paragraph{\rev{Total update count.}}
\rev{Apply \cref{count:lem-positive-product-trajectories} to
\eqref{count:eq-paired-return-value} with}
$S=C\log(2+Q)$ and $\zeta=1/32$. For each observable $g_j$, run a
separate Glauber chain with stationary law $\mu_j$, using independent
randomness across chains. Burn in chain $j$ to TV error $1/(32M)$, and take the
empirical mean $\rev{A_{g_j}}$ of $g_j$ over
\begin{equation}\label{count:eq-counting-trajectory-length}
 L_j=\left\lceil CR_j\log(2+Q)\eps^{-2}\right\rceil
\end{equation}
further observations. \rev{The lemma gives statistical relative error
at most $\eps/2$ except with probability $3/32$, including the event
$A_{E_B}=0$ in its failure allowance.}
By \eqref{count:eq-trajectory-work}, the deterministic number of
burn-in and observation updates is at most
\[
 C\sum_jR_j\left[\log(2+Q)\eps^{-2}+\log^2 n+\log(2+M)\right].
\]
By \eqref{count:eq-counting-sum-costs} and
$\log(2+M)=O(n\delta^2/d+\log n)$, this is at most
\[
 Cn\left(1+\frac nd\right)^{1+\frac1{d-2}}
    \exp\!\left\{\frac{C'n\delta^2}{d}\right\}
    \eps^{-2}\log^4 n.
\]
Fixed powers of $1+n\delta^2/d$ are absorbed by increasing $C'$.
\end{proof}

\subsection{The variance of total occupation}
\label{count:sec-aggregate-counting}

At low fugacities $0\le\lambda\le1/n$, Davies, Sandhu, and Tan bound
the variance of $|X|$ by its value on the edgeless
graph~\cite[Theorem~1.5]{DaviesSandhuTan2025}. We need a bound that
remains uniform through the near-critical regime.

We now prove \cref{count:lem-aggregate-curvature}. Fix
$0<\lambda\le(1+\delta_{\mathrm{up}})\lambdaC(d)$.
Let $G$ be an $n$-vertex graph of maximum degree at most $d\ge3$.
For $v\in V(G)$, write
$p_v=\mu_{G,\lambda}(X_v=1)$ and recall that
$R_{G}(v)=p_v/(1-p_v)$ is its occupation odds.
Varying the fugacity of all vertices together, define the response
\begin{equation}\label{count:eq-common-field-response}
 \begin{aligned}
 w_{G}(v)=\frac{\mathrm d\log R_{G}(v)}{\mathrm d\log\lambda}
 =\E_{\mu_{G,\lambda}}[|X|\mid X_v=1]
   -\E_{\mu_{G,\lambda}}[|X|\mid X_v=0].
 \end{aligned}
\end{equation}
With expectation and variance in the following equation taken under
$\mu_{G,\lambda}$, the law of total variance gives the expected decrease in variance upon
revealing $X_v$
\begin{equation}\label{count:eq-reveal-decrement}
 \Var(|X|)-\E\bigl[\Var(|X|\mid X_v)\bigr]
 =p_v(1-p_v)w_{G}(v)^2.
\end{equation}
We bound this squared response using self-avoiding-walk counts, as in the
connective-constant method of~\cite{SinclairEtAl2017}; such counts have
also been used to control spectral independence~\cite[Section~3]{BezakovaEtAl2024}.

For the path bound below, put
$\delta_+=\max\{\lambda/\lambdaC(d)-1,0\}$: this is zero at and below
criticality, and is the relative excess above criticality.

\begin{lemma}[A self-avoiding-walk bound for the common-field response]
\label{count:lem-saw-response}
Let $G$ be a graph of maximum degree at most $d\ge3$, and let
$0<\lambda\le(1+\delta_{\mathrm{up}})\lambdaC(d)$.
If, for $v\in V(G)$, $S_\ell(G,v)$ counts
self-avoiding walks of length $\ell$ from $v$, then, for an absolute constant $C$,
\begin{equation}\label{count:eq-original-path-response}
 w_{G}(v)^2\le C\left[1+\sum_{j\ge0}(1+\delta_+)^{3j}
       \sqrt{\frac{S_{2j+1}(G,v)}{d(d-1)^{2j}}}\right].
\end{equation}
The sum is finite because $S_\ell(G,v)=0$ for $\ell\ge |V(G)|$.
\end{lemma}

We first use \cref{count:lem-saw-response} to prove
\cref{count:lem-aggregate-curvature}, then return to the proof of
\cref{count:lem-saw-response}.

\paragraph{Independent revelation of vertices.}
\begin{proof}[Proof of \Cref{count:lem-aggregate-curvature}]
Independently of $X$, assign each $v\in V(G)$ an independent random
reveal time $T_v$ with $\Pr(T_v>t)=e^{-t}$ for $t\ge0$.
Unsubscripted expectations in this proof average over both $X$
and the reveal times. At time $t\ge0$, reveal the vertices in
$\mathcal S_t=\{v:T_v\le t\}$, and let $\nu_t$ be the conditional law of $X$
given $\mathcal S_t$ and $X_{\mathcal S_t}$.
This is the coordinate-by-coordinate localization
of~\cite[Section~2.1.1]{ChenEldan2022}, run in continuous time.
Then $\mathcal V(t)=\E[\Var_{\nu_t}(|X|)]$ satisfies
$\mathcal V(0)=\Var_{\mu_{G,\lambda}}(|X|)$ and
$\mathcal V(t)\to0$ as $t\to\infty$.
The finite-sum expression
\[
 \mathcal V(t)=\sum_{S\subseteq V}(1-e^{-t})^{|S|}e^{-t(n-|S|)}
                      \E_{\mu_{G,\lambda}}[\Var_{\mu_{G,\lambda}}(|X|\mid X_S)]
\]
also shows that $\mathcal V$ is differentiable. We compute its rate of
decrease by considering the revelations during $[t,t+h]$.
Let $G'$ be the free residual graph of the current conditional
law $\nu_t$, and write $p_v=\nu_t(X_v=1)$ for its free vertices.
On these vertices, $\nu_t$ is the hard-core distribution $\mu_{G',\lambda}$.
The fixed occupations
add only a constant to $|X|$, so by \eqref{count:eq-reveal-decrement},
revealing a free vertex $v$ decreases the conditional variance in
expectation by $p_v(1-p_v)w_{G'}(v)^2$.
Conditional on the information revealed by time $t$, each unrevealed
vertex is revealed by time $t+h$ with probability $1-e^{-h}=h+O(h^2)$.
For fixed $G$, the probability of two or more revelations is $O(h^2)$.
Since the variance is bounded by $n^2$, these multiple revelations
contribute only $o(h)$ to its expected change. Summing the single-vertex
contributions and averaging over the information at time $t$ gives
\[
 \mathcal V(t+h)-\mathcal V(t)
 =-h\,\E\sum_{v\in V(G')}
       p_v(1-p_v)w_{G'}(v)^2+o(h).
\]
Dividing by $h$ and letting $h\downarrow0$, we obtain
\begin{equation}\label{count:eq-reveal-derivative}
 -\mathcal V'(t)=\E\sum_{v\in V(G')}
                       p_v(1-p_v)w_{G'}(v)^2.
\end{equation}
Conditional on $T_v>t$, a length-$\ell$ self-avoiding walk from $v$ in $G$
can remain in $G'$ only if its other $\ell$ vertices are unrevealed.
For $\ell\ge1$, this event has probability $e^{-\ell t}$ by independence
of their reveal times. Setting $S_\ell(G',v)=0$ when $v\notin V(G')$,
linearity of expectation gives
\[
 \E[S_\ell(G',v)\mid T_v>t]
 \le e^{-\ell t}S_\ell(G,v)
 \le e^{-\ell t}d(d-1)^{\ell-1}.
\]
To bound the expected square root appearing in \cref{count:lem-saw-response},
we use the concavity of the square-root function:
\[
 \E\!\left[\sqrt{\frac{S_\ell(G',v)}{d(d-1)^{\ell-1}}}
       \,\middle|\,T_v>t\right]
 \le\sqrt{\frac{\E[S_\ell(G',v)\mid T_v>t]}{d(d-1)^{\ell-1}}}
 \le e^{-\ell t/2}.
\]
The induced graph $G'$ has maximum degree at most $d$, so
\cref{count:lem-saw-response} applies to it. Since $v$ can be free only
if $T_v>t$, which has probability $e^{-t}$, taking expectations in
\eqref{count:eq-original-path-response} and using the preceding bound
with $\ell=2j+1$ yields, for $t\ge0$,
\begin{equation}\label{count:eq-random-path-survival}
 \E[\one_{\{v\text{ free}\}}w_{G'}(v)^2]
 \le Ce^{-t}\left[1+\sum_{0\le j\le(n-2)/2}(1+\delta_+)^{3j}e^{-(j+1/2)t}\right].
\end{equation}
Set $t_0=3\log(1+\delta_+)$. For $r>0$, bounding the sum in
\eqref{count:eq-random-path-survival} by a geometric series and using
\eqref{count:eq-reveal-derivative} with $p_v\le\lambda$ gives
\[
 -\mathcal V'(t_0+r)\le
 \frac{Cn\lambda e^{-(t_0+r)}}{1-e^{-r}}.
\]
To control this bound near $r=0$, apply \eqref{count:eq-common-field-response}
and the triangle inequality under $\mu_{G',\lambda}$.
Conditional occupation probabilities for $u\ne v$ are at most $\lambda$, so
\[
 |w_{G'}(v)|\le1+\sum_{u\in V(G')\setminus\{v\}}
 \left|\Pr(X_u=1\mid X_v=1)-\Pr(X_u=1\mid X_v=0)\right|
 \le1+n\lambda.
\]
Together with \eqref{count:eq-reveal-derivative}, $p_v\le\lambda$, and
$\Pr(v\in V(G'))\le e^{-t}$, this gives
\[
 -\mathcal V'(t)\le n\lambda e^{-t}(1+n\lambda)^2.
\]
Combining the two bounds, we obtain
\[
 -\mathcal V'(t_0+r)\le Cn\lambda e^{-(t_0+r)}
        \min\{(1+n\lambda)^2,(1-e^{-r})^{-1}\}.
\]
Integrating gives
\begin{equation}
 \mathcal V(t_0)
 \le Cn\lambda\int_0^1\min\{(1+n\lambda)^2,y^{-1}\}\dd y
 \le Cn\lambda\log(2+n\lambda).
 \label{count:eq-pinned-tail-variance}
\end{equation}
If $\lambda\le\lambdaC(d)$, then $\delta_+=0$ and $t_0=0$, proving
\eqref{count:eq-aggregate-critical}.

If $\lambda>\lambdaC(d)$, then $t_0>0$, so we still need to bound
$\mathcal V(0)$ in terms of $\mathcal V(t_0)$. SI ensures that the
variance cannot decrease too quickly, by the argument
of~\cite[Section~3.1]{ChenEldan2022}.
For each conditional law $\nu_t$, restrict to its free coordinates and let
$\Sigma=\Cov_{\nu_t}(X)$ and $W=\diag(p_v(1-p_v))$.
By \cref{thm:main-si} and \eqref{eq:si-covariance-equivalence},
$W^{-1/2}\Sigma W^{-1/2}\preceq C_{\mathrm{SI}}I$ with
$C_{\mathrm{SI}}=C(\sqrt{1+n/d}+(1+n/d)\delta_+)$ for an absolute constant $C$.
The normalized covariance matrix has eigenvalues in $[0,C_{\mathrm{SI}}]$, so
squaring it and undoing the normalization gives
$\Sigma W^{-1}\Sigma\preceq C_{\mathrm{SI}}\Sigma$.
By \eqref{count:eq-common-field-response},
$w=(w_{G'}(v))_v=W^{-1}\Sigma\one$, where $\one$ is the all-ones vector. Hence
\[
 \sum_vp_v(1-p_v)w_v^2
 =\one^{\mathsf T}\Sigma W^{-1}\Sigma\one
 \le C_{\mathrm{SI}}\one^{\mathsf T}\Sigma\one=C_{\mathrm{SI}}\Var_{\nu_t}(|X|).
\]
Taking expectations and using \eqref{count:eq-reveal-derivative} gives $-\mathcal V'(t)\le C_{\mathrm{SI}}\mathcal V(t)$,
and integration from $0$ to $t_0$ yields
\[
 \mathcal V(0)\le e^{C_{\mathrm{SI}}t_0}\mathcal V(t_0).
\]
Finally, $t_0\le3\delta_+$ gives
\[
 C_{\mathrm{SI}}t_0\le C\bigl(\sqrt{1+n/d}\,\delta_++(1+n/d)\delta_+^2\bigr)
       \le C\bigl(1+(1+n/d)\delta_+^2\bigr).
\]
Together with \eqref{count:eq-pinned-tail-variance}, this proves
\eqref{count:eq-aggregate-hot}.
\end{proof}

\paragraph{The two-step recursion and its forcing.}
We turn to the proof of \cref{count:lem-saw-response}.
Recall that $\delta_+=\max\{\lambda/\lambdaC(d)-1,0\}$ and $D=d-1$.

On the self-avoiding-walk tree (\cref{sec:prelim}), let $x_u$ be the
occupation probability of $u$ in its own subtree, and let
$R_u=x_u/(1-x_u)$ be its odds there. At a nonroot node $u$ with
$k\le D$ children, write $x_i$ for the occupation probability of its
$i$th child and $x_{ij}$ for that of this child's $j$th child. Thus
\[
 x_u=\frac{\lambda\prod_i(1-x_i)}{1+\lambda\prod_i(1-x_i)},
 \qquad
 x_i=\frac{\lambda\prod_j(1-x_{ij})}{1+\lambda\prod_j(1-x_{ij})}.
\]
Write $w_u=\mathrm d\log R_u/\mathrm d\log\lambda$ for the response
in the subtree rooted at $u$, with $w_i$ and $w_{ij}$ denoting the
child and grandchild responses. These subtree responses may differ
between copies of the same graph vertex. For the SAW tree rooted at a
graph vertex $v$, preservation of the root marginal gives $w_v=w_G(v)$.
Differentiating the recursion gives
\[
 w_u=1-\sum_i x_iw_i
     =1-\sum_i x_i+\sum_{i,j}x_ix_{ij}w_{ij}.
\]
Write $F_u=1-\sum_i x_i$ for the forcing added at $u$; $x_ix_{ij}$
is the coefficient transmitting a grandchild's response to $u$.
The next lemma gives two estimates: a bound on
$x_u\sum_i x_i=\sum_i(\sqrt{x_ux_i})^2$, a sum of the squared edge weights
used below, and a bound on the forcing in terms of the sum of the two-step coefficients.

\begin{lemma}[Squared weights and common-field forcing]
\label{count:lem-killed-forcing}
At every nonroot node $u$,
\begin{equation}\label{count:eq-squared-edge-budget}
 x_u\sum_i x_i\le\frac{1+\delta_+}{D}.
\end{equation}
Set
\[
 \kappa_u=1-(1+\delta_+)^{-1}\sum_{i,j}x_ix_{ij}.
\]
Then there is an absolute constant $C$ such that
\begin{equation}\label{count:eq-forcing-killing}
 \kappa_u\ge0,\qquad F_u^2\le C(\kappa_u+\delta_+^2).
\end{equation}
\end{lemma}

\begin{proof}
We use the following bound for every nonroot node $u$, proved at the
end of this proof:
\begin{equation}\label{count:eq-child-common-envelope}
 Dx_u\le(1+\delta_+) e^{F_u}.
\end{equation}
Since $F_u\le1$ and $(1-s)e^s\le1$ for $s\le1$, this gives
\[
 x_u\sum_i x_i
 \le\frac{1+\delta_+}{D}(1-F_u)e^{F_u}
 \le\frac{1+\delta_+}{D},
\]
proving \eqref{count:eq-squared-edge-budget}.
\rev{At $\delta_+=0$, this bound can also be obtained from
the $q=2$ maximization in~\cite[Section~4]{SinclairEtAl2017},
restricting the number of children to at most $D$.}

To prove \eqref{count:eq-forcing-killing}, put
$z=(1+\delta_+)^{-1}\sum_i x_i$.
If $z=0$, then $F_u=\kappa_u=1$, so the bound holds.
For $z>0$, applying \eqref{count:eq-child-common-envelope} at each
child gives $\log(Dx_i/(1+\delta_+))\le F_i$.
The $k$ numbers $Dx_i/(1+\delta_+)$, together with $D-k$ zeros, have
average $z$. Jensen's inequality for $s\log s$ (with $0\log0=0$),
followed by the preceding bound, gives
\[
 \begin{aligned}
 z\log z
 &\le\frac1{1+\delta_+}\sum_i x_i
          \log\frac{Dx_i}{1+\delta_+}\\
 &\le\frac1{1+\delta_+}\sum_i x_iF_i
 =z-1+\kappa_u,
 \end{aligned}
\]
where the equality uses $F_i=1-\sum_jx_{ij}$ and the definition of
$\kappa_u$. Since $\kappa_u\le1$, we have $0<z\le e$. Consequently,
\[
 \kappa_u\ge1-z+z\log z\ge\frac{(z-1)^2}{2e}.
\]
In particular, $\kappa_u\ge0$, and
\[
 F_u^2=(1-(1+\delta_+)z)^2
 \le2(1-z)^2+2\delta_+^2z^2
 \le4e\,\kappa_u+2e^2\delta_+^2,
\]
proving \eqref{count:eq-forcing-killing}.

It remains to prove \eqref{count:eq-child-common-envelope}.
Pad the child probabilities $x_1,\ldots,x_k$ with zeros to $D$ entries.
AM--GM gives
\[
 \prod_i(1-x_i)
 \le\left(1-\frac{\sum_i x_i}{D}\right)^D
 =\left(1-\frac{1-F_u}{D}\right)^D.
\]
Define
\[
 f(t)=\frac{D\lambdaC(1-t/D)^D}{1+\lambdaC(1-t/D)^D},
 \qquad 0\le t\le D.
\]
Since $\lambda\le(1+\delta_+)\lambdaC$ and $a/(1+a)$ is increasing
and grows at most linearly in $a$, the tree recursion and the preceding
product bound imply
\[
 Dx_u\le(1+\delta_+)
 \frac{D\lambdaC\prod_i(1-x_i)}{1+\lambdaC\prod_i(1-x_i)}
 \le(1+\delta_+)f(1-F_u).
\]
We have $f(1)=1$ and
\[
 \frac{\mathrm d}{\mathrm d t}\bigl(\log f(t)+t-1\bigr)
       =\frac{f(t)-t}{D-t}.
\]
Since $f$ decreases, this derivative is positive before $1$ and
negative after $1$, so $f(t)\le e^{1-t}$; endpoints follow by
continuity.
Evaluating at $t=1-F_u$ proves \eqref{count:eq-child-common-envelope}.
\end{proof}

\paragraph{A general bound for the response equation.}
The following finite-matrix statement isolates the common argument at
and above criticality. A matrix is substochastic if it is nonnegative
and its row sums are at most one.

\begin{lemma}[A Neumann-series bound]\label{count:lem-killed-resolvent}
Let $K$ be a nonnegative nilpotent matrix, let $\beta\ge1$, and suppose
$P=K/\beta$ is substochastic. If real vectors $w,F$ satisfy $w=F+Kw$
and, coordinatewise,
\[
 F^2\le a(\one-P\one)+b(\beta-1)^2\one
 \qquad(a,b\ge0),
\]
then
\begin{equation}\label{count:eq-killed-resolvent}
 w_v^2\le\bigl[a+b\beta(\beta-1)\bigr]
              \sum_{j\ge0}\beta^{2j}(K^j\one)_v.
\end{equation}
All sums are finite.
\end{lemma}

\begin{proof}
Nilpotence gives $w=\sum_{j\ge0}\beta^jP^jF$. For each $v$, write
\[
 w_v=\sum_{j,u}\beta^j(P^j)_{vu}F_u
     =\sum_{j,u}
       \sqrt{\beta^{3j}(P^j)_{vu}}\,
       \sqrt{\beta^{-j}(P^j)_{vu}}\,F_u.
\]
Cauchy--Schwarz then gives
\[
 w_v^2\le
 \left[\sum_j\beta^{3j}(P^j\one)_v\right]
 \left[\sum_j\beta^{-j}(P^jF^2)_v\right].
\]
Substituting the assumed bound on $F^2$ gives
\[
 \begin{aligned}
 \sum_j\beta^{-j}(P^jF^2)_v
 &\le a\sum_j\beta^{-j}\bigl(P^j(\one-P\one)\bigr)_v
      +b(\beta-1)^2\sum_j\beta^{-j}(P^j\one)_v\\
 &\le a+b\beta(\beta-1).
 \end{aligned}
\]
For the last inequality, the term involving $a$ uses $\beta^{-j}\le1$
and the telescoping identity
\[
 \sum_{j\ge0}P^j(\one-P\one)=\one.
\]
For the term involving $b$, use $P^j\one\le\one$ and
$\sum_{j\ge0}\beta^{-j}=\beta/(\beta-1)$ when $\beta>1$;
when $\beta=1$, this term is zero. Finally,
$\beta^{3j}P^j=\beta^{2j}K^j$, proving the bound.
\end{proof}

\paragraph{From the response equation to self-avoiding walks.}
\begin{proof}[Proof of \Cref{count:lem-saw-response}]
Fix $v\in V(G)$, and let $T$ be its SAW tree, rooted at $v$.
Then $w_v=w_G(v)$.
We first bound the subtree responses at nonroot nodes, which have at
most $D$ children, and then apply the root recursion.
Let $w=(w_u)_{u\ne v}$ and $F=(F_u)_{u\ne v}$, indexed by the nonroot
nodes of $T$. On the same index set, define $K_{uz}=x_yx_z$ for each
path $u\to y\to z$, and zero otherwise; each arrow denotes a step
from a parent to one of its children.
The two-step recursions
\[
 w_u=F_u+\sum_{u\to y\to z}x_yx_zw_z
\]
therefore give $w=F+Kw$. Since $K^j=0$ whenever $2j$ exceeds the height
of $T$, $K$ is nilpotent. For $P=K/(1+\delta_+)$, we have
$(\one-P\one)_u=\kappa_u$. Thus \cref{count:lem-killed-forcing}
verifies the hypotheses of \cref{count:lem-killed-resolvent} with
$\beta=1+\delta_+$ and absolute constants $a=b=C$.
The prefactor in \eqref{count:eq-killed-resolvent} is bounded by an
absolute constant since $\delta_+\le\delta_{\mathrm{up}}$.

Fix a nonroot node $u$. Each multiplication by $K$
extends a path by two edges, so summing over all paths of length $2j$
starting at $u$ gives
\[
 (K^j\one)_u
 =\sum_{u=u_0\to u_1\to\cdots\to u_{2j}}
       \prod_{i=1}^{2j}x_{u_i}.
\]
To bound this sum, assign each parent--child edge $(y,z)$
the weight $\sqrt{x_yx_z}$, as in~\cite{LiLuYin2013,SinclairEtAl2017}.
By \eqref{count:eq-squared-edge-budget}, the squared weights of the
edges leaving any nonroot node $y$ satisfy
\[
 \sum_{z:\,y\to z}\bigl(\sqrt{x_yx_z}\bigr)^2
 =x_y\sum_{z:\,y\to z}x_z\le\frac{1+\delta_+}{D}.
\]
Applying this bound successively at each level gives, for every $\ell\ge0$,
\[
 \sum_{u=u_0\to u_1\to\cdots\to u_\ell}
       \prod_{i=0}^{\ell-1}x_{u_i}x_{u_{i+1}}
 \le\left(\frac{1+\delta_+}{D}\right)^\ell.
\]
The product along a path in the original sum is related to these
edge weights by
\[
 \prod_{i=1}^{\ell}x_{u_i}
   =\sqrt{x_{u_\ell}/x_{u_0}}\,
    \prod_{i=0}^{\ell-1}\sqrt{x_{u_i}x_{u_{i+1}}}.
\]
Equation~\eqref{eq:hc-elementary-residual} gives
$\lambda/(1+\lambda)^{d+1}\le x_y\le\lambda/(1+\lambda)$, so
\[
 \frac{x_{u_\ell}}{x_{u_0}}\le(1+\lambda)^d\le e^{13},
\]
using $d\lambda\le13$.
Let $N_\ell$ count the nodes at depth $\ell$ in the subtree rooted at $u$
(so $N_0=1$). Cauchy--Schwarz therefore gives
\begin{equation}\label{count:eq-path-square-bound}
 \sum_{u=u_0\to u_1\to\cdots\to u_\ell}\ \prod_{i=1}^{\ell}x_{u_i}
 \le C(1+\delta_+)^{\ell/2}\sqrt{N_\ell/D^\ell}.
\end{equation}
We now return to the root $v$, whose children are indexed by $i$.
The one-step recursion $w_v=1-\sum_i x_iw_i$ and Cauchy--Schwarz give
\[
 \begin{aligned}
 w_v^2
 \le2+2\left(\sum_i x_iw_i\right)^2
 \le2+2\left(\sum_i x_i\right)\left(\sum_i x_iw_i^2\right)
 \le2+26\sum_i x_iw_i^2,
 \end{aligned}
\]
where $\sum_i x_i\le d\lambda\le13$, since there are at most $d$
children and each $x_i\le\lambda$.
For each $j\ge0$, let $N_{2j}(i)$ count the depth-$2j$ descendants
of child $i$. Equations~\eqref{count:eq-killed-resolvent} and
\eqref{count:eq-path-square-bound}, the latter with $\ell=2j$, give
\[
 \sum_i x_iw_i^2
 \le C\sum_{j\ge0}(1+\delta_+)^{3j}
           \sum_i x_i\sqrt{N_{2j}(i)/D^{2j}}.
\]
Because $x_i\le\lambda$ and there are at most $d$ children,
Cauchy--Schwarz gives
\[
 \sum_i x_i\sqrt{N_{2j}(i)/D^{2j}}
 \le\lambda\sqrt{\frac{d\sum_iN_{2j}(i)}{D^{2j}}}
 \le d\lambda\sqrt{\frac{S_{2j+1}(G,v)}{dD^{2j}}}.
\]
Here every root-to-node path has distinct original labels and maps
injectively to a self-avoiding walk in $G$. Since $d\lambda\le13$,
the root bound and $w_v=w_G(v)$ prove \eqref{count:eq-original-path-response}.
\end{proof}

\section{Phase gadgets and computational reductions}
\label{sec:lower-main}\label{sec:reduction-main}

We reduce additive-gap cubic MAX-CUT (\cref{prob:cubic-gap}) to hard-core
sampling and counting. Write $H$ for the source graph, $m=|V(H)|$, and
$\tau,\nu$ for its threshold and fixed gap coefficient. The reduction
constructs graphs on $n$ vertices at fugacity $\lambdaC(d)(1+\delta)$,
with
\[
 \frac{n\delta^2}{d}=\Theta(m).
\]
We account explicitly for the number of oracle calls and the work
between calls. Composing this construction with the SAT reduction in
\cref{thm:sat-cubic-source} proves \cref{thm:robust-sat-transfer}.

\Cref{sec:lower-construction,sec:lower-setup} describe the graph
construction and state the guarantees used by the sampling and counting
reductions in \cref{sec:lower-reductions}. We choose the parameters in
\cref{sec:lower-budget} and complete the reduction from SAT in
\cref{sec:robust-sat-proof}. Finally, \cref{sec:glauber-lower-proof} establishes the phase
barrier for local Markov chains, followed by the one-gadget Glauber
and spectral-independence lower bounds. The implementation guarantees are proved
in \cref{sec:square-generator}, and the moment estimates are established
in \cref{sec:gadget-moments,sec:shared-moments}.

\subsection{Graph construction}
\label{sec:lower-construction}

Each vertex $v$ of $H$ is replaced by a bipartite gadget $G_v$ on $2N$
vertices. We connect these graphs by replacing internal edges with edges
between them, using $k$ pairs of matching edges for each edge of $H$.
We first describe the construction for given integers
$N,k$; \cref{sec:lower-budget} will choose them as functions of the
fugacity.
The computational reductions and the Glauber lower bound use mode $1$.
We also analyze mode $0$ to establish the moment estimates before
conditioning on simplicity.

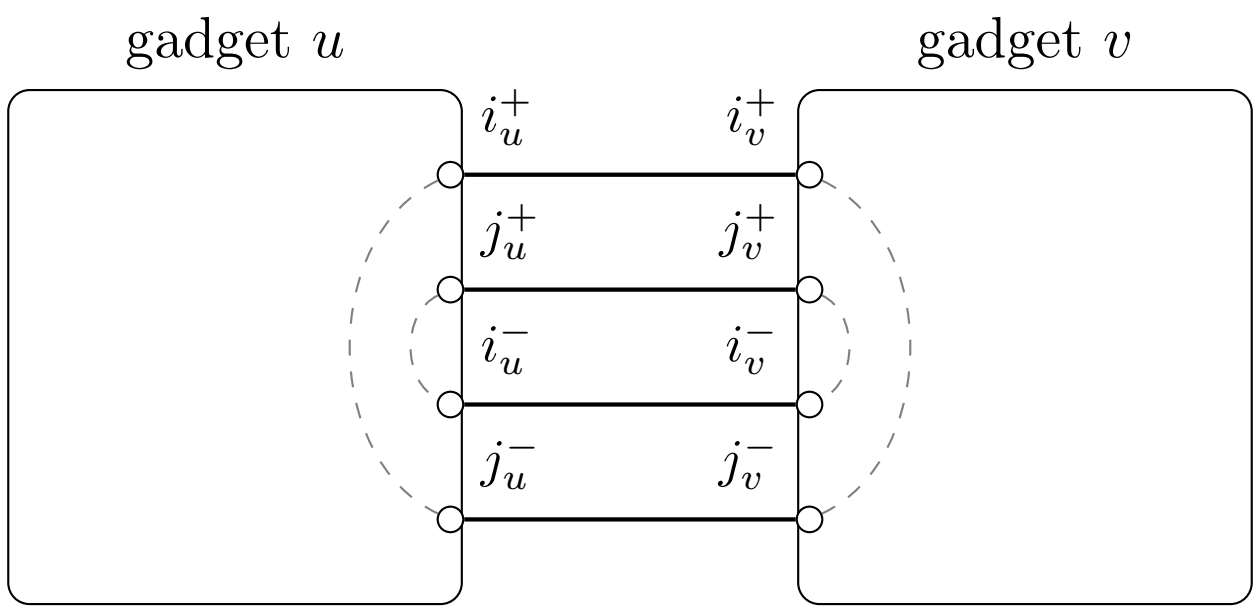
\begin{figure}[t]
\centering
\begin{tikzpicture}[
  scale=0.95,
  gadget/.style={draw,rounded corners,minimum width=30mm,minimum height=34mm,align=center},
  port/.style={circle,draw,fill=white,inner sep=1.7pt},
  ext/.style={thick},
  deleted/.style={dashed,gray}
]
\node[gadget] (u) at (0,0) {};
\node[gadget] (v) at (5.5,0) {};
\node[above] at (u.north) {gadget $u$};
\node[above] at (v.north) {gadget $v$};
\foreach \name/\height/\endpoint/\sign in
  {pa/1.2/i/+,pb/0.4/j/+,ma/-0.4/i/-,mb/-1.2/j/-} {
  \node[port,label={[font=\small]above right:$\endpoint_u^{\sign}$}]
    (u\name) at (1.5,\height) {};
  \node[port,label={[font=\small]above left:$\endpoint_v^{\sign}$}]
    (v\name) at (4,\height) {};
  \draw[ext] (u\name)--(v\name);
}
\draw[deleted] (upa) to[bend right=65] (umb);
\draw[deleted] (upb) to[bend right=65] (uma);
\draw[deleted] (vpa) to[bend left=65] (vmb);
\draw[deleted] (vpb) to[bend left=65] (vma);
\end{tikzpicture}
\caption{One of the $k$ connection blocks for $uv\in E(H)$.
The dashed edges are the original lifts of the two designated matching edges;
the solid edges are the connecting edges that replace them.}
\label{fig:port-wiring}
\end{figure}

\begin{construction}\label{constr:graph-reduction}
Let $H$ be a simple cubic graph on $m$ vertices. Fix $d\ge3$, an even
integer $N$, and a positive integer $k$ with $3k\le N/2$.
In mode $1$, also require $N>d$.
\begin{enumerate}[label=(\arabic*),leftmargin=*]
\item \textbf{Graphs before connections.}
For each $v\in V(H)$, take $d$ perfect matchings
$M_{v,1},\ldots,M_{v,d}$ on $[N]$. Let $B_v$ be their
union, with edge multiplicities and matching indices. In
\emph{mode $1$}, $B_v$ must be simple; \emph{mode $0$} allows parallel edges.
Let $G_v$ be the bipartite double cover of $B_v$: its vertices are
$i_v^+,i_v^-$ for $i\in[N]$, and each copy of an edge $ij$ in $B_v$
gives edges $i_v^+j_v^-$ and $i_v^-j_v^+$. Thus $G_v$ has $g=2N$
vertices. The notation $G_v$ always refers to this graph before connections.
For each $v$, order the edges of $M_{v,1}$ uniformly at random and
orient each edge with either direction equally likely. Conditional
on the matchings, all these orders and orientations are independent.

\item \textbf{Connecting the graphs.}
Start with the disjoint union
\[
 G_{\sqcup}:=\bigsqcup_{v\in V(H)}G_v,\qquad
 Z_{G_{\sqcup}}(\lambda)=\prod_{v\in V(H)}Z_{G_v}(\lambda).
\]
This graph uses the gadgets from the same matching realization, before
any edge replacements.
At each $u\in V(H)$, split the first $3k$ edges of $M_{u,1}$, the
\emph{designated matching edges}, into three consecutive groups of $k$, assigned to the three edges of $H$
incident to $u$ in a fixed order.

For each $uv\in E(H)$, pair the $\ell$-th edges of the groups assigned
to $uv$ in gadgets $u,v$, for $\ell=1,\ldots,k$, using the previously
specified matching order within each group. Each pair of oriented
matching edges $(i_u,j_u),(i_v,j_v)$ determines one \emph{connection block},
so each $uv$ has $k$ blocks\added{; together they form the parallel connection
associated with $uv$}. A block has the four original lifts before
connection; to connect it, replace those lifts by the four connecting edges
shown in \cref{fig:port-wiring}:
\[
 i_u^+i_v^+,\qquad j_u^+j_v^+,\qquad
 i_u^-i_v^-,\qquad j_u^-j_v^-.
\]

\item \textbf{Output.}
In mode $0$, collapse all parallel edges after the replacements above.
In either mode, denote the resulting output graph by $G$.
\end{enumerate}
\end{construction}

In both modes, $G$ has $n=mg=2mN$ vertices. It is simple and
$d$-regular in mode $1$, and simple with maximum degree at most $d$
in mode $0$. Indeed, the designated matching edges at each source vertex lie in one matching, so
rewiring preserves degrees and introduces no loops or parallel edges.
Collapsing existing parallel edges in mode $0$ can only decrease degrees.
In mode $1$, $G_{\sqcup}$ is also simple and $d$-regular on the same
$n$ vertices.

For mode $b\in\{0,1\}$, define a base distribution $\mathbb P_b$
on the random choices in the construction.
Under $\mathbb P_0$, all $M_{v,j}$ are independent uniform perfect
matchings; $\mathbb P_1$ conditions on every $B_v$ being simple.
Both distributions include the order and orientation choices in
\cref{constr:graph-reduction}, and keep different gadgets independent.
Write $\E_b$ for
expectation under $\mathbb P_b$. These distributions are used for the
moment calculations; the reduction actually samples from a reweighted distribution
defined in \cref{sec:lower-setup}.

\subsection{The guarantees needed by the reduction}\label{sec:lower-setup}

Fix a positive rational distance $\delta$ above the uniqueness threshold,
and write
\[
 \lambda=(1+\delta)\lambdaC(d).
\]
In the remainder of this section, Sections~\ref{sec:gadget-moments}--\ref{sec:shared-moments},
and \cref{sec:square-generator}, we suppress this fugacity argument in
partition functions and restricted weights.
For an independent set $I$ of $G$, define the occupation imbalance of
gadget $v$ by
\begin{equation}\label{eq:physical-rho}
 \rho_v=\frac{|\{i\in[N]:i_v^+\in I\}|-
                   |\{i\in[N]:i_v^-\in I\}|}{N}.
\end{equation}
Its \emph{phase} is the sign of $\rho_v$, with zero assigned the positive sign.
Write $\sigma(I)\in\{\pm\}^{V(H)}$ for the gadget phases, which naturally
induce a cut of $H$, as in Sly's reduction~\cite{Sly2010}.
For $\sigma\in\{\pm\}^{V(H)}$, let $Z_\sigma$ be the total weight of
independent sets with phase vector $\sigma$:
\begin{equation}\label{eq:phase-weight-definition}
 Z_\sigma
   :=\sum_{I\in\Ind(G):\,\sigma(I)=\sigma}\lambda^{|I|}.
\end{equation}
Thus $Z_G=\sum_\sigma Z_\sigma$, and a hard-core sample on $G$
has phase vector $\sigma$ with
probability $Z_\sigma/Z_G$.

We compare these weights with $Z_{G_{\sqcup}}>0$, the partition
function of the graph before connections defined in
\cref{constr:graph-reduction}.

For each $v$, the reduction reweights the distribution of
$M_{v,1},\ldots,M_{v,d}$ under $\mathbb P_1$ by the factor
$Z_{G_v}^2/\E_1[Z_{G_v}^2]$, independently across vertices.
The edge orders and orientations remain independent and uniform.
Let $\mathbb Q_1$ denote the resulting distribution of the entire
construction\added{, called the reference-square law}, and let $\omega$ record its matchings, orders, and
orientations. Thus
\begin{equation}\label{eq:reference-square-law}
 \mathbb Q_1(\omega)
 =\frac{Z_{G_{\sqcup}(\omega)}^2\mathbb P_1(\omega)}
        {\E_1 Z_{G_{\sqcup}}^2}.
\end{equation}

The following bounds compare the normalized phase weights with the
cut weights of $H$. The first bounds their expectations; the second gives
a lower bound on the probability that a normalized phase weight exceeds
the stated threshold.
The bounds are proved in
Sections~\ref{sec:gadget-moments}--\ref{sec:shared-moments}.

\begin{theorem}[Bounds for phase weights]
\label{thm:cut-weight-realization}
There is an absolute integer $C_{\rm amp}\ge1$ with the following property.
Fix $d\ge3$, an accuracy $0<\varepsilon\le1$, and positive
upper bounds on $N\delta^2$ and $k\delta$, which may depend on $d$.
There is a constant $C_d>0$ and a rational cutoff
$0<\delta_0(d)\le\min\{\delta_{\mathrm{up}},1/2\}$ such that the
following holds.

For $0<\delta\le\delta_0(d)$, take any even $N$ and positive integer $k$
satisfying the fixed upper bounds, $3k\le N/2$, and
\begin{equation}\label{eq:moment-size-conditions}
 J:=\frac{48(d-1)}{d^3}\,k\delta\ge2,
 \qquad N\delta^2\ge C_{\rm amp}\frac{d^2}{d-2}
                              \frac{J^2}{\varepsilon}.
\end{equation}
Here $J$ is the logarithmic reward for one extra cut edge.
Use mode $1$ of \cref{constr:graph-reduction}.
For every simple cubic $H$ on $m$ vertices and every
$\sigma\in\{\pm\}^{V(H)}$,
\begin{gather}
 \E_{\mathbb Q_1}(Z_\sigma/Z_{G_{\sqcup}})
 \le 2^{-m}e^{-J(|E(H)|-\Cut_H(\sigma))+\varepsilon m},
 \label{eq:main-full-sector-upper}\\
 \mathbb Q_1\left(Z_\sigma/Z_{G_{\sqcup}}\ge
   2^{-m-1}e^{-J(|E(H)|-\Cut_H(\sigma))-\varepsilon m}\right)
 \ge\tfrac14 e^{-C_dm/\sqrt N}.
 \label{eq:main-favorable-probability}
\end{gather}
The constants depend only on $d$, $\varepsilon$,
and the fixed upper bounds.
\end{theorem}

The following proposition gives a bounded-time algorithm for
generating instances from $\mathbb Q_1$.
Its proof is given in \cref{sec:square-generator}.
Write $\bits(t)$ for the binary encoding length of a rational $t$.

\begin{proposition}
\label{prop:reduction-implementation}
Fix $d$ and rational choices of the accuracy and the upper bounds on
$N\delta^2$ and $k\delta$ in \cref{thm:cut-weight-realization}.
For these fixed parameters, there are valid positive rational
values of $C_d,\delta_0(d)$ in that theorem and an integer
threshold $N_{\rm gen}$ for the algorithm below.

Let $H$ be a simple cubic graph on $m$ vertices, and let rational
$\delta$ and integers $N,k$ satisfy that theorem's conditions, with
$N\ge N_{\rm gen}$. Use mode $1$ and fix a number of trials
$\ell\ge1$ and a rational error tolerance $0<\zeta<1$.
A randomized algorithm returns the construction data for $\ell$
mode-$1$ trials, including the matchings,
orders, orientations, and graphs $G_v,G_{\sqcup},G$, or reports an abort.
Its output distribution is within total variation distance $\zeta$
of $\mathbb Q_1^{\otimes\ell}$, the law of $\ell$ independent trials.
An abort is treated as an additional output, assigned probability
zero by the ideal law.
The algorithm has worst-case bit complexity at most
\[
 \ell\,\poly_d\bigl(m,N,\bits(\delta),\bits(\zeta),\log\ell\bigr).
\]
The polynomial depends only on the fixed degree, accuracy, and upper
bounds.
\end{proposition}

\subsection{Sampling and counting reductions}
\label{sec:lower-reductions}

We use the guarantees of \cref{sec:lower-setup} to reduce additive-gap
cubic MAX-CUT to hard-core sampling and counting. Given $(H,\tau)$, with
$m=|V(H)|$, we must distinguish $\MaxCut(H)\ge\tau$ from
$\MaxCut(H)\le\tau-\nu m$.
Use mode $1$ of the construction with parameters satisfying
\cref{thm:cut-weight-realization,prop:reduction-implementation}.
Each algorithm generates $\ell$ trials, with $\ell$ chosen below.

\paragraph{The sampling reduction.}
On each generated graph $G$, call the sampling oracle and decode its
output into a cut using \eqref{eq:physical-rho}. Return YES if any
decoded cut has size at least $\tau-\nu m/2$, and NO otherwise.

\paragraph{The counting reduction.}
In each trial, query both $G$ and $G_{\sqcup}$ from that realization,
at the common fugacity $\lambda=(1+\delta)\lambdaC(d)$.
For each graph, make $O(\log(2\ell))$ independent counting-oracle calls
and take the median of their finite-precision logarithmic estimates,
using the normalization specified in the implementation paragraph below.
Denote the medians by $\widehat L_G$ and $\widehat L_{G_{\sqcup}}$,
and use enough calls that each fails with probability at most $1/(300\ell)$.

With $J,\varepsilon$ as defined in
\cref{thm:cut-weight-realization}, return YES if any trial satisfies
\[
 \widehat L_G-\widehat L_{G_{\sqcup}}>\theta
 :=-J(|E(H)|-\tau+\nu m/4)-(\log2+\varepsilon)m-\tfrac12\log2,
\]
and NO otherwise.

Both algorithms use \cref{prop:reduction-implementation} to generate
the trials; an abort returns NO. The following lemma specifies the
repetition count and accuracy requirements and proves that these
procedures solve the source problem.

\begin{lemma}
\label{lem:square-source-transfer}
Fix a constant $A_*\ge1$. For the construction parameters above,
put $x=n\delta^2/d$ and suppose
\[
 \vartheta:=J\nu/2-\log2-2\varepsilon>0,
 \qquad m\le x\le A_*m.
\]
Choose a rational $\bar a_N$, computable in $\poly_d(N)$ time, with
$C_dN^{-1/2}\le\bar a_N\le2C_dN^{-1/2}$, and assume
$\bar a_N\le\min\{1,\vartheta/16\}$. Set
$\ell=2^{\lceil10+2\bar a_Nm\rceil}$.
For all sufficiently large $m$, each procedure above solves
\cref{prob:cubic-gap} with success probability at least $2/3$,
provided its oracle satisfies the following guarantee in every
possible trial:
\begin{enumerate}[label=(\roman*),leftmargin=*]
\item the sampling oracle on $G$ is within TV distance $1/10$ of the
hard-core distribution;
\item on each of $G$ and $G_{\sqcup}$, the counting oracle approximates
the partition function within a multiplicative factor $e^{c_{\rm app}x}$
with probability at least $2/3$, where $c_{\rm app}>0$ is a fixed rational
satisfying
\begin{equation}\label{eq:reduction-accuracy-margin}
 4c_{\rm app}A_*<J\nu/2.
\end{equation}
\end{enumerate}
If an oracle call takes worst-case time $T$, the sampling algorithm,
given $\delta$, uses at most
\[
 \ell\bigl[T+\poly_d\bigl(n,\bits(\delta)\bigr)\bigr]
\]
bit operations. The counting algorithm uses at most
$\ell\,\poly_d(n,\bits(\delta))(T+1)$ bit operations and
$O(\ell\log(2\ell))$ oracle calls. The polynomials depend only on the
fixed construction and accuracy constants. The sampling algorithm has
one-sided error: a NO source instance is always rejected.
The threshold for ``sufficiently large $m$'' can be chosen independently
of $N,k,\delta$, provided $\vartheta$ and, for counting, the difference
between the right- and left-hand sides of
\eqref{eq:reduction-accuracy-margin} are bounded below by fixed positive
constants.
\end{lemma}
\begin{proof}
We first assume that the $\ell$ trials are independent samples from
$\mathbb Q_1$. The generator's TV error will be added to the failure
probability at the end.
Put
\[
 \theta_{\rm YES}=-J(|E(H)|-\tau)-(\log2+\varepsilon)m-\log2.
\]
If $(H,\tau)$ is a YES instance, fix any cut $\sigma$ with
$\Cut_H(\sigma)\ge\tau$ for the analysis.
The probability bound \eqref{eq:main-favorable-probability} gives
\[
 \mathbb Q_1\left(Z_\sigma/Z_{G_{\sqcup}}\ge e^{\theta_{\rm YES}}\right)
 \ge\tfrac14 e^{-C_dm/\sqrt N}.
\]

\paragraph{Correctness of the sampling reduction.}
If $(H,\tau)$ is a NO instance, the algorithm always returns NO.
For a YES instance, let $Z_{\rm bad}$ be the total weight of configurations whose decoded
cuts have size less than $\tau-\nu m/2$. Summing
\eqref{eq:main-full-sector-upper} over these at most $2^m$ phases gives
\[
 \E_{\mathbb Q_1}(Z_{\rm bad}/Z_{G_{\sqcup}})
 \le e^{-J(|E(H)|-\tau+\nu m/2)+\varepsilon m}
 =2e^{\theta_{\rm YES}-\vartheta m}.
\]
By Markov's inequality,
$Z_{\rm bad}/Z_{G_{\sqcup}}\le2e^{\theta_{\rm YES}-\vartheta m/2}$ except with probability
$e^{-\vartheta m/2}$. Since $C_dN^{-1/2}\le\vartheta/16$, both this bound and
$Z_\sigma/Z_{G_{\sqcup}}\ge e^{\theta_{\rm YES}}$ hold with probability at least
\[
 \tfrac14e^{-C_dm/\sqrt N}-e^{-\vartheta m/2}
 \ge\tfrac18e^{-C_dm/\sqrt N}
\]
for sufficiently large $m$. On any such graph, a hard-core sample
decodes to a bad cut with probability
\[
 \frac{Z_{\rm bad}}{Z_G}
 \le\frac{Z_{\rm bad}/Z_{G_{\sqcup}}}{Z_\sigma/Z_{G_{\sqcup}}}
 \le2e^{-\vartheta m/2}.
\]
Since the oracle's output distribution is within TV distance $1/10$
of the hard-core distribution, its output passes the cut test with
probability at least $1-2e^{-\vartheta m/2}-1/10\ge4/5$.
Thus a trial succeeds with probability at least $e^{-C_dm/\sqrt N}/10$,
and our choice of $\ell$ independent trials makes the failure
probability less than $1/100$.

\paragraph{Correctness of the counting reduction.}
For a YES instance, the common probability bound above and
$Z_G\ge Z_\sigma$ imply that, after $\ell$ independent
trials, some trial satisfies
\[
 \log(Z_G/Z_{G_{\sqcup}})\ge\theta_{\rm YES}
\]
with probability at least $99/100$.

Put
\[
 \theta_{\rm NO}=-J(|E(H)|-\tau+\nu m/2)-(\log2+\varepsilon)m.
\]
If $(H,\tau)$ is a NO instance, every cut has size at most $\tau-\nu m$. Summing
\eqref{eq:main-full-sector-upper} over all $2^m$ phase vectors gives
\[
 \E_{\mathbb Q_1}(Z_G/Z_{G_{\sqcup}})
 \le e^{-J(|E(H)|-\tau+\nu m)+\varepsilon m}
 =e^{\theta_{\rm NO}-\vartheta m}.
\]
Markov's inequality and a union bound give
\[
 \Pr\bigl[\log(Z_G/Z_{G_{\sqcup}})>\theta_{\rm NO}
          \text{ in some trial}\bigr]
 \le\ell e^{-\vartheta m}<1/100
\]
for sufficiently large $m$, by the choice of $\ell$ and
$\bar a_N\le\vartheta/16$.

We now check that estimation errors preserve the threshold decision.
The algorithm uses $\theta=(\theta_{\rm YES}+\theta_{\rm NO})/2$.
As verified in the implementation paragraph below, normalization
increases the error of a successful logarithmic oracle estimate by at
most $1/100$. Each median therefore has logarithmic error at most
$c_{\rm app}x+1/100$ whenever a majority of its oracle calls succeed.
When all estimates satisfy their error guarantees,
\eqref{eq:reduction-accuracy-margin} and
$\theta_{\rm YES}-\theta_{\rm NO}=J\nu m/2-\log2$ give, for sufficiently large $m$,
\[
 \left|\widehat L_G-\widehat L_{G_{\sqcup}}
       -\log\frac{Z_G}{Z_{G_{\sqcup}}}\right|
 \le 2c_{\rm app}x+\frac1{50}
 \le 2c_{\rm app}A_*m+\frac1{50}
 <\frac{\theta_{\rm YES}-\theta_{\rm NO}}2.
\]
The counting reduction therefore fails with probability at most
$1/100$ plus the probability that any estimate violates its error guarantee.

\paragraph{Implementation and running time.}
Generate the $\ell$ trials using \cref{prop:reduction-implementation},
with output distribution within TV distance $1/300$ of
$\mathbb Q_1^{\otimes\ell}$.
For counting, a union bound over the $2\ell$ medians gives
probability at most $2/300$ that any estimate fails.
The TV bound and a union bound,
together with the preceding analysis, give total failure probability
at most $1/100+3/300<1/3$ for either reduction.

Using the rational constants supplied by
\cref{prop:reduction-implementation}, one can take
\[
 \bar a_N=\frac{C_d}{\lfloor \sqrt N\rfloor}.
\]
For counting, choose a rational
$U\in[n\log(1+\lambda),n\log(1+\lambda)+1]$ in polynomial time.
Replace the logarithm $y$ of each positive rational oracle answer by
$\min\{U,\max\{0,y\}\}$, and approximate the result within $1/100$;
map invalid or nonpositive answers to zero.
Both $\log Z_G$ and $\log Z_{G_{\sqcup}}$ lie in $[0,U]$, so
restricting estimates to this interval cannot increase their errors.
Using the bit lengths and sufficiently many leading bits, this
normalization takes time linear in the answer length plus
$\poly_d(n,\bits(\delta))$. The normalized estimates have polynomially
bounded encodings, so taking their median and performing the final
comparison require only polynomial additional work. The strict margin
in \eqref{eq:reduction-accuracy-margin} absorbs the combined $1/50$ error
and fixed absolute comparison precision for sufficiently large $m$.

There is one oracle call per sampling trial and
$O(\log(2\ell))$ calls on each of the two graphs per counting trial, with each answer having length
at most $T$. Together with the costs in
\cref{prop:reduction-implementation}, this proves the stated time bounds,
since $\log\ell=O(m)$ and $m,N\le n$.
The estimates above also give the claimed uniform threshold for $m$.
\end{proof}

\subsection{Choosing the parameters}
\label{sec:lower-budget}

We first calibrate the construction for any fixed positive rational
source gap $\nu$. All constants are independent of $m$ and $\delta$.
The Glauber lower bound uses the same rounding rule with its own fixed reward.

\begin{proposition}[Quantitative reduction from cubic MAX-CUT]
\label{prop:quantitative-cubic-transfer}
Fix a positive rational $\nu$. There are constants $A_*\ge1$
and $c_{\rm app}>0$, depending only on $\nu$, such that,
for every fixed $d\ge3$, there are $L_d\in4\mathbb N$ and a positive
rational $\delta_0(d)$, with $d\le L_d\le A_*d$, for which the following
holds. Given an instance $(H,\tau)$ of Problem~\ref{prob:cubic-gap} with gap
coefficient $\nu$ and $m$ vertices, and a rational distance
$0<\delta\le\delta_0(d)$, the reduction produces simple $d$-regular
oracle graphs of size
\[
 n=4m\left\lceil\frac{L_d}{4\delta^2}\right\rceil,
 \qquad m\le\frac{n\delta^2}{d}\le A_*m.
\]
All oracle queries use fugacity $\lambda=(1+\delta)\lambdaC(d)$.
For sampling, assume that on every possible query graph the oracle's
output distribution is within TV distance $1/10$ of the hard-core distribution.
For counting, assume that on every possible query graph $G$ the oracle
returns, with probability at least $2/3$, a positive estimate
$\widehat Z_G$ satisfying
\[
 e^{-c_{\rm app}n\delta^2/d}Z_G(\lambda)
 \le\widehat Z_G\le
 e^{c_{\rm app}n\delta^2/d}Z_G(\lambda).
\]
Under either oracle assumption, the reduction solves the source instance
with success probability at least $2/3$ using
$\exp\{O_d(1+m\delta)\}$ trials, with the running-time guarantees of
\cref{lem:square-source-transfer}.
The sampling reduction always rejects NO instances.
The constants may depend on $\nu$, which is fixed throughout each application.
\end{proposition}

\begin{proof}
We use mode $1$ of \cref{constr:graph-reduction}, so every oracle query
is a simple $d$-regular graph.

\paragraph{1. Error tolerance and cut reward.}
Choose a rational phase-weight accuracy $0<\varepsilon<1/100$
in \cref{thm:cut-weight-realization}. To cover the $2^m$ phase assignments and
the errors in the weight bounds and estimates, choose a rational target
reward $J_*>2$, independent of $d$, with
$J_*\nu/2>\log2+20\varepsilon$.

\paragraph{2. Connections and gadget sizes.}
The actual reward in \cref{thm:cut-weight-realization} is
$J=48(d-1)k\delta/d^3$, so a fixed target reward calls for
$k=\Theta(d^2/\delta)$. Its other size condition,
\eqref{eq:moment-size-conditions}, is satisfied by taking
$N=\Theta(d/\delta^2)$ with a sufficiently large coefficient.
Using the absolute constant $C_{\rm amp}$ from that theorem, set
$C_*=2C_{\rm amp}(J_*+4)^2/\varepsilon$.
Choose the integer coefficient $L_d$ and construction parameters $k,N$ by
\begin{equation}\label{eq:degree-explicit-lower-parameters}
\begin{split}
 L_d=4\left\lceil\frac{C_*d^2}{4(d-2)}\right\rceil,\quad
 k=\left\lceil\frac{\left\lceil J_*d^3/[48(d-1)]\right\rceil}{\delta}\right\rceil,\quad
 N=2\left\lceil\frac{L_d}{4\delta^2}\right\rceil.
\end{split}
\end{equation}
These choices make $N$ even.
For $0<\delta\le1/10$, the bound $48(d-1)/d^3\le32/9$ gives
\[
 J_*\le J<J_*+4,
 \qquad
 N\delta^2
 \ge \frac{C_*}{2}\frac{d^2}{d-2}
 \ge C_{\rm amp}\frac{d^2}{d-2}\frac{J^2}{\varepsilon}.
\]
The same rounding rules bound $k\delta$ and $N\delta^2$ above by
fixed rational constants depending on
$d$, as required by \cref{thm:cut-weight-realization}.
To bound the output size, set $A_*=3C_*+2$. Since $1\le d/(d-2)\le3$,
\[
 d\le L_d,\qquad (L_d+1)/d\le A_*,\qquad
 L_d\le2N\delta^2<L_d+4\delta^2.
\]
Thus $A_*$ is independent of $d$. Choose the counting-accuracy coefficient
\[
 c_{\rm app}=\varepsilon/A_*.
\]
This coefficient is independent of $d$.
The counting accuracy condition \eqref{eq:reduction-accuracy-margin}
then holds, since $4c_{\rm app}A_*=4\varepsilon<J\nu/2$.
Fixed-relative-error counting satisfies this guarantee for all
sufficiently large $m$.

\paragraph{3. Final distance cutoff.}
Apply \cref{thm:cut-weight-realization,prop:reduction-implementation}
with the accuracy and parameter bounds chosen above.
The gap parameter in \cref{lem:square-source-transfer} satisfies
\[
 \vartheta=J\nu/2-\log2-2\varepsilon>16\varepsilon.
\]
Shrink the cutoff to $\delta_0(d)\le\min\{\delta_{\mathrm{up}},1/10\}$
so that, for every $0<\delta\le\delta_0(d)$, the chosen $N,k$ satisfy
the construction condition $3k\le N/2$, all size thresholds in these two
results, and $\bar a_N\le\varepsilon$.
Such a rational cutoff exists: $k=O_d(\delta^{-1})$,
$N=\Theta_d(\delta^{-2})$, and
$\bar a_N\le2C_d/\sqrt N\to0$.
Thus $\bar a_N\le\min\{1,\vartheta/16\}$, as required by the reduction lemma.

\paragraph{4. Sizes and repetition.}
The construction gives $n=2mN=4m\lceil L_d/(4\delta^2)\rceil$.
The rounding bounds in step~2 imply
\[
 m\le\frac{n\delta^2}{d}\le A_*m,
 \qquad
 \log n=O_d\bigl(\log m+\log(1/\delta)\bigr).
\]
The integer parameters and rational fugacity are computable in
$\poly_d(n,\bits(\delta))$ bit operations.
For the repetition count, we have
\[
 \bar a_N\le2C_d/\sqrt N=O_d(\delta),
 \qquad
 \ell=2^{\lceil10+2\bar a_Nm\rceil}
      =\exp\{O_d(1+m\delta)\}.
\]
Lemma~\ref{lem:square-source-transfer} now gives the claimed success
probability and running-time bounds, and the sampling reduction always
rejects NO instances. Its threshold on $m$ is
uniform in $\delta$; the finitely many smaller source graphs can be
solved directly. This proves the proposition.
\end{proof}

\subsection{Consequences for SAT algorithms}
\label{sec:robust-sat-proof}

We prove \cref{thm:robust-sat-transfer}, completing
the general reduction used to derive \cref{thm:main-transfer} in
\cref{sec:overview-hardness}.
Two padding lemmas then support the proofs of
\cref{cor:near-window-transfer,cor:main-np-hardness,rem:coarse-counting}.
We conclude with extensions to further running-time bounds and distance sequences.

\begin{proof}[Proof of \Cref{thm:robust-sat-transfer}]
The deterministic reduction in \cref{thm:sat-cubic-source} transforms
an input SAT formula of encoding length $s\ge2$ into a cubic MAX-CUT
instance $(H,\tau)$ with a fixed additive gap coefficient and
$m=\Theta(s\lceil\log_2s\rceil^{C_{\rm src}})$ vertices in $H$.
Apply \cref{prop:quantitative-cubic-transfer} to this instance at the
supplied rational distance $\delta$, with this gap coefficient. All oracle graphs
have the same order
\[
 n_*=4m\left\lceil\frac{L_d}{4\delta^2}\right\rceil,
\]
which is even and independent of the random realization.
We may take $\delta_0(d)\le1/10$. Since $L_d\le A_*d$ and $4\delta^2\le d$,
\[
 n_*\le\frac{L_dm}{\delta^2}+4m\le\frac{(A_*+1)dm}{\delta^2},
\]
so $K=A_*+1$ works; it is absolute because the gap coefficient of
\cref{thm:sat-cubic-source} is fixed.
Moreover, \cref{prop:quantitative-cubic-transfer} gives
$n_*\delta^2/d=\Theta(m)$, verifying the size guarantees in
\cref{thm:robust-sat-transfer}.

The number of trials is
\[
 \ell=\exp\{O_d(1+m\delta)\}.
\]
There is one sampling call per trial and
$O(\ell\log(2\ell))$ counting calls in total. Since
$O(\ell\log(2\ell))\le\exp\{O(1+\log\ell)\}$,
both call counts are bounded by $Q$ for a sufficiently large
$C_{\rm rep}(d)$.
The deterministic reduction from SAT to cubic MAX-CUT takes
polynomial time in $s$, hence in $n_*$, since $s=O(m)$ and $m\le n_*$.
The sampling and counting time bounds now follow from their respective
guarantees in \cref{lem:square-source-transfer}, including the cost
of reading oracle answers, with the source reduction absorbed into
$P_d$ and the repetition and call counts absorbed into $Q$.

Unsatisfiable formulas produce NO instances of cubic MAX-CUT,
and hence the sampling reduction always rejects them.
Counting gives a two-sided-error SAT decision algorithm with success
probability at least $2/3$.
\end{proof}

For an algorithm available only at a given distance $\delta_n$
on $n$-vertex graphs, the query size must match the chosen distance.
We first choose a target size $n$ and construct the queries at distance
$\delta_n$. If the graphs produced by the reduction are sufficiently smaller than
$n$, the following lemma pads them to $n$ vertices by adding disjoint
components. This operation is used in the proof of \cref{thm:main-transfer}
and in the consequences proved below.

\begin{lemma}[Regular padding]\label{lem:regular-padding}
Let $H$ be a simple $d$-regular graph on $n_0$ vertices. If $n-n_0$ is
even and at least $2d(d+1)$, one can add disjoint $d$-regular components
in polynomial time to obtain a simple $n$-vertex $d$-regular graph $G$.
For every positive rational $\lambda$, the factor
$Z_G(\lambda)/Z_H(\lambda)$ is explicitly computable in time polynomial
in $n$ and $\bits(\lambda)$. Sampling from $G$ gives sampling from $H$
without increasing TV error, and relative-error counting on $G$ gives
relative-error counting on $H$ with polynomial overhead.
\end{lemma}
\begin{proof}
Since $\gcd(d+1,2d)=\gcd(d+1,2)$ divides $n-n_0$, choose
$0\le u<2d$ such that $u(d+1)\equiv n-n_0\pmod{2d}$, and set
$v=(n-n_0-u(d+1))/(2d)$. The size assumption ensures $v\ge0$.
Add $u$ disjoint copies of $K_{d+1}$ and $v$ disjoint copies of
$K_{d,d}$. The resulting graph is simple and $d$-regular, and
\[
 \frac{Z_G(\lambda)}{Z_H(\lambda)}
 =\bigl(1+(d+1)\lambda\bigr)^u
  \bigl(2(1+\lambda)^d-1\bigr)^v.
\]
This rational factor has polynomial bit length and is computable in
polynomial time. Projecting a sample onto $H$ preserves its target
marginal and cannot increase TV error. Dividing a counting estimate by
the displayed factor preserves relative error, with polynomial overhead
in the graph size, the fugacity encoding length, and the answer length.
\end{proof}

For coarse counting, the allowed logarithmic error grows with the query
size. Replicating the core before padding keeps the error per copy on
the scale of the core.

\begin{lemma}[Replication before padding]\label{lem:replication-padding}
Fix $d\ge3$, a positive rational $\delta$, and a constant $c>0$.
Let $G_0$ be a simple $d$-regular graph of positive even order $n_0$,
and let $n$ be even with $n\ge2(n_0+2d(d+1))$.
One can construct a simple $n$-vertex $d$-regular graph $G$ such that,
at the common fugacity $\lambda=(1+\delta)\lambdaC(d)$,
a logarithmic counting estimate for $G$ with error at most
$cn\delta^2/d$ gives one for $G_0$ with error at most
\[
 2c n_0\delta^2/d+1/100.
\]
The conversion preserves the success event and takes time linear in
the oracle answer length plus $\poly_d(n,\bits(\delta))$;
the construction has the same polynomial overhead.
\end{lemma}

\begin{proof}
Put $r_0=2d(d+1)$ and $t=\lfloor(n-r_0)/n_0\rfloor$.
The number of vertices in the padding graph is $r=n-tn_0$.
It is even and satisfies
$r_0\le r<r_0+n_0$, while $tn_0\ge n/2$.
Take $t$ disjoint copies of the same graph $G_0$ and apply
\cref{lem:regular-padding} to pad their union to $n$ vertices.
Writing $F$ for the padding graph, we have $G=G_0^{\sqcup t}\sqcup F$,
where $Z_F(\lambda)$ is explicitly computable, and
\[
 Z_G(\lambda)=Z_{G_0}(\lambda)^tZ_F(\lambda).
\]
As in the proof of \cref{lem:square-source-transfer}, choose a rational
$U\in[n\log(1+\lambda),n\log(1+\lambda)+1]$
and replace the oracle's logarithmic estimate $y$ by
$\min\{U,\max\{0,y\}\}$. Since $\log Z_G\in[0,U]$, this
cannot increase its error. Subtract $\log Z_F$ and divide by $t$
to obtain $\widehat L_0$, allowing total additional numerical error
at most $1/100$. On every successful oracle call,
\[
 |\widehat L_0-\log Z_{G_0}|
 \le\frac{cn\delta^2}{dt}+\frac1{100}
 \le 2c\frac{n_0\delta^2}{d}+\frac1{100}.
\]
Reading the oracle answer takes time linear in its length; its bit
lengths and sufficiently many leading bits suffice for the required
precision, as in \cref{lem:square-source-transfer}.
The remaining operations, including computing $Z_F(\lambda)$, take
$\poly_d(n,\bits(\delta))$ time.
When applying this construction to a random core, generate $G_0$ once
and then copy it $t$ times. Replication and padding do not change
the distribution of $G_0$.
\end{proof}

For the remaining proofs in this subsection, apply
\cref{thm:sat-cubic-source} to a SAT input of encoding length $s$,
obtaining a cubic MAX-CUT instance with $m$ vertices and a fixed gap
coefficient. For each rational distance $\delta$ in the range of
\cref{prop:quantitative-cubic-transfer}, its core graphs have a common
even order $n_0$, depending on $\delta$. Fix a positive integer $B_0$,
independent of $d$, such that
\begin{equation}\label{eq:sat-consequences-setup}
 m=\Theta(s\lceil\log_2s\rceil^{C_{\rm src}}),
 \qquad n_0\le B_0dm/\delta^2+1,
 \qquad n_0\delta^2/d=\Theta(m).
\end{equation}
For counting, the core queries include both $G$ and $G_{\sqcup}$;
the size and padding arguments below apply to each graph separately.

\paragraph{Consequences stated in the introduction.}

\begin{proof}[Proof of \Cref{cor:near-window-transfer}]
With $m$ and $B_0$ as in \eqref{eq:sat-consequences-setup},
scan $n=2,4,8,\ldots$ and take the first size satisfying
$n\delta_n^2\ge4B_0dm$.
The bound $n\delta_n^2=\Theta((\log n)^r)$ ensures that the
scan terminates and that the values of $n\delta_n^2$ at consecutive
tested sizes differ by at most a fixed factor for sufficiently large $n$.
Thus the failed test
at $n/2$ and the successful test at $n$ give
$n\delta_n^2/d=\Theta(m)$. It follows that
\[
 \log n=\Theta(m^{1/r}),\qquad
 n\delta_n^2/d=\Theta(m),\qquad
 m\delta_n=o(1).
\]
Every unpadded query has even order at most $n/4+1$.
By \cref{lem:regular-padding}, pad each query to exactly $n$ vertices,
preserving its fugacity and accuracy.
The repetition count is bounded, and the scan, exact distance
evaluations, construction, and assumed oracle all have cost
polynomial in $n$. The total running time is therefore
\[
 \exp\{O(m^{1/r})\}
 =\exp\{\widetilde O(s^{1/r})\}.
\]
Finitely many smaller inputs are solved directly.
\end{proof}

We first prove the generalization stated in the introduction,
allowing $c n^{-a}\le\delta_n\le C n^{-a'}$, and then deduce
\cref{cor:main-np-hardness}.

\begin{proposition}[Polynomial-time transfer on distance schedules]
\label{prop:polynomial-schedule-transfer}
Fix $d\ge3$ and constants $0<a'\le a<1/2$.
Let $(\delta_n)$ be positive rationals, exactly evaluable in time
polynomial in $n$, such that, for fixed $c,C>0$ and all
sufficiently large $n$,
\[
 c n^{-a}\le\delta_n\le C n^{-a'}.
\]
If TV-$1/10$ sampling, or counting with relative error $1/10$
and success probability at least $2/3$, is possible in randomized
polynomial time on every sufficiently large simple $n$-vertex
$d$-regular graph at fugacity $(1+\delta_n)\lambdaC(d)$, then
$\mathsf{NP}=\mathsf{RP}$.
For counting, relative error $1/10$ may be replaced by the accuracy
guarantee in \cref{rem:coarse-counting}.
\end{proposition}

\begin{proof}
Use the cubic MAX-CUT instance of size $m$ specified in
\eqref{eq:sat-consequences-setup}.
Choose a fixed integer
\[
 R>\max\{1/a',\,1/(1-2a)\}
\]
and an even target order $n=2\lceil m^R\rceil$.
Evaluate $\delta_n$ and apply
\cref{prop:quantitative-cubic-transfer} at that distance.
Every unpadded query $G_0$ has the same even order
\[
 n_0=O_d(mn^{2a})=o(n),
 \qquad
 m\delta_n=O(mn^{-a'})=o(1).
\]
The first estimate uses $\delta_n\ge c n^{-a}$ and
$R>1/(1-2a)$; the second uses $\delta_n\le C n^{-a'}$ and $R>1/a'$.
Consequently, the number of trials is bounded, and the
construction and exact distance evaluation have polynomial cost.
The distance cutoff and all size requirements hold for
sufficiently large $m$.

Apply \cref{lem:regular-padding} to obtain $n$-vertex queries
for sampling and fixed-relative-error counting.

For coarse counting, choose $c_{\mathrm{coarse}}=c_{\rm app}/4$,
with $c_{\rm app}$ for the fixed MAX-CUT gap coefficient.
Since $n_0=o(n)$ and both orders are even,
\cref{lem:replication-padding} applies for sufficiently large $m$.
For each core query, a successful call gives a logarithmic estimate with error
\[
 \frac{c_{\rm app}}2\frac{n_0\delta_n^2}{d}+\frac1{100}
 \le c_{\rm app}\frac{n_0\delta_n^2}{d},
\]
since $n_0\delta_n^2/d=\Theta(m)\to\infty$.
This meets the accuracy requirement of
\cref{prop:quantitative-cubic-transfer}, with polynomial conversion overhead.

Thus either oracle gives a randomized polynomial-time SAT
algorithm. Sampling already has one-sided error.
For counting, amplify the SAT decision procedure and use it
to find an assignment by successively fixing variables,
accepting only after verifying the resulting assignment.
With inverse-polynomial error per decision call, this succeeds
with probability at least $2/3$ on satisfiable inputs and
never accepts an unsatisfiable input.
Finitely many smaller inputs are solved directly.
Hence SAT belongs to $\mathsf{RP}$.
\end{proof}

\begin{proof}[Proof of \Cref{cor:main-np-hardness}]
Apply \cref{prop:polynomial-schedule-transfer} with
$a'=a=1/2-\xi\in(0,1/2)$.
\end{proof}

\begin{proof}[Proof of \Cref{rem:coarse-counting}]
Set $c_{\mathrm{coarse}}=c_{\rm app}/4$, as in the proof of
\cref{prop:polynomial-schedule-transfer}; this constant is absolute
because the MAX-CUT gap coefficient is fixed.
For \cref{thm:main-transfer,cor:near-window-transfer}, use the same
core graphs and even target sizes as in their proofs.
Every core has even order $n_0\le n/4+1$ and
$n_0\delta_n^2/d=\Theta(m)$, so
$n\ge2(n_0+2d(d+1))$ for sufficiently large $m$.
Lemma~\ref{lem:replication-padding} therefore gives the counting accuracy
required by \cref{thm:robust-sat-transfer}: its additive $1/100$ error
is absorbed since $n_0\delta_n^2/d\to\infty$.
The conversion and counting amplification preserve the stated running-time bounds.
For \cref{cor:main-np-hardness}, apply
\cref{prop:polynomial-schedule-transfer} with $a'=a=1/2-\xi$.
Finitely many smaller inputs are solved directly.
\end{proof}

\paragraph{Further consequences.}

We first extend \cref{cor:near-window-transfer} to distances
closer to the polynomial window.
Fix $d\ge3$ and $b>0$, and let $(\delta_n)$ be positive rationals
exactly evaluable in time polynomial in $n$.
Assume the sampling or counting hypothesis of \cref{cor:near-window-transfer},
with its distance condition replaced by
\[
 \delta_n=\Theta\!\left(
 \sqrt{\frac{\log n\,(\log\log n)^b}{n}}\right).
\]
For $m$ as in \eqref{eq:sat-consequences-setup}, the search over $n=2,4,8,\ldots$
in the proof of \cref{cor:near-window-transfer} gives
\[
 \log n=\Theta\!\left(\frac{m}{(\log m)^b}\right),
 \qquad m\delta_n=o(1).
\]
Consequently, bounded repetition and polynomial work in $n$
give a SAT algorithm with running time
\[
 \exp\!\left\{O\!\left(\frac{m}{(\log m)^b}\right)\right\}
 =
 \exp\!\left\{O\!\left(s(\log s)^{C_{\rm src}-b}\right)\right\}.
\]
In particular, $b>C_{\rm src}$ gives subexponential SAT time.

Returning to power-law distances, the next proposition uses larger
query graphs to extend \cref{thm:main-transfer} to every $0<a<1/2$,
allowing a loss in the power saving.

\begin{proposition}[Power savings with larger query graphs]
\label{prop:padded-power-transfer}
Fix $d\ge3$, $0<a<1/2$, and $0<\gamma<1$.
Let $(\delta_n)$ be positive rationals, exactly evaluable in time
polynomial in $n$, with $\delta_n=\Theta(n^{-a})$.
Suppose that TV-$1/10$ sampling, or relative-error-$1/10$ counting
with success probability at least $2/3$, has a randomized algorithm
with worst-case running time
\[
 \poly(n)\exp\left\{
 O\left((n\delta_n^2/d)^{1-\gamma}\right)\right\}
\]
on all sufficiently large simple $n$-vertex $d$-regular graphs
at fugacity $(1+\delta_n)\lambdaC(d)$.
Put
\[
 \gamma_*=
 \min\left\{\gamma,\,
 \frac{a}{a+(1-2a)(1-\gamma)}\right\}>0.
\]
Then SAT on inputs of binary length $s$ has a randomized algorithm
with success probability at least $2/3$ and running time
\[
 \exp\{\widetilde O(s^{1-\gamma_*})\}.
\]
For $a\ge\gamma/(1+2\gamma)$, we have $\gamma_*=\gamma$.
For counting, relative error $1/10$ may be replaced by the accuracy
guarantee in \cref{rem:coarse-counting}.
\end{proposition}

\begin{proof}
Use the cubic MAX-CUT instance and core bounds in
\eqref{eq:sat-consequences-setup}. Put $b=(1-2a)(1-\gamma)>0$.
At distance $\delta_n$, the unpadded query order is
$n_0=O_d(mn^{2a})$, while the repetition exponent is
$O_d(1+mn^{-a})$.
For a target order $n=\Theta(m^R)$ with
$R\ge1/(1-2a)$, a sufficiently large constant factor ensures
that the core order is at most $n/4$.
Use \cref{lem:regular-padding} to reach exactly $n$ vertices.
For coarse counting, $n_0\le n/4$ implies
$n\ge2(n_0+2d(d+1))$ for sufficiently large $m$.
Apply \cref{lem:replication-padding} with
$c_{\mathrm{coarse}}=c_{\rm app}/4$; since
$n_0\delta_n^2/d=\Theta(m)$, the resulting estimate meets the counting
accuracy required by \cref{prop:quantitative-cubic-transfer}.
The oracle, repetitions, and polynomial overhead then have
total cost
\[
 \poly(m)\exp\{O(m^{bR}+m^{1-aR})\}.
\]
Larger query graphs increase the oracle cost but decrease the
repetition cost. Balancing the exponents $bR$ and $1-aR$,
subject to $R\ge1/(1-2a)$, gives
\[
 R_*=\max\left\{\frac1{1-2a},\frac1{a+b}\right\}.
\]
A direct calculation gives
\[
 \max\{bR_*,1-aR_*\}=1-\gamma_*.
\]

To implement this choice for arbitrary fixed real parameters,
fix an integer $J>R_*$ and enumerate $n=2,4,8,\ldots$ up to $m^J$.
Retain those satisfying the size and distance cutoffs, with core order
at most $n/4$ and $n\ge2(n_0+2d(d+1))$.
There are $L=O(\log m)$ candidates. Evaluating the distances and checking
these conditions for all candidates takes polynomial time in $m$.
Amplify the complete SAT procedure for each candidate to error at most
$1/(6L)$, run them with equal time shares, and return the first answer.
A union bound gives error at most $1/6$, even if finishing times depend
on the answers.
For sufficiently large $m$, a candidate has
$n=\Theta(m^{R_*})$ with a large enough constant factor, and hence
worst-case time $\poly(m)\exp\{O(m^{1-\gamma_*})\}$.
Amplification and time sharing add only polynomial overhead.
Substituting the expression for $m$ proves the claim.
Finitely many smaller inputs are solved directly.
\end{proof}

The next proposition transfers running-time improvements beyond fixed
powers by choosing the query size so that the repetition count is bounded.
For fixed-power bounds, \cref{prop:padded-power-transfer} instead
optimizes the combined oracle and repetition costs.

\begin{proposition}[Transfer of general running-time bounds]
\label{prop:general-profile-transfer}
Fix $d\ge3$ and $0<a<1/2$. Let $(\delta_n)$ be positive
rationals, exactly evaluable in time polynomial in $n$, with
$\delta_n=\Theta(n^{-a})$.
Let $F:[0,\infty)\to[0,\infty)$ be nondecreasing, and put
\[
 \kappa=\max\left\{1,\frac{1-2a}{a}\right\}.
\]
Suppose that TV-$1/10$ sampling, or relative-error-$1/10$ counting
with success probability at least $2/3$, has a randomized algorithm
with worst-case running time
\[
 \poly(n)\exp\left\{O\left(F(n\delta_n^2/d)\right)\right\}
\]
on all sufficiently large simple $n$-vertex $d$-regular graphs
at fugacity $(1+\delta_n)\lambdaC(d)$.
Then there is a constant $B>0$ such that SAT on inputs of binary
length $s$ has a randomized algorithm with success probability
at least $2/3$ and running time
\[
 \poly(s)\exp\left\{
 O\left(F\left(
 B\bigl(s\lceil\log_2s\rceil^{C_{\rm src}}\bigr)^\kappa
 \right)\right)\right\}.
\]
For counting, relative error $1/10$ may be replaced by the accuracy
guarantee in \cref{rem:coarse-counting}.
\end{proposition}

\begin{proof}
With $m$ and $B_0$ as in \eqref{eq:sat-consequences-setup},
scan $n=2,4,8,\ldots$ and take the first size satisfying
\[
 n\delta_n^2\ge4B_0dm,\qquad m\delta_n\le1.
\]
The two-sided bounds $\delta_n=\Theta(n^{-a})$ give
\[
 n=\Theta(m^R),\qquad
 R=\max\left\{\frac1{1-2a},\frac1a\right\},\qquad
 \frac{n\delta_n^2}{d}=\Theta(m^\kappa).
\]
Indeed, each stopping inequality gives the corresponding
lower bound on $n$, and both hold at every sufficiently
large constant multiple of $m^R$.

Every core has even order $n_0\le n/4+1$, so for sufficiently
large $m$ both \cref{lem:regular-padding,lem:replication-padding} apply.
The former handles sampling and fixed-relative-error counting.
For coarse counting, use the latter with
$c_{\mathrm{coarse}}=c_{\rm app}/4$; since $n_0\delta_n^2/d=\Theta(m)$,
its additive $1/100$ error is absorbed, giving the counting accuracy
required by \cref{prop:quantitative-cubic-transfer}.
The repetition count is bounded because $m\delta_n\le1$.
The scan, exact distance evaluations, construction, and padding
have polynomial cost in $m$.
By monotonicity of $F$, the oracle cost is at most
\[
 \poly(m)\exp\{O(F(B_1m^\kappa))\}
\]
for a fixed $B_1>0$.
Substituting the expression for $m$ and enlarging the constant inside
$F$ proves the claim.
Finitely many smaller inputs are solved directly.
\end{proof}

In particular, for every fixed $k\ge1$, a hard-core algorithm
running in $\exp\{O((\log n)^k)\}$ time yields a SAT algorithm
running in $\exp\{O((\log s)^k)\}$ time.
Likewise, a bound
$\poly(n)\exp\{(n\delta_n^2/d)^{o(1)}\}$ yields
$\exp\{s^{o(1)}\}$ SAT time.

For a logarithmic improvement in the exponent, fix $1/3\le a<1/2$
and $q>0$, and take a nondecreasing $F$ that equals
$x/(\log x)^q$ for all sufficiently large $x$.
Here $\kappa=1$, so the resulting SAT running time is
\[
 \exp\{O(s(\log s)^{C_{\rm src}-q})\}.
\]
Thus $q>C_{\rm src}$ gives subexponential SAT time.

The fixed power-law assumption can also be relaxed to nonincreasing
distance sequences satisfying the following upper and lower bounds.

\begin{proposition}[Power savings on monotone schedules]
\label{prop:monotone-schedule-transfer}
Fix $d\ge3$ and $0<\gamma<1$. Let $(\delta_n)$ be nonincreasing
positive rationals, exactly evaluable in time polynomial in $n$.
Suppose that, for fixed $c,C>0$ and all sufficiently large $n$,
\[
 \frac{c(\log n)^{1/[2(1-\gamma)]}}{\sqrt n}
 \le\delta_n\le Cn^{-\gamma/(1+2\gamma)}.
\]
Suppose that TV-$1/10$ sampling, or relative-error-$1/10$ counting
with success probability at least $2/3$, has a randomized algorithm
with worst-case running time
\[
 \poly(n)\exp\!\left(
 O\!\left(\left(\frac{n\delta_n^2}{d}\right)^{1-\gamma}\right)\right)
\]
on all sufficiently large simple $n$-vertex $d$-regular graphs
at fugacity $(1+\delta_n)\lambdaC(d)$. Then SAT on inputs of binary length $s$ is solvable with success
probability at least $2/3$ in time
$\exp\{\widetilde O(s^{1-\gamma})\}$.
\end{proposition}

\begin{proof}
With $m$ and $B_0$ as in \eqref{eq:sat-consequences-setup},
scan $n=2,4,8,\ldots$ and take the first size satisfying
$n\delta_n^2\ge4B_0dm$.
The lower bound on $\delta_n$ implies $n\delta_n^2\to\infty$, so the scan
terminates. For sufficiently large $m$, the preceding size $n/2$
fails the test, and monotonicity of $(\delta_n)$ gives
\[
 4B_0dm\le n\delta_n^2
 \le2(n/2)\delta_{n/2}^2<8B_0dm.
\]
Thus $n\delta_n^2/d=\Theta(m)$.
Since $n\to\infty$ with $m$ and $\delta_n\to0$, all distance
and size cutoffs hold for sufficiently large $m$.

Apply \cref{prop:quantitative-cubic-transfer} at distance $\delta_n$.
Every core has even order $n_0\le n/4+1$, so $n-n_0$ is even
and at least $2d(d+1)$ for sufficiently large $m$.
Apply \cref{lem:regular-padding} to reach exactly $n$ vertices,
preserving the fugacity and oracle accuracy.

The upper bound on $\delta_n$ gives
\[
 m^\gamma\delta_n
 =O\!\left(n^\gamma\delta_n^{1+2\gamma}\right)=O(1),
\]
so the repetition cost is $\exp\{O(m^{1-\gamma})\}$.
The lower bound on $\delta_n$ gives
\[
 (\log n)^{1/(1-\gamma)}=O(n\delta_n^2)=O(m),
 \qquad \log n=O(m^{1-\gamma}).
\]
There are $O(\log n)$ schedule evaluations, each taking time
polynomial in $n$; their output bit lengths are bounded by the same
running-time guarantee. The scan, construction, padding, and other
polynomial overhead therefore have cost
$\exp\{O(m^{1-\gamma})\}$.
The assumed oracle bound and the running-time guarantees of
\cref{prop:quantitative-cubic-transfer} give the same bound for the
complete SAT procedure, including repetitions.
Substituting the expression for $m$ proves the claimed SAT running time
and success probability. Finitely many smaller inputs are solved directly.
\end{proof}

\subsection{Lower bounds for local Markov chains}
\label{sec:glauber-lower-proof}

We first use the phase-weight estimates and a bipartite expander
to prove the local-chain obstruction in \cref{rem:local-chain-obstruction}.
A separate one-gadget analysis gives the Glauber and
spectral-independence lower bounds in
\cref{thm:main-glauber-lower,thm:main-si}.
We begin with elementary bounds for Glauber dynamics and padding.

\begin{lemma}[Elementary bounds for Glauber dynamics]
\label{lem:glauber-linear-lower}
Let $G$ be a graph on $n\ge2$ vertices and let $\lambda>0$.
The Glauber dynamics for $\mu_{G,\lambda}$ satisfies
\[
 t_{\rm rel}\ge n,
 \qquad
 t_{\rm mix}(1/4)\ge(t_{\rm rel}-1)\log2\ge(n-1)\log2.
\]
If $G$ is the disjoint union of a graph $G_0$ on $n_0\ge1$ vertices
and another graph, then, at the same fugacity,
\[
 t_{\rm rel}(G)\ge\frac n{n_0}t_{\rm rel}(G_0).
\]
\end{lemma}
\begin{proof}
For a vertex $v$, its occupation indicator $X_v$ has positive variance.
Only an update at $v$ changes this indicator, so its Dirichlet form is
\[
 \mathcal E(X_v,X_v)
 =\frac1n\E\bigl[\Var(X_v\mid X_{-v})\bigr]
 \le\frac1n\Var(X_v).
\]
The variational characterization of the spectral gap gives
$t_{\rm rel}\ge n$, and the standard spectral lower bound
\cite[Theorem~12.5]{LevinPeres2017} gives
$t_{\rm mix}(1/4)\ge(t_{\rm rel}-1)\log2$.
Finally, the Gibbs law on a disjoint union is a product. For a function
of $G_0$ alone, its variance is unchanged and its Dirichlet form is
multiplied by $n_0/n$. The variational characterization then gives
the last assertion.
\end{proof}

\begin{theorem}[A phase barrier for local Markov chains]
\label{thm:glauber-obstruction}
There are absolute constants $c,A>0$ such that, for every fixed $d\ge3$,
there are $L_d\in4\mathbb N$ and
$0<\delta_{\mathrm{mix}}(d)\le\min\{1/10,\delta_{\mathrm{up}}\}$, with
$d\le L_d$ and $(L_d+1)/d\le A$, for which the following holds.
Let $\mathcal M=\{48\cdot2^j:j\ge0\}$.
For every $m\in\mathcal M$ and every rational
$0<\delta\le\delta_{\mathrm{mix}}(d)$, there is a simple
$d$-regular bipartite graph $G$ on
\[
 n=4m\left\lceil\frac{L_d}{4\delta^2}\right\rceil
\]
vertices with the following property.
At fugacity $\lambda=(1+\delta)\lambdaC(d)$, every irreducible,
aperiodic transition matrix $P$ on $\Ind(G)$ with stationary law
$\mu_{G,\lambda}$ and
\[
 P(I,I')>0
 \quad\Longrightarrow\quad
 |I\triangle I'|\le\lfloor m/40\rfloor
\]
satisfies
\[
 t_{\mathrm{mix}}^P(1/4)\ge\exp\{cn\delta^2/d\}.
\]
The Glauber dynamics on the same graph also satisfies
$t_{\mathrm{rel}}\ge\exp\{cn\delta^2/d\}$.
\end{theorem}

\begin{proof}[Proof of \Cref{thm:glauber-obstruction}]
Fix an infinite family of simple cubic bipartite graphs $H$ with an
absolute edge-expansion constant $h>0$:
\[
 |\partial_H S|\ge h\min\{|S|,m-|S|\}
 \qquad(S\subseteq V(H)),\qquad m=|V(H)|.
\]
Here $\partial_H S$ is the set of edges of $H$ with exactly
one endpoint in $S$.
Start from $K_{3,3}$ and apply the bipartite Ramanujan $2$-lift
construction of Marcus, Spielman, and Srivastava~\cite{MSS2015}.
The lifts remain simple and cubic, and their nontrivial adjacency
eigenvalues have absolute value at most $2\sqrt2$.
The Cheeger inequality~\cite[Theorem~13.10]{LevinPeres2017}
gives $h=(3-2\sqrt2)/2$; discarding the first three sizes leaves
$\mathcal M=\{48\cdot2^j:j\ge0\}$. Choose a bipartition of $H$ in which every edge joins the two
classes, and let $\sigma^*\in\{+,-\}^{V(H)}$ assign $+$ to one class
and $-$ to the other. Each class has size $m/2$, and
$\Cut_H(\sigma^*)=|E(H)|$.
Use mode~$1$ of Construction~\ref{constr:graph-reduction}, with the phase-weight
accuracy in \cref{thm:cut-weight-realization} fixed at $1$.
Mode~$1$ makes $G$ simple and $d$-regular. It is also bipartite:
place a vertex with side label $s\in\{+,-\}$ in gadget $v$
in one class if $s=\sigma_v^*$, and in the other otherwise.
Internal edges join opposite side labels, and external
same-label edges join opposite source classes.

We choose the interaction strength so that the cut loss for changing
a fixed fraction of the gadget phases dominates both the number of
phase assignments and the errors in the phase-weight estimates.
Choose an absolute integer $J_*>2$ so that
$J_*h/5-\log2-3\ge3$. With the absolute constant $C_{\rm amp}$ in
\cref{thm:cut-weight-realization}, set $C_*=2C_{\rm amp}(J_*+4)^2$ and
\[
 L_d=4\left\lceil\frac{C_*d^2}{4(d-2)}\right\rceil,\qquad
 k=\left\lceil
       \frac{\lceil J_*d^3/[48(d-1)]\rceil}{\delta}
     \right\rceil,\qquad
 N=2\left\lceil\frac{L_d}{4\delta^2}\right\rceil.
\]
For $0<\delta\le1/10$, direct rounding gives
\[
 J_*\le J=\frac{48(d-1)}{d^3}k\delta<J_*+4,
 \qquad
 N\delta^2\ge C_{\rm amp}\frac{d^2}{d-2}J^2.
\]
The quantities $N\delta^2$ and $k\delta$ have fixed upper bounds
depending only on $d$. Since $N=\Theta_d(\delta^{-2})$ and
$k=\Theta_d(\delta^{-1})$, choose $0<\delta_{\mathrm{mix}}(d)\le1/10$ below the phase-weight cutoff so that
\cref{thm:cut-weight-realization} applies, $N>d$, $3k\le N/2$, and
$a:=C_d/\sqrt N\le1$. This choice is independent of $m$.
Moreover, with the absolute constant $A=3C_*+2$,
\begin{equation}\label{eq:glauber-size-comparison}
 d\le L_d,\qquad (L_d+1)/d\le A,\qquad
 L_dm\le n\delta^2<(L_d+1)m,
 \qquad m\le n\delta^2/d<Am.
\end{equation}
Here $n=2mN$ is the number of output vertices.

For an independent set $I$ of $G$, put
\[
 q(I)=|\{v:\sigma_v(I)\ne\sigma_v^*\}|,\quad
 r=m/4-1,\quad
 \mathcal A=\{I:q(I)\le r\},\quad
 \mathcal B=\{I:\lceil m/5\rceil\le q(I)\le r\}.
\]
The source sizes make $r$ an integer.
A transition changing at most $\lfloor m/40\rfloor$ occupation
indicators changes at most that many gadget phases.
If it exits $\mathcal A$, its starting configuration therefore satisfies
\[
 q(I)\ge r+1-\lfloor m/40\rfloor
       =m/4-\lfloor m/40\rfloor
       \ge\lceil m/5\rceil.
\]
Thus every such exit starts in $\mathcal B$.
Write $Z_{\mathcal B}$ for the hard-core weight of this band.

For an independent set $I$, let $\widetilde I$ be obtained by exchanging
the $+$ and $-$ labels in every gadget. The paired internal and external
edges ensure that $\widetilde I$ is independent, and $|\widetilde I|=|I|$.
In each of the $m/2$ gadgets with $\sigma_v^*=-$, at least one of
$I,\widetilde I$ has phase $+$, including when the imbalance is zero.
Hence $q(I)+q(\widetilde I)\ge m/2$. Since $2r<m/2$,
$I\in\mathcal A$ implies $\widetilde I\notin\mathcal A$.
Swapping twice recovers $I$, so this gives a weight-preserving injection
from $\mathcal A$ into its complement.
The configurations with $\sigma(I)=\sigma^*$ belong to $\mathcal A$
and have total weight $Z_{\sigma^*}$ by \eqref{eq:phase-weight-definition}
with $\sigma=\sigma^*$. Thus, for $\mu=\mu_{G,\lambda}$,
\[
 0<Z_{\sigma^*}/Z_G\le\mu(\mathcal A)\le1/2.
\]

For a phase vector $\sigma$ represented in $\mathcal B$, put
$S=\{v:\sigma_v\ne\sigma_v^*\}$. Then
$\lceil m/5\rceil\le|S|\le r<m/2$, so
\[
 |E(H)|-\Cut_H(\sigma)
 =|\partial_H S|\ge h|S|\ge hm/5.
\]
Sum \eqref{eq:main-full-sector-upper} over at most $2^m$
phases in $\mathcal B$ to obtain

\[
 \E_{\mathbb Q_1}(Z_{\mathcal B}/Z_{G_{\sqcup}})
 \le e^{-(Jh/5-1)m}.
\]
By \eqref{eq:main-favorable-probability}, the event
\[
 Z_{\sigma^*}/Z_{G_{\sqcup}}\ge 2^{-m-1}e^{-m}
\]
has probability at least $p=\tfrac14e^{-am}$. Markov's inequality
shows that
\[
 Z_{\mathcal B}/Z_{G_{\sqcup}}
 \le8e^{am}e^{-(Jh/5-1)m}
\]
fails with probability at most $p/2$. There is therefore a realization
with both properties. On this realization, since $m\ge48$,
\[
 \frac{Z_{\mathcal B}}{Z_{\sigma^*}}
 \le16e^{-[Jh/5-\log2-2-a]m}
 \le16e^{-3m}\le e^{-2m}.
\]

On this fixed realization,
\[
 \mu(\mathcal A)\le\frac12,\qquad
 \frac{\mu(\mathcal B)}{\mu(\mathcal A)}
 \le\frac{Z_{\mathcal B}}{Z_{\sigma^*}}\le e^{-2m}.
\]
For any transition matrix $P$ in the theorem, every exit from
$\mathcal A$ starts in $\mathcal B$, so
\[
 \frac{\sum_{I\in\mathcal A}\mu(I)P(I,\mathcal A^c)}{\mu(\mathcal A)}
 \le\frac{\mu(\mathcal B)}{\mu(\mathcal A)}\le e^{-2m}.
\]
The bottleneck-ratio bound~\cite[Theorem~7.4]{LevinPeres2017},
valid also for nonreversible chains, gives
$t_{\rm mix}^P(1/4)\ge e^{2m}/4\ge e^m$.
The realization was chosen independently of $P$, so this holds
simultaneously for all the chains in the statement.

For the transition matrix $P$ of Glauber dynamics, testing the
variational characterization of the spectral gap with $1_{\mathcal A}$
and using $\mu(\mathcal A^c)\ge1/2$ gives
\[
 \operatorname{gap}(P)\le2\frac{\mu(\mathcal B)}{\mu(\mathcal A)}
 \le2e^{-2m},\qquad t_{\rm rel}\ge e^{2m}/2\ge e^m.
\]
Together with \eqref{eq:glauber-size-comparison}, these bounds
prove the theorem with the absolute coefficient $c=1/A$.

\end{proof}

We now extend the obstruction to every even size $n\ge2d$,
completing the proof of \cref{rem:local-chain-obstruction}.

\begin{proposition}[Padding to every even size]
\label{prop:glauber-all-sizes}
There are absolute constants $b,c>0$ such that the following holds
for every fixed $d\ge3$.
For every even $n\ge2d$ and rational
$0<\delta\le\delta_{\mathrm{mix}}(d)$ with $n\delta^2/d\ge b$,
there is a simple $d$-regular bipartite graph $G$ on exactly $n$
vertices with the following property.
At fugacity $\lambda=(1+\delta)\lambdaC(d)$, every irreducible,
aperiodic transition matrix $P$ on $\Ind(G)$ with stationary law
$\mu_{G,\lambda}$ and
\[
 P(I,I')>0
 \quad\Longrightarrow\quad
 |I\triangle I'|\le\lfloor n\delta^2/(bd)\rfloor
\]
satisfies
\[
 t_{\rm mix}^P(1/4)\ge\exp\{cn\delta^2/d\}.
\]
The Glauber dynamics on the same graph also satisfies
$t_{\rm rel}\ge\exp\{cn\delta^2/d\}$.
\end{proposition}

\begin{proof}
Use the constants $L_d,A$ and the source sizes $\mathcal M$ from
\cref{thm:glauber-obstruction} and its proof. Set
\[
 g=4\left\lceil\frac{L_d}{4\delta^2}\right\rceil,\qquad
 x=\frac{n\delta^2}{d},\qquad b=160A.
\]
Since $g\delta^2\le L_d+1\le Ad$, the assumption $x\ge b$ gives
$n\ge160g$. Also $g\ge d/\delta^2\ge100d$, so $n\ge4d$.
Choose the largest $m\in\mathcal M$ with $n_0:=mg\le n-2d$.
It exists because $n-2d\ge n/2\ge48g$, and the consecutive sizes
in $\mathcal M$ double. Therefore
\[
 n_0>\frac{n-2d}{2}\ge\frac n4.
\]
Choose the core $G_0$ and sets $\mathcal A,\mathcal B$ from the
proof of \cref{thm:glauber-obstruction}. Its Glauber relaxation
time is at least $e^m$, and
\[
 n_0\delta^2/d<Am,\qquad m>x/(4A).
\]

For padding, take a simple $d$-regular bipartite graph $F$ with
$\ell=(n-n_0)/2\ge d$ vertices on each side: join left vertex $i$
to right vertices $i,i+1,\ldots,i+d-1$ modulo $\ell$.
Let $G=G_0\sqcup F$.

Let $\widehat{\mathcal A}$ and $\widehat{\mathcal B}$ consist of
independent sets of $G$ whose restrictions to $G_0$ belong to
$\mathcal A$ and $\mathcal B$, respectively. Thus
\[
 \widehat{\mathcal A}=\mathcal A\times\Ind(F),\qquad
 \widehat{\mathcal B}=\mathcal B\times\Ind(F).
\]
The Gibbs law $\mu=\mu_{G,\lambda}$ is a product, so
\[
 \mu(\widehat{\mathcal A})\le\frac12,\qquad
 \frac{\mu(\widehat{\mathcal B})}
      {\mu(\widehat{\mathcal A})}\le e^{-2m}.
\]
Moreover,
\[
 \lfloor x/b\rfloor
 =\left\lfloor\frac{x}{160A}\right\rfloor
 \le\lfloor m/40\rfloor.
\]
Every permitted transition therefore changes at most
$\lfloor m/40\rfloor$ occupation indicators in the core, and every
exit from $\widehat{\mathcal A}$ starts in $\widehat{\mathcal B}$.
Applying the bottleneck-ratio bound as in the proof of
\cref{thm:glauber-obstruction}, now on the full state space, gives
\[
 t_{\rm mix}^P(1/4)\ge e^m\ge e^{x/(4A)}.
\]
This holds for every permitted chain on the same graph.

Finally, \cref{lem:glauber-linear-lower} gives the Glauber bound
\[
 t_{\rm rel}(G)
 \ge\frac n{n_0}t_{\rm rel}(G_0)
 \ge e^m\ge e^{x/(4A)}.
\]
Both conclusions follow with $c=1/(4A)$.
\end{proof}

\paragraph{Lower bounds from one gadget.}

We now establish the Glauber and spectral-independence lower bounds
used in the outline proofs of \cref{thm:main-glauber-lower,thm:main-si}.
Following the maxima of the exact one-gadget rate function yields the
$n^{3/2}$ factor in the Glauber lower bound and the matching
spectral-independence lower bound, uniformly throughout the near-critical interval.

\begin{theorem}[Lower bounds from one gadget]
\label{thm:single-gadget-lower}
For every fixed $d\ge3$, there are constants $c_d,C_d,t_d>0$ and $n_d$
such that the following holds.
For every $n\ge n_d$ divisible by four and every $0\le t\le t_d$,
there is a simple $d$-regular bipartite graph $G$ on $n$ vertices
such that, at fugacity $\lambda=\lambdaC(d)e^t$,
\begin{align}
 t_{\rm rel}^G
 &\ge c_d\frac{n^{3/2}}{\sqrt{1+nt^2}}
       \exp\!\left\{n\left(\frac{3(d-2)}{4d^2}t^2-C_dt^3\right)\right\},
 \label{eq:single-gadget-relaxation}\\
 \lambda_{\max}(\Psi_{\mu_{G,\lambda}})
 &\ge c_d(\sqrt n+nt).
 \label{eq:single-gadget-si}
\end{align}
\end{theorem}

\begin{proof}
Let $N=n/2$, and take the symmetric bipartite double cover of
$d$ independent uniform perfect matchings on $N$ base labels.
Condition their union on being simple.
Write $\E_0$ for expectation before this conditioning and $\E_1$
for expectation after it.
Every realization has the automorphism exchanging its two sides
$V^+,V^-$.

For an independent set $I$, put
\[
 M(I)=|I\cap V^+|-|I\cap V^-|,
 \qquad \rho=\frac{M(I)}N=\frac{2M(I)}n.
\]
Use the type $\alpha$ and coordinates $b$ from
\eqref{eq:occupation-type} and \eqref{eq:occupation-coordinates},
and write $\psi_t$ for the rate function in \eqref{eq:one-gadget-rate}
with $\logshift=t$.

\emph{The exact rate and its maxima.}
By \cref{lem:local-quartic-rate,prop:quartic-coercivity},
the stable maximizer $b_t^*(\rho)$ is analytic near $(t,\rho)=(0,0)$.
Define the rate per graph vertex by
\[
 f_t(\rho)=\frac14\psi_t(\rho,b_t^*(\rho)).
\]
Dividing \eqref{eq:optimized-rate-expansion} by four gives
\[
 f_t(\rho)-f_t(0)
 =A_dt\rho^2-B_d\rho^4
   +\rev{O_d(t^2\rho^2+\rho^6)},
\]
where
\[
 A_d=\frac{d(d-2)}{16(d-1)},\qquad
 B_d=\frac{d^4(d-2)}{768(d-1)^2}.
\]
The function $f_t$ is even in $\rho$.
Thus $F_t(u)=f_t(\sqrt u)$ extends analytically near $(t,u)=(0,0)$,
and
\[
 \partial_uF_t(u)=A_dt-2B_du+\rev{O_d(t^2+u^2)},
 \qquad
 \partial_u^2F_t(u)=-2B_d+O_d(t+u).
\]
After decreasing $t_d$, the implicit-function theorem gives
\[
 u_t=\frac{A_d}{2B_d}t+O_d(t^2)\ge0,
 \qquad \rho_t=\sqrt{u_t},
\]
with $\partial_uF_t(u_t)=0$.
Uniform strict concavity in $u$ gives, on a fixed sufficiently
small interval of $\rho$,
\begin{equation}\label{eq:single-gadget-rate-loss}
 c_d(\rho^2-u_t)^2
 \le f_t(\rho_t)-f_t(\rho)
 \le C_d(\rho^2-u_t)^2.
\end{equation}
The rate difference between a maximum and zero imbalance is
\begin{equation}\label{eq:single-gadget-barrier-expansion}
 \mathcal B_d(t):=F_t(u_t)-F_t(0)
 =\frac{3(d-2)}{4d^2}t^2+O_d(t^3).
\end{equation}
The expansion uses $A_d^2/(4B_d)=3(d-2)/(4d^2)$.

\emph{First moments.}
Choose a fixed sufficiently small neighborhood of the critical type
on which the preceding estimates and
\cref{thm:moving-cell-second-moment} hold.
For the type weight $Z_G(\alpha)$ defined in
\eqref{eq:type-partition-function}, \cref{lem:expected-type-weight} gives
\[
 \E_0 Z_G(\alpha)\asymp_d N^{-3/2}e^{N\psi_t(\rho,b)/2}.
\]
The stable Hessian bounds the loss from $b_t^*(\rho)$ above and
below by positive constant multiples of
$\|b-b_t^*(\rho)\|^2$.
By \cref{prop:global-critical-anchor}, the critical rate has a positive
gap outside this fixed neighborhood. Since $\psi_t-\psi_0=O_d(t)$
uniformly over feasible types, the gap persists after decreasing $t_d$.
The global first-moment bound in \cref{prop:quartic-coercivity} therefore
bounds the total expected weight outside the neighborhood by
\[
 n^{C_d}\exp\{nf_t(0)-c_dn\}.
\]
These bounds hold uniformly for $0\le t\le t_d$.

Apply \cref{lem:uniform-cycle-free-transfer} with no exposed matching edges.
The unconditioned simplicity probability is bounded away from zero,
and the simplicity probability under each local one-replica
type law is also bounded away from zero.
Consequently $\E_1Z_G(\alpha)\asymp_d\E_0Z_G(\alpha)$ on the
chosen neighborhood.
Conditioning on simplicity increases the expected weight outside
the chosen neighborhood by a factor of at most
$1/\Pr_0(\text{simple})=O_d(1)$, so the preceding tail bound remains
valid up to a constant factor.

For each integer $j$ with $2j/n$ in a sufficiently small fixed
interval, let $Z_j^{\rm loc}$ be the total weight of independent
sets with $M(I)=j$ and type in the chosen neighborhood, and let
$Z_j^{\rm box}$ additionally impose
$\|b-b_t^*(2j/n)\|_\infty\le K/\sqrt n$, for a fixed $K>0$.
Gaussian summation in the two stable coordinates gives
\begin{equation}\label{eq:single-gadget-level-means}
 c_dn^{-1/2}e^{nf_t(2j/n)}
 \le \E_1Z_j^{\rm box}
 \le \E_1Z_j^{\rm loc}
 \le C_dn^{-1/2}e^{nf_t(2j/n)}.
\end{equation}
Indeed, at fixed $j$ the two stable lattice spacings are $4/N$
and $2/N$, accounting for the parity shift.
A box of width $N^{-1/2}$ therefore contains order $N$ types.
Multiplying by the $N^{-3/2}$ coefficient prefactor leaves
$N^{-1/2}$.
The Gaussian upper sum and the lower box sum are uniform over
these lattice shifts; positivity of the nearby integer
coefficients follows from
\cref{lem:finite-polynomial-coefficients}(i).

\emph{Weights near a maximum and near zero imbalance.}
Put
\[
 w=n^{-1/4}(1+nt^2)^{-1/4},
 \qquad
 A_{n,t}=n^{1/2}w\,e^{nf_t(\rho_t)}.
\]
Uniformly in the stated range, $nw\ge c_d\sqrt n$.
Let $W_+$ consist of configurations with
\[
 \rho\in[\rho_t+2w,\rho_t+3w],
 \qquad
 \|b-b_t^*(\rho)\|_\infty\le K/\sqrt n,
\]
and let $W_-$ be its image under side exchange.
Decreasing $t_d$ and increasing $n_d$ keeps both windows and
their stable-coordinate boxes inside the chosen neighborhood,
uniformly for $n\ge n_d$ and $0\le t\le t_d$.
Write $Z_+$ for the weight of $W_+$, $Z$ for the full partition
function, and $Z_{\rm sad}$ for the total weight of independent
sets with $|\rho|\le w+4/n$.

We claim that
\[
 \begin{aligned}
 c_dA_{n,t}&\le\E_1Z_+\le\E_1Z\le C_dA_{n,t},\\
 \E_1Z_{\rm sad}&\le C_dA_{n,t}e^{-n\mathcal B_d(t)}.
 \end{aligned}
\]
Since $ntw^2\le1$ and $nw^4\le1$,
\eqref{eq:single-gadget-rate-loss} and
\eqref{eq:single-gadget-level-means} give the lower bound on $\E_1Z_+$.
For $Z_{\rm sad}$, the local expansion gives
$n[f_t(\rho)-f_t(0)]=O_d(1)$ for $|\rho|\le w+4/n$;
the global tail is negligible compared with $n^{1/2}w e^{nf_t(0)}$.
To bound $\E_1Z$, summing \eqref{eq:single-gadget-level-means} and
using \eqref{eq:single-gadget-rate-loss} reduces the local contribution to
\[
 \sum_j
 e^{-c_dn((2j/n)^2-u_t)^2}\le C_d nw,
\]
where the sum is over the local imbalance levels.

To bound the corresponding integral, put $a=\sqrt n\,u_t$
and rescale $\rho=n^{-1/4}y$.
For $0\le a\le1$, the inequality
$(y^2-a)^2\ge y^4/2-1$ bounds the rescaled integral by a constant.
For $a\ge1$, the inequality
\[
 |y^2-a|\ge\sqrt a\,\bigl||y|-\sqrt a\bigr|
\]
bounds it by $O_d(a^{-1/2})$, by Gaussian integration on each
half-line. Consequently,
\[
 \int_{\mathbb R}e^{-c_dn(\rho^2-u_t)^2}\dd\rho
 \le C_dn^{-1/4}(1+\sqrt n\,u_t)^{-1/2}
 \le C_dw,
\]
using $u_t\asymp_d t$ for $t>0$ and $u_0=0$.

Extending the sum to all $j\in\mathbb Z$ gives an upper bound.
The function $e^{-c_dn(\rho^2-u_t)^2}$ is bounded by one and is monotone
on at most four intervals, so its sum on the mesh $2/n$ is at most
$n/2$ times the integral plus $O(1)$.
This gives the required $O_d(nw)$ bound, since
$nw\ge c_d\sqrt n$.
The global tail is again negligible.

For the second moment, we have
\[
 \E_1[Z_+^2]\le C_d\E_0[Z_+^2]
 \le C_d(\E_0Z_+)^2\le C_d(\E_1Z_+)^2.
\]
The three inequalities use, respectively, the lower bound on the
simplicity probability, \cref{thm:moving-cell-second-moment} summed over
the types defining $W_+$, and the local first-moment comparison.
The second-moment bound is uniform in $t$, since fugacity factors
cancel at fixed types in \cref{thm:moving-cell-second-moment}.

The Paley--Zygmund inequality gives
$\Pr_1(Z_+\ge\tfrac12\E_1Z_+)\ge p_d>0$.
By the preceding expectation bounds and Markov's inequality,
choosing $C_d$ sufficiently large makes the combined failure
probability of the other two inequalities below less than $p_d$.
Thus, with positive probability,
\begin{equation}\label{eq:single-gadget-good-event}
 Z_+\ge\tfrac12\E_1Z_+,\qquad
 Z\le C_d\E_1Z_+,\qquad
 Z_{\rm sad}\le C_d(\E_1Z_+)e^{-n\mathcal B_d(t)}.
\end{equation}

\emph{Relaxation and spectral independence.}
Fix a graph $G$ satisfying all three inequalities in
\eqref{eq:single-gadget-good-event}, and put $\mu=\mu_{G,\lambda}$.
Set $R=\lfloor nw/2\rfloor$ and
\[
 h(I)=\max\{-1,\min\{1,M(I)/R\}\}.
\]
For sufficiently large $n$, $R\ge1$.
Side exchange gives $\E_\mu h=0$, and $h=\pm1$ on $W_\pm$,
so $\Var_\mu h\ge2Z_+/Z$.
One Glauber update changes $M$ by at most one.
A nonzero change of $h$ requires $|M|\le R+1$ at the starting
configuration, and hence $|\rho|\le w+4/n$.
Thus
\[
 \mathcal E(h,h)\le\frac{Z_{\rm sad}}{2R^2Z},
 \qquad
 t_{\rm rel}\ge\frac{4R^2Z_+}{Z_{\rm sad}}
 \ge c_dR^2e^{n\mathcal B_d(t)}.
\]
Since $R^2\asymp n^{3/2}/\sqrt{1+nt^2}$,
\eqref{eq:single-gadget-barrier-expansion} gives
\eqref{eq:single-gadget-relaxation} for sufficiently large $C_d$.

For spectral independence, side exchange gives $\E_\mu M=0$.
By \eqref{eq:single-gadget-good-event}, $\mu(W_+)\ge c_d$;
the definition of $W_+$ therefore gives
\[
 \Var_\mu M
 \ge c_dn^2(t+w^2)
 \ge c_d(n^{3/2}+n^2t).
\]
The last inequality follows by considering $nt^2\le1$ and
$nt^2\ge1$ separately.
Apply the covariance Rayleigh quotient to the coefficient vector
that is $+1$ on $V^+$ and $-1$ on $V^-$.
Using \eqref{eq:si-covariance-equivalence} and
$\sum_v p_v(1-p_v)\le n/4$, where $p_v=\mu(X_v=1)$, we obtain
\[
 \lambda_{\max}(\Psi_\mu)
 \ge\frac{\Var_\mu M}{\sum_vp_v(1-p_v)}
 \ge c_d(\sqrt n+nt).
\]
This proves \eqref{eq:single-gadget-si}.
\end{proof}

\begin{corollary}[Lower bounds at every size]
\label{cor:glauber-superlinear-lower}
After decreasing $\delta_{\mathrm{mix}}(d)$ if necessary, there is
an absolute constant $c>0$ such that, for every fixed $d\ge3$,
there are constants $a_d,b_d>0$ with the following property.
For every $n\ge d+1$ and every
$0\le\delta\le\delta_{\mathrm{mix}}(d)$, there is a simple
$n$-vertex graph $G$ of maximum degree $d$ such that, at fugacity
$(1+\delta)\lambdaC(d)$,
\[
 \min\{t_{\rm rel}^G,t_{\rm mix}^G(1/4)\}
 \ge a_dn^{3/2}\exp\!\left(c\frac{n\delta^2}{d}\right),
 \qquad
 \lambda_{\max}(\Psi_{\mu_G})
 \ge b_d(\sqrt n+n\delta).
\]
For even $n\ge2d$, the graph may be chosen $d$-regular and bipartite.
\end{corollary}

\begin{proof}
With $C_d,t_d$ from \cref{thm:single-gadget-lower}, decrease
$\delta_{\mathrm{mix}}(d)$ so that $t=\log(1+\delta)\le t_d$ and
\[
 \frac{3(d-2)}{4d^2}t^2-C_dt^3
 \ge\frac{3(d-2)}{8d^2}t^2\ge\frac{t^2}{8d}.
\]
For sufficiently large $n$, put $n_0=4\lfloor n/4\rfloor$,
apply \cref{thm:single-gadget-lower} on $n_0$ vertices, and add
$n-n_0$ isolated vertices.
The graph still has maximum degree $d$.
By \cref{lem:glauber-linear-lower}, padding cannot decrease
the relaxation time.
Put $x=n\delta^2/d$.
Since $n_0\ge n/2$ and $t\ge\delta/2$ in the stated interval,
\[
 t_{\rm rel}^G
 \ge c_d n^{3/2}(1+x)^{-1/2}e^{x/64}
 \ge a_dn^{3/2}e^{x/128}.
\]
The last inequality absorbs $(1+x)^{-1/2}$ into the exponential.
The mixing bound follows from
\cref{lem:glauber-linear-lower}, after decreasing $a_d$.

The influence matrix of a disjoint union is block diagonal.
Hence padding also preserves the core's SI lower bound, and
$n_0\asymp n$, $t\asymp\delta$ give the asserted estimate.

For the regular bipartite class, instead take
$n_0=4\lfloor(n-2d)/4\rfloor$.
The even remainder has at least $2d$ vertices, so use the bipartite
padding graph from the proof of \cref{prop:glauber-all-sizes}.
For sufficiently large $n$, again $n_0\ge n/2$, so the relaxation
and SI bounds above hold with the same absolute exponent coefficient.

Graphs of maximum degree $d$ exist for every $n\ge d+1$;
simple $d$-regular bipartite graphs exist for every even $n\ge2d$.
For fixed $d$, the remaining sizes are finite and $\delta$ lies in a
bounded interval. Thus \cref{lem:glauber-linear-lower} and
$\lambda_{\max}(\Psi_\mu)\ge1$ cover them after uniformly decreasing
$a_d,b_d$.
\end{proof}

\paragraph{Above the polynomial mixing window.}
Fix $d\ge3$ and a sequence of rationals
$0<\delta_n\le\delta_{\mathrm{mix}}(d)$, and put $x_n=n\delta_n^2/d$.
If $x_n/\log n\to\infty$, then at fugacity
$(1+\delta_n)\lambdaC(d)$,
\[
 \max_G t_{\mathrm{mix}}^G(1/4)
 =\exp\{\Theta(x_n)\}=n^{\Theta(x_n/\log n)},
\]
where the maximum is over simple $n$-vertex graphs of maximum degree $d$.
Indeed, \cref{cor:glauber-superlinear-lower} and
\cref{thm:glauber-upper} at $\eta=1/8$, with
$t_{\mathrm{mix}}(1/4)\le t_{\mathrm{mix}}(1/8)$, give the two bounds:
the logarithms of their polynomial and polylogarithmic prefactors
are $O_d(\log n)=o(x_n)$.
The same scale holds over simple $d$-regular bipartite graphs at even sizes.
The ratio $x_n/\log n$ can tend to infinity arbitrarily slowly while
$\delta_n\to0$, so the exponent of $n$ in this mixing-time scale can
also diverge arbitrarily slowly.

\section{Mean weights of near-critical gadgets}
\label{sec:gadget-moments}\label{sec:quartic-unmarked-main}

We begin the proof of \cref{thm:cut-weight-realization} by showing that
mean phase weights follow the cut weights of the source graph $H$.
Throughout this section, $\E_0$ denotes expectation under the construction
law $\mathbb P_0$: all gadget matchings are independent and uniform,
and, for each $v$, $M_{v,1}$ has a uniform edge order and independent
uniform edge orientations as in \cref{constr:graph-reduction}, before conditioning on
simplicity. Recall that $Z_\sigma$ is the partition function of the \rev{output}
graph $G$ restricted to phase vector $\sigma$, and $Z_{G_{\sqcup}}$ is the
partition function of the disjoint union of the same gadgets before
connections. We estimate ratios of expectations such as
$\E_0Z_\sigma/\E_0Z_{G_{\sqcup}}$.
Section~\ref{sec:shared-moments} will use mixed moments to transfer these
estimates to the reweighted construction law $\mathbb Q_1$ required by
the theorem.

In Sections~\ref{sec:gadget-moments}--\ref{sec:shared-moments}, an edge's
\emph{color} is the index of its matching $M_{v,j}$. Put
\[
 L=2N\delta^2,\qquad \logshift=\log(1+\delta).
\]
Fix $d,\varepsilon$ and the upper bounds in \cref{thm:cut-weight-realization}.
Its assumptions bound both $N\delta^2$ and $k\delta$ above and away from
zero: the positive lower bounds follow from $J\ge2$ and the gadget-size
condition. 
Unless stated otherwise, the estimates below are uniform over
these allowed parameter choices, which may vary with the instance.

\paragraph{Occupation counts and restricted weights.}
Section~\ref{sec:shared-moments} uses a second-moment estimate for a
restricted weight $X_\sigma$ to prove the probability lower bound in
\cref{thm:cut-weight-realization}. We therefore need a lower bound on
$\E_0X_\sigma$ as well as an upper bound on $\E_0Z_\sigma$.
We now specify which independent sets contribute to $X_\sigma$.

For an independent set $I$ of $G$, define
\[
 S_v=\{i\in[N]:i_v^+\in I\},\qquad
 T_v=\{i\in[N]:i_v^-\in I\}.
\]
Suppress the subscript $v$ while
considering one gadget, and define
\[
 n_{10}=|S\setminus T|,\quad n_{01}=|T\setminus S|,\quad
 n_{11}=|S\cap T|,\quad n_{00}=|[N]\setminus(S\cup T)|.
\]
These count labels occupied only on the $+$ side, only on the $-$ side,
on both sides, and on neither side, respectively.
The state $a=(a_+,a_-)\in\Omega:=\{00,10,01,11\}$ of a label records
these two memberships, with $1$ meaning occupied and $0$ unoccupied. Let
\begin{equation}\label{eq:occupation-type}
 \alpha_a=n_a/N\quad(a\in\Omega).
\end{equation}
The vector $\alpha=(\alpha_{00},\alpha_{10},\alpha_{01},\alpha_{11})$
records the fraction of labels in each state; we call it the \emph{type}
of the configuration in this gadget. In type sums, the counts $N\alpha_a$
are nonnegative integers summing to $N$; for analytic expansions of the
rate function, we allow continuous proportions $\alpha$.
We use the imbalance $\rho$ and the pair $b=(b_1,b_2)$, defined by
\begin{equation}\label{eq:occupation-coordinates}
 \rho=\alpha_{10}-\alpha_{01}=\frac{|S|-|T|}{N},\qquad
 b_1=2(\alpha_{10}+\alpha_{01}),\qquad b_2=2\alpha_{11}.
\end{equation}
Thus $b_1/2$ is the fraction of labels occupied on exactly one side,
and $b_2/2$ is the fraction occupied on both sides. The phase
is the sign of $\rho$, with zero assigned the positive sign, as in
\eqref{eq:physical-rho}.

To define $X_\sigma$, we will restrict the occupation proportions to a
region carrying most of the expected one-gadget weight. We first describe
how this expected weight depends on the occupation proportions.

For the gadget $G_v$ before connections, define
\begin{equation}\label{eq:type-partition-function}
 Z_{G_v}(\alpha):=\sum_{I\in\Ind(G_v):\,\alpha(I)=\alpha}
                 \lambda^{|I|},
\end{equation}
where $\alpha(I)$ is computed from the occupation counts in
\eqref{eq:occupation-type}.

Let $\alpha$ be determined by $(\rho,b)$ through the preceding relations.
Let $\mathcal P(\alpha)$ be the set of symmetric couplings of $\alpha$
with itself satisfying
\[
 P_{aa'}=0\quad\text{unless }a_+a'_-=a_-a'_+=0.
\]
Intuitively, these couplings describe how states with frequencies $\alpha$
can be paired along base edges without violating the independent-set
constraint.

For continuous proportions $\alpha_a\ge0$, $\sum_a\alpha_a=1$, with
$\mathcal P(\alpha)\ne\varnothing$, define the rate function below, writing
$\mathsf H(p)=-\sum_xp_x\log p_x$ with $0\log0=0$ and using
$\log\lambda=\log\lambdaC(d)+\logshift$:
\begin{equation}\label{eq:one-gadget-rate}
 \begin{aligned}
 \psi_{\logshift}(\rho,b)
 =2(1-d)\mathsf H(\alpha)
  +(b_1+2b_2)\bigl(\log\lambdaC(d)+\logshift\bigr)
 +d\max_{P\in\mathcal P(\alpha)}\mathsf H(P).
 \end{aligned}
\end{equation}
At the critical fugacity $\lambda=\lambdaC(d)$, equivalently
$\logshift=0$, the rate $\psi_0(\rho,b)$ is uniquely maximized at
$(\rho,b)=(0,b_c)$, where
\[
 b_c=\left(\frac{4(d-1)}{d^2},\frac2{d^2}\right).
\]
This follows from the one-replica case of
\cref{prop:global-critical-anchor}.

For types near $(0,b_c)$ with $\E_0Z_{G_v}(\alpha)>0$,
\cref{lem:expected-type-weight} gives the precise connection to the
expected partition function at type $\alpha$:
\[
 \log\E_0Z_{G_v}(\alpha)
 =\frac N2\psi_{\logshift}(\rho,b)+O_d(\log N).
\]

For sufficiently small $\rho,\logshift$, let $b_{\logshift}^{*}(\rho)$
be the unique local maximizer of $b\mapsto\psi_{\logshift}(\rho,b)$ in a
sufficiently small fixed neighborhood of $b_c$.
Its existence and uniqueness are proved in
\cref{lem:local-quartic-rate}.
\begin{definition}[Restricted phase weight]\label{def:restricted-phase-weight}
\rev{Fix an interior neighborhood $\mathcal U$ of $(0,b_c)$
satisfying \cref{prop:quartic-coercivity}, small enough that
equation~\eqref{eq:connection-type-tilt} is solvable and
the local connection estimate in \cref{prop:marked-local-coercivity}
holds throughout $\mathcal U$.
On this neighborhood, define the rescaled variables}
\begin{equation}\label{eq:exact-stable-center}
 r=\rho/\sqrt{\logshift},\qquad
 s=\sqrt{N/2}\,[b-b_{\logshift}^{*}(\rho)].
\end{equation}
For fixed cutoffs $R,W>0$ and a phase vector $\sigma$, define the restricted phase weight
\begin{equation}\label{eq:restricted-phase-weight}
 X_\sigma:=
 \sum_{\substack{I\in\Ind(G):\ \sigma(I)=\sigma\\
                  (\rho_v,b_v)\in\mathcal U,\ |r_v|\le R,\
                  \|s_v\|_\infty\le W\ \text{for all }v}}
       \lambda^{|I|}.
\end{equation}
Here $\rho_v,b_v,r_v,s_v$ are the occupation variables of $I$ in gadget $v$.
In particular, $0\le X_\sigma\le Z_\sigma$.
\end{definition}

Thus $R$ bounds the scaled imbalance, and $W$ bounds both components of
the scaled deviation from $b_{\logshift}^{*}(\rho)$.

\begin{proposition}[First-moment bounds for phase weights]
\label{prop:uniform-annealed-soft-spin}
Fix $d\ge3$, $0<\varepsilon\le1$, and positive upper bounds on
$N\delta^2$ and $k\delta$.
There exist finite cutoffs $R,W>0$ and $\delta_0>0$, depending only on
$d,\varepsilon$ and these fixed upper bounds, with the following property.
For any $0<\delta\le\delta_0$, take an even $N$ and a positive integer $k$
satisfying $3k\le N/2$, the fixed upper bounds, and
\[
 J=\frac{48k\delta(d-1)}{d^3}\ge2,\qquad
 L=2N\delta^2\ge3480\frac{d^2}{d-2}\frac{J^2}{\varepsilon}.
\]
Then for every simple cubic graph $H$ on $m$ vertices and every phase vector
$\sigma\in\{\pm\}^{V(H)}$,
\begin{equation}\label{eq:annealed-restricted-lower}
 \begin{aligned}
 2^{-m}e^{-J(|E(H)|-\Cut_H(\sigma))-\varepsilon m}
 \le \frac{\E_0X_\sigma}{\E_0Z_{G_{\sqcup}}}
 \le \frac{\E_0Z_\sigma}{\E_0Z_{G_{\sqcup}}}
 \le 2^{-m}e^{-J(|E(H)|-\Cut_H(\sigma))+\varepsilon m}.
 \end{aligned}
\end{equation}
\synlabelalias{eq:uniform-annealed-sign-upper}{eq:annealed-restricted-lower}
\end{proposition}

Section~\ref{sec:soft-spin} derives \cref{prop:uniform-annealed-soft-spin}
from an integral representation of the expected phase weights.
Sections~\ref{sec:quartic}--\ref{sec:first-moment-restoration} prove this
representation by analyzing individual gadgets, their connections,
and the contribution outside the truncation windows.

\subsection{Integral representation of \rev{mean phase weights}}
\label{app:reduction-details}\label{sec:soft-spin}

We introduce an integral $\mathcal I_{H,\sigma}(R)$ that averages the
connection weights over the scaled gadget imbalances with phases $\sigma$.
For sufficiently large cutoffs, $2^{-m}\mathcal I_{H,\sigma}(R)$
approximates the normalized mean phase weights, as quantified in
\cref{lem:normalized-amplitude-representation}.
To define the auxiliary distribution, put
\begin{equation}\label{eq:Vd}
 V_d(r)=A_dr^2-B_dr^4,\qquad
 A_d=\frac{d(d-2)}{16(d-1)},\qquad
 B_d=\frac{d^4(d-2)}{768(d-1)^2}.
\end{equation}
\synlabelalias{eq:AB-constants}{eq:Vd}
Its two maxima are at $\pm r_d$, where $r_d>0$ and
\begin{equation}\label{eq:rd}
 r_d^2=\frac{24(d-1)}{d^3},\qquad
 V_d(r_d)-V_d(r)=B_d(r^2-r_d^2)^2.
\end{equation}
\synlabelalias{eq:well-identity}{eq:rd}
Recall that $r_v=\rho_v/\sqrt{\logshift}$ is the scaled occupation imbalance
in gadget $v$. For a fixed phase vector $\sigma$ and $2r_d\le R\le\infty$,
let $\nu_{\sigma_v,R}$ be the probability measure on $\mathbb R$ given by
\[
 \nu_{\sigma_v,R}(dr)
 \propto e^{LV_d(r)}\mathbf 1_{\{0\le\sigma_v r\le R\}}\,dr.
\]
\Cref{prop:uniform-normalized-tails} shows that this measure approximates
the expected one-gadget weights after restriction to the same phase and
window and normalization. The comparison in \cref{lem:compact-connection-comparison}
motivates averaging the following interaction factor
under the product measure:
\begin{equation}\label{eq:amplitude-model}
 \mathcal I_{H,\sigma}(R)
 :=\E_{r\sim\bigotimes_{v\in V(H)}\nu_{\sigma_v,R}}
 \exp\left\{-\frac{k\delta}{2}\sum_{uv\in E(H)}
                    (r_u+r_v)^2\right\}.
\end{equation}
At $r_v=\sigma_v r_d$ for every vertex, the exponent is
$-J(|E(H)|-\Cut_H(\sigma))$, since $J=2k\delta r_d^2$.
The first lemma relates this integral to the mean phase weights; the
second bounds the logarithmic error from fixing $r_v=\sigma_v r_d$
at every vertex.

\begin{lemma}[Integral representation]
\label{lem:normalized-amplitude-representation}
Fix $d\ge3$ and positive lower and upper bounds on $L=2N\delta^2$ and
$k\delta$. For every $\eta>0$, there exist finite $R\ge2r_d$,
$W>0$, and $\delta_0>0$, depending only on $d,\eta$ and these
bounds, with the following property. For every $0<\delta\le\delta_0$,
even $N$ and positive integer $k$ satisfying $3k\le N/2$ and the fixed
bounds on $L,k\delta$, every simple cubic graph $H$ on $m$ vertices
and every phase vector $\sigma$ satisfy
\[
 2^{-m}e^{-\eta m}\mathcal I_{H,\sigma}(R)
 \le\frac{\E_0X_\sigma}{\E_0Z_{G_{\sqcup}}}
 \le\frac{\E_0Z_\sigma}{\E_0Z_{G_{\sqcup}}}
 \le2^{-m}e^{\eta m}\mathcal I_{H,\sigma}(R).
\]
Here $X_\sigma$ is the restricted weight of
\cref{def:restricted-phase-weight} with these cutoffs $R,W$.
\end{lemma}

The proof is completed in \cref{sec:first-moment-restoration}, after
establishing the one-gadget and connection estimates.

\begin{lemma}[Cut-weight estimate]
\label{lem:amplitude-integration}\label{lem:amplitude-moments}
For $J=2k\delta r_d^2\ge2$ and
$L=2N\delta^2\ge12\frac{d^2}{d-2}J^2$, every cubic $H$ and every phase vector $\sigma$
satisfy, uniformly for $2r_d\le R\le\infty$,
\begin{equation}\label{eq:amp-integration}
 \left|\log\mathcal I_{H,\sigma}(R)
           +J(|E(H)|-\Cut_H(\sigma))\right|
 \le174\frac{d^2}{d-2}\frac{J^2}{L}\,m.
\end{equation}
\end{lemma}

\begin{proof}[Proof of \Cref{prop:uniform-annealed-soft-spin}]
Apply \cref{lem:normalized-amplitude-representation} with
$\eta=\varepsilon/2$ to choose $R,W,\delta_0$ such that
\[
 2^{-m}e^{-\varepsilon m/2}\mathcal I_{H,\sigma}(R)
 \le\frac{\E_0X_\sigma}{\E_0Z_{G_{\sqcup}}}
 \le\frac{\E_0Z_\sigma}{\E_0Z_{G_{\sqcup}}}
 \le2^{-m}e^{\varepsilon m/2}\mathcal I_{H,\sigma}(R).
\]
The assumed lower bound on $L$ ensures that
\cref{lem:amplitude-integration} applies and gives
\[
 \left|\log\mathcal I_{H,\sigma}(R)
          +J(|E(H)|-\Cut_H(\sigma))\right|
 \le\frac{\varepsilon m}{20}.
\]
Exponentiating this bound and substituting into the first display proves
the proposition, since $\varepsilon m/2+\varepsilon m/20\le\varepsilon m$.
\end{proof}

\begin{proof}[Proof of \Cref{lem:amplitude-integration}]
Set $Y_v=\sigma_v r_v/r_d-1$, the relative deviation from the maximum
within the phase $\sigma_v$. Expanding the edge costs in
\eqref{eq:amplitude-model} gives
\[
 \mathcal I_{H,\sigma}(R)
 =e^{-J(|E(H)|-\Cut_H(\sigma))}\E e^{-\Delta},
\]
where
\[
 \Delta=J\sum_{uv\ {\rm uncut}}(Y_u+Y_v)
  +\frac J4\sum_{uv\ {\rm uncut}}(Y_u+Y_v)^2
  +\frac J4\sum_{uv\ {\rm cut}}(Y_u-Y_v)^2.
\]
It therefore suffices to bound $\log\E e^{-\Delta}$.
We use exponential moments for the upper bound and Jensen's inequality
for the lower bound.

The product measure in \eqref{eq:amplitude-model} makes the $Y_v$
independent, and the evenness of $V_d$ gives them the same distribution.
Let $Y$ have this common law and put
$t=LB_dr_d^4=3L(d-2)/(4d^2)\ge9J^2$.
By \eqref{eq:well-identity}, the density of $Y$ at $y$ is proportional to
$e^{-t y^2(2+y)^2}$ on $[-1,R/r_d-1]$, whose upper endpoint is at least $1$.
The following estimates, proved below, hold for $t\ge1$ and
$|h|\le\sqrt t$, uniformly in that endpoint:
\begin{equation}\label{eq:dimensionless-amplitude-moments}
 |\E Y|\le\frac{27}{t},\qquad
 \E[Y^2e^{|hY|}]\le\frac{20}{t},\qquad
 \log\E e^{hY}\le\frac{27|h|+10h^2}{t}.
\end{equation}

Let $d_\sigma(v)\le3$ count the uncut edges incident to $v$, so that
$Jd_\sigma(v)\le\sqrt t$. Discarding the nonnegative quadratic terms
in $\Delta$ and using independence gives
\[
 \log\E e^{-\Delta}
 \le\sum_v\log\E e^{-Jd_\sigma(v)Y_v}
 \le\frac{81J+90J^2}{t}\,m.
\]
For the lower bound, use $(Y_u\pm Y_v)^2\le2Y_u^2+2Y_v^2$
and the fact that every vertex has degree three to obtain
\[
 \E\Delta
 \le3J\sum_v|\E Y_v|+\frac{3J}{2}\sum_v\E Y_v^2
 \le\frac{111J}{t}\,m.
\]
Jensen's inequality then gives
\[
 \log\E e^{-\Delta}\ge-\E\Delta\ge-\frac{111J}{t}\,m.
\]
Since $J\ge2$, both bounds are at most $261J^2m/(2t)$ in absolute value.
Substitute $t=3L(d-2)/(4d^2)$ to obtain \eqref{eq:amp-integration}.

It remains to prove \eqref{eq:dimensionless-amplitude-moments}.
The normalizing integral satisfies
\[
 \int_{-1}^{R/r_d-1}e^{-t y^2(2+y)^2}\dd y
 \ge\frac{e^{-25/16}}{\sqrt t},
\]
by integration on $|y|\le1/(2\sqrt t)$.
Throughout the domain, $y^2(2+y)^2\ge y^2$.
Using $|hy|\le ty^2/2+h^2/(2t)$ and integrating a Gaussian gives
\[
 \E[Y^2e^{|hY|}]
 \le\frac{e^{33/16}\sqrt{2\pi}}{t}<\frac{20}{t}.
\]
The second-moment bound alone gives only $|\E Y|=O(t^{-1/2})$;
to obtain $O(t^{-1})$, we use cancellation between $y$ and $-y$
on $[0,1]$. Since $y^2(2+y)^2-y^2(2-y)^2=8y^3$,
using $1-e^{-x}\le x$ for $x\ge0$ gives
\[
 \left|\int_{-1}^{1}y e^{-ty^2(2+y)^2}\dd y\right|
 \le 8t\int_0^\infty y^4e^{-ty^2}\dd y
 =3\sqrt\pi\,t^{-3/2}.
\]
The unpaired upper tail contributes at most
$\int_1^\infty ye^{-ty^2}\dd y=e^{-t}/(2t)$.
Dividing by the normalizing integral and using its lower bound gives
$|\E Y|\le27/t$.
Finally, $e^z\le1+z+z^2e^{|z|}/2$ yields
\[
 \E e^{hY}
 \le1+h\E Y+\frac{h^2}{2}\E[Y^2e^{|hY|}]
 \le1+\frac{27|h|+10h^2}{t}.
\]
Taking logarithms and using $\log(1+x)\le x$ proves the last bound
in \eqref{eq:dimensionless-amplitude-moments}.
\end{proof}

\subsection{The normalized one-gadget law}
\label{app:quartic-details}\label{sec:quartic}

We estimate the fraction of $\E_0Z_{G_v}$ contributed by fixed
ranges of the scaled coordinates $r,s$ in \eqref{eq:exact-stable-center},
and bound the contribution outside these ranges,
as stated precisely in \cref{prop:uniform-normalized-tails} below. Together with the
connection estimates in \cref{sec:ports}, these bounds will yield the
integral representation in \cref{lem:normalized-amplitude-representation}.

Fix one gadget $G_v$ before connections. Recall that $Z_{G_v}(\alpha)$
in \eqref{eq:type-partition-function} sums the weights of independent sets
of type $\alpha$. For a set $\mathcal S$ of types, define
\[
 Z_{G_v}(\mathcal S):=\sum_{\alpha\in\mathcal S}Z_{G_v}(\alpha).
\]
Removing the type restriction gives the full partition function $Z_{G_v}$.

The proposition below uses a Gaussian density proportional to $e^{-s^{\mathsf T}H_ds/2}$,
where the positive-definite matrix $H_d$ is specified by
\begin{align}\label{eq:H-d}
 (y,z)H_d(y,z)^{\mathsf T}
 &=\frac{d^2}{2(d-2)}(y+2z)^2+\frac{d^2(d-2)}{2(d-1)(d^2-2d+2)}((d-2)z-y)^2,\\*
 \det H_d&=\frac{d^6}{4(d-1)(d^2-2d+2)}.
 \nonumber
\end{align}

Recall that $\mathcal U$ is the fixed neighborhood of $(0,b_c)$ used
to define the coordinates in \eqref{eq:exact-stable-center}.

\begin{proposition}[One-gadget law]
\label{prop:uniform-normalized-tails}\label{prop:bin-integral-mean}
As $N\to\infty$ with $L=2N\delta^2$ between the fixed positive lower
and upper bounds, the following estimates hold uniformly over the
allowed parameter choices. For each fixed bounded interval $\mathcal J$
and fixed $W>0$, restricting the sum to types in $\mathcal U$, where $s$
is defined, we have
\begin{equation}\label{eq:bin-integral-mean}
 \frac{\E_0 Z_{G_v}(\{r\in\mathcal J,\ \|s\|_\infty\le W\})}{\E_0Z_{G_v}}
 =\frac{\int_{\mathcal J}e^{LV_d(r)}\dd r}{\int_{\mathbb R}e^{LV_d(r)}\dd r}
   \frac{\sqrt{\det H_d}}{2\pi}
       \int_{[-W,W]^2}e^{-s^{\mathsf T}H_ds/2}\dd s+o_N(1).
\end{equation}
If we restrict only $r$, the corresponding formula is, uniformly over
all intervals $\mathcal J\subseteq\mathbb R$,
\begin{equation}\label{eq:one-gadget-imbalance-marginal}
 \frac{\E_0 Z_{G_v}(\{r\in\mathcal J\})}{\E_0Z_{G_v}}
 =\frac{\int_{\mathcal J}e^{LV_d(r)}\dd r}
       {\int_{\mathbb R}e^{LV_d(r)}\dd r}+o_N(1).
\end{equation}
If $\mathcal J$ has positive length, the ratio in \eqref{eq:bin-integral-mean}
has a positive lower bound for all sufficiently large $N$, depending only
on $d,\mathcal J,W$ and the fixed bounds on $L$.
For some fixed $c,C>0$, every type $\alpha$ with $(\rho,b)\in\mathcal U$ satisfies
\begin{equation}\label{eq:normalized-cell-envelope}
 \frac{\E_0Z_{G_v}(\alpha)}{\E_0Z_{G_v}}
 \le\frac{C}{N^2\sqrt{\logshift}}
              e^{Cr^2-cr^4-c\|s\|^2}.
\end{equation}
In the following tail bound, both events involving $r,s$
are restricted to types in $\mathcal U$, where $s$ is defined.
For $R,W\ge2$ and sufficiently large $N$,
\begin{equation}\label{eq:uniform-integrated-tail}
 \frac{\E_0 Z_{G_v}(\{|r|>R\}\cup\{\|s\|_\infty>W\}\cup\mathcal U^c)}
      {\E_0Z_{G_v}}
 \le C\left[\int_{|u|>R-1}e^{Cu^2-cu^4}\dd u+e^{-cW^2}\right]+e^{-cN}.
\end{equation}
\end{proposition}

We prove the local expansion first, then establish a dominating bound
and sum over the lattice to prove this proposition.

\subsubsection{Exact coefficients and the quartic expansion}

We obtain the local type-weight estimate by combining a coefficient
estimate with a Taylor expansion of $\psi_{\logshift}$.

\begin{proposition}[Quartic approximation]
\label{prop:quartic-expansion}
There exists $c_N=c_N(d,N,\delta)$, independent of the type $\alpha$,
such that, for each fixed $R,W$, uniformly over types with
$\E_0Z_{G_v}(\alpha)>0$, $|r|\le R$, and $\|s\|_\infty\le W$,
\begin{equation}\label{eq:quartic-normal-form}
 \log\E_0Z_{G_v}(\alpha)
 =c_N+LV_d(r)-\frac12s^{\mathsf T}H_ds+O_{d,R,W}(N^{-1/4}).
\end{equation}
\end{proposition}

We derive \cref{prop:quartic-expansion} from the following coefficient
estimate and rate expansion. The type corresponding to $(0,b_c)$ is
\begin{equation}\label{eq:critical-type}
 \alpha_c=d^{-2}((d-1)^2,d-1,d-1,1).
\end{equation}

\begin{lemma}[Expected type weights]\label{lem:expected-type-weight}
There is a positive analytic function $h_d(\alpha)$, independent of
$N,\delta$, defined for continuous proportions $\alpha$ with $(\rho,b)$
in a fixed neighborhood of $(0,b_c)$ and bounded above and away from zero there.
As $N\to\infty$ through even values, uniformly over types in this neighborhood with $\E_0Z_{G_v}(\alpha)>0$,
and uniformly over all $\lambda>0$, with $\logshift=\log(\lambda/\lambdaC(d))$,
\begin{equation}\label{eq:one-gadget-coefficient-estimate}
 \E_0Z_{G_v}(\alpha)
 =N^{-3/2}e^{N\psi_{\logshift}(\rho,b)/2}
   h_d(\alpha)\bigl(1+O_d(N^{-1/2})\bigr).
\end{equation}
\end{lemma}

\begin{lemma}\label{lem:local-quartic-rate}
The function $(\rho,b,\logshift)\mapsto\psi_{\logshift}(\rho,b)$
is analytic near $(0,b_c,0)$ and satisfies
$\nabla_b^2\psi_0(0,b_c)=-H_d$.
For each sufficiently small $\rho,\logshift$, the function
$b\mapsto\psi_{\logshift}(\rho,b)$ has a unique local maximizer in a
fixed neighborhood of $b_c$. This maximizer $b_{\logshift}^{*}(\rho)$
depends analytically on $(\rho,\logshift)$.
For fixed $R,W$, uniformly on
$|r|\le R$, $\|s\|_\infty\le W$,
\[
 \frac N2\bigl[\psi_{\logshift}(\rho,b)
   -\psi_{\logshift}(0,b_{\logshift}^{*}(0))\bigr]
 =2N\logshift^2V_d(r)-\frac12s^{\mathsf T}H_ds
   +O_{d,R,W}(N^{-1/4}).
\]
The error is uniform for $L$ between its fixed positive bounds.
\end{lemma}

\begin{proof}[Proof of \Cref{prop:quartic-expansion}]
Set
\[
 c_N:=\frac N2\psi_{\logshift}(0,b_{\logshift}^{*}(0))
       -\frac32\log N+\log h_d(\alpha_c).
\]
For fixed $R,W$ and sufficiently large
$N$, taking logarithms in
\cref{lem:expected-type-weight} and using \cref{lem:local-quartic-rate}
gives, uniformly on the stated window,
\[
 \begin{aligned}
 \log\E_0Z_{G_v}(\alpha)-c_N-LV_d(r)+\frac12s^{\mathsf T}H_ds
 =\log\frac{h_d(\alpha)}{h_d(\alpha_c)}
      +(2N\logshift^2-L)V_d(r)+O_{d,R,W}(N^{-1/4}).
 \end{aligned}
\]
Since $L=2N\delta^2$ stays between its fixed bounds,
$\delta=O(N^{-1/2})$; \eqref{eq:exact-stable-center} and analyticity of
$b_{\logshift}^{*}(\rho)$, with $b_0^{*}(0)=b_c$, give
$\|\alpha-\alpha_c\|=O_{d,R,W}(N^{-1/4})$ on the stated window.
Positivity and analyticity of $h_d$, together with
$\logshift=\delta+O(\delta^2)$, imply
\[
 \log\frac{h_d(\alpha)}{h_d(\alpha_c)}=O_{d,R,W}(N^{-1/4}),\qquad
 (2N\logshift^2-L)V_d(r)=O_{d,R}(N\delta^3)=O_{d,R}(N^{-1/2}).
\]
Substitution proves \eqref{eq:quartic-normal-form}.
\end{proof}

\begin{proof}[Proof of \Cref{lem:expected-type-weight}]
\emph{Exact counting.}
Recall $n_a=N\alpha_a$ from \eqref{eq:occupation-type}.
For a base edge $ij$, let $a,a'\in\{00,10,01,11\}$ record the occupation
states of $(i^+,i^-)$ and $(j^+,j^-)$, respectively, in $G_v$.
The crossed edges $i^+j^-$ and $i^-j^+$ obey the independent-set constraint
precisely when $a_+a'_-=a_-a'_+=0$. Let $B_{aa'}$ be $1$ for these
allowed pairs and $0$ otherwise. In the state order $00,10,01,11$,
\begin{equation}\label{eq:base-compatibility}
 B=\begin{pmatrix}
 1&1&1&1\\1&1&0&0\\1&0&1&0\\1&0&0&0
 \end{pmatrix}.
\end{equation}
Assign the monomials $1,x,y,z$ to states $00,10,01,11$, respectively.
Multiplying the two monomials for each allowed ordered pair and summing
gives the generating polynomial for one base edge:
\begin{equation}\label{eq:P-poly}
 P(x,y,z)=1+2x+2y+2z+x^2+y^2.
\end{equation}
The exponents count endpoints in states $10,01,11$. Since a perfect
matching has $N/2$ edges with disjoint endpoints,
$[x^{n_{10}}y^{n_{01}}z^{n_{11}}]P^{N/2}$ counts its compatible state
assignments of type $\alpha$.

There are $\binom{N}{n_{00},n_{10},n_{01},n_{11}}$ state assignments of
this type. Fix one and generate a uniform matching by uniformly
permuting the labels and pairing consecutive positions. The states at
these positions are uniform over all assignments of type $\alpha$,
so the compatibility probability is the coefficient above divided by
this multinomial count. Independence of the $d$ matchings raises this
ratio to the $d$th power. Summing over the state assignments, each with
weight $\lambda^{n_{10}+n_{01}+2n_{11}}$, gives~\cite[Equation~(51)]{CCYZ2026}
\begin{equation}\label{eq:alpha-coefficient}
  \E_0Z_{G_v}(\alpha)
  =\lambda^{n_{10}+n_{01}+2n_{11}}
   \frac{([x^{n_{10}}y^{n_{01}}z^{n_{11}}]P^{N/2})^d}
        {\binom{N}{n_{00},n_{10},n_{01},n_{11}}^{d-1}}.
\end{equation}

\emph{Identifying the entropy objective.}
For each monomial $x^{j_1}y^{j_2}z^{j_3}$ appearing in $P$, write
$j=(j_1,j_2,j_3)$ and let $c_j$ be its coefficient.
Let $p=(p_j)$ range over probability distributions on these six monomials
satisfying $\sum_jp_jj=2(\alpha_{10},\alpha_{01},\alpha_{11})$, the mean
constraint in \cref{lem:finite-polynomial-coefficients} for $M=N/2$
and target $(n_{10},n_{01},n_{11})$.
Recall that $\mathcal P(\alpha)$ consists of symmetric distributions $Q$
on allowed ordered state pairs, with each endpoint distributed as $\alpha$:
\[
 Q(a,a')=Q(a',a),\qquad \sum_{a'}Q(a,a')=\alpha_a,\qquad
 Q(a,a')=0\ \text{if }B_{aa'}=0.
\]
We claim that the entropy objective in that lemma satisfies
\[
 \max_p\left\{\mathsf H(p)+\sum_jp_j\log c_j\right\}
 =\max_{Q\in\mathcal P(\alpha)}\mathsf H(Q).
\]
Given $p$, assign probability $p_j/c_j$ to each ordered pair producing
the monomial with exponent $j$. For example, the two pairs $(00,10)$ and $(10,00)$
producing $x$ each receive half its probability.
Each monomial comes from one diagonal pair or two reversed pairs, so
the resulting distribution $Q$ is symmetric. Since $j$ counts the
two endpoint states, the mean constraint gives
\[
 \left(\sum_{a'}Q(10,a'),\sum_{a'}Q(01,a'),\sum_{a'}Q(11,a')\right)
 =\frac12\sum_jp_jj=(\alpha_{10},\alpha_{01},\alpha_{11}).
\]
The remaining marginal probability is $\alpha_{00}$, so both endpoint
marginals are $\alpha$ and $Q\in\mathcal P(\alpha)$.
Conversely, every $Q\in\mathcal P(\alpha)$ is obtained this way by
summing its probabilities over pairs producing each monomial;
symmetry makes the two probabilities equal when $c_j=2$.
For these corresponding distributions,
\[
 \mathsf H(Q)
 =-\sum_j c_j\frac{p_j}{c_j}\log\frac{p_j}{c_j}
 =\mathsf H(p)+\sum_jp_j\log c_j,
\]
which proves the claimed equality of maxima.

\emph{Coefficient estimates and assembly.}
We verify the hypotheses of \cref{lem:finite-polynomial-coefficients}(i)
for $P^{N/2}$. Tilt $P$ at
$((d-2)^{-1},(d-2)^{-1},(d-2)^{-2})$: give the monomial with exponent $j$ probability
proportional to $c_j(d-2)^{-j_1-j_2-2j_3}$, and let $X$ be its exponent
vector. Its three coordinates count states $10,01,11$ among the two
endpoints of an allowed base edge. Normalizing gives
\begin{equation}\label{eq:XN-law}
 \E[x^{X_1}y^{X_2}z^{X_3}]
 =\frac{(d-2)^2+2(d-2)(x+y)+2z+x^2+y^2}{d^2}.
\end{equation}
Thus $\E X=2(\alpha_{c,10},\alpha_{c,01},\alpha_{c,11})$, as required
at the critical type. The support contains the origin and the three
unit vectors, so its difference lattice is $\mathbb Z^3$ and its
covariance is positive definite. Since the derivative of the tilted
mean with respect to the logarithmic tilt is this covariance matrix,
the inverse-function theorem gives analytic tilts realizing
$2(\alpha_{10},\alpha_{01},\alpha_{11})$ for all $\alpha$ in a fixed
small neighborhood of $\alpha_c$. These tilts stay in a compact set.
Let $\Sigma_\alpha$ be the covariance matrix of the tilted exponent
with mean $2(\alpha_{10},\alpha_{01},\alpha_{11})$.
Applying \cref{lem:finite-polynomial-coefficients}(i) with $h=3$,
$M=N/2$, and target $(n_{10},n_{01},n_{11})$ gives, uniformly in this neighborhood,
\[
 [x^{n_{10}}y^{n_{01}}z^{n_{11}}]P(x,y,z)^{N/2}
 =\frac{\exp\!\left\{\frac N2\max_{Q\in\mathcal P(\alpha)}\mathsf H(Q)\right\}}
 {(\pi N)^{3/2}\sqrt{\det\Sigma_\alpha}}
 \bigl(1+O_d(N^{-1/2})\bigr).
\]

For the denominator, Stirling's formula gives, uniformly in this neighborhood,
\[
 \binom{N}{n_{00},n_{10},n_{01},n_{11}}
 =\frac{e^{N\mathsf H(\alpha)}}{(2\pi N)^{3/2}\sqrt{\prod_{a\in\Omega}\alpha_a}}
   \bigl(1+O_d(N^{-1})\bigr),
\]
since all four proportions are bounded away from zero.
Substituting these two estimates into \eqref{eq:alpha-coefficient},
the exponential terms combine as
\[
 \frac{dN}{2}\max_{Q\in\mathcal P(\alpha)}\mathsf H(Q)
 -(d-1)N\mathsf H(\alpha)+\frac N2(b_1+2b_2)\log\lambda
 =\frac N2\psi_{\logshift}(\rho,b).
\]
Since $N^{-3d/2}N^{3(d-1)/2}=N^{-3/2}$, the prefactor is
$N^{-3/2}h_d(\alpha)$, where
\[
 h_d(\alpha)=\frac{2^{3(d-1)/2}}{\pi^{3/2}}
 \frac{(\prod_{a\in\Omega}\alpha_a)^{(d-1)/2}}
 {(\det\Sigma_\alpha)^{d/2}}.
\]
For fixed $d$, the combined relative error is $O_d(N^{-1/2})$,
which proves \eqref{eq:one-gadget-coefficient-estimate}.
The analytic tilts depend only on $\alpha$, so $\det\Sigma_\alpha$
and $h_d(\alpha)$ are analytic and independent of $N,\delta$.
On a smaller fixed neighborhood of $\alpha_c$, the determinant and all
four proportions are bounded above and away from zero; the displayed
formula gives the same bounds for $h_d$.
{The fugacity enters \eqref{eq:alpha-coefficient} only through
an exact monomial, so the relative error is uniform over all
positive fugacities.}
\end{proof}

\begin{proof}[Proof of \Cref{lem:local-quartic-rate}]
We express $\psi_{\logshift}$, defined in \eqref{eq:one-gadget-rate},
using two Legendre transforms, then expand
in coordinates that diagonalize their quadratic terms. The cancellation
in the imbalance direction leaves the required quartic term.

\emph{Expressing $\psi_{\logshift}$ through Legendre transforms.}
Let $(\omega_+,\omega_-)\in\Omega=\{00,10,01,11\}$ have probability law
$\alpha_c$ from \eqref{eq:critical-type}. Each bit records whether the
corresponding vertex of one base label is occupied ($1$) or unoccupied ($0$).
The product form of $\alpha_c$ makes the bits independent Bernoulli
variables of mean $1/d$. Put $D=d-1$ and define
\[
 \xi_+=\frac{d\omega_+-1}{\sqrt D},\quad
 \xi_-=\frac{d\omega_--1}{\sqrt D},\qquad
 U=\frac{\xi_+-\xi_-}{\sqrt2},\quad V=\frac{\xi_++\xi_-}{\sqrt2},\quad U_0=\xi_+\xi_-.
\]
The functions $1,U,V,U_0$ form an orthonormal basis under $\alpha_c$.
Recall the compatibility matrix $B$ from \eqref{eq:base-compatibility}:
$B_{ij}=1$ exactly when $i_+j_-=i_-j_+=0$.
Using the one-edge generating polynomial $P$ from \eqref{eq:P-poly},
define the critical distribution of the two endpoint states $i,j\in\Omega$ by
\[
 P_c(i,j)=\frac{B_{ij}t_it_j}{P(t_{10},t_{01},t_{11})},
 \qquad
 t=(1,(d-2)^{-1},(d-2)^{-1},(d-2)^{-2}).
\]
Here $P(t_{10},t_{01},t_{11})=d^2/(d-2)^2$, and the row sums of $P_c$ are $\alpha_c$.
For $c\in\mathbb R$ and $\zeta\in\mathbb R^2$, define
\[
 \begin{aligned}
 \rev{\kappa_{\mathrm D}}(c,\zeta)
   &=\log\E_{\alpha_c}e^{cU+\zeta_1V+\zeta_2U_0},\\
 \rev{\kappa_{\mathrm N}}(c,\zeta)
   &=\log\E_{P_c}\exp\!\left\{
       \sum_{i=1}^2(cU_i+\zeta_1V_i+\zeta_2U_{0,i})\right\}.
 \end{aligned}
\]
Here the unindexed functions are evaluated at a state sampled from
$\alpha_c$; $U_i,V_i,U_{0,i}$ are their values at endpoint $i=1,2$
of a state pair sampled from $P_c$. All are scalar random variables.
The subscripts $\rev{\mathrm D,\mathrm N}$ refer to the
multinomial denominator and matching numerator in \eqref{eq:alpha-coefficient}.
Let $\rev{I_{\mathrm D}},\rev{I_{\mathrm N}}$ be their Legendre transforms: for either pair $(\kappa,I)$,
\[
 I(x,w)=\sup_{c\in\mathbb R,\,\zeta\in\mathbb R^2}
       \{cx+\zeta^{\mathsf T}w-\kappa(c,\zeta)\},
 \qquad x\in\mathbb R,\quad w\in\mathbb R^2.
\]

For a type $\alpha$ with coordinates $(\rho,b)$ from
\eqref{eq:occupation-coordinates}, let $(u,v_1,v_2)$ be the mean of
$(U,V,U_0)$ under $\alpha$, and write $v=(v_1,v_2)$. Explicitly,
\begin{equation}\label{eq:diagonal-mixed-coordinates}
 u=\frac{d\rho}{\sqrt{2D}},\qquad
 v_1=\frac{d(b_1+2b_2)-4}{2\sqrt{2D}},\qquad
 v_2=\frac{d((d-2)b_2-b_1)+2}{2D}.
\end{equation}
The mean of the sum over both endpoints is twice this vector.
We now relate $\rev{I_{\mathrm D}},\rev{I_{\mathrm N}}$ to the entropy terms in \eqref{eq:one-gadget-rate}.
The basis property of $1,U,V,U_0$, and $P_c(00,j)>0$ for every $j\in\Omega$,
show that both random vectors have full-dimensional support.
Their covariance matrices are nonsingular, so the inverse-function theorem
gives exponential tilts realizing every nearby mean.
The relative-entropy identity in the proof of
\cref{lem:finite-polynomial-coefficients} then expresses each Legendre
transform as the minimum KL divergence subject to its mean constraint,
attained at the corresponding tilt.
For $\rev{I_{\mathrm D}}$, the constraint fixes the state law to $\alpha$;
for $\rev{I_{\mathrm N}}$, it fixes the average of the endpoint marginals to $\alpha$.
Symmetrizing a law $Q$ on allowed state pairs as $(Q+Q^{\mathsf T})/2$
preserves this constraint and cannot increase $\KL(Q\|P_c)$.
The symmetric law has both marginals equal to $\alpha$, so the minimum
can be taken over $\mathcal P(\alpha)$. Consequently,
\[
 \begin{aligned}
 \rev{I_{\mathrm D}}(u,v)
 &=\KL(\alpha\|\alpha_c)
   =-\mathsf H(\alpha)-\sum_{i\in\Omega}\alpha_i\log\alpha_{c,i},\\
 \rev{I_{\mathrm N}}(2u,2v)
 &=\min_{Q\in\mathcal P(\alpha)}\KL(Q\|P_c)\\
 &=-\max_{Q\in\mathcal P(\alpha)}\mathsf H(Q)
   -2\sum_{i\in\Omega}\alpha_i\log t_i
   +\log P(t_{10},t_{01},t_{11}).
 \end{aligned}
\]
The last equality follows by expanding $\log P_c(i,j)$ and using the
marginals of $Q$. Substituting into \eqref{eq:one-gadget-rate}, the
critical coefficient of each singly occupied proportion is twice
\[
 \log\lambdaC+d\log(d-2)-(d-1)\log(d-1)=0,
\]
and that of $\alpha_{11}$ is four times this value. The remaining fugacity term is
$\logshift[(b_1-b_{c,1})+2(b_2-b_{c,2})]=2\sqrt{2D}\,\logshift v_1/d$,
where $\logshift=\log(1+\delta)$.
Subtracting the value at $(0,b_c)$ therefore gives
\begin{equation}\label{eq:local-rate-legendre}
 \psi_{\logshift}(\rho,b)-\psi_{\logshift}(0,b_c)
 =2D \rev{I_{\mathrm D}}(u,v)-d\rev{I_{\mathrm N}}(2u,2v)
   +\frac{2\sqrt{2D}}d\logshift v_1.
\end{equation}

\emph{Computing the local expansion.}
We expand the two Legendre transforms in \eqref{eq:local-rate-legendre}
to obtain the quartic term in the imbalance and the quadratic terms in
the remaining coordinates. All error bounds below are uniform for
$|u|\le CN^{-1/4}$ and $\|v\|\le CN^{-1/2}$, with $C$ fixed.
For either $\kappa\in\{\rev{\kappa_{\mathrm D}},\rev{\kappa_{\mathrm N}}\}$, write
\[
 \kappa_2=\partial_c^2\kappa(0),\qquad
 \Sigma=\nabla_\zeta^2\kappa(0),\qquad
 \gamma=\partial_c^2\nabla_\zeta\kappa(0),\qquad
 \kappa_4=\partial_c^4\kappa(0).
\]
Exchanging the two occupation bits negates $U$ and preserves $V,U_0$,
so $\kappa$ is even in $c$. For $c=O(N^{-1/4})$ and
$\zeta=O(N^{-1/2})$, its Taylor expansion is
\[
 \kappa(c,\zeta)=\tfrac12\kappa_2c^2+\tfrac12\zeta^{\mathsf T}\Sigma\zeta
       +\tfrac12c^2\gamma^{\mathsf T}\zeta+\tfrac1{24}\kappa_4c^4+O(N^{-5/4}).
\]
Analyticity and evenness give this remainder; differentiating the omitted
Taylor terms gives a $c$-derivative $O(N^{-1})$ and a $\zeta$-gradient
$O(N^{-3/4})$.
The local inverse established above gives the unique maximizing tilt
$(\tilde c,\tilde\zeta)$ with $\nabla\kappa(\tilde c,\tilde\zeta)=(u,v)$.
The mixed covariance $\partial_c\nabla_\zeta\kappa(0)$ vanishes
by evenness in $c$; this symmetry and the analytic inverse give
$\tilde c=O(N^{-1/4})$ and $\tilde\zeta=O(N^{-1/2})$.
The stationary equations are thus
\[
 \kappa_2\tilde c+\tilde c\gamma^{\mathsf T}\tilde\zeta
       +\frac{\kappa_4}{6}\tilde c^3+O(N^{-1})=u,
 \qquad
 \Sigma\tilde\zeta+\frac12\tilde c^2\gamma+O(N^{-3/4})=v.
\]
Solving to the required orders yields
\[
 \tilde c=\frac{u}{\kappa_2}+O(N^{-3/4}),\qquad
 \tilde\zeta=\Sigma^{-1}\left(v-\frac{u^2\gamma}{2\kappa_2^2}\right)
             +O(N^{-3/4}).
\]
Since
\[
 u\tilde c-\frac{\kappa_2}{2}\tilde c^2
 =\frac{u^2}{2\kappa_2}
 -\frac{\kappa_2}{2}\left(\tilde c-\frac u{\kappa_2}\right)^2
 =\frac{u^2}{2\kappa_2}+O(N^{-3/2}),
\]
we need no further correction to $\tilde c$. Substituting in
$I=u\tilde c+v^{\mathsf T}\tilde\zeta-\kappa(\tilde c,\tilde\zeta)$ gives
\[
 I(u,v)=\frac{u^2}{2\kappa_2}
 +\frac12\left(v-\frac{u^2\gamma}{2\kappa_2^2}\right)^{\mathsf T}
 \Sigma^{-1}\left(v-\frac{u^2\gamma}{2\kappa_2^2}\right)
 -\frac{\kappa_4u^4}{24\kappa_2^4}+O(N^{-5/4}).
\]

We now compute the coefficients for $\rev{\kappa_{\mathrm D}}$ and $\rev{\kappa_{\mathrm N}}$.
The binary relation $\xi_+^2=1+(D-1)\xi_+/\sqrt D$ gives
\begin{equation}\label{eq:binary-quartic-identity}
 U^2=1+aV-U_0,\qquad a=\frac{D-1}{\sqrt{2D}}.
\end{equation}
\rev{For the edge law $P_c$, define the transition matrix $T_c$ by
$T_c(i,j)=P_c(i,j)/\alpha_c(i)$. Its $i$-th row is the conditional distribution
of the second endpoint state given that the first is $i$.} This conditional distribution
factors over the two crossed pairs of occupation bits:
\begin{equation}\label{eq:critical-binary-channel}
 T_c(i,j)=(\rev{T_{\rm bin}})_{i_+,j_-}(\rev{T_{\rm bin}})_{i_-,j_+},\qquad
 \rev{T_{\rm bin}}=\begin{pmatrix}(D-1)/D&1/D\\1&0\end{pmatrix}.
\end{equation}
The rows and columns of $\rev{T_{\rm bin}}$ are indexed by $0,1$.
\rev{For a state function $f:\Omega\to\mathbb R$, $T_c$ acts by taking the
conditional mean:}
\[
 (T_cf)(i)=\sum_{j\in\Omega}T_c(i,j)f(j).
\]
Since $\rev{T_{\rm bin}}$ preserves $(D/d,1/d)$ and multiplies the centered binary
spin by $-1/D$, for $F\in\{U,V,U_0\}$ we have
\[
 T_cF=\theta_FF,\qquad
 (\theta_U,\theta_V,\theta_{U_0})=(D^{-1},-D^{-1},D^{-2}).
\]
These identities give the following coefficients:
\[
\begin{array}{c|cc}
 &\rev{\kappa_{\mathrm D}}&\rev{\kappa_{\mathrm N}}\\ \hline
 \kappa_2&1&2d/D\\[1mm]
 \Sigma&I_2&\operatorname{diag}\!\left(\frac{2(D-1)}D,
                                 \frac{2(D^2+1)}{D^2}\right)\\[1mm]
 \gamma&(a,-1)^{\mathsf T}&(2da/D,-2d^2/D^2)^{\mathsf T}\\[1mm]
 \kappa_4&a^2-1&d^2(D-5)/D^2
\end{array}
\]
To verify the table, recall that the coordinates $U,V,U_0$
are centered and orthonormal under $\alpha_c$. For $(i,j)\sim P_c$,
write $F_1=F(i)$ and $F_2=F(j)$. Conditioning on the first endpoint gives
\[
 \E[F_2\mid i]=\theta_FF_1,\qquad
 \E[F_1F_2]=\theta_F,\qquad
 \Var(F_1+F_2)=2(1+\theta_F).
\]
For distinct $F,G\in\{U,V,U_0\}$, orthogonality gives
$\E[F_1G_2]=\theta_G\E[F_1G_1]=0$; the other three terms likewise vanish,
so $\Cov(F_1+F_2,G_1+G_2)=0$. This proves the first two rows.
The identity \eqref{eq:binary-quartic-identity}
gives $\E[U^2V]=a$, $\E[U^2U_0]=-1$, and $\E U^4=2+a^2$.
Expanding the cubic moment for $\rev{\kappa_{\mathrm N}}$ and conditioning one endpoint
on the other gives, for $Z=V,U_0$,
\[
 \E[(U_1+U_2)^2(Z_1+Z_2)]
 =2(1+\theta_Z+2/D)\E[U^2Z].
\]
Finally, $\E[U_1^3U_2]=(2+a^2)/D$ and
$\E[U_1^2U_2^2]=1-a^2/D+D^{-2}$. Hence the fourth cumulant for $\rev{\kappa_{\mathrm N}}$ is
\[
 2(2+a^2)+\frac8D(2+a^2)
 +6(1-a^2/D+D^{-2})-3(2d/D)^2
 =\frac{d^2(D-5)}{D^2}.
\]
For $\rev{\kappa_{\mathrm D}}$, the fourth cumulant is $\E U^4-3=a^2-1$.

Apply the common expansion to $\rev{I_{\mathrm D}}(u,v)$ and $\rev{I_{\mathrm N}}(2u,2v)$ using
these two columns, and substitute into \eqref{eq:local-rate-legendre},
with $\logshift=O(N^{-1/2})$ as in the stated parameter range.
The quadratic imbalance terms cancel exactly:
\[
 2D\frac{u^2}{2}-d\frac{(2u)^2}{2(2d/D)}=0.
\]
Completing squares in $v_1,v_2$ gives
\begin{align}\label{eq:diagonal-quartic-polynomial}
 &\psi_{\logshift}(\rho,b)-\psi_{\logshift}(0,b_c)
 =\frac{D-1}{d^2}\logshift^2+\frac{D-1}{2d}\logshift u^2
     -\frac{D-1}{48}u^4\nonumber\\*
 &\quad-\frac{2D}{D-1}[v_1-a(u^2/4+\logshift/d)]^2
       -\frac{D(D-1)}{D^2+1}(v_2+u^2/2)^2
       +O(N^{-5/4}).
\end{align}
In a fixed neighborhood of $(u,v)=(0,0)$, the omitted analytic terms
have total degree at least three in $(u^2,v_1,v_2)$, by evenness in $u$
and the computed coefficients. They are $O_d((u^2+\|v\|)^3)$,
with $v$-gradient $O_d((u^2+\|v\|)^2)$.
The $\logshift$ term in \eqref{eq:local-rate-legendre} is exact.

\emph{The exact center and the uniform error.}
We now replace the centers of the squares in \eqref{eq:diagonal-quartic-polynomial}
by the exact maximizer and express the expansion in $r,s$.
The analytic inverse maps above make $\rev{I_{\mathrm D}},\rev{I_{\mathrm N}}$, and hence
$\psi_{\logshift}$, analytic near the critical point.
The coordinate relations \eqref{eq:diagonal-mixed-coordinates} give
\[
 \frac{4D}{D-1}v_1^2+\frac{2D(D-1)}{D^2+1}v_2^2
 =(b-b_c)^{\mathsf T}H_d(b-b_c),
\]
with $H_d$ as in \eqref{eq:H-d}. Thus the expansion gives
$\nabla_b\psi_0(0,b_c)=0$ and $\nabla_b^2\psi_0(0,b_c)=-H_d$.
Since $H_d$ is positive definite, the
implicit-function theorem gives a unique analytic solution
$b_{\logshift}^{*}(\rho)$ of
$\nabla_b\psi_{\logshift}(\rho,b_{\logshift}^{*}(\rho))=0$
in a fixed neighborhood of $b_c$. Shrinking this neighborhood so that
the Hessian remains negative definite makes it the unique local maximizer.
Symmetry makes the center even in $\rho$, so
$b_{\logshift}^{*}(\rho)-b_c=O_d(\rho^2+|\logshift|)$.
Since $s=\sqrt{N/2}(b-b_{\logshift}^{*}(\rho))$ by
\eqref{eq:exact-stable-center}, bounded $r,s$ and the fixed bounds on
$L=2N\delta^2$ give
$u=O_{d,R}(N^{-1/4})$, $v=O_{d,R,W}(N^{-1/2})$, and
$\logshift=O(N^{-1/2})$, as required above.

At $b=b_{\logshift}^{*}(\rho)$, we have
$\nabla_v\psi_{\logshift}=0$ and $v=O_d(u^2+|\logshift|)$.
The gradient bound following \eqref{eq:diagonal-quartic-polynomial} therefore gives
\begin{equation}\label{eq:w-star}
 \bigl(v_1-a(u^2/4+\logshift/d),\ v_2+u^2/2\bigr)
 =O_d((u^2+|\logshift|)^2)=O_{d,R}(N^{-1}).
\end{equation}
Since \eqref{eq:diagonal-mixed-coordinates} is a fixed invertible affine
map from $b$ to $v$, the two centers also differ by $O_{d,R}(N^{-1})$
in $b$, or $O_{d,R}(N^{-1/2})$ after scaling by $\sqrt{N/2}$.
Since $s$ is bounded, the two
negative quadratic terms in \eqref{eq:diagonal-quartic-polynomial},
after multiplication by $N/2$, sum to
$-\frac12s^{\mathsf T}H_ds+O_{d,R,W}(N^{-1/2})$.
For the remaining terms, $r=\rho/\sqrt{\logshift}$ from
\eqref{eq:exact-stable-center} and \eqref{eq:diagonal-mixed-coordinates}
identify $V_d$ in \eqref{eq:Vd}:
\[
 \frac N2\left[\frac{D-1}{2d}\logshift u^2-\frac{D-1}{48}u^4\right]
 =2N\logshift^2V_d(r).
\]
Subtracting \eqref{eq:diagonal-quartic-polynomial} at
$(\rho,b)=(0,b_{\logshift}^{*}(0))$, where $r=s=0$, cancels its constant term.
Multiplying by $N/2$ therefore yields
\[
 \frac N2[\psi_{\logshift}(\rho,b)-\psi_{\logshift}(0,b_{\logshift}^{*}(0))]
 =2N\logshift^2V_d(r)-\frac12s^{\mathsf T}H_ds
   +O_{d,R,W}(N^{-1/4}).
\]
The center replacement and Taylor remainder contribute
$O(N^{-1/2})+N\,O(N^{-5/4})=O(N^{-1/4})$, proving the lemma.
\end{proof}

\subsubsection{Tail bounds and normalization}

We bound the weight outside compact windows in $r,s$, then sum the local
expansion over the type lattice to obtain the normalized one-gadget law.

\begin{proposition}\label{prop:quartic-coercivity}\label{lem:uniform-cell-prefactor}
There are a fixed
interior neighborhood $\mathcal U$ of $(0,b_c)$ and positive constants $c,C,\logshift_0$ such that, for
${0\le\logshift\le\logshift_0}$ and on $\mathcal U$,
\begin{align}
 \psi_{\logshift}(\rho,b)&\le\psi_{\logshift}(\rho,b_{\logshift}^{*}(\rho))
                   -c\|b-b_{\logshift}^{*}(\rho)\|^2,\label{eq:stable-rate-loss}\\
 \psi_{\logshift}(\rho,b_{\logshift}^{*}(\rho))-\psi_{\logshift}(0,b_c)
 &\le C\logshift^2+C\logshift\rho^2-c\rho^4.\label{eq:quartic-rate-loss}
\end{align}
For types in $\mathcal U$ with $\E_0Z_{G_v}(\alpha)>0$,
\begin{equation}\label{eq:uniform-cell-prefactor}
 \log\E_0Z_{G_v}(\alpha)
 =\frac N2\psi_{\logshift}(\rho,b)-\frac32\log N+O_d(1).
\end{equation}
For every type in the domain of \eqref{eq:one-gadget-rate},
$\E_0Z_{G_v}(\alpha)\le N^{C_d}e^{N\psi_{\logshift}(\rho,b)/2}$
for some $C_d>0$. Outside $\mathcal U$,
$\psi_{\logshift}(\rho,b)\le\psi_{\logshift}(0,b_c)-c$
after a possibly smaller $\logshift_0$.
\end{proposition}

\begin{proof}
By \cref{lem:local-quartic-rate}, the Hessian in $b$ is uniformly negative
definite in a sufficiently small fixed neighborhood. Taylor expansion
about the local maximizer gives \eqref{eq:stable-rate-loss}.
On a fixed neighborhood of $(\rho,\logshift)=(0,0)$,
evaluate \eqref{eq:diagonal-quartic-polynomial} at the center using
\eqref{eq:w-star} and the fixed-neighborhood analytic remainder bound
following \eqref{eq:diagonal-quartic-polynomial}.
\rev{Evenness in $\rho$ and
$2|\logshift|\rho^4\le\logshift^2\rho^2+\rho^6$ give}
\begin{equation}\label{eq:optimized-rate-expansion}
 \begin{aligned}
 \psi_{\logshift}(\rho,b_{\logshift}^{*}(\rho))
   -\psi_{\logshift}(0,b_{\logshift}^{*}(0))
 =4A_d\logshift\rho^2-4B_d\rho^4
   +\rev{O_d(\logshift^2\rho^2+\rho^6)},
 \end{aligned}
\end{equation}
where $A_d,B_d>0$ are the coefficients of $V_d$ in \eqref{eq:Vd}.
Also, $b_{\logshift}^{*}(0)-b_c=O_d(\logshift)$ and
$\nabla_b\psi_{\logshift}(0,b_c)=O_d(\logshift)$, so
$\psi_{\logshift}(0,b_{\logshift}^{*}(0))-\psi_{\logshift}(0,b_c)
=O_d(\logshift^2)$. Shrinking the neighborhood and $\logshift_0$
absorbs the remainder and proves \eqref{eq:quartic-rate-loss}.

\Cref{lem:expected-type-weight} gives
\eqref{eq:uniform-cell-prefactor}. For the global upper bound, use
\cref{lem:finite-polynomial-coefficients}(ii) and Stirling's polynomial
lower bound for the multinomial denominator in \eqref{eq:alpha-coefficient};
these also cover zero state counts.

\rev{The local estimates remain valid on a smaller fixed
neighborhood $\mathcal U$ of $(0,b_c)$.
By \cref{prop:global-critical-anchor} and compactness,
there is a positive critical gap outside each such neighborhood.
The fugacity perturbation is uniformly $O(\logshift)$ on the
domain of \eqref{eq:one-gadget-rate}, so half this gap persists
for sufficiently small $\logshift$.
Thus $\mathcal U$ may also meet the local requirements in
\cref{def:restricted-phase-weight}, after adjusting the constants
and $\logshift_0$.}
\end{proof}

\begin{proof}[Proof of \Cref{prop:uniform-normalized-tails}]
Recall from the proof of \cref{prop:quartic-expansion} that
\[
 e^{c_N}=N^{-3/2}h_d(\alpha_c)
       e^{N\psi_{\logshift}(0,b_{\logshift}^{*}(0))/2}.
\]
The prefactor bounds and
\eqref{eq:stable-rate-loss}--\eqref{eq:quartic-rate-loss} imply
\[
 \E_0Z_{G_v}(\alpha)\le C e^{c_N}e^{Cr^2-cr^4-c\|s\|^2}
 \qquad((\rho,b)\in\mathcal U),
\]
since $N\logshift^2$ is bounded above and away from zero.
The gap outside $\mathcal U$ absorbs the polynomial prefactor and the
polynomial number of types, giving, after decreasing $c$,
\[
 \sum_{\alpha:\,(\rho,b)\notin\mathcal U}\E_0Z_{G_v}(\alpha)
 \le e^{c_N}e^{-cN}.
\]

We sum first over $s$ at each fixed $r$, then over $r$.
At fixed $n_{10}-n_{01}$, parity requires
$n_{10}+n_{01}\equiv n_{10}-n_{01}\pmod2$.
Thus consecutive values of $b_1,b_2$ are spaced $4/N,2/N$ apart.
By \eqref{eq:exact-stable-center}, the allowed $s$-values lie on a rectangular
lattice with spacings $2\sqrt{2/N},\sqrt{2/N}$. Subtracting
$b_{\logshift}^{*}(\rho)$ shifts this lattice without changing the spacings.
Each lattice cell has area $4/N$, so converting a sum over $s$ to an
integral gives a factor $N/4$. Consecutive $r$-values are spaced
$(N\sqrt{\logshift})^{-1}$ apart, giving the total factor
$N^2\sqrt{\logshift}/4$ when summing over both $r$ and $s$.
For $R,W\ge2$, comparing the Gaussian and quartic sums with integrals over their cells,
uniformly in the lattice's position, bounds the tail sets specified in
\eqref{eq:uniform-integrated-tail} by
\[
 \frac{\E_0Z_{G_v}(\{|r|>R\}\cup\{\|s\|_\infty>W\}\cup\mathcal U^c)}
      {e^{c_N}N^2\sqrt{\logshift}/4}
 \le C\left[\int_{|u|>R-1}e^{Cu^2-cu^4}\dd u+e^{-cW^2}\right]+e^{-cN}.
\]
Here the $\mathcal U^c$ term uses the exponential bound above.

On a fixed compact box, \eqref{eq:quartic-normal-form} approximates the
type weights by $e^{c_N}e^{LV_d(r)}e^{-s^{\mathsf T}H_ds/2}$.
All lattice types in this box have positive coefficients for large $N$
by \cref{lem:finite-polynomial-coefficients}(i).
For each allowed $r$ in this box, the corresponding sum over $s$ satisfies
\[
 \frac4N\sum_{\|s\|_\infty\le W}e^{-s^{\mathsf T}H_ds/2}
 =\int_{[-W,W]^2}e^{-s^{\mathsf T}H_ds/2}\dd s+O_{d,W}(N^{-1/2}),
\]
uniformly in the lattice's position. Multiplying the subsequent sum over
$r$ by its spacing gives the one-dimensional Riemann sum.
For any given accuracy, first choose the box using the unnormalized tail bound above,
then choose $N$ to control the remaining errors. This yields
\[
 \E_0Z_{G_v}=e^{c_N}\frac{N^2\sqrt{\logshift}}4\,
       \frac{2\pi}{\sqrt{\det H_d}}
       \left[\int_{\mathbb R}e^{LV_d(r)}\dd r+o_N(1)\right].
\]
The integral is uniformly bounded away from zero for the allowed $L$, so
$e^{c_N}N^2\sqrt{\logshift}/(4\E_0Z_{G_v})=O(1)$.
Using this bounded factor in the type and tail estimates proves
\eqref{eq:normalized-cell-envelope} and
\eqref{eq:uniform-integrated-tail}.
Dividing the restricted sum by this expression proves
\eqref{eq:bin-integral-mean}; the common factor
$e^{c_N}N^2\sqrt{\logshift}/4$ cancels.
For $\mathcal J$ of positive length and $W>0$, the limiting factors are
positive and continuous in $L$, so its compact range gives the claimed
positive lower bound.

The tail bound removes the restrictions on $s$ and $\mathcal U$ to give
\eqref{eq:one-gadget-imbalance-marginal}. On a fixed compact $r$ interval,
the endpoints of $\mathcal J$ cut at most two intervals between consecutive
$r$-values. Their normalized contribution is
$O((N\sqrt{\logshift})^{-1})$, uniformly in the endpoints.
The tail bound extends this uniformity to all intervals, including
half-lines. At $r=0$, summing the type bound over $s$ and adding the
exponentially small contribution from $\mathcal U^c$ gives
\[
 \frac{\E_0Z_{G_v}(\rho=0)}{\E_0Z_{G_v}}
 =O\!\left(\frac1{N\sqrt{\logshift}}\right)=O(N^{-3/4}),
\]
so assigning zero imbalance to the positive phase does not affect the limit.
After fixing cutoffs via \eqref{eq:uniform-integrated-tail}, the quantitative bounds in
\eqref{eq:quartic-normal-form} and
\cref{lem:finite-polynomial-coefficients}(i), together with the compact-box
Riemann-sum estimates, give a computable threshold for $N$ at any given accuracy.
\end{proof}

For $G_v$ before connections, with $g=2N$ vertices, let
$Z_{G_v}(\rho\in\mathcal J)$ denote its partition function at fugacity
$(1+\delta)\lambdaC$, restricted to imbalances in $\mathcal J$.
Recall that $\logshift=\log(1+\delta)$ and $\E_0$ denotes expectation
before conditioning on simplicity.

\begin{proposition}[Quartic imbalance law]
\label{prop:imbalance-law}
Fix $d\ge3$ and a compact interval $\mathcal L\subset(0,\infty)$, and put
$\Phi_\delta(\rho)=A_d\logshift\,\rho^2-B_d\rho^4$, where
$A_d,B_d$ are defined in \eqref{eq:AB-constants}.
As $\delta\to0$ with
$g\logshift^2\in\mathcal L$, uniformly over intervals
$\mathcal J\subseteq\mathbb R$,
\[
  \frac{\E_0 Z_{G_v}(\rho\in\mathcal J)}{\E_0 Z_{G_v}}
 =\frac{\int_{\mathcal J}e^{g\Phi_\delta(\rho)}\dd\rho}
       {\int_{\mathbb R}e^{g\Phi_\delta(\rho)}\dd\rho}+o(1).
\]
\end{proposition}

\begin{proof}[Proof of \Cref{prop:imbalance-law}]
Recall that $g=2N$. Apply \eqref{eq:one-gadget-imbalance-marginal} to
$\mathcal J/\sqrt{\logshift}$ and substitute
$\rho=\sqrt{\logshift}\,r$. The exact identity is
$g\Phi_\delta(\rho)=2N\logshift^2V_d(r)$.
Since $2N\logshift^2=L+O(\delta)$, replacing $L$ by this coefficient
changes the normalized integrals by $o_N(1)$, uniformly in the interval:
use convergence on a compact set and the common quartic tails outside it.
The factor $\sqrt{\logshift}$ in $\dd\rho=\sqrt{\logshift}\,\dd r$
cancels from the ratio of integrals. This gives the claimed
uniform interval law.
\end{proof}

\subsection{The effect of gadget connections}
\label{app:port-details}\label{sec:ports}

We compute how connecting the gadgets changes their expected weights at
fixed occupation types. Let $Z_H(\boldsymbol\alpha)$ be the sum of
$\lambda^{|I|}$ over independent sets $I$ of the output graph $G$ whose
type in each gadget $v$ is $\alpha_v$.

\begin{lemma}[Connection weights]
\label{lem:compact-connection-comparison}
Fix $d\ge3$, positive lower and upper bounds on $L=2N\delta^2$ and
$k\delta$, and finite $R,W>0$. There exists $\delta_0>0$ such that,
for every $0<\delta\le\delta_0$, even $N$ and positive integer $k$
satisfying these bounds and $3k\le N/2$,
the following holds for every simple cubic graph $H$ on $m$ vertices.
For every type assignment $\boldsymbol\alpha$ with $\E_0Z_{G_v}(\alpha_v)>0$
at each $v$ and in the window $|r_v|\le R$, $\|s_v\|_\infty\le W$
of \eqref{eq:exact-stable-center},
\begin{equation}\label{eq:connection-comparison-goal}
 \frac{\E_0Z_H(\boldsymbol\alpha)}
      {\prod_v\E_0Z_{G_v}(\alpha_v)}
 =\exp\left\{-\frac{k\delta}{2}\sum_{uv\in E(H)}(r_u+r_v)^2
                  +O(\logshift)m\right\}.
\end{equation}
The error constant and $\delta_0$ depend only on
$d,R,W$ and the fixed parameter bounds.
\end{lemma}
We first compute the change for one connection block, then compare the
coefficients for all blocks simultaneously: evaluating polynomials at
positive arguments does not itself enforce the occupation counts. The coefficient
comparison also supplies a one-sided bound for tails and a general form
used for mixed moments.

Here $N\logshift^2=\Theta(1)$, $k\logshift=k\delta+O(\logshift)$,
and $k=\Theta(\sqrt N)$.

\subsubsection{The weight of one connection block}

For each gadget $u$, assign the monomials $1,x_u,y_u,z_u$ to the states
$00,10,01,11$, respectively, and write $P(u)=P(x_u,y_u,z_u)$,
with $P$ defined in \eqref{eq:P-poly}.
Take designated matching edges $(i_u,j_u)$ and $(i_v,j_v)$, oriented as in
\cref{constr:graph-reduction}, with the wiring in \cref{fig:port-wiring}.
The new edges $i_u^+i_v^+$ and $i_u^-i_v^-$
forbid the states at labels $i_u,i_v$ from both occupying $+$ or both
occupying $-$. \rev{If neither label is in state $00$, only $10/01$ and
$01/10$ are allowed.} Thus their generating polynomial is
\begin{equation}\label{eq:S-poly}
  S(u,v)=1+x_u+y_u+z_u+x_v+y_v+z_v+x_uy_v+y_ux_v.
\end{equation}
The labels $j_u,j_v$ give the same factor, so the connected block has
polynomial $S(u,v)^2$. Keeping or restoring its original lifts gives
$P(u)P(v)$.

For positive $x_u,y_u,z_u$, define coordinates $\eta_u,c_u,\zeta_u$ by
\begin{equation}\label{eq:scq-coords}
 x_u=\frac{e^{\eta_u+c_u}}{d-2},\qquad
 y_u=\frac{e^{\eta_u-c_u}}{d-2},\qquad
 z_u=\frac{e^{\zeta_u}}{(d-2)^2},
\end{equation}
and similarly at $v$.
The variable $c_u$ changes the weights of states
$10$ and $01$ in opposite directions, $\eta_u$ changes them by the same
factor, and $\zeta_u$ changes the weight of $11$. Write
$c=(c_u,c_v)$, $\eta=(\eta_u,\eta_v)$, and $\zeta=(\zeta_u,\zeta_v)$.
The origin corresponds to the critical type $\alpha_c$ in
\eqref{eq:critical-type}, as computed in the proof of \cref{lem:expected-type-weight}.

\begin{proposition}\label{prop:port-hessian}
Let
\begin{equation}\label{eq:connection-log-ratio}
  h:=\log\frac{S(u,v)^2}{P(u)P(v)}.
\end{equation}
At $c=\eta=\zeta=0$,
\[
  h=0,
  \qquad
  \nabla h=0.
\]
Here the gradient is with respect to all six coordinates. At the same point,
the Hessian in $(c_u,c_v)$ is
\[
  \nabla^2_{(c_u,c_v)}h
  =-\frac{4}{d^2}
    \begin{pmatrix}1&1\\1&1\end{pmatrix},
\]
and every mixed second derivative between $c$ and $(\eta,\zeta)$ vanishes.
In a fixed neighborhood of zero,
\begin{equation}\label{eq:h-port-local}
  h=-\frac{2}{d^2}(c_u+c_v)^2
    +O_d\bigl(\|c\|^4+\|(\eta,\zeta)\|^2\bigr).
\end{equation}
\end{proposition}

\begin{proof}
Write $a=1/(d-2)$. At this point, $P(u)=P(v)=S(u,v)=(1+2a)^2$.
The derivatives of $P(u)$ there with respect to $(\eta_u,\zeta_u)$
are $(4a(1+a),2a^2)$, and similarly for $P(v)$.
The derivatives of $S$ with respect to
$(\eta_u,\zeta_u)$ or $(\eta_v,\zeta_v)$ are $(2a(1+a),a^2)$.
Thus $h=0$ and its gradient in $(\eta,\zeta)$ vanishes. Its gradient
in $c$ vanishes because interchanging the two sides of both gadgets
sends $c$ to $-c$ and leaves $h$ unchanged.

With $\eta=\zeta=0$, direct factorization gives
\[
 P(u)=(1+2a\cosh c_u)^2,\qquad P(v)=(1+2a\cosh c_v)^2,
\]
\[
 S(u,v)=(1+2a\cosh c_u)(1+2a\cosh c_v)
           -4a^2\sinh^2\!\left(\frac{c_u+c_v}{2}\right).
\]
Consequently
\[
 h=2\log\left[1-
 \frac{4a^2\sinh^2((c_u+c_v)/2)}
      {(1+2a\cosh c_u)(1+2a\cosh c_v)}\right]
 =-\frac{2}{d^2}(c_u+c_v)^2+O_d(\|c\|^4).
\]
This gives the stated Hessian in $c$. Interchanging both sides gives
$h(c,\eta,\zeta)=h(-c,\eta,\zeta)$, so the mixed second derivatives
between $c$ and $(\eta,\zeta)$ and the cubic terms involving only $c$ vanish.
Taylor's theorem and
$\|c\|^2\|(\eta,\zeta)\|\le(\|c\|^4+\|(\eta,\zeta)\|^2)/2$
now give \eqref{eq:h-port-local}.
\end{proof}

We now express the block ratio in occupation coordinates.
For each type with $(\rho_v,b_v)\in\mathcal U$, choose positive
$x_v,y_v,z_v$ and define $\theta_v:=(\log x_v,\log y_v,\log z_v)$,
with the values chosen to satisfy
\begin{equation}\label{eq:connection-type-tilt}
 \nabla_{\theta_v}\log P(e^{\theta_v})
 =2(\alpha_{v,10},\alpha_{v,01},\alpha_{v,11}).
\end{equation}
The solution is unique near $c_v=\eta_v=\zeta_v=0$
and depends analytically on the type, as shown in the proof of
\cref{lem:expected-type-weight}. We write $h(\theta_u,\theta_v)$ for
\eqref{eq:connection-log-ratio} in these logarithmic coordinates.
In the coordinates of \eqref{eq:scq-coords}, the mean constraint and
\eqref{eq:XN-law} give, suppressing the vertex subscript,
\[
 \partial_c\log P=2\rho,\qquad
 \left.\partial_c^2\log P\right|_{c=\eta=\zeta=0}
 =\Var(X_1-X_2)=\frac4d.
\]
Interchanging $+$ and $-$ makes $c$ odd and $\eta,\zeta$ even in $\rho$.
Differentiating the first identity at $(\rho,b)=(0,b_c)$ therefore gives
$\partial_\rho c=d/2$.
Using $b-b_{\logshift}^{*}(\rho)=\sqrt{2/N}\,s$ from
\eqref{eq:exact-stable-center} and
$b_{\logshift}^{*}(\rho)-b_c=O_d(\logshift+\rho^2)$ gives
\begin{equation}\label{eq:c-rho}
 c=\frac d2\rho+O_d\bigl((\logshift+\rho^2+\|s\|/\sqrt N)|\rho|\bigr).
\end{equation}
The other two variables satisfy
$\eta,\zeta=O_d(\logshift+\rho^2+\|s\|/\sqrt N)$.

\begin{proposition}
\label{prop:marked-local-coercivity}
\rev{For types in the fixed neighborhood $\mathcal U$ chosen in
\cref{def:restricted-phase-weight}, and for the allowed parameters,}
let $\theta_u,\theta_v$ satisfy
\eqref{eq:connection-type-tilt} for their respective types. Then
\begin{equation}\label{eq:marked-local-coercive}
 \left|kh(\theta_u,\theta_v)+\frac{k\delta}{2}(r_u+r_v)^2\right|
 \le C\logshift\left(1+r_u^4+r_v^4+\|s_u\|^2+\|s_v\|^2\right).
\end{equation}
The constants depend only on $d$ and the fixed parameter bounds.
\end{proposition}

\begin{proof}
The coordinate relation \eqref{eq:c-rho} and the bounds
$\eta,\zeta=O_d(\logshift+\rho^2+\|s\|/\sqrt N)$, inserted in
\eqref{eq:h-port-local}, give
\[
 \left|h+\tfrac12(\rho_u+\rho_v)^2\right|
 \le C\left(\logshift^2+\rho_u^4+\rho_v^4+(\|s_u\|^2+\|s_v\|^2)/N\right).
\]
Here $\logshift\rho^2$ and $\rho^2\|s\|/\sqrt N$ are bounded by squares.
Multiply by $k=O(\logshift^{-1})$ and use $\rho=\sqrt{\logshift}\,r$,
$k/N=O(\logshift)$.
Replacing $k\logshift$ by $k\delta$ contributes
$O(\logshift(r_u^2+r_v^2))$, which is absorbed into the displayed bound.
\end{proof}

\subsubsection{Coefficient comparison}

We first express the expected number of independent sets with given
occupation counts as a coefficient, then compare coefficients under
block replacement.

For the following matching identity, use the ordered matchings and block
pairing of \cref{constr:graph-reduction} under $\mathbb P_0$.
Allow $H$ to have maximum degree at most three. At each $v$, use the
first $\deg_H(v)k$ edges of $M_{v,1}$, with $\deg_H(v)k\le N/2$ and $N$ even.

Fix $F\subseteq E(H)$. Define $G^{(1)}(F)$ by the same construction as
$G$, keeping all vertices but using only $M_{v,1}$ instead of all $d$
matchings in each $G_v$. Use the same matching realizations, orders,
orientations, and block pairing. For $uv\in F$, connect all its $k$
blocks; for $uv\notin F$, keep their original edges.
The corresponding one-block polynomial is $Q_{uv}=S(u,v)^2$ for
$uv\in F$ and $Q_{uv}=P(u)P(v)$ otherwise.
For a vertex set $I$, let $n_v(I)$ be the vector counting labels $i$ for which
$(\mathbf1_{\{i_v^+\in I\}},\mathbf1_{\{i_v^-\in I\}})$ equals
$10,01,11$, respectively. For counts $n_v=(n_{v,10},n_{v,01},n_{v,11})$,
write $\mathbf z_v=(x_v,y_v,z_v)$ and
$\mathbf z_v^{n_v}=x_v^{n_{v,10}}y_v^{n_{v,01}}z_v^{n_{v,11}}$.
State $00$ has weight $1$; use $P$ from \eqref{eq:P-poly}.

\begin{lemma}
\label{lem:ordered-slot-factorization}
For occupation counts $\mathbf n=(n_v)_v$ with $n_v\in\mathbb Z_{\ge0}^3$
and $n_{v,10}+n_{v,01}+n_{v,11}\le N$,
\begin{equation}\label{eq:global-marked-coefficient}
  \E_0\left|\left\{I\in\Ind(G^{(1)}(F)):
                       n_v(I)=n_v\ \forall v\right\}\right|
  =
  \left[\prod_v\mathbf z_v^{n_v}\right]
    \prod_v P(\mathbf z_v)^{N/2-\deg_H(v)k}
    \prod_{uv\in E(H)}Q_{uv}^{k}.
\end{equation}
\end{lemma}

\begin{proof}
As in the proof of \cref{lem:expected-type-weight}, generate each matching
by uniformly permuting the labels and pairing consecutive positions;
the permutation also gives the required edge order and orientations.
At each $v$, fix a labeled state assignment with counts $n_v$, including
$n_{v,00}=N-n_{v,10}-n_{v,01}-n_{v,11}$. Its state sequence in the ordered
endpoint positions is uniform among sequences with those counts, each
having probability $\prod_{a\in\Omega}n_{v,a}!/N!$, where
$\Omega=\{00,10,01,11\}$.

Each matching edge outside the blocks contributes $P$, and each block
contributes $Q_{uv}$. Hence the coefficient in
\eqref{eq:global-marked-coefficient} counts state sequences that place
no two occupied vertices at the ends of an edge of $G^{(1)}(F)$.
For a fixed vertex set $I$ with these counts, the probability
that $I\in\Ind(G^{(1)}(F))$ is this coefficient multiplied by
$\prod_v(\prod_a n_{v,a}!/N!)$.
Summing over the $\prod_v(N!/\prod_a n_{v,a}!)$ assignments cancels this
factor and proves the identity.
\end{proof}

The next lemma separates the evaluated block ratios from the correction
needed to enforce the given counts.

\begin{lemma}[Coefficient comparison]
\label{lem:normalized-block-replacement}\label{lem:conditional-lclt-marked}\label{lem:residual-targets-square-root}
Let $H$ have bounded maximum degree and $m$ vertices. Fix a dimension $p$
and variables $\mathbf z_v=(z_{v,1},\ldots,z_{v,p})$ at each vertex.
Let $P_v(\mathbf z_v)$ and $Q_e(\mathbf z_u,\mathbf z_v)$, $e=uv$,
range over fixed finite families of nonzero polynomials with nonnegative
coefficients. Assume each $P_v$ has positive coefficients at
$1,z_{v,1},\ldots,z_{v,p}$.
Let $N$ be even and $k$ a nonnegative integer with $k/\sqrt N$ bounded
above by a fixed constant. Consider integer targets $n_v\in\mathbb Z^p$
admitting $\theta_v$ in a fixed compact set with
\[
 n_v=\frac N2\mu_v,\qquad \mu_v=\nabla\log P_v(e^{\theta_v}).
\]
Let $\mathfrak C_Q$ be the coefficient on the right-hand side of
\eqref{eq:global-marked-coefficient}, using $P_v$ in place of $P$ and
blocks $Q_e$. Replacing each $Q_e$ by $P_uP_v$ gives the reference coefficient
\[
 \mathfrak C_0(\mathbf n)
 =\prod_v[\mathbf z_v^{n_v}]P_v(\mathbf z_v)^{N/2}.
\]
For $e=uv$, put
\[
 h_e=\log\frac{Q_e(e^{\theta_u},e^{\theta_v})}
                       {P_u(e^{\theta_u})P_v(e^{\theta_v})},\qquad
 b_{e\to v}=\nabla_{\theta_v}h_e.
\]
There is a constant $C$ such that, for all sufficiently large $N$,
both coefficients are positive and
\begin{equation}\label{eq:normalized-block-replacement}
 -C\left(\frac m{\sqrt N}
       +\frac{k^2}{N}\sum_v\left\|\sum_{e\ni v}b_{e\to v}\right\|^2\right)
 \le \log\frac{\mathfrak C_Q(\mathbf n)}{\mathfrak C_0(\mathbf n)}
       -k\sum_e h_e
 \le \frac{Cm}{\sqrt N}.
\end{equation}
The constant $C$ and the lower bound on $N$ depend only on the fixed
polynomial families, compact tilt set, degree bound, and bound on $k/\sqrt N$.
\end{lemma}

\begin{proof}
All probabilities and expectations below are for independent exponent
vectors with the following tilted laws.
Give each exponent $a$ of $P_v$ probability
$[\mathbf z_v^a]P_v\,e^{\langle\theta_v,a\rangle}/P_v(e^{\theta_v})$,
with mean $\mu_v$ and covariance $\Sigma_v$.
There are $\ell_v:=N/2-k\deg_H(v)$ such factors at $v$.
For each of the $k$ copies of $Q_e$, let
$Y_{e,i}=(Y_{e,i}^u,Y_{e,i}^v)$ have law
\[
 \Pp\bigl(Y_{e,i}=(a,b)\bigr)
 =\frac{[\mathbf z_u^a\mathbf z_v^b]Q_e\,
          e^{\langle\theta_u,a\rangle+\langle\theta_v,b\rangle}}
        {Q_e(e^{\theta_u},e^{\theta_v})}.
\]
Then $\E Y_{e,i}^v-\mu_v=b_{e\to v}$:
\added{this is the change in the mean exponent vector at $v$
when $P_uP_v$ is replaced by $Q_e$ under these tilted laws.}
Put
\[
 U_v=\sum_{e\ni v}\sum_{i=1}^k(Y_{e,i}^v-\mu_v),\qquad
 B(Y)=\sum_v\frac{U_v^{\mathsf T}\Sigma_v^{-1}U_v}{2\ell_v}.
\]
After conditioning on all $Y$, the counts still required at $v$ form the integer vector
$\ell_v\mu_v-U_v=n_v-\sum_{e\ni v}\sum_{i=1}^kY_{e,i}^v$, and
$\|U_v\|\le C\sqrt N$ deterministically. Here $\ell_v\asymp N$.
The support assumption gives
difference lattice $\mathbb Z^p$ and covariance eigenvalues bounded above
and away from zero over the compact set of $\theta_v$.
Thus \cref{lem:moving-lclt}, after undoing the tilt, gives
\[
 [\mathbf z_v^{\ell_v\mu_v-U_v}]P_v(\mathbf z_v)^{\ell_v}
 =
 \frac{P_v(e^{\theta_v})^{\ell_v}
       e^{-\langle\theta_v,\ell_v\mu_v-U_v\rangle}}
      {(2\pi\ell_v)^{p/2}\sqrt{\det\Sigma_v}}
 e^{-U_v^{\mathsf T}\Sigma_v^{-1}U_v/(2\ell_v)}
 \bigl(1+O(N^{-1/2})\bigr),
\]
uniformly in $Y$; hence these coefficients are positive for sufficiently
large $N$.
Let $B_0$ be $B(Y)$ under the same construction with each $Q_e$ replaced
by $P_uP_v$.
Summing over the block exponents combines the tilt factors into
$e^{-\sum_v\langle\theta_v,n_v\rangle}$.
This factor, the Gaussian prefactors, and
$\prod_v P_v(e^{\theta_v})^{\ell_v}$ cancel in the ratio.
The block normalizers give $e^{k\sum_eh_e}$, so multiplying the uniform
relative errors gives
\[
 \frac{\mathfrak C_Q(\mathbf n)}{\mathfrak C_0(\mathbf n)}
 =e^{k\sum_eh_e}
   \frac{\E e^{-B}}{\E e^{-B_0}}\,e^{O(m/\sqrt N)}.
\]
Since $B,B_0\ge0$, Jensen's inequality yields
\[
 -\E B\le\log\frac{\E e^{-B}}{\E e^{-B_0}}\le\E B_0.
\]
Independence of the bounded block copies gives
$\Var U_v\preceq CkI$ and
$\E U_v=k\sum_{e\ni v}b_{e\to v}$. The bounds
$\|\Sigma_v^{-1}\|\le C$ and $\ell_v\asymp N$ imply
\[
 \E B\le C\left(\frac{km}{N}
       +\frac{k^2}{N}\sum_v\left\|\sum_{e\ni v}b_{e\to v}\right\|^2\right).
\]
For the reference blocks every $b_{e\to v}$ is zero, so
$\E B_0=O(km/N)$. Substituting these bounds in the coefficient ratio
gives \eqref{eq:normalized-block-replacement},
using $k/N=O(N^{-1/2})$.
\end{proof}

\subsubsection{From coefficients to expected weights}

The tail estimates in \cref{sec:first-moment-restoration} require an
upper bound throughout $\mathcal U$; on fixed windows we also obtain
a matching lower bound.

\begin{lemma}
\label{lem:local-marked-coefficient-bound}
Under this section's parameter bounds, let $H$ have maximum degree
three and $m$ vertices. For every type
assignment $\boldsymbol\alpha$ with $(\rho_v,b_v)\in\mathcal U$ for every $v$,
take $\theta_v$ satisfying \eqref{eq:connection-type-tilt} and put
\begin{equation}\label{eq:comparison-type-weight}
 W_H(\boldsymbol\alpha)
 :=\prod_v\E_0Z_{G_v}(\alpha_v)
     \exp\left\{k\sum_{uv\in E(H)}h(\theta_u,\theta_v)\right\}.
\end{equation}
For sufficiently large $N$,
\begin{equation}\label{eq:local-marked-upper}
 \E_0Z_H(\boldsymbol\alpha)
       \le e^{Cm/\sqrt N}W_H(\boldsymbol\alpha).
\end{equation}
For fixed $R,W>0$, if the coordinates in \eqref{eq:exact-stable-center}
satisfy $|r_v|\le R$, $\|s_v\|_\infty\le W$ for every $v$, then
\[
 e^{-C_{R,W}m/\sqrt N}W_H(\boldsymbol\alpha)
 \le \E_0Z_H(\boldsymbol\alpha).
\]
The constants and size thresholds depend only on $d$ and the fixed
parameter bounds, and additionally on $R,W$ for the lower bound.
\end{lemma}

\begin{proof}
Put $n_v=N(\alpha_{v,10},\alpha_{v,01},\alpha_{v,11})$ and use
$\mathfrak C_Q,\mathfrak C_0$ from \cref{lem:normalized-block-replacement}
with $P_v=P$ and $Q_{uv}=S(u,v)^2$, so $h_e=h(\theta_u,\theta_v)$.
Equation~\eqref{eq:connection-type-tilt} gives the required mean condition.
The exponent support of $P$ contains the origin and the three unit vectors,
and the chosen $\theta_v$ lie in a compact set. Thus that lemma applies,
and both coefficients are positive for large $N$.
By \cref{lem:ordered-slot-factorization},
\[
 \frac{\E_0Z_H(\boldsymbol\alpha)}
      {\prod_v\E_0Z_{G_v}(\alpha_v)}
 =\frac{\mathfrak C_Q(\mathbf n)}{\mathfrak C_0(\mathbf n)}.
\]
For fixed types, the other matching colors, assignment counts, and fugacity
factors are identical before and after connections and cancel.
The upper estimate in \eqref{eq:normalized-block-replacement} gives
\eqref{eq:local-marked-upper}.

On the fixed window, \eqref{eq:c-rho}
and the bounds following it give $c_v=O(N^{-1/4})$ and
$\eta_v,\zeta_v=O(N^{-1/2})$.
\rev{The vanishing gradient in \cref{prop:port-hessian}
means that, at the critical tilt, replacing $P(u)P(v)$ by $S(u,v)^2$
preserves the mean exponent vector at each endpoint gadget under the
tilted laws of \cref{lem:normalized-block-replacement}.
Analyticity and the affine change of logarithmic coordinates in
\eqref{eq:scq-coords} therefore give
$b_{e\to v}=\nabla_{\theta_v}h_e=O(N^{-1/4})$.}
Since $k^2/N=O(1)$ and $\deg_H(v)\le3$,
\[
 \frac{k^2}{N}\sum_v\left\|\sum_{e\ni v}b_{e\to v}\right\|^2
 =O(m/\sqrt N).
\]
The lower estimate in \eqref{eq:normalized-block-replacement}
therefore gives the claimed lower bound.
\end{proof}

\begin{proof}[Proof of \Cref{lem:compact-connection-comparison}]
On the fixed window, \cref{lem:local-marked-coefficient-bound} and
\eqref{eq:marked-local-coercive} give
\[
 \begin{aligned}
 \log\frac{\E_0Z_H(\boldsymbol\alpha)}
               {\prod_v\E_0Z_{G_v}(\alpha_v)}
 &=k\sum_{uv\in E(H)}h(\theta_u,\theta_v)+O(m/\sqrt N)\\
 &=-\frac{k\delta}{2}\sum_{uv\in E(H)}(r_u+r_v)^2
      +O(\logshift m).
 \end{aligned}
\]
Indeed, each edge contributes $O(\logshift)$ to the error, there are
$3m/2$ edges, and $N^{-1/2}=O(\logshift)$.
\end{proof}

\subsection{Controlling first-moment tails}
\label{sec:first-moment-restoration}

We bound $\E_0Z_\sigma/\E_0X_\sigma$ and use this bound to complete
\cref{lem:normalized-amplitude-representation}. We treat separately types
outside $\mathcal U$ and types inside $\mathcal U$ but outside the \rev{fixed window
$|r_v|\le R$, $\|s_v\|_\infty\le W$ at every vertex, as in
Definition~\ref{def:restricted-phase-weight}}.
Throughout this subsection, $o_N(1)$ tends to zero uniformly in
$H,m,\sigma$ and the allowed parameters, with the cutoffs fixed.

\begin{proposition}[First-moment tails]
\label{prop:wired-first-moment-tails}
Let $X_\sigma$ be the restricted phase weight in
\eqref{eq:restricted-phase-weight} and $Z_\sigma$ the full phase weight.
There is a nonnegative function $\xi_{\rm tail}(R,W)\to0$ as
$\min\{R,W\}\to\infty$ such that, for every fixed sufficiently large
$R,W$, as $N\to\infty$,
\begin{equation}\label{eq:wired-first-moment-tails}
 \E_0Z_\sigma
 \le \exp\{[\xi_{\rm tail}(R,W)+o_N(1)]m\}
       \E_0X_\sigma.
\end{equation}
This holds for every source graph of maximum degree three and every
phase vector.
\rev{The same bound holds after summing $Z_\sigma$ and $X_\sigma$ over any set of phase vectors.}
\end{proposition}

With $W_H$ defined in \eqref{eq:comparison-type-weight}, set
\begin{equation}\label{eq:comparison-weight-sums}
 \mathcal L_{H,\sigma}:=\sum_{\boldsymbol\alpha}W_H(\boldsymbol\alpha),\qquad
 \mathcal L_{H,\sigma}^{R,W}
 :=\sum_{\boldsymbol\alpha:\,\forall v,\ |r_v|\le R,\ \|s_v\|_\infty\le W}
 W_H(\boldsymbol\alpha).
\end{equation}
Both sums run over type assignments with phases $\sigma$ and
$(\rho_v,b_v)\in\mathcal U$ for every $v$.

\begin{proof}[Proof of \Cref{prop:wired-first-moment-tails}]
The upper bound in \cref{lem:local-marked-coefficient-bound} and
\cref{cor:marked-macroscopic-tail} give
\[
 \E_0Z_\sigma\le e^{Cm/\sqrt N}(1+e^{-cN/2})^m
                    \mathcal L_{H,\sigma}.
\]
Fix constants $c,C>0$ as in \cref{lem:local-saddle-tail-clipping} and set
\[
 \xi_{\rm tail}(R,W)
 =\log\left(1+C\int_{|u|>R-1}e^{Cu^2-cu^4}\dd u+Ce^{-cW^2}\right).
\]
That lemma gives
\[
 \mathcal L_{H,\sigma}
 \le e^{m[\xi_{\rm tail}(R,W)+o_N(1)]}\mathcal L_{H,\sigma}^{R,W}.
\]
The lower bound in \cref{lem:local-marked-coefficient-bound} gives
$\mathcal L_{H,\sigma}^{R,W}\le e^{C_{R,W}m/\sqrt N}\E_0X_\sigma$.
Combining the three bounds proves \eqref{eq:wired-first-moment-tails};
positivity preserves the same bound \rev{under sums over phase vectors}.
\end{proof}

\begin{proof}[Proof of \Cref{lem:normalized-amplitude-representation}]
Let $(r,s)\in\mathbb R\times\mathbb R^2$ have probability density
proportional to $e^{LV_d(r)-s^{\mathsf T}H_ds/2}$, the continuous
approximation in \cref{prop:uniform-normalized-tails}.
For fixed $R\ge2r_d$ and $W>0$, define $p_{R,W}$ to be the conditional probability
of $r\le R$ and $\|s\|_\infty\le W$ given $r\ge0$.
Since the density factors into a function of $r$ and a function of $s$,
\[
 p_{R,W}=
 \frac{\int_0^R e^{LV_d(a)}\dd a}{\int_0^\infty e^{LV_d(a)}\dd a}
 \frac{\sqrt{\det H_d}}{2\pi}
     \int_{[-W,W]^2}e^{-s^{\mathsf T}H_ds/2}\dd s.
\]
We suppress the dependence of $p_{R,W}$ on $L$.
By symmetry, this distribution assigns probability $p_{R,W}/2$ to
either phase restricted to the window.
Since $\E_0Z_{G_{\sqcup}}=\prod_v\E_0Z_{G_v}$, combining the normalized
one-gadget law \eqref{eq:bin-integral-mean} with the connection comparison
in \cref{lem:compact-connection-comparison} gives
\begin{equation}\label{eq:compact-normalized-integral}
 \frac{\E_0X_\sigma}{\E_0Z_{G_{\sqcup}}}
 =2^{-m}p_{R,W}^m\mathcal I_{H,\sigma}(R)e^{m\,o_N(1)}.
\end{equation}
To justify the error in \eqref{eq:compact-normalized-integral} uniformly
in $m$, partition the magnitude interval $[0,R]$ into finitely many
intervals of positive length, with maximum length $\ell$.
For each fixed partition, the continuous weights in
\eqref{eq:bin-integral-mean} of its finitely many intervals are bounded
away from zero uniformly over the allowed parameters. Thus the lattice sums
have relative error $o_N(1)$. The zero-imbalance types assigned to the
positive phase contribute normalized weight $O(N^{-3/4})$ to the interval
containing zero, by the proof of \cref{prop:uniform-normalized-tails};
this is included in the relative errors.
With phases fixed, the exponent in \eqref{eq:amplitude-model} varies
within each product of intervals by at most $C\ell$ per edge, hence
$Cm\ell$ in total. Multiplying the one-gadget estimates and summing
positive contributions gives
\[
 \left|\log\frac{\E_0X_\sigma}
 {2^{-m}p_{R,W}^m\mathcal I_{H,\sigma}(R)\E_0Z_{G_{\sqcup}}}\right|
 \le m\bigl(C\ell+o_N(1)\bigr).
\]
Here $C$ is independent of the partition; $o_N(1)$ is for each fixed
partition and includes the coefficient-comparison error.
Taking $\ell$ arbitrarily small proves \eqref{eq:compact-normalized-integral}.

Combining this estimate with \cref{prop:wired-first-moment-tails} and
$p_{R,W}\le1$, there is a nonnegative $\varepsilon_N\to0$ for fixed
$R,W$ such that
\[
 \begin{aligned}
 \frac{\E_0X_\sigma}{\E_0Z_{G_{\sqcup}}}
 &\ge 2^{-m}\mathcal I_{H,\sigma}(R)
       e^{-m[-\log p_{R,W}+\varepsilon_N]},\\
 \frac{\E_0Z_\sigma}{\E_0Z_{G_{\sqcup}}}
 &\le 2^{-m}\mathcal I_{H,\sigma}(R)
       e^{m[\xi_{\rm tail}(R,W)+\varepsilon_N]}.
 \end{aligned}
\]
Choose $R,W$ sufficiently large that
$-\log p_{R,W},\xi_{\rm tail}(R,W)\le\eta/3$ uniformly over the allowed
parameters. Since $L=2N\delta^2$ is bounded away from zero, choosing
$\delta_0$ sufficiently small ensures $\varepsilon_N\le\eta/3$.
These bounds and $X_\sigma\le Z_\sigma$ prove the lemma.
\end{proof}

\subsubsection{Types outside the fixed neighborhood}

We first bound the contribution from vertices whose types lie outside $\mathcal U$.

\begin{corollary}
\label{cor:marked-macroscopic-tail}
Let $\mathcal L_{H,\sigma}$ be as in \eqref{eq:comparison-weight-sums}.
There are $c,C>0$, depending only on $d$ and the fixed parameter bounds,
such that, for all sufficiently large $N$, every phase vector $\sigma$
and nonempty set $S\subseteq V(H)$ satisfy
\[
 \sum_{\boldsymbol\alpha:\,\{v:(\rho_v,b_v)\notin\mathcal U\}=S}
 \E_0Z_H(\boldsymbol\alpha)
 \le e^{-cN|S|+C\sqrt N\log N|S|+Cm/\sqrt N}\,\mathcal L_{H,\sigma},
\]
where the sum is over type assignments with phases $\sigma$.
Summing over all nonempty $S$ gives at most
$e^{Cm/\sqrt N}[(1+e^{-cN/2})^m-1]\mathcal L_{H,\sigma}$.
\end{corollary}

For every block on an edge of $H$ incident to $S$, delete its connecting
edges and restore the lifts of its two matching edges, as in
Figure~\ref{fig:port-wiring}. This lets us apply the one-gadget global bounds.
The following lemma controls the resulting shifts of the coefficient targets.

\begin{lemma}[Restoring connection blocks]
\label{lem:shifted-cell-domination}
Recall $\mathfrak C_Q$ and the polynomial assumptions from
\cref{lem:normalized-block-replacement}. Let $N$ be even and $k$ a
nonnegative integer with $k\deg_H(v)\le N/2$ for every $v$.
For a set $E_{\rm rest}\subseteq E(H)$ of edges whose blocks are to be
restored, define the modified family
\[
 Q^{\rm rest}_{uv}=
 \begin{cases}
 P_uP_v,&uv\in E_{\rm rest},\\
 Q_{uv},&uv\notin E_{\rm rest}.
 \end{cases}
\]
Let $\mathfrak C_{Q^{\rm rest}}(\mathbf n)$ be the coefficient obtained
from $\mathfrak C_Q(\mathbf n)$ by replacing each block polynomial $Q_e$
with $Q_e^{\rm rest}$, leaving all other factors unchanged.
There are a constant $C>0$, depending only on the fixed polynomial families,
and a probability distribution $\omega_{E_{\rm rest}}$ on nonnegative integer
vectors $\mathbf a=(a_v)_v$ satisfying
$\|a_v\|_1\le Ck\deg_{E_{\rm rest}}(v)$ for every $v$, such that,
for every integer target $\mathbf n$,
\begin{equation}\label{eq:partial-restoration-coefficient}
 \mathfrak C_Q(\mathbf n)
 \le e^{Ck|E_{\rm rest}|}\sum_{\mathbf a}\omega_{E_{\rm rest}}(\mathbf a)
                   \mathfrak C_{Q^{\rm rest}}(\mathbf n-\mathbf a).
\end{equation}
\end{lemma}

\begin{proof}
Expand $\prod_{e\in E_{\rm rest}}Q_e^k=\sum_{\mathbf a}c_{E_{\rm rest}}(\mathbf a)\mathbf z^{\mathbf a}$.
Here $a_v$ records the exponents of $\mathbf z_v$. Only
$k\deg_{E_{\rm rest}}(v)$ factors can involve these variables, each of bounded
degree, so $\|a_v\|_1\le Ck\deg_{E_{\rm rest}}(v)$.
Each term contributes $c_{E_{\rm rest}}(\mathbf a)$ times the coefficient
at $\mathbf n-\mathbf a$ of the remaining factors.
In the restored polynomial, those factors are multiplied by
$\prod_v P_v^{k\deg_{E_{\rm rest}}(v)}$. Taking the constant terms from
these added factors shows that $\mathfrak C_{Q^{\rm rest}}(\mathbf n-\mathbf a)$
is at least $\prod_v P_v(0)^{k\deg_{E_{\rm rest}}(v)}$ times the remaining
coefficient, since all other contributions are nonnegative.
Dividing by this positive product and summing with weights
$c_{E_{\rm rest}}(\mathbf a)$ gives
\[
 \mathfrak C_Q(\mathbf n)
 \le\left(\prod_v P_v(0)^{-k\deg_{E_{\rm rest}}(v)}\right)
      \sum_{\mathbf a}c_{E_{\rm rest}}(\mathbf a)\mathfrak C_{Q^{\rm rest}}(\mathbf n-\mathbf a).
\]
Both the prefactor and the coefficient sum
$\sum_{\mathbf a}c_{E_{\rm rest}}(\mathbf a)=\prod_{e\in E_{\rm rest}}Q_e(\mathbf1)^k$
are at most $e^{Ck|E_{\rm rest}|}$. Normalize by this positive sum to define
$\omega_{E_{\rm rest}}(\mathbf a)=c_{E_{\rm rest}}(\mathbf a)/\prod_{e\in E_{\rm rest}}Q_e(\mathbf1)^k$.
\end{proof}

We also use the following identity: for an integer $j\ge1$ and
$q_1,\ldots,q_j\ge0$, summing over pairwise disjoint sets
$\mathcal D_1,\ldots,\mathcal D_j\subseteq V(H)$ gives
\begin{equation}\label{eq:exceptional-set-summation}
 \sum_{\mathcal D_1,\ldots,\mathcal D_j}
 \prod_{i=1}^j q_i^{|\mathcal D_i|}
 =\left(1+\sum_{i=1}^jq_i\right)^m.
\end{equation}
Each vertex belongs either to none of the sets (weight $1$) or to one
$\mathcal D_i$ (weight $q_i$).

\begin{proof}[Proof of \Cref{cor:marked-macroscopic-tail}]\label{par:exceptional-restoration}
Put $E_{\rm rest}=\{e\in E(H):e\cap S\ne\varnothing\}$, so $|E_{\rm rest}|\le3|S|$, and apply
\cref{lem:shifted-cell-domination} with $P_v=P$ and $Q_{uv}=S(u,v)^2$.
Fix $\mathbf a$ with $\omega_{E_{\rm rest}}(\mathbf a)>0$. Only the coefficient targets for $M_{v,1}$
change, from $n_v$ to $n_v-a_v$; the assignment counts, the factors
for the other matchings, and the fugacity factors are evaluated at the original types.

At $v\notin S$, keep the original tilt $\theta_v$ from
\eqref{eq:connection-type-tilt}. Fix the monomial contributed by each block
on $uv\notin E_{\rm rest}$. The remaining coefficient at $v$ comes from
$P(\mathbf z_v)^{N/2-k\deg_{E(H)\setminus E_{\rm rest}}(v)}$,
and its target differs from its tilted mean by $O(\sqrt N)$.
Apply \cref{lem:moving-lclt} as in the proof of
\cref{lem:normalized-block-replacement}, bounding the Gaussian factor by $1$.
Relative to the original one-gadget weights and the factors $e^{kh}$
on edges outside $E_{\rm rest}$, the prefactor ratios multiply to
$e^{O(m/\sqrt N)}$. The shifts $n_v\mapsto n_v-a_v$ multiply the tilt factors by
$\exp\{\sum_{v\notin S}\langle\theta_v,a_v\rangle\}\le e^{Ck|E_{\rm rest}|}$,
since the tilts are bounded and only endpoints of $E_{\rm rest}$ have nonzero shifts.

Put $f_{\logshift}^*=\tfrac12\max_{\rho,b}\psi_{\logshift}(\rho,b)$,
where the maximum is over the domain of \eqref{eq:one-gadget-rate}.
At $v\in S$, all incident blocks are restored, so the first matching
color contributes $[\mathbf z_v^{n_v-a_v}]P(\mathbf z_v)^{N/2}$.
Writing $F_P$ for the coefficient entropy function in
\cref{lem:finite-polynomial-coefficients}(ii), that lemma and
$\|a_v\|_1=O(\sqrt N)$ give, for original types with $\mathcal P(\alpha_v)\ne\varnothing$,
\[
 [\mathbf z_v^{n_v-a_v}]P(\mathbf z_v)^{N/2}
 \le N^C\exp\left\{\frac N2F_P(2n_v/N)+C\sqrt N\log N\right\}.
\]
Also, \cref{prop:quartic-coercivity} gives
$\psi_{\logshift}(\rho,b)/2\le f_{\logshift}^*-c$ outside $\mathcal U$.
Combining these bounds with the other matching colors and the assignment
and fugacity factors in \eqref{eq:alpha-coefficient}, then summing over
at most $(N+1)^3$ types, gives at most
$N^C e^{Nf_{\logshift}^*-cN+C\sqrt N\log N}$ per vertex.

To compare with $\mathcal L_{H,\sigma}$, complete each type assignment
outside $S$ by summing, at each $v\in S$, over types satisfying
$r_d/2\le\sigma_v r\le3r_d/2$ and $\|s\|_\infty\le1$.
On these windows, \cref{lem:local-quartic-rate,prop:quartic-coercivity}
and $N\logshift^2=\Theta(1)$ give
\[
 0\le Nf_{\logshift}^*-\frac N2\psi_{\logshift}(\rho,b)\le C.
\]
The prefactor estimate \eqref{eq:uniform-cell-prefactor} therefore bounds
their total one-gadget expected weight below by $N^{-C}e^{Nf_{\logshift}^*}$.
All completed types lie in $\mathcal U$, so the tilts at both endpoints
of every reinstated edge lie in a fixed compact set, on which $h$ is bounded.
Reinstating at most $3|S|$ factors $e^{kh}$ therefore multiplies the weights
by at least $e^{-Ck|S|}$.
Including restoration and target-shift costs, the factor per vertex in $S$
is at most
\[
 \frac{N^C e^{Nf_{\logshift}^*-cN+C\sqrt N\log N}}
      {N^{-C}e^{Nf_{\logshift}^*}}\,e^{O(k)}
 =e^{-cN+O(\sqrt N\log N)}.
\]
Multiplying over $S$, tracking the factor $e^{O(m/\sqrt N)}$, and
averaging over $\omega_{E_{\rm rest}}$ proves the first bound. Apply
\eqref{eq:exceptional-set-summation} with $j=1$ and omit the empty set
for the second.
\end{proof}

\subsubsection{Types outside the \rev{fixed window}}

The following lemma compares the two sums in \eqref{eq:comparison-weight-sums},
bounding the contribution from types outside the \rev{fixed window}.

\begin{lemma}
\label{lem:local-saddle-tail-clipping}
There are $c,C,R_0,W_0>0$, depending only on $d$ and the fixed parameter
bounds, such that, for fixed $R\ge R_0,W\ge W_0$ and sufficiently large $N$,
\begin{equation}\label{eq:local-saddle-tail-clipping}
 \mathcal L_{H,\sigma}
 \le e^{m\,o_N(1)}
 \left(1+C\int_{|u|>R-1}e^{Cu^2-cu^4}\dd u+Ce^{-cW^2}\right)^m
 \mathcal L_{H,\sigma}^{R,W}.
\end{equation}
\end{lemma}

\begin{proof}
We compare each type assignment with assignments inside the window and
with the same phases, then average over the replacements. For $\boldsymbol\alpha$ in $\mathcal U$
with phases $\sigma$, set
\[
 S_r=\{v:|r_v|>R\},\qquad
 S_s=\{v\notin S_r:\|s_v\|_\infty>W\}.
\]
Take $R_0\ge\max\{2,2r_d\}$ and $W_0\ge2$. At $v\in S_r$, replace the type
by any type satisfying $r_d/2\le\sigma_v r'_v\le3r_d/2$ and
$\|s'_v\|_\infty\le1$. At $v\in S_s$, keep $r'_v=r_v$ and replace only
the stable coordinate, with $\|s'_v\|_\infty\le1$.
Leave other types unchanged, obtaining $\boldsymbol\alpha'$.
All replacements are types arising from nonnegative integer counts
summing to $N$.
They lie in $\mathcal U$ and have positive expected weight for large $N$
by \cref{lem:finite-polynomial-coefficients}(i).

For each replacement, Taylor expansion at $b_{\logshift}^{*}(\rho'_v)$
using \cref{lem:local-quartic-rate} and $\|s'_v\|_\infty\le1$ gives
$0\le\frac N2[\psi_{\logshift}(\rho'_v,b_{\logshift}^{*}(\rho'_v))
-\psi_{\logshift}(\rho'_v,b'_v)]\le C$.
For $v\in S_r$, $r'_v$ is also in a fixed interval, so the same lemma
and $N\logshift^2=\Theta(1)$ give
\[
 \left|\frac N2\bigl[\psi_{\logshift}(\rho'_v,b'_v)
       -\psi_{\logshift}(0,b_c)\bigr]\right|\le C.
\]
Together with \cref{lem:uniform-cell-prefactor} and
\eqref{eq:stable-rate-loss}--\eqref{eq:quartic-rate-loss}, these bounds imply
\[
 \frac{\E_0Z_{G_v}(\alpha_v)}{\E_0Z_{G_v}(\alpha'_v)}
 \le\begin{cases}
 C\exp\{Cr_v^2-cr_v^4-c\|s_v\|^2\},&v\in S_r,\\
 C\exp\{-c\|s_v\|^2\},&v\in S_s.
 \end{cases}
\]
In the second line, $r'_v=r_v$, so the terms
$\psi_{\logshift}(\rho_v,b_{\logshift}^{*}(\rho_v))$ cancel.

In \eqref{eq:comparison-type-weight}, each edge $uv$ contributes
$e^{kh(\theta_u,\theta_v)}$.
By \eqref{eq:marked-local-coercive}, its exponent is
$-k\delta(r_u+r_v)^2/2$ plus the remainder bounded there.
Since $r'_v=r_v$ outside $S_r$ and $|r'_v|\le3r_d/2$ on $S_r$,
\[
 \frac{k\delta}{2}\left|\sum_{uv\in E(H)}
 \bigl[(r'_u+r'_v)^2-(r_u+r_v)^2\bigr]\right|
 \le Ck\delta\sum_{v\in S_r}(1+r_v^2).
\]
Indeed, for $v\in S_r$ a neighbor $u\notin S_r$ satisfies
$|r_u|\le R<|r_v|$. For edges within $S_r$, use
$2|r_ur_v|\le r_u^2+r_v^2$; at most three edges meet each vertex.

It remains to compare the remainders in \eqref{eq:marked-local-coercive}.
Edges disjoint from $S_r\cup S_s$ give no difference, and the sum of
remainder differences is bounded in absolute value by
\[
 C\logshift\sum_{v\in S_r}r_v^4
 +C\logshift\sum_{v\in S_r\cup S_s}\|s_v\|^2
 +o_N(1)|S_r\cup S_s|.
\]
Here $|r_v|\le R$ on $S_s$, and unchanged neighbors and replacement types
lie in fixed windows; their contributions are included in $o_N(1)$.
For sufficiently small $\logshift$, the first two terms are bounded
by half the magnitudes of $-cr_v^4$ and $-c\|s_v\|^2$ in the exponents
of the one-gadget bounds. Since $k\delta$ is bounded, the quadratic contribution above is
at most $C\sum_{v\in S_r}(1+r_v^2)$. Enlarging $C$ and decreasing $c$ gives
\[
 \frac{W_H(\boldsymbol\alpha)}{W_H(\boldsymbol\alpha')}
 \le e^{m\,o_N(1)}
 \prod_{v\in S_r}Ce^{Cr_v^2-cr_v^4-c\|s_v\|^2}
 \prod_{v\in S_s}Ce^{-c\|s_v\|^2}.
\]

Fix $S_r,S_s$, the unchanged types, and the imbalances at $S_s$.
By the proof of \cref{prop:uniform-normalized-tails}, consecutive $r$-values
are spaced $(N\sqrt{\logshift})^{-1}$ apart. The replacement interval has
length $r_d$, so it contains $\Theta(N\sqrt{\logshift})$ such values.
At fixed $r$, the two $s$-coordinates have spacings
$2\sqrt{2/N}$ and $\sqrt{2/N}$, giving lattice cells of area $4/N$.
Thus $[-1,1]^2$ contains $\Theta(N)$ allowed $s$-values.
There are therefore at least $cN^2\sqrt{\logshift}$ replacement types
at each vertex in $S_r$, and at least $cN$ at each vertex in $S_s$,
where $r$ is kept fixed. Averaging over replacements and then summing
the original tail types gives the sums below; their prefactors are the
reciprocals of these replacement counts, up to a constant.
Fix positive constants $c,C$, depending only on $d$ and the parameter bounds,
for which the same lattice estimates apply, and set
\[
 q_R=C\int_{|u|>R-1}e^{Cu^2-cu^4}\dd u,\qquad q_W=Ce^{-cW^2}.
\]
Then
\[
 \frac{C}{N^2\sqrt{\logshift}}
 \sum_{|r|>R,\,s}e^{Cr^2-cr^4-c\|s\|^2}\le q_R,
 \qquad
 \frac CN\sum_{\|s\|_\infty>W}e^{-c\|s\|^2}\le q_W.
\]
Here both sums run over the type lattice in $\mathcal U$; the second is
at fixed $r$.
For fixed $S_r,S_s$, each replacement assignment uniquely determines
the unchanged types and the imbalances kept at $S_s$.
Summing over these data therefore counts
each replacement assignment at most once, giving a sum bounded by
$\mathcal L_{H,\sigma}^{R,W}$. Hence
\[
 \sum_{\boldsymbol\alpha:\,S_r(\boldsymbol\alpha)=S_r,\ S_s(\boldsymbol\alpha)=S_s}
 W_H(\boldsymbol\alpha)
 \le e^{m\,o_N(1)}q_R^{|S_r|}q_W^{|S_s|}\mathcal L_{H,\sigma}^{R,W}.
\]
The sum on the left is restricted to $\mathcal U$ and phases $\sigma$.
Apply \eqref{eq:exceptional-set-summation} with $q_1=q_R$ and $q_2=q_W$
to obtain \eqref{eq:local-saddle-tail-clipping}.
\end{proof}

\section{Shared fluctuations and \rev{mixed moments}}
\label{sec:shared-moments}\label{sec:marked-main}\label{app:square-planting}

Recall that $\mathbb P_0$ is the unconditioned construction law and
$\mathbb P_1$ conditions every base graph $B_v$ on being simple.
Write $\E_b$ for expectation under $\mathbb P_b$, for $b\in\{0,1\}$;
$\mathbb Q_1$ reweights $\mathbb P_1$ by $Z_{G_{\sqcup}}^2$, as in
\eqref{eq:reference-square-law}. We use the fixed parameter bounds
from \cref{sec:gadget-moments}.

Section~\ref{sec:gadget-moments} bounds ratios of expected phase weights.
We show that the four normalized mixed moments below have a common
factor, which cancels in Lemma~\ref{lem:sqplant-cancellation}. Together with
a comparison of normalized means under $\mathbb P_0$ and $\mathbb P_1$,
this gives the phase-weight bounds under $\mathbb Q_1$ in
\cref{thm:cut-weight-realization}.

\begin{theorem}[Mixed moments]
\label{thm:sqplant-full-moments}
Recall that $Z_{G_{\sqcup}}=\prod_vZ_{G_v}$ is the partition function
before connections. Let $Z_\sigma$ be the full phase-sector weight and,
for fixed cutoffs $R\ge2r_d$ and $W>0$, let $X_\sigma$ be the restricted
weight of Definition~\ref{def:restricted-phase-weight}.
There are constants $\mathfrak c_0,\mathfrak c_1>0$ depending only on $d$,
and constants $C_d,N_0>0$ computable from $d$, the fixed parameter bounds,
and the cutoffs $R,W$,
such that the following holds for each mode $b\in\{0,1\}$.
For every simple cubic graph $H$ on $m$ vertices, every phase vector
$\sigma\in\{\pm\}^{V(H)}$, and allowed $N,k,\delta$
with $N\ge N_0$, each of
\begin{equation}\label{eq:sqplant-full-moments}
 \frac{\E_b Z_{G_{\sqcup}}^2}{(\E_b Z_{G_{\sqcup}})^2},\qquad
 \frac{\E_b X_\sigma^2}{(\E_b X_\sigma)^2},\qquad
 \frac{\E_b(X_\sigma Z_{G_{\sqcup}})}{\E_b X_\sigma\E_b Z_{G_{\sqcup}}},\qquad
 \frac{\E_b(Z_\sigma Z_{G_{\sqcup}})}{\E_b Z_\sigma\E_b Z_{G_{\sqcup}}}
\end{equation}
lies in $[\mathfrak c_b^m e^{-C_dm/\sqrt N},
\mathfrak c_b^m e^{C_dm/\sqrt N}]$.
\end{theorem}

The following lemma transfers the normalized means from $\mathbb P_0$ to $\mathbb P_1$;
its proof is at the end of \cref{sec:full-moment-restoration}.

\begin{lemma}[First moments after conditioning]\label{lem:conditioned-first-means}
Under the assumptions of \cref{thm:sqplant-full-moments}, there is a
constant $C>0$, computable from $d$, the fixed parameter bounds, and
the cutoffs $R,W$, such that, for $U\in\{X_\sigma,Z_\sigma\}$,
\begin{equation}\label{eq:conditioned-first-means}
 e^{-Cm/\sqrt N}\le
 \frac{\E_1U/\E_1Z_{G_{\sqcup}}}{\E_0U/\E_0Z_{G_{\sqcup}}}
 \le e^{Cm/\sqrt N}.
\end{equation}
\end{lemma}

\begin{lemma}[Change of measure]
\label{lem:sqplant-cancellation}
Let $A>0$ and $X\ge0$ be random variables on a finite probability
space with law $\mathbb P$, with $\E_{\mathbb P}X>0$.
Define $\mathbb Q$ by $\dd\mathbb Q/\dd\mathbb P=A^2/\E_{\mathbb P}A^2$.
All unmarked expectations below are under $\mathbb P$.
Suppose that, for some $C_0>0$ and $a\ge0$, both
\[
 \frac{\E A^2}{(\E A)^2},\qquad
 \frac{\E(XA)}{\E X\E A}
\]
lie in $[C_0e^{-a},C_0e^a]$. Then
\begin{equation}\label{eq:sqplant-means}
 e^{-2a}\frac{\E X}{\E A}
 \le\E_{\mathbb Q}(X/A)
 \le e^{2a}\frac{\E X}{\E A}.
\end{equation}
Moreover, if $\E X^2/(\E X)^2\le C_0e^a$, then
\begin{equation}\label{eq:sqplant-relative-second}
 \frac{\E_{\mathbb Q}[(X/A)^2]}
      {[\E_{\mathbb Q}(X/A)]^2}\le e^{4a}.
\end{equation}
\end{lemma}

\begin{proof}
By the definition of $\mathbb Q$,
\[
 \frac{\E_{\mathbb Q}(X/A)}{\E X/\E A}
 =\frac{\E(XA)/(\E X\E A)}{\E A^2/(\E A)^2},
 \qquad
 \frac{\E_{\mathbb Q}[(X/A)^2]}{[\E_{\mathbb Q}(X/A)]^2}
 =\frac{\E X^2\,\E A^2}{[\E(XA)]^2}.
\]
The first ratio lies in $[e^{-2a},e^{2a}]$; under the additional
assumption, the second is at most $\frac{(C_0e^a)^2}{(C_0e^{-a})^2}=e^{4a}$.
\end{proof}

\begin{proof}[Proof of \Cref{thm:cut-weight-realization}]
Take $C_{\rm amp}=3480$ and choose the cutoffs $R,W$ from
Proposition~\ref{prop:uniform-annealed-soft-spin} at accuracy $\varepsilon/2$;
use these cutoffs to define $X_\sigma$. Its proof takes them from
\cref{lem:normalized-amplitude-representation}, so $R\ge2r_d$. The proposition's size condition follows
from \eqref{eq:moment-size-conditions}, since $L=2N\delta^2$.
With $C$ from Lemma~\ref{lem:conditioned-first-means}, choose $\delta_0$
no larger than that proposition's cutoff and small
enough that the size condition forces $N\ge N_0$ and
$(C+2C_d)/\sqrt N\le\varepsilon/2$.
Equation~\eqref{eq:annealed-restricted-lower}, with $\varepsilon$ replaced
by $\varepsilon/2$, gives the mean bounds under $\mathbb P_0$.
Passing to $\mathbb P_1$ changes each normalized mean by a factor in
$[e^{-Cm/\sqrt N},e^{Cm/\sqrt N}]$.

Apply the mean bound in Lemma~\ref{lem:sqplant-cancellation}
under $\mathbb P_1$, with $A=Z_{G_{\sqcup}}$ and
$X\in\{X_\sigma,Z_\sigma\}$, and its second-moment bound with
$X=X_\sigma$. In both applications, take $C_0=\mathfrak c_1^m$
and $a=C_dm/\sqrt N$; the hypotheses follow from
\cref{thm:sqplant-full-moments} with $b=1$. Thus, for
$X\in\{X_\sigma,Z_\sigma\}$,
\[
 \frac{\E_{\mathbb Q_1}(X/Z_{G_{\sqcup}})}{\E_1 X/\E_1Z_{G_{\sqcup}}}
 \in[e^{-2C_dm/\sqrt N},e^{2C_dm/\sqrt N}],\qquad
 \frac{\E_{\mathbb Q_1}[(X_\sigma/Z_{G_{\sqcup}})^2]}
 {[\E_{\mathbb Q_1}(X_\sigma/Z_{G_{\sqcup}})]^2}
 \le e^{4C_dm/\sqrt N}.
\]
Combining the mean bounds gives \eqref{eq:main-full-sector-upper} and
\[
 \E_{\mathbb Q_1}(X_\sigma/Z_{G_{\sqcup}})
 \ge2^{-m}e^{-J(|E(H)|-\Cut_H(\sigma))-\varepsilon m}.
\]
Applying the Paley--Zygmund inequality to $X_\sigma/Z_{G_{\sqcup}}$ gives
\[
 \mathbb Q_1\left(X_\sigma/Z_{G_{\sqcup}}\ge\tfrac12
             \E_{\mathbb Q_1}(X_\sigma/Z_{G_{\sqcup}})\right)
 \ge\tfrac14e^{-4C_dm/\sqrt N}.
\]
Use $X_\sigma\le Z_\sigma$ and enlarge $C_d$ to obtain
\eqref{eq:main-favorable-probability}. All cutoffs depend only on the
fixed parameters in the theorem.
\end{proof}

We prove the mixed-moment estimates by computing the second moment of
one gadget at fixed types, then accounting for connections and simplicity.
Finally, we sum the fixed-type estimates and bound the contributions
outside the neighborhoods where those estimates apply.

\subsection{Second moments for one gadget}
\label{app:unmarked-details}

We estimate the mixed second moment at two fixed types, normalized by the
corresponding first moments, for one gadget before connections.
A product of two partition functions sums over two configurations on
the same random graph; we call these configurations \emph{replicas}.
Recall the state set $\Omega=\{00,10,01,11\}$ and types from
\eqref{eq:occupation-type}. \rev{If the two replicas in a gadget have types $\alpha,\beta$, their}
\emph{overlap} is the $4\times4$ matrix
\[
 \Gamma_{ab}=N^{-1}|\{i:\text{label }i\text{ has state }a
       \text{ in the first replica and }b\text{ in the second}\}|.
\]
Its row sums are $\alpha$ and its column sums are $\beta$.
For overlap matrices and their differences, $\|\cdot\|_2$ is the
Euclidean norm after viewing the matrix as a vector of its 16 entries.
For actual configurations, $N\Gamma_{ab}$ counts base labels with
joint state $(a,b)$, so it is a nonnegative integer. The product
matrix $\alpha\otimes\beta$ means that the two state assignments are
independent at a uniformly chosen label. We obtain the second-moment
factor by summing the expected pair weights near this product matrix.

Recall the type-restricted partition function $Z_{G_v}(\alpha)$ from
\eqref{eq:type-partition-function}.
\begin{theorem}[One-gadget second moments]\label{thm:moving-cell-second-moment}
There is a positive smooth function $\cC_d^{\rm pm}(\alpha,\beta)$,
computed in \eqref{eq:moving-pm-constant}, and a compact neighborhood
$\mathcal U_d$ of $\alpha_c$ from \eqref{eq:critical-type}, within the
positive probability vectors on $\Omega$, such that the following holds.
The coordinates $(\rho,b)$ of every $\alpha\in\mathcal U_d$ lie in
\rev{$\mathcal U$ fixed in \cref{def:restricted-phase-weight}.}
Uniformly for integer types $\alpha,\beta\in\mathcal U_d$, as even $N\to\infty$,
\begin{equation}\label{eq:moving-cell-pm-asymptotic}
  \frac{\E_0[Z_{G_v}(\alpha)Z_{G_v}(\beta)]}
       {\E_0Z_{G_v}(\alpha)\,\E_0Z_{G_v}(\beta)}
  =\cC_d^{\rm pm}(\alpha,\beta)
   \left[1+O_d(N^{-1/2})\right].
\end{equation}
{The estimate holds at every positive fugacity, without a
restriction on $N\delta^2$, since the fugacity factors cancel
from the ratio at fixed types.}
\end{theorem}

The type weights in the denominator are positive for sufficiently large
$N$, by Lemma~\ref{lem:finite-polynomial-coefficients}(i).

\subsubsection{The overlap sum and edge distributions}
\label{sec:critical-cell-second-moment}\label{sec:moving-pointwise-second-moment}

We express the second moment as a sum over overlap matrices and define
the edge distributions used to estimate this sum.

Recall $B$ from \eqref{eq:base-compatibility}. Suppose two base labels
have states $a,a'$ in the first replica and $b,b'$ in the second.
Pairing these labels is allowed in both replicas exactly when
\[
 (B\otimes B)_{(a,b),(a',b')}=B_{aa'}B_{bb'}=1.
\]
Fix the states of all $N$ base labels in both replicas, with
$N\Gamma_{ab}$ labels having states $a,b$, respectively.
Following Lemma~\ref{lem:one-color-matching-asymptotic}, let
$\mathcal M_N^{B\otimes B}(\Gamma)$ be the number of perfect matchings
satisfying the displayed condition at every pair. Relabeling shows
that this number depends only on $\Gamma$.
A uniform perfect matching therefore satisfies both replicas' constraints
with probability $\mathcal M_N^{B\otimes B}(\Gamma)/(N-1)!!$.
There are $N!/\prod_{a,b}(N\Gamma_{ab})!$ state assignments with these
counts. Summing over $\Gamma$, multiplying by the fugacity weight, and
using independence of the $d$ matching colors gives
\begin{equation}\label{eq:exact-overlap-sum-new}
 \E_0[Z_{G_v}(\alpha)Z_{G_v}(\beta)]
 =\lambda^{N(\alpha_{10}+\alpha_{01}+2\alpha_{11}
                   +\beta_{10}+\beta_{01}+2\beta_{11})}
   \sum_\Gamma\frac{N!}{\prod_{a,b}(N\Gamma_{ab})!}
       \left[\frac{\mathcal M_N^{B\otimes B}(\Gamma)}{(N-1)!!}\right]^d.
\end{equation}
Here the sum is over matrices $\Gamma$ with row sums $\alpha$, column
sums $\beta$, and $N\Gamma_{ab}\in\mathbb Z_{\ge0}$ for all $a,b$.
Normalizing the expected pair weights in \eqref{eq:exact-overlap-sum-new}
defines the distribution $\mu_{\alpha,\beta}$ on overlap matrices:
\begin{equation}\label{eq:unconnected-overlap-law}
 \mu_{\alpha,\beta}(\Gamma)\propto
 \frac{[\mathcal M_N^{B\otimes B}(\Gamma)]^d}
      {\prod_{a,b}(N\Gamma_{ab})!}.
\end{equation}
The following exponential-moment bound will control the connection
errors in Section~\ref{app:marked-details}.

\begin{lemma}\label{lem:overlap-exponential-moment}
There are constants $c,C>0$, depending only on $d$, such that for all
sufficiently large even $N$ and integer types $\alpha,\beta\in\mathcal U_d$,
a random matrix $\Gamma\sim\mu_{\alpha,\beta}$ satisfies
\begin{equation}\label{eq:overlap-exponential-moment}
 \E_{\mu_{\alpha,\beta}}e^{cN\|\Gamma-\alpha\otimes\beta\|_2^2}\le C.
\end{equation}
For each fixed sufficiently small $\eta>0$, the same bound holds
under $\mu_{\alpha,\beta}$ conditioned on
$\|\Gamma-\alpha\otimes\beta\|_2\le\eta$, for sufficiently large
$N$ depending also on $\eta$.
\end{lemma}
Theorem~\ref{thm:moving-cell-second-moment} and
Lemma~\ref{lem:overlap-exponential-moment} are proved in
Section~\ref{sec:unmarked-gaussian}.

To estimate the overlap sum, we express its exponential rate using edge
distributions and compute its curvature using their transition matrices.
Put $D=d-1$. Recall that
$\mathcal P(\alpha)$ in \eqref{eq:one-gadget-rate} consists of symmetric
couplings with marginal $\alpha$ and $P_{ab}=0$ whenever $B_{ab}=0$.
For $\alpha$ near $\alpha_c$, define
\begin{equation}\label{eq:moving-edge-law}
 P_\alpha:=\mathop{\rm arg\,max}_{P\in\mathcal P(\alpha)}\mathsf H(P),
 \qquad T_\alpha(a,b):=P_\alpha(a,b)/\alpha_a.
\end{equation}
The positive exponential tilt constructed in the proof of
Lemma~\ref{lem:expected-type-weight} gives this unique optimizer and its
analytic dependence on $\alpha$.
For each $a$, $T_\alpha(a,\cdot)$ is the conditional distribution of the
second endpoint state given that the first is $a$.
The matrix $T_\alpha$ is reversible with stationary law $\alpha$.
Acting on functions by $(T_\alpha f)(a)=\sum_bT_\alpha(a,b)f(b)$,
it preserves $L^2_0(\alpha)=\{f:\sum_a\alpha_af(a)=0\}$, the space
orthogonal to constant functions in $L^2(\alpha)$.
Write $T_{\alpha,\perp}$ for its restriction to this space.
At $\alpha=\alpha_c$, the critical edge distribution $P_c$ constructed
in Lemma~\ref{lem:local-quartic-rate} has the required marginal and
exponential-tilt form, so $P_{\alpha_c}=P_c$.
Put $T_c=T_{\alpha_c}$. By \eqref{eq:critical-binary-channel},
the spectrum (the set of eigenvalues) of $T_{c,\perp}$ is
\begin{equation}\label{eq:critical-channel-spectrum-new}
  \spec(T_{c,\perp})=\{D^{-1},-D^{-1},D^{-2}\}.
\end{equation}

\subsubsection{The overlap rate and its maximizer}\label{sec:unmarked-localization}

We show that, for fixed types $\alpha,\beta$ sufficiently close to
$\alpha_c$, the exponential rate of the overlap summands is uniquely maximized at
$\alpha\otimes\beta$, and compute its quadratic decrease there.

Define the entropy expressions
\begin{equation}\label{eq:overlap-rates}
 \begin{aligned}
 f(\alpha)&=(1-d)\mathsf H(\alpha)+\frac d2\mathsf H(P_\alpha),\\
 \cF_{\alpha,\beta}(\Gamma)&=(1-d)\mathsf H(\Gamma)
                         +\frac d2\max_Y\mathsf H(Y).
 \end{aligned}
\end{equation}
Here $\Gamma$ ranges over all nonnegative real matrices with row sums
$\alpha$ and column sums $\beta$, without requiring $N\Gamma$ to be
integral. The maximum is over symmetric probability matrices
$Y=(Y_{(a,b),(a',b')})$ on pairs of joint states, with $Y\one=\Gamma$
and $\supp Y\subseteq\supp(B\otimes B)$.
Set $\cF_{\alpha,\beta}(\Gamma)=-\infty$ when no such $Y$ exists.
For integer overlaps in a fixed small neighborhood of
$\alpha\otimes\beta$ with positive matching count,
Lemma~\ref{lem:one-color-matching-asymptotic} gives
\[
 \log\frac{\mathcal M_N^{B\otimes B}(\Gamma)}{(N-1)!!}
 =N\left[\frac12\max_Y\mathsf H(Y)-\mathsf H(\Gamma)\right]+O_d(1).
\]
The ratio on the left is the probability that one uniform perfect
matching satisfies both replicas' constraints. The $d$ matching colors
are independent, so their log probabilities add. Stirling's formula gives
$\log[N!/\prod_{a,b}(N\Gamma_{ab})!]=N\mathsf H(\Gamma)+O_d(\log N)$.
Consequently, the summand in \eqref{eq:exact-overlap-sum-new}, with its
fugacity factor removed, satisfies
\[
 \log\left\{\frac{N!}{\prod_{a,b}(N\Gamma_{ab})!}
 \left[\frac{\mathcal M_N^{B\otimes B}(\Gamma)}{(N-1)!!}\right]^d\right\}
 =N\cF_{\alpha,\beta}(\Gamma)+O_d(\log N).
\]
The same calculation for one replica gives, for positive-weight integer
types $\alpha\in\mathcal U_d$,
\[
 \log\frac{\E_0Z_{G_v}(\alpha)}
 {\lambda^{N(\alpha_{10}+\alpha_{01}+2\alpha_{11})}}
 =Nf(\alpha)+O_d(\log N).
\]
These estimates are uniform in the stated neighborhoods. For arbitrary
feasible overlaps, including boundary matrices,
Lemma~\ref{lem:finite-polynomial-coefficients}(ii) and Stirling's bounds
give the upper bound $N^{O_d(1)}e^{N\cF_{\alpha,\beta}(\Gamma)}$
on each summand with the fugacity factor removed.
\begin{lemma}\label{lem:moving-relative-rate}
For every $\Gamma$ for which the constraints on $Y$ in
\eqref{eq:overlap-rates} are feasible,
\begin{equation}\label{eq:moving-relative-rate}
  \cF_{\alpha,\beta}(\Gamma)-f(\alpha)-f(\beta)
  =(d-1)\KL(\Gamma\|\alpha\otimes\beta)
   -\frac d2
    \inf_{\substack{Y=Y^{\mathsf T}\ge0,\,Y\one=\Gamma\\
                    \supp Y\subseteq\supp(P_\alpha\otimes P_\beta)}}
      \KL(Y\|P_\alpha\otimes P_\beta).
\end{equation}
At $\Gamma=\alpha\otimes\beta$ the minimizing matrix is $Y=P_\alpha\otimes P_\beta$, and the right-hand side vanishes.
\end{lemma}

\begin{proof}
The fixed marginals give
\[
 \mathsf H(\Gamma)=\mathsf H(\alpha)+\mathsf H(\beta)-\KL(\Gamma\|\alpha\otimes\beta).
\]
The entropy-maximizing matrix has the form
$P_\alpha(a,a')=B_{aa'}r_ar_{a'}$ for positive $r_a$, as in
Lemma~\ref{lem:one-color-matching-asymptotic}; the same holds for $P_\beta$.
Thus, when $B_{aa'}=1$, $\log P_\alpha(a,a')=\log r_a+\log r_{a'}$.
Using $Y\one=\Gamma$ and the row and column sums $\alpha,\beta$ of
$\Gamma$ gives
\[
 \sum_{\substack{B_{aa'}=1\\B_{bb'}=1}}Y_{(a,b),(a',b')}\log\!\bigl(P_\alpha(a,a')P_\beta(b,b')\bigr)
 =\sum_{B_{aa'}=1}P_\alpha(a,a')\log P_\alpha(a,a')
  +\sum_{B_{bb'}=1}P_\beta(b,b')\log P_\beta(b,b').
\]
Consequently,
\[
 \mathsf H(Y)=\mathsf H(P_\alpha)+\mathsf H(P_\beta)
              -\KL(Y\|P_\alpha\otimes P_\beta).
\]
Substituting these identities into \eqref{eq:overlap-rates} and
maximizing over $Y$ proves \eqref{eq:moving-relative-rate}.
\end{proof}

To compute the quadratic decrease, choose eigenbases of
$T_{\alpha,\perp}$ and $T_{\beta,\perp}$, orthonormal in
$L^2_0(\alpha)$ and $L^2_0(\beta)$,
\[
  T_{\alpha,\perp}f_i=t_if_i,
  \qquad
  T_{\beta,\perp}g_j=s_jg_j,
  \qquad 1\le i,j\le3.
\]
For each $i,j$, perturb the product overlap by
\[
 \Gamma_z(a,b)=\alpha_a\beta_b\bigl(1+zf_i(a)g_j(b)\bigr).
\]
For sufficiently small $z$, this is a probability matrix with the same
marginals, since $f_i,g_j$ have mean zero. The nine functions
$f_i\otimes g_j$ form an orthonormal basis for such marginal-preserving
perturbations relative to $\alpha\otimes\beta$; they have eigenvalues $t_is_j$ under
$T_\alpha\otimes T_\beta$.

\begin{lemma}[Overlap rate bounds]\label{lem:moving-product-localization}
After shrinking $\mathcal U_d$ if necessary, there is $c_d>0$ such that for all
$\alpha,\beta\in\mathcal U_d$ and all overlap matrices $\Gamma$ with these marginals,
\begin{equation}\label{eq:moving-global-gap}
  \cF_{\alpha,\beta}(\Gamma)
  \le f(\alpha)+f(\beta)
      -c_d\min\{\|\Gamma-\alpha\otimes\beta\|_2^2,1\}.
\end{equation}
For each perturbation $\Gamma_z$ defined above,
\begin{equation}\label{eq:moving-mode-curvature}
 -\left.\frac{d^2}{dz^2}\cF_{\alpha,\beta}(\Gamma_z)\right|_{z=0}
 =\gamma(t_is_j):=\frac{1-Dt_is_j}{1+t_is_j}.
\end{equation}
\end{lemma}

\begin{proof}
\emph{Quadratic decrease near the product matrix.}
Put $h=f_i\otimes g_j$. For $(X_1,X_2)\sim P_\alpha\otimes P_\beta$,
orthonormality and the eigenfunction relation give
\[
 \E[h(X_1)^2]=1,\qquad \E[h(X_2)\mid X_1]=t_is_jh(X_1).
\]
For $x,y\in\Omega^2$, set $a(x,y)=z[h(x)+h(y)]/(1+t_is_j)$.
Then $\E[a(X_1,X_2)\mid X_1]=zh(X_1)$, and likewise at the other endpoint.
Thus $(P_\alpha\otimes P_\beta)(x,y)(1+a(x,y))$ has both marginals
$\Gamma_z$ and is nonnegative for sufficiently small $z$.
For any symmetric $Y\ge0$ with $Y\one=\Gamma_z$ and
$\supp Y\subseteq\supp(B\otimes B)$, write
$Y(x,y)=(P_\alpha\otimes P_\beta)(x,y)(1+q(x,y))$ on this support.
Then $q-a$ has conditional mean zero given either $X_1$ or $X_2$ and is
orthogonal to $a$, which is a sum of a function of $X_1$ and a function of $X_2$.
Consequently, under the same distribution of $(X_1,X_2)$,
\[
 \E[q(X_1,X_2)^2]
 =\frac{2z^2}{1+t_is_j}+\E[(q-a)(X_1,X_2)^2].
\]
Thus $a$ uniquely minimizes the squared norm. Relative entropy has
quadratic term one half of this norm squared.

To obtain the constrained KL minimizer, tilt the positive entries by
\[
 Y_\theta(x,y)\propto(P_\alpha\otimes P_\beta)(x,y)e^{\theta_x+\theta_y},
 \qquad \theta_{(00,00)}=0.
\]
By symmetry, $Y_\theta\one=\Gamma_z$ is equivalent to
\[
 \E_{Y_\theta}[\mathbf1_{\{X_1=x\}}+\mathbf1_{\{X_2=x\}}]
 =2\Gamma_z(x),\qquad x\in\Omega^2.
\]
The vector of these endpoint counts, with the $(00,00)$ coordinate
omitted, takes the value $0$ and every unit vector with positive
probability, since $(00,00)$ can pair with every state, including itself. Lemma~\ref{lem:finite-polynomial-coefficients}(i)
therefore gives a tilt analytic in the marginal $\Gamma$ near
$\alpha\otimes\beta$, hence in $z$ along $\Gamma_z$.
Its entropy variational identity identifies $Y_\theta$ as the minimizer
under the stated constraints.
The first-order perturbation minimizes the quadratic form just computed,
so the relative perturbation is $a+O_d(z^2)$. Consequently,
\[
 \KL(\Gamma_z\|\alpha\otimes\beta)=\frac{z^2}{2}+O_d(|z|^3),\qquad
 \inf_{Y\one=\Gamma_z}\KL(Y\|P_\alpha\otimes P_\beta)
 =\frac{z^2}{1+t_is_j}+O_d(|z|^3),
\]
where the infimum is still subject to symmetry, nonnegativity, and the support constraint.
Substitution into \eqref{eq:moving-relative-rate} gives curvature
$d/(1+t_is_j)-(d-1)=\gamma(t_is_j)$.
The orthogonality of the functions $f_i\otimes g_j$ and the diagonal
action of $T_\alpha\otimes T_\beta$ imply that their corresponding
$h(X_1)+h(X_2)$ are orthogonal; hence there are no mixed quadratic terms.
At criticality, $|t_is_j|\le D^{-2}<1$ and $Dt_is_j\le D^{-1}<1$,
so every $\gamma(t_is_j)$ is positive.
The KL expansions have no linear terms, and the entries of $\alpha,\beta$
are uniformly bounded away from zero on $\mathcal U_d$, so the weighted
coordinate norm and the entrywise Euclidean norm are uniformly equivalent.
By continuity of the Hessian and analyticity of $\cF_{\alpha,\beta}$,
there are $\eta,c>0$ such that
$\cF_{\alpha,\beta}(\Gamma)\le f(\alpha)+f(\beta)-c\|\Gamma-\alpha\otimes\beta\|_2^2$
whenever $\alpha,\beta\in\mathcal U_d$ and
$\|\Gamma-\alpha\otimes\beta\|_2\le\eta$, after shrinking $\mathcal U_d$.

\emph{A uniform gap outside this neighborhood.}
Proposition~\ref{prop:global-critical-anchor} gives a unique critical
maximizer for two replicas; its fugacity term is constant at fixed
replica marginals. If no uniform gap persisted as
$\alpha,\beta\to\alpha_c$, choose a sequence of matrices $\Gamma$ with
$\|\Gamma-\alpha\otimes\beta\|_2\ge\eta$ whose rates approach the
critical maximum. For each $\Gamma$, choose $Y$ attaining the maximum
in \eqref{eq:overlap-rates}. Compactness gives a convergent subsequence
of $(\Gamma,Y)$; continuity of entropy, including at zero entries with
$0\log0=0$, makes its limit attain the critical maximum away from
$\alpha_c\otimes\alpha_c$, contradicting uniqueness.
Combining this gap with the local bound proves \eqref{eq:moving-global-gap}.
\end{proof}

\subsubsection{Gaussian summation}\label{sec:unmarked-gaussian}

We evaluate the overlap sum to prove the second-moment estimate, then
prove the exponential-moment bound for the overlap.

\begin{proof}[Proof of \Cref{thm:moving-cell-second-moment}]
Divide \eqref{eq:exact-overlap-sum-new} by
$\E_0Z_{G_v}(\alpha)\,\E_0Z_{G_v}(\beta)$; we estimate the resulting
normalized summands.
The supports of $B$ and $B\otimes B$ are connected, with loops at
$00$ and $(00,00)$. The proof of Lemma~\ref{lem:moving-product-localization}
establishes analytic tilts and the full difference lattice near
$\Gamma=\alpha\otimes\beta$. There, $P_\alpha\otimes P_\beta$ is positive
at every $(x,y)$ with $(B\otimes B)_{xy}=1$.
Fix a sufficiently small $\eta>0$. Let $Y$ maximize entropy under the
constraints in \eqref{eq:overlap-rates}. Draw $N/2$ independent pairs of
joint states with law $Y$; their endpoint-count vectors have mean $2\Gamma$.
After deleting the $(00,00)$ coordinate from these vectors and the target
$N\Gamma$, Lemma~\ref{lem:moving-lclt} gives positive probability at this
target for large $N$, uniformly for $\|\Gamma-\alpha\otimes\beta\|_2\le\eta$.
Thus the matching count is positive, and
Lemma~\ref{lem:one-color-matching-asymptotic} applies uniformly.

The summand estimates in Section~\ref{sec:unmarked-localization} and
\eqref{eq:moving-global-gap} bound the total normalized contribution
outside this neighborhood by $N^Ce^{-cN}$: the fixed rate gap dominates
the polynomial number of overlap matrices.

Put $\mathcal X=\sqrt N(\Gamma-\alpha\otimes\beta)$.
Using the eigenfunctions $f_i,g_j$ from Lemma~\ref{lem:moving-product-localization},
define $x=(x_{ij})$ by
\[
 \mathcal X_{ab}=\alpha_a\beta_b\sum_{i,j=1}^3x_{ij}f_i(a)g_j(b).
\]
The norms $\|x\|_2$ and $\|\mathcal X\|_2$ are uniformly equivalent
for $\alpha,\beta\in\mathcal U_d$.
Analyticity of the entropy-maximizing matrix $Y$ and
\eqref{eq:moving-mode-curvature} give, uniformly in a fixed small neighborhood,
\begin{equation}\label{eq:moving-local-taylor}
 N[\cF_{\alpha,\beta}(\alpha\otimes\beta+\mathcal X/\sqrt N)-\cF_{\alpha,\beta}(\alpha\otimes\beta)]
 =-\frac12\sum_{i,j=1}^3\gamma(t_is_j)x_{ij}^2
       +O_d(\|\mathcal X\|^3/\sqrt N).
\end{equation}

We evaluate the leading prefactors at $\Gamma=\alpha\otimes\beta$.
On the mean-zero subspace of $L^2(\alpha\otimes\beta)$, the eigenfunctions
$f_i\otimes1$, $1\otimes g_j$, and $f_i\otimes g_j$ of
$T_\alpha\otimes T_\beta$ have eigenvalues $t_i$, $s_j$, and $t_is_j$.
Lemma~\ref{lem:one-color-matching-asymptotic} gives the following ratio
of matching prefactors for two replicas and two separate single replicas,
with exponent $-d/2$ accounting for all $d$ matching colors:
\begin{equation}\label{eq:moving-matching-prefactor}
 \left[\frac{\det(I+(T_\alpha\otimes T_\beta)_\perp)}
 {\det(I+T_{\alpha,\perp})\det(I+T_{\beta,\perp})}\right]^{-d/2}
 =\prod_{i,j=1}^3(1+t_is_j)^{-d/2}.
\end{equation}
The factors $1+t_i$ and $1+s_j$ cancel in this ratio.
For the assignment counts $N!/\prod_{a,b}(N\Gamma_{ab})!$ in
\eqref{eq:exact-overlap-sum-new} and the two single-replica counts,
Stirling's formula gives the prefactor ratio
\begin{equation}\label{eq:moving-multinomial-prefactor}
 \frac{(2\pi N)^{-15/2}\prod_{a,b}(\alpha_a\beta_b)^{-1/2}}
 {(2\pi N)^{-3}(\prod_a\alpha_a\prod_b\beta_b)^{-1/2}}
 =(2\pi N)^{-9/2}
 \left(\prod_a\alpha_a\prod_b\beta_b\right)^{-3/2}.
\end{equation}
Differences of count matrices $N\Gamma$ with fixed marginals lie in
the lattice of integer $4\times4$ matrices with zero row and column sums.
With $e_a$ the unit vector for state $a\in\Omega$, an integer basis is
$E_{ab}=(e_a-e_{11})(e_b-e_{11})^{\mathsf T}$,
$a,b\in\Omega\setminus\{11\}$: the nine entries with $a,b\ne11$
determine the rest through the row and column sums.
Before the $N^{-1/2}$ scaling, count increments $M,M'$ have inner
product $\sum_{a,b}M_{ab}M'_{ab}/(\alpha_a\beta_b)$ in these coordinates.
The Gram matrix of this basis is therefore $G_\alpha\otimes G_\beta$, where
\[
 G_\alpha=\diag(\alpha_{00}^{-1},\alpha_{10}^{-1},\alpha_{01}^{-1})
             +\alpha_{11}^{-1}\one\one^{\mathsf T},
 \qquad \det G_\alpha=(\prod_a\alpha_a)^{-1}.
\]
Scaling each of the nine coordinates by $N^{-1/2}$, the lattice in $x$
has fundamental cell volume
$N^{-9/2}(\prod_a\alpha_a\prod_b\beta_b)^{-3/2}$.
It is generally translated, since $N(\alpha\otimes\beta)$ need not be an integer matrix.
Equation~\eqref{eq:moving-multinomial-prefactor} is this cell volume
times $(2\pi)^{-9/2}$. The resulting integral is
\[
 (2\pi)^{-9/2}\int_{\mathbb R^9}
 e^{-\frac12\sum_{i,j=1}^3\gamma(t_is_j)x_{ij}^2}\,dx
 =\prod_{i,j=1}^3\gamma(t_is_j)^{-1/2}.
\]
Combining it with \eqref{eq:moving-matching-prefactor}, each pair eigenvalue contributes
\[
 (1+t_is_j)^{-d/2}\gamma(t_is_j)^{-1/2}
 =[(1-Dt_is_j)(1+t_is_j)^D]^{-1/2}.
\]
Let $A_{\rm col}=\one\one^{\mathsf T}-I_d$ be the $d\times d$ matrix
with zero diagonal and all off-diagonal entries equal to one.
Its eigenvalues are $D$ and $-1$, the latter with multiplicity $D$.
Thus the product of the nine contributions gives the factor
\begin{equation}\label{eq:moving-pm-constant}
  \cC_d^{\rm pm}(\alpha,\beta)
  :=\det\!\left(
      I-A_{\rm col}\otimes T_{\alpha,\perp}\otimes T_{\beta,\perp}
    \right)^{-1/2}.
\end{equation}
The determinant is positive in a sufficiently small neighborhood of
$\alpha_c$, because at the critical law the spectral radii of $A_{\rm col}$
and $T_{c,\perp}$ are $D$ and $D^{-1}$, so that
$A_{\rm col}\otimes T_{c,\perp}\otimes T_{c,\perp}$ has spectral radius
$D\cdot D^{-2}=D^{-1}<1$.
Analyticity of $P_\alpha$ and $P_\beta$ then makes this factor smooth.

For the error estimate, the leading Gaussian weight is the product of
\eqref{eq:moving-matching-prefactor},
\eqref{eq:moving-multinomial-prefactor}, and
$\exp\{-\tfrac12\sum_{i,j}\gamma(t_is_j)x_{ij}^2\}$.
The uniform matching asymptotic, Stirling's formula and
\eqref{eq:moving-local-taylor} show that the exact normalized summand
is this weight multiplied by
\[
 [1+O_d((1+\|x\|_2)/\sqrt N)]e^{R_N(x)},
 \qquad |R_N(x)|\le C\|x\|_2^3/\sqrt N.
\]
For sufficiently small fixed $\eta>0$, on
$\|\Gamma-\alpha\otimes\beta\|_2\le\eta$ the cubic remainder is bounded by
one quarter of $\sum_{i,j}\gamma(t_is_j)x_{ij}^2$, since
$\|x\|_2\le C\eta\sqrt N$ and the curvatures have a uniform
positive lower bound. Consequently,
\[
 e^{-\frac12\sum_{i,j}\gamma(t_is_j)x_{ij}^2}|e^{R_N(x)}-1|
 \le C\|x\|_2^3e^{-c\|x\|_2^2}/\sqrt N.
\]
The prefactor error is bounded by
$CN^{-1/2}(1+\|x\|_2)e^{-c\|x\|_2^2}$ after including the exponential.
The integer basis above gives translated lattice cells in $x$ of
diameter $O(N^{-1/2})$ and the computed volume, uniformly over
$\mathcal U_d$. The polynomial times Gaussian bounds above remain
integrable when maximized over each cell, so their sums times the cell
volume are $O(N^{-1/2})$.
The Gaussian gradient is bounded by $C(1+\|x\|_2)e^{-c\|x\|_2^2}$;
cellwise comparison with the integral gives the same error, uniformly
in the lattice translation. The tails outside the fixed neighborhood
are exponentially small. This gives an absolute error $O_d(N^{-1/2})$;
the uniform positive lower bound on $\cC_d^{\rm pm}$ converts it to the
relative error in \eqref{eq:moving-cell-pm-asymptotic}.
\synlabelalias{eq:sqplant-unmarked-precision}{eq:moving-cell-pm-asymptotic}
\end{proof}

\begin{proof}[Proof of Lemma~\ref{lem:overlap-exponential-moment}]
The estimates \eqref{eq:moving-local-taylor}--\eqref{eq:moving-multinomial-prefactor}
and the remainder bounds in the preceding proof control each summand
normalized by the two first moments. Their sum is the second-moment
ratio \eqref{eq:moving-cell-pm-asymptotic}, uniformly bounded away from zero.
Dividing by this sum gives, for $\mathcal X=\sqrt N(\Gamma-\alpha\otimes\beta)$
and sufficiently small fixed $\eta>0$,
\[
 \mu_{\alpha,\beta}(\Gamma)\le C N^{-9/2}e^{-c_0\|\mathcal X\|_2^2},
 \qquad \|\Gamma-\alpha\otimes\beta\|_2\le\eta.
\]
For $0<c<c_0$, multiplying by $e^{c\|\mathcal X\|_2^2}$ and summing
over the nine-dimensional lattice, whose cell volume is $\Theta_d(N^{-9/2})$,
bounds the local contribution uniformly.
Outside this neighborhood, \eqref{eq:moving-global-gap} and the bounded
diameter of overlap matrices show that decreasing $c$ preserves half
the fixed rate gap, making the weighted contribution exponentially small.
This proves \eqref{eq:overlap-exponential-moment}.
By Markov's inequality,
\[
 \mu_{\alpha,\beta}(\|\Gamma-\alpha\otimes\beta\|_2>\eta)
 \le C e^{-cN\eta^2}.
\]
For fixed $\eta>0$, the conditioning event therefore has probability
at least $1/2$ for large $N$. Dividing by this probability proves the
conditional bound after doubling $C$.
\end{proof}

The logarithm of the second-moment factor also admits an expansion
over cycle lengths. Apply $-\log(1-z)=\sum_{i\ge1}z^i/i$ to the
tensor-product eigenvalues in \eqref{eq:moving-pm-constant}.
Their moduli are less than one, and $\tr A_{\rm col}=0$ removes the
$i=1$ term, giving
\begin{equation}\label{eq:pm-cycle-series}
  \log\cC_d^{\rm pm}(\alpha,\beta)
  =\sum_{i\ge2}\frac{\tr(A_{\rm col}^i)}{2i}\,
     \tr(T_{\alpha,\perp}^i)\,\tr(T_{\beta,\perp}^i).
\end{equation}
Here $\tr(A_{\rm col}^i)/(2i)$ is the limiting mean number of
$i$-cycles of the base graph $B_v$ under $\mathbb P_0$.
Such cycle expansions also arise in small-subgraph
conditioning~\cite{RobinsonWormald1992,RobinsonWormald1994,Janson1995,Wormald1999}.
Related fixed-type second moments for bipartite models appear
in~\cite[Theorem~3.3 and Lemma~7.6]{MosselWeitzWormald2009}
and~\cite[Lemmas~A.8 and~A.9]{GSV2016}.

\subsection{The effect of connections on mixed moments}
\label{app:marked-details}\label{app:high-girth-details}

We show that connecting gadgets preserves their common second-moment
factor, up to an error that can be integrated over the occupation types.
Let $H$ be a simple graph on $m$ vertices with maximum degree at most three,
and let $G$ be its graph from Construction~\ref{constr:graph-reduction}.
For subcubic $H$, use the first $\deg_H(v)k$ edges of $M_{v,1}$ at each
$v$, as for \cref{lem:ordered-slot-factorization}.
For $F\subseteq E(H)$, define $G(F)$ by keeping all connection blocks on
edges in $F$ and undoing all $k$ blocks on each edge $uv\notin F$:
for each block formed from $(i_u,j_u),(i_v,j_v)$, remove the four
connecting edges in Figure~\ref{fig:port-wiring} and restore
$i_g^+j_g^-,i_g^-j_g^+$ for $g=u,v$. All other edges remain unchanged.
Thus $G(E(H))=G$ and $G(\varnothing)=G_{\sqcup}$.
Use the same random matchings, edge orders, and orientations for all
graphs $G(F)$. For each $F$, define
\begin{equation}\label{eq:partial-graph-type-weight}
 Z_F(\boldsymbol\alpha):=
 \sum_{I\in\Ind(G(F)):\,\alpha_v(I)=\alpha_v\ \forall v}\lambda^{|I|}.
\end{equation}
Thus $Z_\varnothing(\boldsymbol\alpha)=\prod_vZ_{G_v}(\alpha_v)$ and
$Z_{E(H)}(\boldsymbol\alpha)=Z_H(\boldsymbol\alpha)$.

Use the type neighborhood $\mathcal U_d$ from
Theorem~\ref{thm:moving-cell-second-moment} and the coordinates in
\eqref{eq:exact-stable-center}. To record the type-dependent error, put
\begin{equation}\label{eq:type-error-weight}
 \mathsf w(\alpha)=1+r^4+\|s\|^2,\qquad
 \mathsf W(\boldsymbol\alpha)=\sum_v\mathsf w(\alpha_v).
\end{equation}
Fix a sufficiently small $\eta>0$, depending only on $d$, as permitted by
Lemma~\ref{lem:overlap-exponential-moment}, and write
$\mathcal V(\alpha,\beta)$ for the overlap matrices with these marginals
and $\|\Gamma-\alpha\otimes\beta\|_2\le\eta$.
For fixed $F_1,F_2\subseteq E(H)$, let $\Gamma_v(I,J)$ be the overlap
of $I,J$ in gadget $v$, and define
\[
 M^{\rm loc}_{F_1,F_2;b}(\boldsymbol\alpha,\boldsymbol\beta)
 :=\E_b\!\left[
 \sum_{\substack{I\in\Ind(G(F_1)),\,J\in\Ind(G(F_2))\\
                 \alpha_v(I)=\alpha_v,\,\alpha_v(J)=\beta_v\ \forall v}}
 \lambda^{|I|+|J|}
 \prod_{v\in V(H)}\mathbf1_{\{\Gamma_v(I,J)\in\mathcal V(\alpha_v,\beta_v)\}}
 \right].
\]
We first treat mode $b=0$. Take $\mathfrak c_0$ in
Theorem~\ref{thm:sqplant-full-moments} to be
$\cC_d^{\rm pm}(\alpha_c,\alpha_c)$ from \eqref{eq:moving-pm-constant}.

\begin{lemma}[Local mixed moments]\label{lem:sqplant-weighted-kernel}
For all sufficiently large $N$, all $F_1,F_2\subseteq E(H)$ and integer
marginal types in $\mathcal U_d$,
\begin{equation}\label{eq:sqplant-local-weighted-kernel}
 \left|\log\frac{M^{\rm loc}_{F_1,F_2;0}(\boldsymbol\alpha,\boldsymbol\beta)}
 {\mathfrak c_0^m\E_0Z_{F_1}(\boldsymbol\alpha)\E_0Z_{F_2}(\boldsymbol\beta)}\right|
 \le\frac{C}{\sqrt N}[\mathsf W(\boldsymbol\alpha)+\mathsf W(\boldsymbol\beta)].
\end{equation}
The constant $C$ and the required lower bound on $N$ depend only on
$d$ and the fixed parameter bounds.
\end{lemma}

The proof combines the one-gadget second moment with the coefficient
comparison of Lemma~\ref{lem:normalized-block-replacement}. The key step
expresses the first overlap derivative through products of endpoint
marginal deviations, one from each replica.

\subsubsection{Endpoint marginals and overlap derivatives}
\label{sec:marked-fixed-source}

A connection block pairs $(i_u,j_u)\in M_{u,1}$ with
$(i_v,j_v)\in M_{v,1}$, as in Construction~\ref{constr:graph-reduction}.
In one replica, let $\omega_\ell=(\omega_{\ell,+},\omega_{\ell,-})\in\Omega$
record occupation of $\ell^+,\ell^-$ for each of these four base labels,
and write $\omega=(\omega_{i_u},\omega_{j_u},\omega_{i_v},\omega_{j_v})$.
Choose $e=0$ to keep the original lifts or $e=1$ to replace them by
connecting edges. Set $w_e(\omega)=1$ if no edge of this block has both
endpoints occupied, and $w_e(\omega)=0$ otherwise. With $B$ from
\eqref{eq:base-compatibility},
\[
 w_0(\omega)=B_{\omega_{i_u}\omega_{j_u}}B_{\omega_{i_v}\omega_{j_v}},\qquad
 w_1(\omega)=\prod_{\epsilon\in\{+,-\}}
 (1-\omega_{i_u,\epsilon}\omega_{i_v,\epsilon})
 (1-\omega_{j_u,\epsilon}\omega_{j_v,\epsilon}).
\]
Give a label in state $00,10,01,11$ the monomial $1,x_g,y_g,z_g$,
respectively, in gadget $g$. Summing the four-label monomials with
indicator $w_e$ gives
\begin{equation}\label{eq:Qsrc-Qref}
 Q_0=P(u)P(v),\qquad Q_1=S(u,v)^2,
\end{equation}
using \eqref{eq:P-poly} and \eqref{eq:S-poly}.
Let $\theta=(\theta_u,\theta_v)$ be the logarithmic tilts from
\eqref{eq:connection-type-tilt}, with entries indexed by $10,01,11$;
set $\theta_{g,00}=0$. Define
\begin{equation}\label{eq:marked-local-law}
 \nu_e^\theta(\omega)=
 \frac{w_e(\omega)\exp\{\sum_{g\in\{u,v\}}\sum_{\ell\in\{i_g,j_g\}}\theta_{g,\omega_\ell}\}}
      {Q_e(e^\theta)},\qquad e=0,1.
\end{equation}
For types $\boldsymbol\alpha=(\alpha_u,\alpha_v)$, write
$\nu_{e,\ell}^{\boldsymbol\alpha}$ for the marginal of $\omega_\ell$
under \eqref{eq:marked-local-law} at the tilts for these types,
where $\ell\in\{i_g,j_g\}$ in gadget $g=u,v$.

\begin{lemma}\label{lem:critical-endpoint-centering}
At $\alpha_u=\alpha_v=\alpha_c$,
\begin{equation}\label{eq:endpoint-marginal-equality}
 \nu_{1,\ell}^{(\alpha_c,\alpha_c)}
 =\nu_{0,\ell}^{(\alpha_c,\alpha_c)}=\alpha_c.
\end{equation}
For types in $\mathcal U_d$, each label under $\nu_0$ has marginal
$\nu_{0,\ell}^{\boldsymbol\alpha}=\alpha_g$, and
\begin{equation}\label{eq:moving-endpoint-defect}
 \max_{g\in\{u,v\},\,\ell\in\{i_g,j_g\}}\|\nu_{1,\ell}^{\boldsymbol\alpha}-\alpha_g\|_2
 \le C_d(\|\alpha_u-\alpha_c\|_2+\|\alpha_v-\alpha_c\|_2).
\end{equation}
\end{lemma}
\begin{proof}
Since $Q_0=P(u)P(v)$, under $\nu_0$ the two matching edges have laws
$P_{\alpha_u},P_{\alpha_v}$ from \eqref{eq:moving-edge-law}, so
$\nu_{0,\ell}^{\boldsymbol\alpha}=\alpha_g$.
The coordinates in \eqref{eq:scq-coords} are related to the logarithmic
tilts by an invertible affine map, so the gradient in
Proposition~\ref{prop:port-hessian} also vanishes in these tilts. Thus the expected label-state counts under
$\nu_0$ and $\nu_1$ agree at criticality. Both laws are invariant under swapping
$i_g,j_g$ simultaneously in the two gadgets, giving equality at each label
and hence \eqref{eq:endpoint-marginal-equality}.
Analyticity of the tilts and \eqref{eq:marked-local-law} now gives
\eqref{eq:moving-endpoint-defect}.
\end{proof}

To compare pair weights under connections, record each label's state
in both replicas. The following polynomials count assignments of these
joint states to a matching edge and to a connection block.
For a joint state $(a,b)\in\Omega^2$, use a variable $z_{g,ab}$ and
fix $z_{g,00,00}=1$. The two-replica polynomial for a matching edge is
\begin{equation}\label{eq:pair-edge-polynomial}
 P_2(\mathbf z_g)=\sum_{a,a',b,b'\in\Omega}
 B_{aa'}B_{bb'}z_{g,ab}z_{g,a'b'}.
\end{equation}
Write $\Theta_{g,ab}=\log z_{g,ab}$ on the other fifteen coordinates,
with $\Theta_{g,00,00}=0$.
Fix $\boldsymbol\alpha=(\alpha_u,\alpha_v)$ and
$\boldsymbol\beta=(\beta_u,\beta_v)$ with all four types in $\mathcal U_d$.
For overlaps $\Gamma_g$ near $\alpha_g\otimes\beta_g$ with marginals
$\alpha_g,\beta_g$, choose the unique tilt satisfying
\begin{equation}\label{eq:pair-type-tilt}
 \partial_{\Theta_{g,ab}}\log P_2(e^{\Theta_g})=2\Gamma_{g,ab}
 \quad((a,b)\ne(00,00)).
\end{equation}
The omitted mean follows because every edge has two endpoints.
At $\Gamma_g=\alpha_g\otimes\beta_g$, the solution is
$\Theta_{g,ab}=\theta^{\alpha_g}_{g,a}+\theta^{\beta_g}_{g,b}$, using
one-replica tilts for the indicated types: the normalized weights in
\eqref{eq:pair-edge-polynomial} then factor as $P_{\alpha_g}\otimes P_{\beta_g}$
and have marginal $\alpha_g\otimes\beta_g$.
The exponent support of $P_2$ contains $0,e_1,\ldots,e_{15}$, since $(00,00)$ is compatible
with every joint state. Thus the covariance is positive definite, and
the same inverse-function argument as in Lemma~\ref{lem:expected-type-weight}
gives analytic tilts on a fixed neighborhood of the product overlaps.
Choose $\eta$ small enough that this neighborhood contains every
$\mathcal V(\alpha_g,\beta_g)$.

For two block choices $e,f\in\{0,1\}$, define
\begin{equation}\label{eq:pair-block-polynomial}
 Q^{(2)}_{ef}(\mathbf z_u,\mathbf z_v)
 =\sum_{\omega,\omega'\in\Omega^4}w_e(\omega)w_f(\omega')
       \prod_{g\in\{u,v\}}\prod_{\ell\in\{i_g,j_g\}} z_{g,\omega_\ell\omega'_\ell}.
\end{equation}
Write $Q^{(2)}_{ef}(\Gamma_u,\Gamma_v)$ for its value at the tilts
in \eqref{eq:pair-type-tilt}.
Write $Q_e(\boldsymbol\alpha)$ for $Q_e$ evaluated at the tilts for
$\boldsymbol\alpha$, and similarly for $\boldsymbol\beta$.
The additive form of the joint tilt at product overlap gives
\begin{equation}\label{eq:pair-block-factorization}
 Q^{(2)}_{ef}(\alpha_u\otimes\beta_u,\alpha_v\otimes\beta_v)=Q_e(\boldsymbol\alpha)Q_f(\boldsymbol\beta).
\end{equation}
The logarithmic difference between the two-replica connection ratio
and the product of the two single-replica ratios is
\begin{equation}\label{eq:Psi-definition}
 \Psi_{ef}(\Gamma_u,\Gamma_v)
 =\log\frac{Q^{(2)}_{ef}(\Gamma_u,\Gamma_v)}{Q^{(2)}_{00}(\Gamma_u,\Gamma_v)}
 -\log\frac{Q_e(\boldsymbol\alpha)}{Q_0(\boldsymbol\alpha)}
 -\log\frac{Q_f(\boldsymbol\beta)}{Q_0(\boldsymbol\beta)}.
\end{equation}
It vanishes at $(\alpha_u\otimes\beta_u,\alpha_v\otimes\beta_v)$.
For $\ell\in\{i_g,j_g\}$, put $\Delta_{e,\ell}^{\boldsymbol\alpha}
=\nu_{e,\ell}^{\boldsymbol\alpha}-\alpha_g$, so $\Delta_{0,\ell}^{\boldsymbol\alpha}=0$.

\begin{lemma}[Endpoint cancellation]
\label{lem:pair-saddle-derivative}\label{lem:one-leg-derivative}
For $g=u,v$, consider the perturbation
\[
 \Gamma_g(t)=\alpha_g\otimes\beta_g+t\dot\Gamma_g,
\]
where $\dot\Gamma_g$ has zero row and column sums, so the marginal
types remain fixed. With $T_\alpha$ from \eqref{eq:moving-edge-law},
the first-order change in $\Psi_{ef}$ is
\begin{equation}\label{eq:one-leg-formula-correct}
 \left.\frac{d}{dt}\Psi_{ef}(\Gamma_u(t),\Gamma_v(t))\right|_{t=0}
 =\sum_{g\in\{u,v\}}\sum_{\ell\in\{i_g,j_g\}}
 \left\langle
 \frac{\Delta_{e,\ell}^{\boldsymbol\alpha}}{\alpha_g}
 \otimes\frac{\Delta_{f,\ell}^{\boldsymbol\beta}}{\beta_g},
 (I+T_{\alpha_g}\otimes T_{\beta_g})^{-1}
 \frac{\dot\Gamma_g}{\alpha_g\otimes\beta_g}
 \right\rangle_{\alpha_g\otimes\beta_g}.
\end{equation}
The inner product is weighted by $\alpha_g\otimes\beta_g$ and quotients
are entrywise. Each term contains an endpoint marginal deviation from
each replica, so the derivative vanishes if $e=0$ or $f=0$.
For types in $\mathcal U_d$, its operator norm on these marginal-preserving
directions, using Euclidean norms, satisfies
\begin{equation}\label{eq:Psi-gradient-bound}
 \|\mathrm D\Psi_{ef}(\alpha_u\otimes\beta_u,\alpha_v\otimes\beta_v)\|
 \le C_d\!\sum_{g=u,v}\|\alpha_g-\alpha_c\|\!
           \sum_{g=u,v}\|\beta_g-\alpha_c\|.
\end{equation}
Thus the derivative is $O(N^{-1/2})$ on each fixed $r,s$ window,
with a constant depending on the window and the fixed parameter bounds.
Second derivatives along these directions are uniformly bounded for
$\Gamma_g\in\mathcal V(\alpha_g,\beta_g)$, $g=u,v$.
\end{lemma}
\begin{proof}
Suppress the gadget index and set $K=T_\alpha\otimes T_\beta$.
Put $u=\dot\Gamma/(\alpha\otimes\beta)$. The zero row and column sums
of $\dot\Gamma$ imply $u\in L^2_0(\alpha)\otimes L^2_0(\beta)$.
Let $(X,Y)$ have law $P_\alpha\otimes P_\beta$: $X,Y$ are the
joint states at the two endpoints of a matching edge, and $e_X,e_Y$
are their unit vectors. Allowing all sixteen logarithmic weights to vary temporarily,
the Hessian of $\log P_2$ at product is
\[
 \Cov(e_X+e_Y)=2\diag(\alpha\otimes\beta)(I+K)-4(\alpha\otimes\beta)(\alpha\otimes\beta)^{\mathsf T}.
\]
On perturbations with mean zero under $\alpha\otimes\beta$, the last
term vanishes. Differentiating \eqref{eq:pair-type-tilt} with
$\Theta_{00,00}=0$ gives, for $(a,b)\in\Omega^2$,
\[
 \mathrm D\Theta_{ab}(\alpha\otimes\beta)[\dot\Gamma]
 =[(I+K)^{-1}u]_{ab}-[(I+K)^{-1}u]_{00,00}.
\]
A common shift of all coordinates leaves
$\log(Q^{(2)}_{ef}/Q^{(2)}_{00})$ unchanged, since both blocks have two
labels per gadget. We may therefore compute its derivative using
the zero-mean vector $(I+K)^{-1}u$. The critical spectrum in
\eqref{eq:critical-channel-spectrum-new} and continuity give a uniformly
bounded inverse near criticality; it preserves
$L^2_0(\alpha)\otimes L^2_0(\beta)$.

Restore the gadget indices. At product overlap the replicas are
independent in the block law. Differentiating the finite sum
\eqref{eq:pair-block-polynomial} therefore gives, in all sixteen coordinates,
\[
 \nabla_{\Theta_g}\log
 \frac{Q^{(2)}_{ef}(e^{\Theta_u},e^{\Theta_v})}
      {Q^{(2)}_{00}(e^{\Theta_u},e^{\Theta_v})}
 =\sum_{\ell\in\{i_g,j_g\}}\bigl(
 \nu_{e,\ell}^{\boldsymbol\alpha}\otimes\nu_{f,\ell}^{\boldsymbol\beta}
 -\alpha_g\otimes\beta_g\bigr).
\]
Each summand equals
\[
 \Delta_{e,\ell}^{\boldsymbol\alpha}\otimes\beta_g
 +\alpha_g\otimes\Delta_{f,\ell}^{\boldsymbol\beta}
 +\Delta_{e,\ell}^{\boldsymbol\alpha}\otimes\Delta_{f,\ell}^{\boldsymbol\beta}.
\]
Pair each summand with
$(I+T_{\alpha_g}\otimes T_{\beta_g})^{-1}[\dot\Gamma_g/(\alpha_g\otimes\beta_g)]$.
Its mean in either replica's state with the other fixed is zero,
so the first two terms vanish. The last term, summed over $g,\ell$,
gives \eqref{eq:one-leg-formula-correct}; the single-replica terms in
\eqref{eq:Psi-definition} are constant because the marginals are fixed.
The bounded inverse and
\eqref{eq:moving-endpoint-defect} give \eqref{eq:Psi-gradient-bound};
on fixed windows, each type differs from $\alpha_c$ by $O(N^{-1/4})$.
The second-derivative bound follows from analyticity on the compact
neighborhood of positive tilts.
\end{proof}

\subsubsection{Proof of the mixed-moment comparison}

\begin{proof}[Proof of Lemma~\ref{lem:sqplant-weighted-kernel}]
For each $uv\in E(H)$, set $e=\mathbf1_{\{uv\in F_1\}}$ and
$f=\mathbf1_{\{uv\in F_2\}}$.
We first define the remainder in the connection comparison.
In \eqref{eq:global-marked-coefficient}, write
$\mathfrak C_F(N\boldsymbol\alpha)$ for the coefficient with $P_v=P$
and $Q_{uv}=Q_{\mathbf1_{\{uv\in F\}}}$, and
$\mathfrak C^{(2)}_{F_1,F_2}(N\boldsymbol\Gamma)$ for the coefficient
with $P_v=P_2$ and $Q_{uv}=Q^{(2)}_{ef}$.
The targets omit the $00$ and $(00,00)$ coordinates, respectively.

For the two-replica identity, fix an assignment of joint states
$(a,b)\in\Omega^2$ to the labels, with counts $N\Gamma_{v,ab}$.
As in the proof of Lemma~\ref{lem:ordered-slot-factorization}, uniformly
permute the labels in each gadget and pair consecutive positions.
At $v$, each joint-state sequence with these counts has probability
$\prod_{a,b}(N\Gamma_{v,ab})!/N!$.
By their definitions, $P_2,Q^{(2)}_{ef}$ count sequences satisfying the
edge constraints in both replicas, so the probability that the fixed
assignment satisfies these constraints is $\mathfrak C^{(2)}_{F_1,F_2}(N\boldsymbol\Gamma)$ times
the product of these probabilities over $v$.
There are $N!/\prod_{a,b}(N\Gamma_{v,ab})!$ assignments at each $v$;
summing over assignments cancels these factors. Thus
$\mathfrak C^{(2)}_{F_1,F_2}(N\boldsymbol\Gamma)$ is the expected number
of pairs $I\in\Ind(G^{(1)}(F_1))$, $J\in\Ind(G^{(1)}(F_2))$
with overlaps $\Gamma_v$ at every $v$.

The exponent support of $P_2$ contains $0,e_1,\ldots,e_{15}$, and its
tilts lie in a fixed compact set. Thus
Lemma~\ref{lem:normalized-block-replacement} applies to both versions;
in particular, these coefficients are positive for sufficiently large $N$.
For integer overlaps $\Gamma_v\in\mathcal V(\alpha_v,\beta_v)$, define
\[
 E(\boldsymbol\Gamma):=
 \log\frac{\mathfrak C^{(2)}_{F_1,F_2}(N\boldsymbol\Gamma)
             \mathfrak C_\varnothing(N\boldsymbol\alpha)
             \mathfrak C_\varnothing(N\boldsymbol\beta)}
            {\mathfrak C^{(2)}_{\varnothing,\varnothing}(N\boldsymbol\Gamma)
             \mathfrak C_{F_1}(N\boldsymbol\alpha)
             \mathfrak C_{F_2}(N\boldsymbol\beta)}
 -k\sum_{uv\in E(H)}\Psi_{ef}(\Gamma_u,\Gamma_v).
\]
We will bound this remainder in \eqref{eq:pair-coefficient-error}.
Fixing the overlaps in the two-replica comparison, and the marginal
types in each one-replica comparison, makes the assignment counts,
other matching colors, and fugacity factors cancel in each
connected-to-unconnected ratio. The block normalizers combine as
$\Psi_{ef}$ in \eqref{eq:Psi-definition}. Summing over integer overlaps
therefore gives, with $\mu_{\alpha_v,\beta_v}$ from \eqref{eq:unconnected-overlap-law},
\begin{equation}\label{eq:local-overlap-decomposition}
\begin{aligned}
 \frac{M^{\rm loc}_{F_1,F_2;0}(\boldsymbol\alpha,\boldsymbol\beta)}
 {\E_0Z_{F_1}(\boldsymbol\alpha)\E_0Z_{F_2}(\boldsymbol\beta)}
 &=\prod_v
 \frac{\E_0[Z_{G_v}(\alpha_v)Z_{G_v}(\beta_v)]}
 {\E_0Z_{G_v}(\alpha_v)\E_0Z_{G_v}(\beta_v)}\\
 &\quad\times\sum_{\Gamma_v\in\mathcal V(\alpha_v,\beta_v)\ \forall v}
 e^{k\sum_{uv\in E(H)}\Psi_{ef}(\Gamma_u,\Gamma_v)+E(\boldsymbol\Gamma)}
 \prod_v\mu_{\alpha_v,\beta_v}(\Gamma_v).
\end{aligned}
\end{equation}

\emph{One-gadget factors.}
Exchanging states $10,01$ preserves $B$ and, by uniqueness in
\eqref{eq:moving-edge-law}, conjugates $T_\alpha$ by the same permutation.
Thus the determinant in \eqref{eq:moving-pm-constant} is even in each
replica's imbalance. Analyticity gives
\[
 |\log\cC_d^{\rm pm}(\alpha,\beta)-\log\mathfrak c_0|
 \le C(\rho_\alpha^2+\rho_\beta^2+
             \|b_\alpha-b_c\|+\|b_\beta-b_c\|).
\]
Using \eqref{eq:exact-stable-center}, the estimate
$b_{\logshift}^{*}(\rho)-b_c=O(\logshift+\rho^2)$ from
Lemma~\ref{lem:local-quartic-rate}, and $\logshift=O(N^{-1/2})$
in the fixed parameter range yields, throughout $\mathcal U_d$,
\begin{equation}\label{eq:sqplant-physical-weight}
 \|\alpha-\alpha_c\|^2+\rho^2+\|b-b_c\|
 \le\frac C{\sqrt N}\mathsf w(\alpha).
\end{equation}
Consequently,
\[
 \left|\log\frac{\prod_v\cC_d^{\rm pm}(\alpha_v,\beta_v)}{\mathfrak c_0^m}\right|
 \le\frac C{\sqrt N}[\mathsf W(\boldsymbol\alpha)+\mathsf W(\boldsymbol\beta)].
\]
By \eqref{eq:sqplant-unmarked-precision}, the first product in the
decomposition satisfies the same bound, since $\mathsf W\ge m$.

\emph{Coefficient errors.}
Write $a_v=\|\alpha_v-\alpha_c\|$, $a'_v=\|\beta_v-\alpha_c\|$, and
$\mathcal X_v=\sqrt N(\Gamma_v-\alpha_v\otimes\beta_v)$.
The block's mean endpoint-count vector at $v$, minus that for a
matching edge, is
\[
 \nabla_{\Theta_v}\log
 \frac{Q^{(2)}_{ef}(e^{\Theta_u},e^{\Theta_v})}
      {P_2(e^{\Theta_u})P_2(e^{\Theta_v})}.
\]
At critical product types it vanishes by Lemma~\ref{lem:critical-endpoint-centering} and
\eqref{eq:pair-block-factorization}. Analyticity therefore bounds its norm by
$C[a_u+a_v+a'_u+a'_v+(\|\mathcal X_u\|+\|\mathcal X_v\|)/\sqrt N]$.
The one-replica defects satisfy the same bound without the other
replica or overlap terms. Since $k/N=O(N^{-1/2})$, $k^2/N=O(1)$
and degrees are bounded, applying
\eqref{eq:normalized-block-replacement} to the three coefficient ratios
in the definition of $E$ gives
\begin{equation}\label{eq:pair-coefficient-error}
 |E(\boldsymbol\Gamma)|\le
 C m/\sqrt N+C\sum_v(a_v^2+(a'_v)^2)
                 +\frac C N\sum_v\|\mathcal X_v\|^2.
\end{equation}
The $m/\sqrt N$ term includes local coefficient errors and the
$O(mk/N)$ variance contribution. In \eqref{eq:normalized-block-replacement},
$k^2/N$ multiplies squared mean shifts, giving the other terms.

\emph{Overlap integration.}
For a fixed sufficiently large $C$, set
$t_v=C\sum_{u:\,uv\in F_1\cap F_2}(a_u+a_v)(a'_u+a'_v)$.
By \eqref{eq:pair-block-factorization}, each $\Psi_{ef}$ vanishes at
product overlap. Taylor expansion and Lemma~\ref{lem:pair-saddle-derivative} give
\begin{equation}\label{eq:pair-block-taylor-bound}
 \left|k\sum_{uv}\Psi_{ef}(\Gamma_u,\Gamma_v)\right|
 \le\sum_v t_v\|\mathcal X_v\|+
                   \frac C{\sqrt N}\sum_v\|\mathcal X_v\|^2,
 \qquad
 \sum_vt_v\le C\sum_v(a_v^2+(a'_v)^2).
\end{equation}
Here blocks with $e=0$ or $f=0$ have zero first derivative;
$k/\sqrt N=O(1)$ and $k/N=O(N^{-1/2})$ control the linear and
quadratic terms, respectively. Bounded degree gives the bound on
$\sum_vt_v$ and, since $\mathcal U_d$ is compact, a uniform bound $t_v\le T$.

Under $\mu_{\alpha_v,\beta_v}$, or its conditioning on
$\mathcal V(\alpha_v,\beta_v)$, \eqref{eq:overlap-exponential-moment}
gives, uniformly for $0\le t\le T$ and sufficiently small $z\ge0$,
\[
 -C(t+z)\le\log\E e^{-t\|\mathcal X_v\|-z\|\mathcal X_v\|^2}
 \le0\le\log\E e^{t\|\mathcal X_v\|+z\|\mathcal X_v\|^2}\le C(t+z).
\]
For the upper bound, differentiate the log moment along the segment from
$(0,0)$ to $(t,z)$; its derivatives are uniformly bounded by the
quadratic exponential moment. Jensen's inequality gives the lower bound.
Lemma~\ref{lem:overlap-exponential-moment} and Markov's inequality give
$\mu_{\alpha_v,\beta_v}(\mathcal V(\alpha_v,\beta_v)^c)\le Ce^{-cN\eta^2}$.
For fixed $\eta$ and sufficiently large $N$,
\[
 e^{-Cm e^{-c'N}}\le
 \prod_v\mu_{\alpha_v,\beta_v}(\mathcal V(\alpha_v,\beta_v))\le1.
\]
Apply the exponential bounds with $t=t_v$ and $z=C/\sqrt N$, also
covering the $C/N$ term in \eqref{eq:pair-coefficient-error}.
On the restricted overlaps, \eqref{eq:pair-coefficient-error} and
\eqref{eq:pair-block-taylor-bound} bound the exponent from both sides.
For the lower bound, condition on every $\Gamma_v\in\mathcal V(\alpha_v,\beta_v)$
and multiply by this event's probability. For the upper bound, extend
the sum of the positive exponential upper bound to all overlaps.
Independence lets us apply the one-vertex exponential bounds separately.
The resulting errors satisfy, by \eqref{eq:sqplant-physical-weight}
and \eqref{eq:pair-block-taylor-bound},
\[
 \sum_vt_v+\frac m{\sqrt N}+\sum_v(a_v^2+(a'_v)^2)
 \le\frac C{\sqrt N}[\mathsf W(\boldsymbol\alpha)+\mathsf W(\boldsymbol\beta)].
\]
The restriction adds at most $Cm e^{-c'N}$ to the logarithmic error,
absorbed using $\mathsf W\ge m$. Thus the logarithm of the overlap sum
has absolute value at most
$CN^{-1/2}[\mathsf W(\boldsymbol\alpha)+\mathsf W(\boldsymbol\beta)]$.
Combining with the one-gadget factors proves
\eqref{eq:sqplant-local-weighted-kernel}.
\end{proof}

\subsection{Conditioning on simplicity}\label{sec:marked-cycle-conditioning}

We extend the local mixed-moment comparison to the simple-graph mode.
Let $\mathcal E_v=\{B_v\text{ is simple}\}$, so
$\mathbb P_1=\mathbb P_0(\cdot\mid\cap_v\mathcal E_v)$.
This event concerns the original base graphs: no two matching colors
may share an edge, or equivalently form a cycle of length two.
Recall that $Z_F(\boldsymbol\alpha)$ in \eqref{eq:partial-graph-type-weight}
is the independent-set weight on $G(F)$ with gadget types
$\boldsymbol\alpha$. As in \cref{app:marked-details},
$M^{\rm loc}_{F_1,F_2;b}$ restricts the overlap in each gadget to
$\mathcal V(\alpha_v,\beta_v)$; use $\mathsf W$ from \eqref{eq:type-error-weight}.
With $\mathfrak c_0=\cC_d^{\rm pm}(\alpha_c,\alpha_c)$ from
\eqref{eq:moving-pm-constant} and $T_\alpha$ from \eqref{eq:moving-edge-law},
take the remaining constant in Theorem~\ref{thm:sqplant-full-moments} to be
\begin{equation}\label{eq:common-multipliers}
 \mathfrak c_1=\mathfrak c_0e^{-\frac{d(d-1)}4[\tr(T_{\alpha_c}^2)-1]^2}.
\end{equation}
Here \eqref{eq:critical-channel-spectrum-new} gives
$\tr(T_{\alpha_c}^2)-1=2(d-1)^{-2}+(d-1)^{-4}$.
Since $\tr(A_{\rm col}^2)=d(d-1)$ and
$\tr(T_{c,\perp}^2)=\tr(T_{\alpha_c}^2)-1$, the exponent in
\eqref{eq:common-multipliers} is minus the $i=2$ term of
\eqref{eq:pm-cycle-series} at $\alpha=\beta=\alpha_c$. Hence
\[
 \log\mathfrak c_1
 =\sum_{i\ge3}\frac{\tr(A_{\rm col}^i)}{2i}\,\tr(T_{c,\perp}^i)^2>0,
\]
since $\tr(A_{\rm col}^i)=D^i+(-1)^iD>0$ for $i\ge3$ and
$\tr(T_{c,\perp}^3)=D^{-6}$ by \eqref{eq:critical-channel-spectrum-new}.
Thus $\mathfrak c_1>1$.

\begin{lemma}[Conditioned mixed moments]\label{lem:conditioned-local-moments}
Under the parameter, graph, and type assumptions of
Lemma~\ref{lem:sqplant-weighted-kernel}, for all sufficiently large $N$,
for every $F_1,F_2\subseteq E(H)$,
\begin{equation}\label{eq:both-local-kernels}
 \left|\log\frac{M^{\rm loc}_{F_1,F_2;1}(\boldsymbol\alpha,\boldsymbol\beta)}
 {\mathfrak c_1^m\E_1Z_{F_1}(\boldsymbol\alpha)\E_1Z_{F_2}(\boldsymbol\beta)}\right|
 \le\frac{C_d}{\sqrt N}[\mathsf W(\boldsymbol\alpha)+\mathsf W(\boldsymbol\beta)].
\end{equation}
The constant $C_d$ and the required lower bound on $N$ depend only on
$d$ and the fixed parameter bounds.
\end{lemma}

Together with Lemma~\ref{lem:sqplant-weighted-kernel}, this gives the
local mixed-moment bound for both $b=0,1$. To estimate simplicity
probabilities, reweight $\mathbb P_0$ as follows:
\[
 \frac{d\mathbb P^{\rm pl}_{F;\boldsymbol\alpha}}{d\mathbb P_0}
 =\frac{Z_F(\boldsymbol\alpha)}{\E_0Z_F(\boldsymbol\alpha)},\qquad
 \frac{d\mathbb P^{\rm pl}_{F_1,F_2;\boldsymbol\alpha,\boldsymbol\beta}}
      {d\mathbb P_0}
 =\frac{Z_{F_1}(\boldsymbol\alpha)Z_{F_2}(\boldsymbol\beta)}
       {\E_0[Z_{F_1}(\boldsymbol\alpha)Z_{F_2}(\boldsymbol\beta)]}.
\]
Given the construction, sample independent sets in the respective graphs
independently, with probabilities proportional to $\lambda^{|I|}$ and
subject to their specified types. Use the same notation for these joint
planted laws. For given overlap matrices
$\boldsymbol\Gamma=(\Gamma_v)_v$, condition the pair law additionally
on having overlap $\Gamma_v$ in every gadget $v$.

Define the simplicity probabilities
\begin{equation}\label{eq:planted-simplicity-probabilities}
 p_F(\boldsymbol\alpha):=\mathbb P^{\rm pl}_{F;\boldsymbol\alpha}
 \!\left(\bigcap_v\mathcal E_v\right),\quad
 p_{F_1,F_2}(\boldsymbol\Gamma):=
 \mathbb P^{\rm pl}_{F_1,F_2;\boldsymbol\alpha,\boldsymbol\beta}
 \!\left(\bigcap_v\mathcal E_v\,\middle|\,
 \Gamma_v(I,J)=\Gamma_v\ \forall v\right),
\end{equation}
where the conditioning event must have positive probability.
\added{Under $\mathbb P_0$, the events $\mathcal E_v$ are independent
with a common probability, so
$\mathbb P_0(\bigcap_v\mathcal E_v)=\mathbb P_0(\mathcal E_v)^m$.}
The definition of $\mathbb P^{\rm pl}_{F;\boldsymbol\alpha}$ gives
\begin{equation}\label{eq:first-moment-conditioning-identity}
 \E_1Z_F(\boldsymbol\alpha)
 =\E_0Z_F(\boldsymbol\alpha)\,
   \frac{p_F(\boldsymbol\alpha)}{\mathbb P_0(\mathcal E_v)^m}.
\end{equation}
At fixed overlap $\boldsymbol\Gamma$, the analogous identity uses
$p_{F_1,F_2}(\boldsymbol\Gamma)$ in place of $p_F(\boldsymbol\alpha)$.

The factorial-moment and Poisson method is classical
\cite{Janson1995,Wormald1999,JohnsonPaquette2013}; here we control its error
uniformly after exposing $O(\sqrt N)$ edges of $M_{v,1}$ and their
endpoint states.
For one gadget $v$, fix the occupation bits at every base label
for one vertex set ($j=1$, $\xi=\alpha$) or two vertex sets
($j=2$, $\xi=\Gamma$). A matching edge is compatible if neither of
its two lifts has both endpoints in any one of the fixed sets;
this is the condition $B_{aa'}=1$ for one set and
$B_{aa'}B_{bb'}=1$ for two sets. For $j=0$, impose no restriction
and put $\xi=(1)$.

Let $\mathcal R$ specify disjoint edges of the original $M_{v,1}$,
including their endpoint labels and states. Define
$\mathbb P^{(j)}_\xi(\cdot\mid\mathcal R)$ as the following distribution
on the $d$ matchings: keep these edges fixed, choose a uniform compatible
perfect matching on the remaining labels to complete $M_{v,1}$,
and independently choose each other matching uniformly among compatible
perfect matchings on all labels. Assume these choices are feasible.
The fixed edges need not be compatible, since their lifts may have been
removed by connections; they still count when testing simplicity of $B_v$.
\rev{In the joint planted laws, the partition-function weights cancel
the normalizing factors for sampling the independent sets.
Given all occupation states and designated matching edges, the fugacity
weights and all cross-gadget constraints are fixed. The remaining
matchings therefore have laws $\mathbb P^{(j)}_\xi(\cdot\mid\mathcal R)$,
independently across gadgets.}

The simplicity probabilities will be expressed using $T_\alpha$ from
\eqref{eq:moving-edge-law}. Define $P_\Gamma^{(2)}$ and $T_\Gamma^{(2)}$
by the same formula with $B\otimes B,\Gamma$ in place of $B,\alpha$.
Both $B$ and $B\otimes B$ satisfy the support condition in
\cref{lem:finite-incidence-transfer}: state $00$, or $(00,00)$,
is compatible with every state, so the support is connected and has a loop.
The coupling $P_\alpha\otimes P_\beta$ has marginal
$\alpha\otimes\beta$ and the required exponential-tilt form, so
uniqueness gives $T_{\alpha\otimes\beta}^{(2)}=T_\alpha\otimes T_\beta$.
Taking the trace of its square yields
\begin{equation}\label{eq:pair-cycle-product-factor}
 \tr[(T_{\alpha\otimes\beta}^{(2)})^2]
 =\tr(T_\alpha^2)\tr(T_\beta^2).
\end{equation}

\begin{lemma}[Simplicity probabilities]
\label{lem:uniform-cycle-free-transfer}\label{lem:uniform-cycle-poisson-complete}
\label{cor:effective-cycle-free-lower}\label{lem:square-fixed-cycles}
Fix $d\ge3$, $K>0$ and $j\in\{0,1,2\}$. For $j=1,2$, let
$\xi=\alpha$ or $\Gamma$, respectively, range over a fixed compact set
where all state proportions and all coupling entries corresponding to
$B_{aa'}=1$ or $B_{aa'}B_{bb'}=1$, respectively, are positive.
For $j=0$, use $\xi=(1)$. Let $\mathcal R$ specify
$q_N\le K\sqrt N$ disjoint edges of $M_{v,1}$ and their endpoint states,
and use the residual law $\mathbb P^{(j)}_\xi(\cdot\mid\mathcal R)$.
Let $X$ count cycles of length two, each specified by a base edge
and two matching colors containing it; exposed original edges are included.
Define the limiting means
\[
 \Lambda_0=\frac{d(d-1)}4,\quad
 \Lambda_1(\alpha)=\frac{d(d-1)}4\tr(T_\alpha^2),\quad
 \Lambda_2(\Gamma)=\frac{d(d-1)}4\tr[(T_\Gamma^{(2)})^2],
\]
and put $\mathfrak a_j(\xi)=e^{-\Lambda_j(\xi)}$.
There are constants $C,N_0>0$, depending only on $d,K$ and the fixed compact set, such that, for even
$N\ge N_0$ and uniformly after every admissible exposure $\mathcal R$,
\begin{equation}\label{eq:uniform-no-cycle-quantitative}
 \left|\log\frac{\mathbb P^{(j)}_\xi(X=0\mid\mathcal R)}{\mathfrak a_j(\xi)}\right|
 \le\frac C{\sqrt N}.
 \synlabelalias{eq:cycle-free-transfer}{eq:uniform-no-cycle-quantitative}
\end{equation}
Moreover, the simplicity probability satisfies
$\mathbb P^{(j)}_\xi(X=0\mid\mathcal R)\ge\tfrac12e^{-4d(d-1)}$.
Each $\mathfrak a_j(\xi)$ is analytic in $\xi$ on a neighborhood
of the fixed compact set.
\added{For rational $K$ and a compact type set specified by rational
coordinate bounds, $C,N_0$ can be computed from $d$, $K$, and these bounds.}
\end{lemma}

\begin{proof}

All expectations below use $\mathbb P^{(j)}_\xi(\cdot\mid\mathcal R)$.
In this proof write $P_\xi,T_\xi$ for $P_\alpha,T_\alpha$ if $j=1$,
$P_\Gamma^{(2)},T_\Gamma^{(2)}$ if $j=2$, and $(1),(1)$ if $j=0$.
Put $w(a,b)=P_\xi(a,b)/(\xi(a)\xi(b))$, which is zero for
incompatible states and uniformly bounded otherwise.

\emph{Cycles meeting exposed endpoints.}
Deleting the $2q_N\le2K\sqrt N$ exposed endpoints changes the state
proportions by $O(N^{-1/2})$. By compactness and continuity of the couplings,
for sufficiently large $N$ the original and residual proportions lie in
a common fixed compact set on which all state proportions and permitted
coupling entries have uniform positive lower bounds.
Lemma~\ref{lem:finite-incidence-transfer}, applied on this set, shows that any $\ell\le N^{1/10}$ additional
disjoint compatible edges in one color, away from its exposed vertices,
with endpoint states $(a_i,b_i)$, have joint probability
\[
 N^{-\ell}\prod_{i=1}^{\ell}w(a_i,b_i)
 \left[1+\rev{O\left(\frac\ell{\sqrt N}\right)}\right].
\]
Let $U$ be the set of $2q_N$ exposed endpoints, and let $X_0$ count
cycles outside $U$. A cycle meeting $U$ and using the first color must
contain one of the exposed matching edges; the other edge has probability $O(N^{-1})$.
For other color pairs, there are $O(q_NN)$ placements meeting $U$,
each of probability $O(N^{-2})$. Thus
\begin{equation}\label{eq:simplicity-exposure-cost}
 \E(X-X_0)=O(q_N/N),\qquad
 |\Pr(X=0)-\Pr(X_0=0)|=O(N^{-1/2}).
\end{equation}
An exposed matching edge with incompatible endpoint states cannot be repeated by a compatible color.

\emph{Factorial moments.}
\rev{For integers $t\ge1$ with $2t\le N^{1/[4(d+1)]}$, we estimate}
$(X_0)_t=X_0(X_0-1)\cdots(X_0-t+1)$.
A cycle specifies an unordered label pair and two distinct colors.
For endpoint states $a,b$, the probability that both colors contain
this pair has leading term $N^{-2}w(a,b)^2$.
The sum $\Lambda_N$ of these terms over cycles outside $U$ satisfies
\[
 \Lambda_N=\frac12\binom d2
       \sum_{a,b}\xi(a)\xi(b)w(a,b)^2+O((q_N+1)/N)
       =\Lambda_j(\xi)+O(N^{-1/2}).
\]
Indeed, symmetry of $P_\xi$ gives
$\sum_{a,b}P_\xi(a,b)^2/(\xi(a)\xi(b))=\tr(T_\xi^2)$;
the error deletes coincident and exposed labels. The sum of these leading terms over
cycles through any specified vertex is $O(N^{-1})$. Removing tuples sharing a vertex
from $\Lambda_N^t$ therefore costs at most \rev{$C^tt^2/N\le C^tt/\sqrt N$}.
Lemma~\ref{lem:finite-incidence-transfer}, applied in each matching color, compares every remaining
joint probability with the product of these leading terms to relative error
$\rev{O(t/\sqrt N)}$. Since
$|\Lambda_N^t-\Lambda_j^t|\le C^tt/\sqrt N$, the vertex-disjoint tuples
have total error at most \rev{$C^tt/\sqrt N$}.

For tuples of cycles sharing vertices, form the union of their distinct
colored edges. Let $p,v,e$ count its distinct endpoint pairs, vertices, and
colored edges. For endpoint pair $a$, let $h_a$ count its colors and
$t_a$ its selected cycles. The cycles in a factorial moment are distinct,
so
\[
 2\le h_a\le d,\qquad 1\le t_a\le\binom{h_a}{2},\qquad
 \sum_a h_a=e,\quad\sum_a t_a=t.
\]
Write $r=e-v$ and $j'=2t-v$. Since
$2(t_a-1)\le(d+1)(h_a-2)$ and $2p-v\ge0$,
\begin{equation}\label{eq:cycle-edge-excess}
 r=\sum_a(h_a-2)+(2p-v),\qquad
 j'=2\sum_a(t_a-1)+(2p-v)\le(d+1)r.
\end{equation}
Moreover $r=0$ holds exactly for vertex-disjoint cycles.
Each of the $t$ cycles contributes two endpoint positions.
A partition of these $2t$ positions records which represent the same
base label. There are at most $(2t)^{2j'}$ such partitions into
$2t-j'$ vertices: choose the $j'$ nonminimal positions and assign each
the least position in its part. Choosing two colors and ordering the two
endpoints of each cycle gives at most $[2\binom d2]^t\le C^t$ choices.
For each $r\ge1$, \eqref{eq:cycle-edge-excess} therefore bounds the
number of patterns by $2C^t(2t)^{2(d+1)r}$. A pattern has at most
$N^v$ labelings, each with probability at most $C^tN^{-e}$.
If two distinct edges of the same color share an endpoint, the probability
is zero. \rev{Consequently, the total contribution from tuples sharing vertices is}
at most
\[
 C^t\sum_{r=1}^{2t}
       \left(\frac{(2t)^{2(d+1)}}N\right)^r
 \le \frac{C^t(2t)^{2(d+1)}}N
 \rev{\le\frac{C^t}{\sqrt N}}.
\]
The geometric ratio is at most $N^{-1/2}$, and the range also ensures
$2t\le N^{1/10}$ as required by Lemma~\ref{lem:finite-incidence-transfer}. Combining the two
kinds of tuples gives
\begin{equation}\label{eq:square-cycle-factorial}
 |\E(X_0)_t-\Lambda_j(\xi)^t|
 \le \rev{\frac{C^tt}{\sqrt N}}.
\end{equation}

\emph{Probability of no cycles.}
Let $L_N=2\lfloor\log N/(2\log\log N)\rfloor$.
The Bonferroni inequalities give
\[
 \sum_{t=0}^{L_N+1}\frac{(-1)^t}{t!}\E(X_0)_t
 \le\Pr(X_0=0)\le
 \sum_{t=0}^{L_N}\frac{(-1)^t}{t!}\E(X_0)_t,
\]
where $(X_0)_0=1$. Both truncation orders lie in the range of
\eqref{eq:square-cycle-factorial} for sufficiently large $N$.
Replacing $\E(X_0)_t$ by $\Lambda_j(\xi)^t$ in either sum costs at most
\[
 \sum_{t=1}^{L_N+1}\frac{|\E(X_0)_t-\Lambda_j(\xi)^t|}{t!}
 =O(N^{-1/2}),
\]
\rev{since $\sum_{t\ge1}C^tt/t!<\infty$ for fixed $C$.}
The resulting sums are truncations of $e^{-\Lambda_j(\xi)}$.
Each transition matrix is stochastic on at most $16$ states, so
$0\le\Lambda_j(\xi)\le\Lambda_*:=4d(d-1)$ for $j=0,1,2$.
For sufficiently large $N$, both exponential-series remainders are at most
$\Lambda_*^{L_N+1}/(L_N+1)!=N^{-1+o(1)}$.
Thus $\Pr(X_0=0)=e^{-\Lambda_j}+O(N^{-1/2})$.
Equation~\eqref{eq:simplicity-exposure-cost} gives the same estimate for $\Pr(X=0)$;
$e^{-\Lambda_j}\ge e^{-\Lambda_*}>0$ converts the error to the logarithmic
bound in \eqref{eq:uniform-no-cycle-quantitative}. Taking
$C/\sqrt{N_0}\le\log 2$ gives the stated lower bound.
Analyticity follows from that of the couplings.
In our applications, the type ranges can be chosen as closed boxes
with rational endpoints satisfying the positivity conditions above.
Bounds on the tilts and their derivatives can then be computed from
the finite coupling equations. Together with the Fourier bounds in
the proof of \cref{lem:moving-lclt}, these give computable $C,N_0$.

\end{proof}

\begin{proof}[Proof of Lemma~\ref{lem:conditioned-local-moments}]
\rev{Applying \eqref{eq:first-moment-conditioning-identity} and its fixed-overlap version}
to the pair weights and the two first moments
in \eqref{eq:local-overlap-decomposition} gives
\[
\begin{aligned}
 &\frac{M^{\rm loc}_{F_1,F_2;1}(\boldsymbol\alpha,\boldsymbol\beta)}
 {\E_1Z_{F_1}(\boldsymbol\alpha)\E_1Z_{F_2}(\boldsymbol\beta)}
 =\frac{\mathbb P_0(\mathcal E_v)^m}
 {p_{F_1}(\boldsymbol\alpha)p_{F_2}(\boldsymbol\beta)}
 \prod_v\frac{\E_0[Z_{G_v}(\alpha_v)Z_{G_v}(\beta_v)]}
 {\E_0Z_{G_v}(\alpha_v)\E_0Z_{G_v}(\beta_v)}\\
 &\qquad\times\sum_{\Gamma_v\in\mathcal V(\alpha_v,\beta_v)\ \forall v}
 e^{k\sum_{uv\in E(H)}\Psi_{ef}(\Gamma_u,\Gamma_v)+E(\boldsymbol\Gamma)}
 p_{F_1,F_2}(\boldsymbol\Gamma)\prod_v\mu_{\alpha_v,\beta_v}(\Gamma_v).
\end{aligned}
\]
Under $\mathbb P^{\rm pl}_{F_1;\boldsymbol\alpha}$, condition on all occupation
states and expose the designated matching edges. The remaining matchings
are independent across gadgets with laws
$\mathbb P^{(1)}_{\alpha_v}(\cdot\mid\mathcal R_v)$.
At most $3k=O(\sqrt N)$ matching edges are exposed per gadget.
The compact tilt ranges in Section~\ref{sec:marked-fixed-source} bound
the state proportions and positive entries of $P_\alpha,P_\Gamma^{(2)}$
away from zero, uniformly for $\alpha,\beta\in\mathcal U_d$ and
$\Gamma\in\mathcal V(\alpha,\beta)$.
Thus Lemma~\ref{lem:square-fixed-cycles} applies uniformly, and averaging gives
\[
 p_{F_1}(\boldsymbol\alpha)
 =\E_{\mathbb P^{\rm pl}_{F_1;\boldsymbol\alpha}}
   \left[\prod_v\mathbb P^{(1)}_{\alpha_v}(\mathcal E_v\mid\mathcal R_v)\right],
\]
and hence
\begin{equation}\label{eq:one-replica-simplicity-probability}
 \left|\log\frac{p_{F_1}(\boldsymbol\alpha)}{\prod_v\mathfrak a_1(\alpha_v)}\right|
 \le\frac{C_dm}{\sqrt N}.
\end{equation}
For pairs, the same argument conditional on $\Gamma_v(I,J)=\Gamma_v$ for all $v$ gives
\[
 p_{F_1,F_2}(\boldsymbol\Gamma)
 =\E_{\mathbb P^{\rm pl}_{F_1,F_2;\boldsymbol\alpha,\boldsymbol\beta}}
 \!\left[\prod_v\mathbb P^{(2)}_{\Gamma_v}(\mathcal E_v\mid\mathcal R_v)
 \,\middle|\,\Gamma_v(I,J)=\Gamma_v\ \forall v\right],
\]
and hence
\begin{equation}\label{eq:pair-simplicity-probability}
 \left|\log\frac{p_{F_1,F_2}(\boldsymbol\Gamma)}{\prod_v\mathfrak a_2(\Gamma_v)}\right|
 \le\frac{C_dm}{\sqrt N}.
\end{equation}
Under $\mathbb P_0$, Lemma~\ref{lem:square-fixed-cycles} with $j=0$ gives
$|\log[\mathbb P_0(\mathcal E_v)/\mathfrak a_0]|\le C_d/\sqrt N$.
Thus
\begin{equation}\label{eq:finite-conditioning-factor}
 \left|\log\frac{\mathbb P_0(\mathcal E_v)^m p_{F_1,F_2}(\boldsymbol\Gamma)}
 {p_{F_1}(\boldsymbol\alpha)p_{F_2}(\boldsymbol\beta)}
 -\sum_v\log\frac{\mathfrak a_0\mathfrak a_2(\Gamma_v)}
 {\mathfrak a_1(\alpha_v)\mathfrak a_1(\beta_v)}\right|
 \le\frac{Cm}{\sqrt N}.
\end{equation}
At product overlap, \eqref{eq:pair-cycle-product-factor} gives
\begin{equation}\label{eq:critical-simplicity-cancellation}
 \frac{\mathfrak a_0\mathfrak a_2(\alpha\otimes\beta)}
      {\mathfrak a_1(\alpha)\mathfrak a_1(\beta)}
 =e^{-\frac{d(d-1)}4[\tr(T_\alpha^2)-1][\tr(T_\beta^2)-1]}.
\end{equation}
At $\alpha=\beta=\alpha_c$, this factor equals
$\mathfrak c_1/\mathfrak c_0$ by \eqref{eq:common-multipliers}.
We bound its variation with the marginal types and then with the overlap.

Exchanging states $10,01$ conjugates $T_\alpha$ by the same permutation,
as in the proof of \cref{lem:sqplant-weighted-kernel}. Hence
$\tr(T_\alpha^2)$ is even in $\rho$. Analyticity and
\eqref{eq:sqplant-physical-weight} give
\begin{equation}\label{eq:cycle-multiplier-evenness}
 |\tr(T_\alpha^2)-\tr(T_{\alpha_c}^2)|
 \le C_d(\rho^2+\|b-b_c\|)
 \le \frac{C_d}{\sqrt N}\mathsf w(\alpha).
\end{equation}

Substituting into \eqref{eq:critical-simplicity-cancellation} gives
\[
 \left|\log\frac{\mathfrak c_0\mathfrak a_0\mathfrak a_2(\alpha\otimes\beta)}
 {\mathfrak c_1\mathfrak a_1(\alpha)\mathfrak a_1(\beta)}\right|
 \le\frac C{\sqrt N}[\mathsf w(\alpha)+\mathsf w(\beta)].
\]

For $\Gamma_v=\alpha_v\otimes\beta_v+\mathcal X_v/\sqrt N$
in $\mathcal V(\alpha_v,\beta_v)$, positivity and analyticity of
$\mathfrak a_2$ on the fixed compact set give
\[
 \left|\log\frac{\mathfrak a_2(\Gamma_v)}
                       {\mathfrak a_2(\alpha_v\otimes\beta_v)}\right|
 \le\frac C{\sqrt N}\|\mathcal X_v\|.
\]
Combining these bounds with \eqref{eq:finite-conditioning-factor} and
$\mathsf W\ge m$ yields
\begin{equation}\label{eq:conditioning-overlap-error}
 \left|\log\frac{\mathfrak c_0^m\mathbb P_0(\mathcal E_v)^m
 p_{F_1,F_2}(\boldsymbol\Gamma)}
 {\mathfrak c_1^m p_{F_1}(\boldsymbol\alpha)p_{F_2}(\boldsymbol\beta)}\right|
 \le\frac C{\sqrt N}\left[\mathsf W(\boldsymbol\alpha)
 +\mathsf W(\boldsymbol\beta)+\sum_v\|\mathcal X_v\|\right].
\end{equation}
In the conditioned overlap sum, factor out
$(\mathfrak c_1/\mathfrak c_0)^m$. The displayed bound contributes
$C[\mathsf W(\boldsymbol\alpha)+\mathsf W(\boldsymbol\beta)]/\sqrt N$
independently of the overlaps and adds $C/\sqrt N$ to each coefficient
$t_v$ from \eqref{eq:pair-block-taylor-bound} in the upper and lower
exponential bounds in the proof of Lemma~\ref{lem:sqplant-weighted-kernel}.
By \eqref{eq:sqplant-physical-weight} and $\mathsf W\ge m$,
\[
 \sum_v(t_v+C/\sqrt N)
 \le\frac C{\sqrt N}[\mathsf W(\boldsymbol\alpha)+\mathsf W(\boldsymbol\beta)].
\]
The same exponential-moment estimates bound the absolute value of the
logarithm of the remaining overlap sum by
$C[\mathsf W(\boldsymbol\alpha)+\mathsf W(\boldsymbol\beta)]/\sqrt N$.
Combining this with the one-gadget factors from that proof gives
\eqref{eq:both-local-kernels}, since
$\mathfrak c_0^m(\mathfrak c_1/\mathfrak c_0)^m=\mathfrak c_1^m$.
\end{proof}

\subsection{Integrating type-dependent errors}

The preceding mixed-moment bounds have multiplicative factors in
$[e^{-C\mathsf W/\sqrt N},e^{C\mathsf W/\sqrt N}]$, with $\mathsf W$ defined in
\eqref{eq:type-error-weight}. We average these factors with probabilities
proportional to the expected type weights on $\mathcal U_d$, obtaining
bounds in $[e^{-C'm/\sqrt N},e^{C'm/\sqrt N}]$. Types outside $\mathcal U_d$ are treated
in the next subsection.

\begin{lemma}[Exponential moment]
\label{lem:sqplant-coupled-moment}
Use the fixed parameter bounds and graphs of
Lemma~\ref{lem:sqplant-weighted-kernel}. Fix $b\in\{0,1\}$ and
$F\subseteq E(H)$. All type assignments below satisfy
$\alpha_v\in\mathcal U_d$ and have integer counts $N\alpha_{v,a}$ at every vertex.
Choose $\mathcal A$ to be one of three sets: all such assignments;
those with phase vector $\sigma$; or those with phase vector $\sigma$
and the fixed $R,W$ window in \eqref{eq:restricted-phase-weight}.
For each choice of $\mathcal A$, define the corresponding probability distribution by
\begin{equation}\label{eq:annealed-type-law}
 \nu(\boldsymbol\alpha)=
 \frac{\mathbf1_{\{\boldsymbol\alpha\in\mathcal A\}}
           \E_bZ_F(\boldsymbol\alpha)}
      {\sum_{\boldsymbol\beta\in\mathcal A}\E_bZ_F(\boldsymbol\beta)}.
\end{equation}
For each of these three choices, there are constants $c,C>0$ such that, for all sufficiently large $N$,
\begin{equation}\label{eq:sqplant-coupled-moment}
 \E_\nu e^{c\mathsf W}\le e^{Cm}.
\end{equation}
The constants $c,C$ and the required lower bound on $N$ depend only on
$d$ and the fixed parameter bounds, and additionally on $R,W$ in the window case.
\end{lemma}

\begin{proof}
For the choice of $\mathcal A$ with the $R,W$ window, every vertex satisfies
$|r_v|\le R$ and $\|s_v\|_\infty\le W$. Hence
$\mathsf W\le m(1+R^4+2W^2)$, proving the claim in both modes.
It remains to treat the two choices without a window.
For $b=0$, apply Lemma~\ref{lem:local-marked-coefficient-bound} with
connections only on $F$, and use \eqref{eq:marked-local-coercive} to obtain
\[
 \E_0Z_F(\boldsymbol\alpha)
 \le\prod_v\E_0Z_{G_v}(\alpha_v)
 \exp\left\{\frac{Cm}{\sqrt N}-\frac{k\delta}{2}
       \sum_{uv\in F}(r_u+r_v)^2+C\logshift\mathsf W(\boldsymbol\alpha)\right\}.
\]
Drop the nonpositive quadratic term. Choose $c$ smaller than half the quartic
and stable decay constants in \eqref{eq:normalized-cell-envelope}, and
then take $N$ large enough that $C\logshift\le c$.
That bound and the lattice summation in the proof of
Proposition~\ref{prop:uniform-normalized-tails} give
\[
 \sum_{\alpha\in\mathcal U_d}\E_0Z_{G_v}(\alpha)e^{2c\mathsf w(\alpha)}
 \le C\E_0Z_{G_v}.
\]
Enlarging the type sum to all $\alpha_v\in\mathcal U_d$ and multiplying
the one-gadget bounds gives
\[
 \sum_{\boldsymbol\alpha\in\mathcal A}
 \E_0Z_F(\boldsymbol\alpha)e^{c\mathsf W(\boldsymbol\alpha)}
 \le\E_0Z_{G_{\sqcup}}e^{Cm}.
\]

For the denominator, restrict to $|r_v-\sigma_vr_d|\le r_d/2$ and
$\|s_v\|_\infty\le1$ at every vertex, choosing any fixed $\sigma$ if phases
are unrestricted. By \eqref{eq:bin-integral-mean}, each one-gadget
restriction has expected weight at least $c\E_0Z_{G_v}$.
On these boxes, \eqref{eq:marked-local-coercive} bounds the logarithmic
connection factor below by a constant per edge, and
Lemma~\ref{lem:local-marked-coefficient-bound} gives a coefficient comparison
factor of at least $e^{-Cm/\sqrt N}$.
Since $|F|\le3m/2$, this gives
\[
 \sum_{\boldsymbol\alpha\in\mathcal A}\E_0Z_F(\boldsymbol\alpha)
 \ge\E_0Z_{G_{\sqcup}}e^{-Cm}.
\]
Dividing proves the claim for $b=0$.

\rev{For $b=1$, use $p_F$ from
\eqref{eq:planted-simplicity-probabilities}.
Equation~\eqref{eq:first-moment-conditioning-identity} cancels the common factor}
$\mathbb P_0(\mathcal E_v)^{-m}$, giving
\[
 \E_\nu e^{c\mathsf W}
 =\frac{\sum_{\boldsymbol\alpha\in\mathcal A}
 \E_0Z_F(\boldsymbol\alpha)p_F(\boldsymbol\alpha)e^{c\mathsf W(\boldsymbol\alpha)}}
 {\sum_{\boldsymbol\alpha\in\mathcal A}
 \E_0Z_F(\boldsymbol\alpha)p_F(\boldsymbol\alpha)}.
\]
Since $p_F\le1$, the numerator has the same upper bound as before.
Equation~\eqref{eq:one-replica-simplicity-probability} and the uniform positive lower bound in
Lemma~\ref{lem:square-fixed-cycles} give $p_F(\boldsymbol\alpha)\ge e^{-Cm}$
on the boxes used for the denominator. The denominator is therefore
at least $\E_0Z_{G_{\sqcup}}e^{-Cm}$ after enlarging $C$, proving the claim.
\end{proof}

We will also use the following elementary integration rule. If a probability
law $\nu$ satisfies $\E_\nu e^{c\mathsf W}\le e^{Cm}$ for $\mathsf W\ge0$, then, for
$0\le t\le c$,
\begin{equation}\label{eq:sqplant-holder}
 e^{-Ctm/c}\le\E_\nu e^{-t\mathsf W}\le1
 \le\E_\nu e^{t\mathsf W}\le e^{Ctm/c}.
\end{equation}
The upper bound follows from H\"older's inequality. Jensen's inequality gives
$c\E\mathsf W\le\log\E e^{c\mathsf W}\le Cm$, and another application gives the lower bound.
For $t=O(N^{-1/2})$, both expectations lie in
$[e^{-C'm/\sqrt N},e^{C'm/\sqrt N}]$ for a suitable constant $C'$.

\subsection{Removing type and overlap restrictions}
\label{sec:full-moment-restoration}

We now include types outside $\mathcal U_d$ and overlaps outside
$\mathcal V(\alpha_v,\beta_v)$. The following bounds complete the
mixed-moment and conditioned first-moment estimates stated at the
start of this section.

For $U\in\{Z_{G_{\sqcup}},X_\sigma,Z_\sigma\}$, define $U_{\mathcal U_d}$
by restricting the independent-set sum defining $U$ to sets $I$ satisfying
$\alpha_v(I)\in\mathcal U_d$ for every $v\in V(H)$.

\begin{lemma}[Removing restrictions]
\label{lem:remove-moment-restrictions}
Use the fixed parameter bounds and graphs of
Lemma~\ref{lem:sqplant-weighted-kernel}, the cutoffs $R\ge2r_d$ and $W>0$ of
\cref{thm:sqplant-full-moments}, and the neighborhoods
$\mathcal U_d,\mathcal V(\alpha,\beta)$ from
\cref{thm:moving-cell-second-moment,app:marked-details}.
In either mode $b$ and for sufficiently large $N$, the following hold.
\begin{enumerate}[label=(\roman*),leftmargin=*]
\item For integer marginals in $\mathcal U_d$ and any $F_1,F_2\subseteq E(H)$,
\[
 1\le\frac{\E_b[Z_{F_1}(\boldsymbol\alpha)Z_{F_2}(\boldsymbol\beta)]}
 {M^{\rm loc}_{F_1,F_2;b}(\boldsymbol\alpha,\boldsymbol\beta)}
 \le\exp\left\{\frac C{\sqrt N}
 [\mathsf W(\boldsymbol\alpha)+\mathsf W(\boldsymbol\beta)]\right\}.
\]
\item For
\[
 (U,V)\in\{(Z_{G_{\sqcup}},Z_{G_{\sqcup}}),(X_\sigma,X_\sigma),
 (X_\sigma,Z_{G_{\sqcup}}),(Z_\sigma,Z_{G_{\sqcup}})\},
\]
choose $F_1,F_2$ and the corresponding type sets $\mathcal A_1,\mathcal A_2$ as in
Lemma~\ref{lem:sqplant-coupled-moment} so that
$U_{\mathcal U_d}=\sum_{\boldsymbol\alpha\in\mathcal A_1}Z_{F_1}(\boldsymbol\alpha)$
and $V_{\mathcal U_d}=\sum_{\boldsymbol\beta\in\mathcal A_2}Z_{F_2}(\boldsymbol\beta)$.
Define their local pair contribution by
\[
 \mathcal M_{\mathrm{loc},b}(U,V)
 :=\sum_{\boldsymbol\alpha\in\mathcal A_1,\,\boldsymbol\beta\in\mathcal A_2}
 M^{\rm loc}_{F_1,F_2;b}(\boldsymbol\alpha,\boldsymbol\beta).
\]
Then $1\le\E_b(UV)/\mathcal M_{\mathrm{loc},b}(U,V)\le e^{Cm/\sqrt N}$.
\item For $U\in\{Z_{G_{\sqcup}},X_\sigma,Z_\sigma\}$,
$1\le\E_bU/\E_bU_{\mathcal U_d}\le e^{Cm/\sqrt N}$.
\end{enumerate}
The constant $C$ and the required lower bound on $N$ depend only on
$d$, the fixed parameter bounds and $R,W$.
\end{lemma}

We first control the cost of restoring blocks while the types and overlaps
remain in these neighborhoods. Fix $F_1,F_2\subseteq E(H)$ and choose
$\mathcal A_1,\mathcal A_2$ as in Lemma~\ref{lem:sqplant-coupled-moment}.
For $F\subseteq E(H)$, define
\[
 \mathcal M_{F,b}
 :=\sum_{\boldsymbol\alpha\in\mathcal A_1,\,\boldsymbol\beta\in\mathcal A_2}
 M^{\rm loc}_{F_1\setminus F,F_2\setminus F;b}(\boldsymbol\alpha,\boldsymbol\beta).
\]
For any integer overlap assignment, also write
\[
 \mathcal M_{F,b}(\boldsymbol\Gamma)
 :=\E_b\!\sum_{I\in\Ind(G(F_1\setminus F)),\,J\in\Ind(G(F_2\setminus F)):\,\Gamma_v(I,J)=\Gamma_v\ \forall v}
 \lambda^{|I|+|J|}.
\]
Thus $\mathcal M_{F,b}$ sums these weights over overlaps in
$\mathcal V(\alpha_v,\beta_v)$ whose marginals belong to
$\mathcal A_1,\mathcal A_2$. On this set, define a probability law by
$\Pr_{\mathrm{loc},b}(\boldsymbol\Gamma)
=\mathcal M_{\varnothing,b}(\boldsymbol\Gamma)/\mathcal M_{\varnothing,b}$,
and denote expectation by $\mathbb E_{\mathrm{loc},b}$.
Put $\mathcal X_v=\sqrt N(\Gamma_v-\alpha_v\otimes\beta_v)$ and
$\mathcal W=\sum_v[\mathsf w(\alpha_v)+\mathsf w(\beta_v)+\|\mathcal X_v\|^2]$.

\begin{lemma}\label{lem:local-pair-restoration}
For some $c,C>0$ and all sufficiently large $N$,
\begin{equation}\label{eq:sqplant-joint-local-moment}
 \mathbb E_{\mathrm{loc},b}e^{c\mathcal W}\le e^{Cm},
 \qquad b=0,1,
\end{equation}
and, for every $F\subseteq E(H)$,
\begin{equation}\label{eq:sqplant-restored-local-comparison}
 e^{-C\sqrt N|F|-Cm/\sqrt N}
 \le\frac{\mathcal M_{F,b}}{\mathcal M_{\varnothing,b}}
 \le e^{C\sqrt N|F|+Cm/\sqrt N}.
\end{equation}
For fixed integer marginals $\boldsymbol\alpha,\boldsymbol\beta$ in $\mathcal U_d$,
\[
 \left|\log\frac{M^{\rm loc}_{F_1\setminus F,F_2\setminus F;b}(\boldsymbol\alpha,\boldsymbol\beta)}
 {M^{\rm loc}_{F_1,F_2;b}(\boldsymbol\alpha,\boldsymbol\beta)}\right|
 \le C\sqrt N|F|+\frac C{\sqrt N}
 [\mathsf W(\boldsymbol\alpha)+\mathsf W(\boldsymbol\beta)].
\]
The constants and required lower bound on $N$ depend only on $d$,
the fixed parameter bounds and any imposed $R,W$.
\end{lemma}
\begin{proof}
For fixed marginals, insert $e^{c\sum_v\|\mathcal X_v\|^2}$ into the overlap
sum in \eqref{eq:local-overlap-decomposition}; dividing by the sum without
this factor gives the conditional exponential moment. The bounds
\eqref{eq:pair-coefficient-error}--\eqref{eq:pair-block-taylor-bound}
control the numerator using products of expectations under
$\mu_{\alpha_v,\beta_v}$ conditioned on $\mathcal V(\alpha_v,\beta_v)$.
By Lemma~\ref{lem:overlap-exponential-moment}, for small fixed $c>0$,
\[
 \E_{\mu_{\alpha_v,\beta_v}(\cdot\mid\mathcal V(\alpha_v,\beta_v))}
 e^{t_v\|\mathcal X_v\|+(c+C/\sqrt N)\|\mathcal X_v\|^2}\le C.
\]
Here $t_v$ from \eqref{eq:pair-block-taylor-bound} is uniformly bounded
on $\mathcal U_d$, so $t_v\|\mathcal X_v\|$ is bounded by a constant plus a small
multiple of $\|\mathcal X_v\|^2$. Divide the resulting upper bound
by the lower local-moment bound in Lemma~\ref{lem:sqplant-weighted-kernel}.
For $b=1$, \eqref{eq:conditioning-overlap-error} adds $C/\sqrt N$ to $t_v$;
use the lower local-moment bound in Lemma~\ref{lem:conditioned-local-moments}. Thus
\begin{equation}\label{eq:conditional-local-overlap-moment}
 \mathbb E_{\mathrm{loc},b}
 [e^{c\sum_v\|\mathcal X_v\|^2}\mid\boldsymbol\alpha,\boldsymbol\beta]
 \le e^{Cm+C[\mathsf W(\boldsymbol\alpha)+\mathsf W(\boldsymbol\beta)]/\sqrt N}.
\end{equation}
Let $\nu_1,\nu_2$ be the type laws \eqref{eq:annealed-type-law} for
$G(F_1),G(F_2)$ with sets $\mathcal A_1,\mathcal A_2$ in mode $b$.
Summing over overlaps gives the marginal-type weight
$M^{\rm loc}_{F_1,F_2;b}(\boldsymbol\alpha,\boldsymbol\beta)$.
The local-moment bounds compare this weight with
$\mathfrak c_b^m\E_bZ_{F_1}(\boldsymbol\alpha)\E_bZ_{F_2}(\boldsymbol\beta)$.
Normalizing over types cancels $\mathfrak c_b^m$ and gives
\[
 \mathbb E_{\mathrm{loc},b}e^{c\mathcal W}
 \le e^{Cm}
 \frac{\E_{\nu_1\otimes\nu_2}e^{(c+C/\sqrt N)[\mathsf W(\boldsymbol\alpha)+\mathsf W(\boldsymbol\beta)]}}
 {\E_{\nu_1\otimes\nu_2}e^{-C[\mathsf W(\boldsymbol\alpha)+\mathsf W(\boldsymbol\beta)]/\sqrt N}}
 \le e^{C'm}.
\]
The last inequality follows from Lemma~\ref{lem:sqplant-coupled-moment}
and \eqref{eq:sqplant-holder}, after decreasing $c$ and increasing the
lower bound on $N$. This proves \eqref{eq:sqplant-joint-local-moment}.

At fixed local $\boldsymbol\Gamma$, apply
Lemma~\ref{lem:normalized-block-replacement} at the tilts
\eqref{eq:pair-type-tilt} to both choices of edge sets. Each changed
block ratio has bounded logarithm, and there are at most $k|F|$ changed
blocks. The mean-deviation estimates used in
\eqref{eq:pair-coefficient-error}, together with
\eqref{eq:sqplant-physical-weight}, bound both coefficient errors by
$C\mathcal W/\sqrt N$. Assignment counts, other matching colors and
fugacity factors cancel. Consequently,
\[
 \left|\log\frac{\mathcal M_{F,0}(\boldsymbol\Gamma)}
 {\mathcal M_{\varnothing,0}(\boldsymbol\Gamma)}\right|
 \le C\sqrt N|F|+C\mathcal W/\sqrt N.
\]
For $b=1$, the fixed-overlap version of
\eqref{eq:first-moment-conditioning-identity}, applied to
$G(F_1\setminus F)$ and $G(F_2\setminus F)$, multiplies
$\mathcal M_{F,0}(\boldsymbol\Gamma)$ by
$p_{F_1\setminus F,F_2\setminus F}(\boldsymbol\Gamma)/\mathbb P_0(\mathcal E_v)^m$.
Substituting \eqref{eq:pair-simplicity-probability} gives, uniformly in $F$,
\[
 \left|\log\frac{\mathcal M_{F,1}(\boldsymbol\Gamma)}{\mathcal M_{F,0}(\boldsymbol\Gamma)}
 -\sum_v\log\frac{\mathfrak a_2(\Gamma_v)}{\mathbb P_0(\mathcal E_v)}\right|
 \le\frac{Cm}{\sqrt N}.
\]
The sum is independent of $F$ and cancels when comparing edge sets, so the same logarithmic bound
holds for $b=1$, after enlarging $C$ since $\mathcal W\ge2m$.
Averaging the pointwise ratio under $\Pr_{\mathrm{loc},b}$ and applying
\eqref{eq:sqplant-joint-local-moment} and \eqref{eq:sqplant-holder} proves
\eqref{eq:sqplant-restored-local-comparison}.
With marginals fixed, use \eqref{eq:conditional-local-overlap-moment} and
\eqref{eq:sqplant-holder}: the overlap part contributes at most
$C m/\sqrt N+C[\mathsf W(\boldsymbol\alpha)+\mathsf W(\boldsymbol\beta)]/N$
to the logarithmic error. Together with the fixed marginal part and
$\mathsf W\ge m$, this proves the last assertion.
\end{proof}

\begin{proof}[Proof of \Cref{lem:remove-moment-restrictions}]
We prove (ii) first, then specialize to fixed marginals for (i) and
one replica for (iii). Fix $F_1,F_2$ and the phase and window restrictions;
for (ii), choose those representing $(U,V)$.
\added{\emph{Restoring blocks at exceptional vertices.}}
For an arbitrary pair type define the disjoint sets
\[
 S=\{v:\alpha_v\notin\mathcal U_d\text{ or }\beta_v\notin\mathcal U_d\},
 \qquad
 T=\{v\notin S:\Gamma_v\notin\mathcal V(\alpha_v,\beta_v)\}.
\]
Put $\mathcal D=S\cup T$ and
$E_{\rm rest}=\{uv\in F_1\cup F_2:\{u,v\}\cap\mathcal D\ne\varnothing\}$.
We will restore these blocks, obtaining $G(F_i\setminus E_{\rm rest})$
for $i=1,2$; $|E_{\rm rest}|\le3|\mathcal D|$.
\added{The rate loss at each vertex of $\mathcal D$ will dominate
the restoration cost; exponential-moment bounds control the
accumulated error on the remaining vertices.}
Before restoration, in mode $1$ condition
$\mathbb P^{\rm pl}_{F_1,F_2;\boldsymbol\alpha,\boldsymbol\beta}$ from
Section~\ref{sec:marked-cycle-conditioning} on the original overlap
assignment $\boldsymbol\Gamma$ and expose the designated matching edges
and endpoint states. Conditional on this exposure and all occupation states,
the residual matchings are independent across gadgets. At
$v\notin\mathcal D$, Lemma~\ref{lem:square-fixed-cycles} places the
simplicity probability uniformly in
$[\mathfrak a_2(\Gamma_v)e^{-C/\sqrt N},\mathfrak a_2(\Gamma_v)e^{C/\sqrt N}]$;
at $v\in\mathcal D$,
use the upper bound $1$. Averaging and conditioning on simplicity therefore give
\[
 \mathcal M_{\varnothing,1}(\boldsymbol\Gamma)
 \le\frac{\mathcal M_{\varnothing,0}(\boldsymbol\Gamma)}
 {\mathbb P_0(\mathcal E_v)^m}
 \prod_{v\notin\mathcal D}\mathfrak a_2(\Gamma_v)e^{Cm/\sqrt N}.
\]
We apply coefficient restoration to the unconditioned weight in this
bound, leaving the displayed simplicity factors at the original types.

Equation~\eqref{eq:partial-restoration-coefficient} bounds the first-color
coefficient by $e^{C\sqrt N|\mathcal D|}$ times an average of restored
coefficients at targets $N\Gamma_v-a_v$.
Fix $\mathbf a$ in the support of $\omega_{E_{\rm rest}}$, so
$\|a_v\|_1\le Ck\deg_{E_{\rm rest}}(v)$; the bounds below are uniform in this choice.
Targets omit the $(00,00)$ coordinate.
The multinomial counts and fugacity weights keep their original types;
at $v\notin\mathcal D$, keep the original tilts from \eqref{eq:pair-type-tilt}.
At $v\in\mathcal D$, every incident block is restored.
For a nonzero original summand, the unchanged matching colors ensure
that $2\Gamma_v$ lies in the convex hull of the exponent support of $P_2$.
For $P_2$ in \eqref{eq:pair-edge-polynomial}, the coefficient entropy $F_{P_2}$ in
Lemma~\ref{lem:coefficient-entropy-modulus}(ii) gives
\[
 [\mathbf z^{N\Gamma_v-a_v}]P_2(\mathbf z)^{N/2}
 \le N^C\exp\{\tfrac N2F_{P_2}(2\Gamma_v)+C\sqrt N\log N\},
\]
where $\Gamma_v$ also omits that coordinate in the argument of $F_{P_2}$.
The other matching colors keep the original targets.
At $S$, \cref{prop:unrestricted-critical-pair-anchor} gives a fixed
critical rate gap; the fugacity change is $O(N\logshift)=O(\sqrt N)$.
At $T$, \cref{lem:moving-product-localization} gives a fixed gap
with the marginals held fixed. Thus in either case the rate loss is
$cN$ and the coefficient shifts and polynomial type count cost at most
$C\sqrt N\log N$ per exceptional vertex. The global coefficient bound
also applies when some state counts are zero.

\added{\emph{Comparison with local replacements.}}
Outside $\mathcal D$, first fix the monomial contributed by each
unrestored block. At each vertex, the remaining factor is a power of
$P_2$ with exponent $N/2-O(\sqrt N)$; its target is $N\Gamma_v-a_v$
minus these monomial exponents, hence differs from its tilted mean by
$O(\sqrt N)$. Apply Lemma~\ref{lem:moving-lclt} at the original tilts,
bound the Gaussian factor by one, and sum over the block monomials.
The prefactor ratios contribute $O(m/\sqrt N)$
to the logarithmic error, and the target shifts contribute at most
$\sum_{v\notin\mathcal D}|\langle\Theta_v,a_v\rangle|
\le C\sqrt N|\mathcal D|$.
The replacement configurations below keep $\Gamma_v$ unchanged outside
$\mathcal D$. On these vertices, bound the restored graph's coefficient
at the unshifted targets $N\Gamma_v$ below using
\eqref{eq:normalized-block-replacement}.
Every unrestored block has the same endpoint types in both coefficients,
so its evaluated polynomial factor cancels.
The squared mean-deviation term is controlled by the estimate in the
proof of \eqref{eq:pair-coefficient-error} and \eqref{eq:sqplant-physical-weight},
bounding the additional logarithmic cost by
$CN^{-1/2}\sum_{v\notin\mathcal D}[\mathsf w(\alpha_v)+\mathsf w(\beta_v)+\|\mathcal X_v\|^2]$.

At $v\in S$, in each replica sum replacement integer types over
\[
 r_d/2\le\sigma_v r\le3r_d/2,\qquad \|s\|_\infty\le\min\{1,W\},
\]
where $\sigma_v$ is that replica's required phase, or $+$ if unrestricted.
Since $R\ge2r_d$ (\cref{thm:sqplant-full-moments}), these
regions satisfy the window restriction. Sum over local integer overlaps
for these replacement types.
These replacement regions lie in $\mathcal U_d$ for large $N$ by
\eqref{eq:exact-stable-center} and Lemma~\ref{lem:local-quartic-rate}.
The type lattice from Proposition~\ref{prop:uniform-normalized-tails}
and Lemma~\ref{lem:finite-polynomial-coefficients}(i) give positive weights there.
Uniformly for integer types $\alpha,\beta\in\mathcal U_d$,
Lemma~\ref{lem:overlap-exponential-moment} gives
\[
 \mu_{\alpha,\beta}\bigl(\mathcal V(\alpha,\beta)\bigr)
 \ge1-Ce^{-cN\eta^2}\ge\tfrac12.
\]
Thus restricting overlaps preserves a fixed fraction of the one-gadget
second moment. Write
$f_{\logshift}^*=\frac12\max\psi_{\logshift}$, with the maximum over
the domain of \eqref{eq:one-gadget-rate}. On the replacement regions,
$N\psi_{\logshift}(\rho,b)/2\ge Nf_{\logshift}^*-C$, as in the proof of
Corollary~\ref{cor:marked-macroscopic-tail}.
Lemma~\ref{lem:expected-type-weight} and Theorem~\ref{thm:moving-cell-second-moment}
therefore give total replacement weight at least $N^{-C}e^{2Nf_{\logshift}^*}$.
The ratio of the upper bound for the omitted weights to this lower bound is
\[
 \frac{N^Ce^{2Nf_{\logshift}^*-cN+C\sqrt N\log N}}
      {N^{-C}e^{2Nf_{\logshift}^*}}
 \le e^{-cN+C'\sqrt N\log N}.
\]
At $T$, keep both marginals and sum over local overlaps.
The same mass bound and Theorem~\ref{thm:moving-cell-second-moment} give
replacement weight at least $c\E_0Z_{G_v}(\alpha_v)\E_0Z_{G_v}(\beta_v)$.
Divide both the shifted far-overlap weight and the replacement weight
by their common fugacity factor. With $f$ from \eqref{eq:overlap-rates},
the former is at most $N^Ce^{N[f(\alpha_v)+f(\beta_v)]-cN+C\sqrt N\log N}$,
and the latter is at least $N^{-C}e^{N[f(\alpha_v)+f(\beta_v)]}$.
Their ratio is again at most $e^{-cN+C'\sqrt N\log N}$.
In mode $1$, all replacement types are local, so their simplicity
probabilities satisfy Lemma~\ref{lem:square-fixed-cycles}.
The factors at $v\notin\mathcal D$ and $\mathbb P_0(\mathcal E_v)^{-m}$
cancel against those in the displayed upper bound. The factors at
$v\in\mathcal D$ are bounded below by a fixed positive constant,
costing at most $e^{C|\mathcal D|}$.

\added{\emph{Summing over exceptional sets.}}
Fix $S,T$, all joint types outside $\mathcal D$, and the marginals at $T$.
Replacement matrices satisfy
\[
 \Gamma'_v=\Gamma_v\quad(v\notin\mathcal D),\qquad
 \Gamma'_v\one=\alpha_v,\quad(\Gamma'_v)^{\mathsf T}\one=\beta_v\quad(v\in T).
\]
At $S$, sum over the replacement regions above; at $T$, sum over local overlaps.
After restoration, no edge joins a gadget indexed by $\mathcal D$ to
another gadget, so their weights factor. The preceding bounds compare each vertex's total omitted
weight with its total replacement weight. Multiply these comparisons and
use the coefficient comparison outside $\mathcal D$.
The bounds are uniform in $\mathbf a$, so averaging in
\eqref{eq:partial-restoration-coefficient} preserves them.
Now sum over the fixed joint types and the marginal types kept at $T$.
For fixed $S,T$, each replacement assignment determines these data uniquely,
so it is counted at most once. Enlarging the replacement sum to all local
types gives
\[
 \sum_{\boldsymbol\Gamma:\,S(\boldsymbol\Gamma)=S,\,T(\boldsymbol\Gamma)=T}
 \mathcal M_{\varnothing,b}(\boldsymbol\Gamma)
 \le e^{-cN|\mathcal D|+C\sqrt N\log N|\mathcal D|}
 \sum_{\boldsymbol\Gamma'}
 \mathcal M_{E_{\rm rest},b}(\boldsymbol\Gamma')
 e^{C\mathcal W(\boldsymbol\Gamma')/\sqrt N}.
\]
Both sums obey the original phase and window restrictions; the second
also restricts the marginals to $\mathcal U_d$ and overlaps to $\mathcal V$.
Apply \eqref{eq:sqplant-joint-local-moment} and \eqref{eq:sqplant-holder}
on the restored graphs to bound the second sum by
$e^{Cm/\sqrt N}\mathcal M_{E_{\rm rest},b}$.
Then \eqref{eq:sqplant-restored-local-comparison} and
$|E_{\rm rest}|\le3|\mathcal D|$ give
\[
 \frac{\sum_{\boldsymbol\Gamma:\,S(\boldsymbol\Gamma)=S,\,T(\boldsymbol\Gamma)=T}
 \mathcal M_{\varnothing,b}(\boldsymbol\Gamma)}{\mathcal M_{\varnothing,b}}
 \le e^{-cN(|S|+|T|)+C\sqrt N\log N(|S|+|T|)+Cm/\sqrt N}.
\]
Apply \eqref{eq:exceptional-set-summation} with the two exceptional types
$S,T$. The total factor is
$e^{Cm/\sqrt N}(1+2e^{-cN/2})^m=e^{O(m/\sqrt N)}$.
Nonnegativity gives the opposite comparison, proving (ii).

\added{\emph{Fixed marginals and first moments.}}
For (i), fix integer marginals $\alpha_v,\beta_v\in\mathcal U_d$ for every $v$
and take $S=\varnothing$. Apply
\eqref{eq:conditional-local-overlap-moment} with $F_i$ replaced by
$F_i\setminus E_{\rm rest}$, $i=1,2$. For the local law on these graphs,
\eqref{eq:sqplant-holder} gives
\[
 \mathbb E_{\mathrm{loc},b}
 [e^{C\mathcal W/\sqrt N}\mid\boldsymbol\alpha,\boldsymbol\beta]
 \le \exp\{C'[\mathsf W(\boldsymbol\alpha)+\mathsf W(\boldsymbol\beta)]/\sqrt N\}.
\]
Use the fixed-marginal restoration bound in
Lemma~\ref{lem:local-pair-restoration} and sum over $T$ as above.
Since $\mathsf W\ge m$, the factor $(1+e^{-cN/2})^m$ is absorbed into
the displayed error, proving (i).

For (iii), use $F=\varnothing$ for $U=Z_{G_{\sqcup}}$ and $F=E(H)$
for $U=X_\sigma,Z_\sigma$, with their original phase and window restrictions.
Put $S=\{v:\alpha_v\notin\mathcal U_d\}$ and restore
$E_{\rm rest}=\{e\in F:e\cap S\ne\varnothing\}$.
Apply the same shifted-target bound with $P$ in place of $P_2$,
the one-replica critical rate gap from
Proposition~\ref{prop:global-critical-anchor}, and the same replacement
region in the single replica. Its one-gadget weight has the lower bound
used in Corollary~\ref{cor:marked-macroscopic-tail}.
For $b=1$, the preceding conditioning argument uses
$\mathfrak a_1(\alpha_v)$ in place of $\mathfrak a_2(\Gamma_v)$.
Equations~\eqref{eq:normalized-block-replacement} and
\eqref{eq:sqplant-physical-weight}, followed by
Lemma~\ref{lem:sqplant-coupled-moment} and \eqref{eq:sqplant-holder}, bound
the restoration cost by $e^{C\sqrt N|E_{\rm rest}|+Cm/\sqrt N}$.
Since $|E_{\rm rest}|\le3|S|$, the first term is absorbed by the
$cN|S|$ rate loss. Summing over $S$ yields
\[
 1\le\frac{\E_bU}{\E_bU_{\mathcal U_d}}
 \le e^{Cm/\sqrt N}(1+e^{-cN/2})^m\le e^{C'm/\sqrt N}.
\]

\end{proof}

\begin{corollary}[Fixed-type mixed moments]
\label{cor:sqplant-conditioned-moments}
Under the assumptions of Lemma~\ref{lem:remove-moment-restrictions},
for $b=0,1$, all $F_1,F_2\subseteq E(H)$ and integer marginal types
in $\mathcal U_d$,
\begin{equation}\label{eq:sqplant-weighted-kernel}
 \left|\log\frac{\E_b[Z_{F_1}(\boldsymbol\alpha)Z_{F_2}(\boldsymbol\beta)]}
 {\mathfrak c_b^m\E_b Z_{F_1}(\boldsymbol\alpha)\E_b Z_{F_2}(\boldsymbol\beta)}\right|
 \le\frac{C_d}{\sqrt N}[\mathsf W(\boldsymbol\alpha)+\mathsf W(\boldsymbol\beta)].
\end{equation}
Here all overlaps are included. The bound is uniform in $H,m$ and the
type assignments.
\end{corollary}

\begin{proof}
Combine the local bound in Lemma~\ref{lem:sqplant-weighted-kernel}
($b=0$) or Lemma~\ref{lem:conditioned-local-moments} ($b=1$) with the
fixed-marginal comparison in \cref{lem:remove-moment-restrictions}.
Increase $C_d$ to absorb the latter's constant.
\end{proof}

\begin{proof}[Proof of \Cref{thm:sqplant-full-moments}]
Let $(U,V)$ be one of the four pairs listed in
Lemma~\ref{lem:remove-moment-restrictions}(ii).
Exponentiate \eqref{eq:sqplant-weighted-kernel} and average against
the product of the type laws
\eqref{eq:annealed-type-law} corresponding to $U_{\mathcal U_d},V_{\mathcal U_d}$.
Lemma~\ref{lem:sqplant-coupled-moment} and \eqref{eq:sqplant-holder} give
\begin{equation}\label{eq:sqplant-local-integrated}
 \left|\log\frac{\E_b(U_{\mathcal U_d}V_{\mathcal U_d})}
 {\mathfrak c_b^m\E_bU_{\mathcal U_d}\E_bV_{\mathcal U_d}}\right|
 \le\frac{Cm}{\sqrt N}.
\end{equation}
Since $\mathcal M_{\mathrm{loc},b}(U,V)\le
\E_b(U_{\mathcal U_d}V_{\mathcal U_d})$, Lemma~\ref{lem:remove-moment-restrictions}(ii)
and nonnegativity give
\[
 1\le\frac{\E_b(UV)}{\E_b(U_{\mathcal U_d}V_{\mathcal U_d})}
 \le e^{Cm/\sqrt N}.
\]
Part~(iii) gives the same bounds for $\E_bU/\E_bU_{\mathcal U_d}$ and
$\E_bV/\E_bV_{\mathcal U_d}$. Substituting these three bounds into
\eqref{eq:sqplant-local-integrated} proves the theorem after enlarging $C_d$.

\end{proof}

\begin{proof}[Proof of \Cref{lem:conditioned-first-means}]
Let $U\in\{Z_{G_{\sqcup}},X_\sigma,Z_\sigma\}$ and first restrict its
types to $\mathcal U_d$. Use $F=\varnothing$ for $U=Z_{G_{\sqcup}}$ and
$F=E(H)$ for $U=X_\sigma,Z_\sigma$, with their original phase and window
restrictions. At fixed types, \eqref{eq:first-moment-conditioning-identity}
and \eqref{eq:one-replica-simplicity-probability}, together with
\eqref{eq:uniform-no-cycle-quantitative} for $j=0$, give
\[
 \left|\log\frac{\E_1Z_F(\boldsymbol\alpha)}{\E_0Z_F(\boldsymbol\alpha)}
 -\sum_v\log\frac{\mathfrak a_1(\alpha_v)}{\mathfrak a_0}\right|
 \le\frac{Cm}{\sqrt N}.
\]
Since $\log\mathfrak a_1(\alpha)=-\frac{d(d-1)}4\tr(T_\alpha^2)$,
\eqref{eq:cycle-multiplier-evenness} gives
\[
 |\log\mathfrak a_1(\alpha)-\log\mathfrak a_1(\alpha_c)|
 \le \frac C{\sqrt N}\mathsf w(\alpha).
\]
Exponentiate these bounds and average against the type law
\eqref{eq:annealed-type-law} in mode $0$.
By \cref{lem:sqplant-coupled-moment} and \eqref{eq:sqplant-holder},
\[
 \left|\log\frac{\E_1U_{\mathcal U_d}}{\E_0U_{\mathcal U_d}}
 -m\log\frac{\mathfrak a_1(\alpha_c)}{\mathfrak a_0}\right|
 \le\frac{Cm}{\sqrt N}.
\]
Part~(iii) of \cref{lem:remove-moment-restrictions} removes the marginal
restriction in both modes with error $e^{O(m/\sqrt N)}$.
Subtracting the logarithmic ratios for $U=X_\sigma$ or $Z_\sigma$ and
for $Z_{G_{\sqcup}}$ cancels the common term; the triangle inequality
proves \eqref{eq:conditioned-first-means} after enlarging $C$.\qedhere
\end{proof}

\section{Further questions}\label{sec:discussion}

\paragraph{Optimal critical mixing.}
For fixed $d\ge3$, the worst-case mixing time of the Glauber dynamics
for $\mu_{G,\lambdaC(d)}$ over simple $n$-vertex graphs of maximum
degree $d$ lies between $\Omega_d(n^{3/2})$ and
$O_d(n^{2+1/(d-2)}\log^2 n)$
(\cref{cor:glauber-superlinear-lower,thm:glauber-upper}).
The lower bound uses our analysis of the construction
in~\cite[Section~5.2]{CCYZ2026}.
What is the optimal worst-case mixing time for each fixed $d$?

\paragraph{Deterministic approximate counting.}
Can deterministic approximate counting at fixed $d$ attain the same
$\exp\{O_d(n\delta^2/d)\}\poly(n,\varepsilon^{-1})$ scale as
\cref{thm:main-upper}, uniformly for $0\le\delta\le\delta_{\mathrm{up}}$?
At $\delta=0$, this is the polynomial-time counting
question posed in~\cite[Section~7]{CCYZ2026}, whose Theorem~6.4 gives
a deterministic bound of $\exp\{O(\sqrt n\log^2(n/\varepsilon))\}$.

\paragraph{Sharper quantitative reductions.}
For the schedules covered by \cref{thm:main-transfer}, the reduction
preserves a fixed power improvement in the exponent, up to
polylogarithmic factors. Can the polylogarithmic size loss in the
constructive PCP be reduced or avoided?
The $O_d(N^{-1/2})$ precision of the gadget moment estimates gives
the repetition bound $\exp\{O_d(1+m\delta)\}$ in
\cref{thm:robust-sat-transfer}, where $m$ is the number of vertices
in the cubic MAX-CUT instance.
Can one reduce this bound and preserve the full power improvement
at $\delta_n=\Theta(n^{-a})$ for every fixed $0<a<1/2$,
strengthening \cref{prop:padded-power-transfer}, while keeping the
quadratic dependence of the graph size on $\delta^{-1}$?

\paragraph{Other antiferromagnetic two-spin systems.}
A nondegenerate quartic bifurcation suggests the same quadratic
dependence of the worst-case Glauber mixing exponent on the distance
from the tree-uniqueness threshold.
Does this behavior extend to other antiferromagnetic two-spin systems
on bounded-degree graphs as parameters approach the threshold from
the non-uniqueness side?
Can one also obtain analogous power-saving transfers from sampling
or counting to SAT?

\paragraph{Criticality and post-ETH fine-grained complexity}
What principles can organize fine-grained complexity without relying on an exponential-time lower bound for SAT as a fixed benchmark? Computational phase transitions suggest studying how computational complexity scales jointly with input size and distance from a tractability threshold. Our results provide one example: for fixed $d$, the parameter $n\delta^2/d$ governs the exponential scale of worst-case Glauber mixing and, under the conditions of our main transfer theorem, serves as an effective SAT instance size for transferring fixed-power improvements in sampling or counting algorithms to SAT. Can analogous scaling parameters be identified for other computational phase transitions? 
More broadly, can we develop a general theory of critical scaling in computational complexity, uncovering common scaling laws across seemingly unrelated problems and understanding how fine-grained reductions relate their near-critical behavior? Can such a theory distinguish critical slowing down of particular algorithms from intrinsic computational barriers, while remaining informative as benchmark algorithms improve?

\section*{AI methodology}
We interacted with GPT 5.6 Pro extensively in obtaining both the upper and lower bounds. We provided high-level guidance, and GPT 5.6 Pro helped us quickly explore and rule out many false candidates. For example, the first hint of the current shape of the bound came from asking GPT 5.6 to analyze the mixing time of Glauber dynamics on a few variants of the symmetric bipartite double cover of $d$ independent uniform perfect matchings, all of which came out with the same exponent. %
GPT 6 Astra was used in polishing the current writeup. We have checked and re-assembled the texts generated by AI, and we take full responsibility for their correctness.

\bibliographystyle{alpha}
\bibliography{references}

\appendix
\crefalias{section}{appendix}
\crefalias{subsection}{subappendix}
\section{Supporting results for the upper bounds}
\label{app:upper-bound-support}

\subsection{Auxiliary estimates for Glauber dynamics}
\label{app:glauber-estimates}

Throughout this subsection, $D=d-1$ and $\lambdaC=\lambdaC(d)$.

\subsubsection{Subcritical spectral stability}
\label{app:glauber-subcritical}

\begin{proof}[Proof of \Cref{lem:glauber-occupation-bound}]
Write $\theta=\theta_d(\lambda)$, so
$\lambda=\theta/(1-\theta)^{D+1}$.
Positive pinnings leave
hard-core distributions on induced subgraphs of maximum degree at most $d$,
and each $a\le\lambda$ is subcritical, so it suffices to prove
\eqref{eq:glauber-stability-covariance} with $\mathcal C(\lambda)=1/(1-D\theta)$
for $\mu_{G,\lambda}$ itself.

Fix a vertex $v$ and use its SAW tree from \cref{sec:prelim}.
For each node $u$, let $\Phi_u$ be the sum of the absolute derivatives
of $\log R_u$ with respect to the log fugacities at all nodes of its
subtree, treating these fugacities as independent variables.
For a child $w$, \eqref{eq:glauber-deletion-recursion} gives
\[
 \frac{\partial\log R_u}{\partial\log R_w}=-p_w,
 \qquad \Phi_u=1+\sum_{w\text{ child of }u}p_w\Phi_w.
\]

To prove $\Phi_u\le U(p_u)/(1-D\theta)$ by induction at nonroot
nodes, define the potential $U$ and the contribution $B(z)$ of a child
with vacancy probability $z$ by
\[
\begin{aligned}
 U(p)&=\min\left\{\frac{1-\theta}{1-p},
                \frac{\theta(1-p)}{(1-\theta)p}\right\},
 \qquad U(0)=1-\theta,\\
 B(z)&=(1-z)U(1-z)
 =\min\left\{\frac{(1-\theta)(1-z)}z,
              \frac{\theta z}{1-\theta}\right\}.
\end{aligned}
\]
For $D$ children with vacancy probabilities $z_i$, put $P=\prod_i z_i$,
so the parent occupation probability is $p=\lambda P/(1+\lambda P)$.
We will prove
\begin{equation}\label{eq:glauber-potential-drift}
 U(p)\ge 1-D\theta+\sum_i B(z_i).
\end{equation}

We use the elementary inequality
$\sum_{i=1}^k x_i\le k-1+\prod_{i=1}^k x_i$ for $0\le x_i\le1$.
First, for $k\ge1$ and $z_i\in(0,1]$, we claim
\begin{equation}\label{eq:glauber-potential-union}
 \sum_i B(z_i)\le\theta\left(k-1+
                         \frac{\prod_i z_i}{(1-\theta)^k}\right).
\end{equation}
Decreasing any $z_i>1-\theta$ to $1-\theta$ increases the left-hand side
and decreases the right-hand side. It therefore suffices to assume
$z_i\le1-\theta$ for all $i$. Then $B(z_i)=\theta z_i/(1-\theta)$,
and the product inequality applied to $z_i/(1-\theta)$ proves the claim.

For the parent with $D$ children, substitution gives
\[
 U(p)=\min\left\{1-\theta+\frac{\theta P}{(1-\theta)^D},
                         \frac{(1-\theta)^D}{P}\right\}.
\]
The first branch satisfies \eqref{eq:glauber-potential-drift} by
\eqref{eq:glauber-potential-union} with $k=D$.
For the second branch, increasing any $z_i<1-\theta$ to $1-\theta$
decreases $(1-\theta)^D/P$ and increases $B(z_i)$, so it suffices to
assume $z_i\ge1-\theta$ for all $i$.
Then $B(z_i)=(1-\theta)/z_i-(1-\theta)$, and the second branch minus
the right-hand side of \eqref{eq:glauber-potential-drift} is
\[
 \prod_i\frac{1-\theta}{z_i}-\sum_i\frac{1-\theta}{z_i}+D-1\ge0
\]
by the same product inequality. This proves
\eqref{eq:glauber-potential-drift} for both branches.
For fewer than $D$ children, pad with $z_i=1$; these entries leave
the product unchanged and contribute $B(1)=0$.
Induction from the leaves using \eqref{eq:glauber-potential-drift}
and the recursion for $\Phi_u$ now gives
$\Phi_u\le U(p_u)/(1-D\theta)$ at every nonroot node; at a leaf,
$\Phi_u=1$ and the drift inequality has an empty sum.

The root $v$ has at most $D+1$ children. Padding as above and applying
\eqref{eq:glauber-potential-union} with $k=D+1$ gives
\[
 \begin{aligned}
 (1-D\theta)\Phi_v
 \le1-D\theta+\sum_{i=1}^{D+1}p_iU(p_i)
 \le1+\lambda\prod_i(1-p_i)=\frac1{1-p_v}.
 \end{aligned}
\]
Apply Weitz's marginal identity from \cref{sec:prelim} and
differentiate with respect to an original vertex fugacity $\lambda_w$,
at homogeneous fugacity $\lambda$:
\[
 \frac{\partial\log R_v}{\partial\log\lambda_w}
 =\frac{\Cov(X_v,X_w)}{p_v(1-p_v)}.
\]
Each derivative is the signed sum over copies of $w$. The triangle
inequality bounds the sum over original labels by $\Phi_v$. Keeping
the root vacancy factor therefore gives
\begin{equation}\label{eq:glauber-subcritical-row-sum}
 \frac1{p_v}\sum_w|\Cov(X_v,X_w)|
 \le (1-p_v)\Phi_v\le\frac1{1-D\theta}.
\end{equation}
The matrix $\diag(p_v)^{-1}\Cov(X)$ has absolute row sums at most
$1/(1-D\theta)$ and is similar to the symmetric positive semidefinite
matrix $\diag(p_v)^{-1/2}\Cov(X)\diag(p_v)^{-1/2}$. Its eigenvalues
are consequently at most $1/(1-D\theta)$, proving
\eqref{eq:glauber-stability-covariance} with
$\mathcal C(\lambda)=1/(1-(d-1)\theta_d(\lambda))$.
Finally, convexity of $\theta\mapsto\theta/(1-\theta)^{D+1}$ gives
$1-\lambda/\lambdaC\le\frac{2D}{D-1}(1-D\theta)$~\cite[Lemma~3.7]{ChenJiang2025}, proving the upper bound
in \eqref{eq:glauber-subcritical-input}.
\end{proof}

The row bound \eqref{eq:glauber-subcritical-row-sum} is sharp on the
infinite $d$-regular tree: its unique subcritical Gibbs distribution has
$p_v=\theta/(1+\theta)$ and absolute influence $\theta^k$ along each
path of length $k$. There are $dD^{k-1}$ vertices at distance $k\ge1$,
so the row sum in \eqref{eq:glauber-subcritical-row-sum} equals
\[
 (1-p_v)\left(1+\sum_{k\ge1}dD^{k-1}\theta^k\right)
 =\frac1{1+\theta}\left(1+\frac{d\theta}{1-D\theta}\right)
 =\frac1{1-D\theta}.
\]

The root vacancy factor matters for the mixing exponent.
Using only $pU(p)\le\theta$ gives
$\Phi_v\le1+d\theta/(1-D\theta)=(1+\theta)/(1-D\theta)$.
Bounding the variance-normalized row sums this way and using
$\diag(p_v(1-p_v))\preceq\diag(p_v)$ would give the weaker
spectral-stability rate $(1+\theta)/(1-D\theta)$, the constant
of~\cite[Theorem~3.1]{ChenJiang2025}, yielding the
polynomial exponent $1+2/(d-2)$ instead of $1+1/(d-2)$ in
\cref{prop:glauber-integrated-profile}.

\subsubsection{The critical two-step contraction and marginal response constants}
\label{app:glauber-cubic}

\begin{proof}[Proof of \Cref{lem:glauber-cubic-deficit}]
We prove both claims using the strict concavity of $\log g'$.
The univariate derivative bound also follows from the proof of
\cite[Lemma~3.6, Equation~(3)]{ChenJiang2025} by taking the critical limit.
The proof of~\cite[Theorem~5.5]{SongYinZhao2019} shows the cubic
contraction of $g$ near its fixed point; we quantify it uniformly here.
Put $z=z(x)=e^x/(1+e^x)$ and $W=W(x)=\lambdaC(1-z)^D$.
Then $g'(x)=D^2zW/(1+W)>0$, and $L(x)=\log g'(x)$ satisfies
\[
\begin{aligned}
 L'&=1-z-\frac{Dz}{1+W},\\
 L''&=-z(1-z)\left(1+\frac{D}{1+W}\right)
       -\frac{D^2z^2W}{(1+W)^2}<0.
\end{aligned}
\]
At $x_*=-\log(D-1)$, we have $z(x_*)=1/D$ and
$W(x_*)=1/(D-1)$, so $L(x_*)=L'(x_*)=0$.
Strict concavity therefore gives $L(x)\le0$, with equality only at
$x_*$; hence $0<g'\le1$ and $g'$ is unimodal.

Since $\frac{\mathrm d}{\mathrm dx}\log(z(1-z))=1-2z\in[-1,1]$
and $z(x_*)=1/D$, on $|x-x_*|\le1$ we have
$z(1-z)\ge(D-1)/(3D^2)$.
Also $W\le\lambdaC\le4$, and hence
$1+D/(1+W)\ge(D+1)/5$. Thus, on this interval,
\[
 L''(x)\le-\frac{1-D^{-2}}{15}\le-\frac1{90}.
\]
Using $L(x_*)=L'(x_*)=0$ and unimodality outside the interval yields
\[
 g'(x)\le
 \exp\left\{-\frac{\min\{(x-x_*)^2,1\}}{180}\right\}
 \le1-\frac{\min\{(x-x_*)^2,1\}}{360}.
\]
The last inequality uses $e^{-a}\le1-a/2$ for $0\le a\le1$.
Since $\min\{y^2,1\}$ is even and nondecreasing in $|y|$, its integral
over any interval of length $s\le1$ is at least
$\int_{-s/2}^{s/2}y^2\dd y=s^3/12$.
Thus \eqref{eq:glauber-integrated-deficit} holds with $c=1/4320$:
\[
 g(x+s)-g(x)
 \le s-\frac1{360}\int_x^{x+s}\min\{(y-x_*)^2,1\}\dd y
 \le s-\frac{s^3}{4320}.\qedhere
\]
\end{proof}

\paragraph{Constants for the marginal response bound.}
\label{app:glauber-response-constants}
Here we verify the choices used in the proof of
\cref{lem:glauber-field-response}. Take
\[
 c=\frac1{4320},\qquad C_1=2,\qquad C_2=34,\qquad
 A=128.
\]
For $t_+\le\delta_{\mathrm{up}}=2^{-18}$ and $h\le h_0=2^{-27}$,
we have $t+h<\log2$, $e^{t_+}-1\le2t_+$, and
$2+4D\lambdaC\le34$ since $D\lambdaC\le8$.
Moreover,
\[
 s=128(\sqrt{t_+}+h^{1/3})\le\frac12,\qquad
 \frac2{128^2}+\frac{34}{128^3}<\frac1{4320}.
\]
Since $h\le h^{1/3}$ and $A\ge C_2$, we have $C_2h\le s$.

\subsection{Implementation of counting and sampling}
\label{count:sec-implementation}

\begin{proof}[Completion of the proof of \Cref{thm:counting-upper}]
We use the computational conventions of \cref{sec:computational-conventions}
and resume the schedule notation of \cref{count:sec-improved-counting},
with $Q=1+n/d$. Write $0=b_0<\cdots<b_K=B$ for the
annealing endpoints, so there are $M=2K+1$ observables.

\paragraph{Schedule and run lengths.}
Compute $\lambdaC(d)$ by repeated squaring in $O(\log n)$ operations.
To obtain computable upper bounds for $V_{\mathcal B}$ and $R_j$, replace
$z^{\gamma_d}$ by $2^{\lceil\gamma_d\lceil\log_2z\rceil\rceil}$ for
$z\ge1$, where $\gamma_d=1+1/(d-2)\in[1,2]$, and replace $e^t$ by
$2^{\lceil2t\rceil}$ for $t\ge0$. Find logarithmic upper bounds by
doubling. These replacements preserve the bounds after enlarging
absolute constants. Round $R_j$ upward and choose the observation
lengths according to \eqref{count:eq-counting-trajectory-length}.

Round each $V_{\mathcal B}$ upward to a power of two
and use \eqref{count:eq-counting-band-mesh}. Let $h_*$ be the smallest
step length. Since all band lengths are at least $w_0$ and
$V_{\mathcal B}\le CQ\log(2+Q)e^{Cn\delta^2/d}$,
$\log(1/h_*)=O(n\delta^2/d+\log n)$.
To keep the fugacities and paired factors rational, choose a
common integer $P$ with
$2^{-P}<c_0h_*2^{-\lceil B\rceil}\le c_0h_*e^{-B/2}$.
Approximate each $e^{-b_i/2}$ by a rational with absolute error at most
$2^{-P-2}$, then round it to a nearest multiple $a_i$ of $2^{-P}$,
keeping $a_0=1$. The total error is at most
$2^{-P-2}+2^{-P-1}<2^{-P}$.
For small absolute $c_0$, the offsets
$\widehat b_i=-2\log a_i=b_i+O(c_0h_*)$ remain strictly increasing,
and step lengths change by only constant factors.
The initial band's variance and relaxation bounds hold for all $b\ge0$.
On a rounded cold band, $b-\delta$ stays within constant factors of its
scale $w$, since $h_*\le w$; on the rounded late band, $R(b)=O(n)$
and $\int f''(b)\dd b\le n\lambda\le CQ$ still hold.
The relaxation-time estimates are therefore preserved. Each band's
largest step times its integrated variance remains $O(c_*)$, so
\cref{count:lem-band-budget} still gives $O(1)$ total relative variance
per band.

Set $\widehat\lambda_i=\lambda a_i^2$. The final endpoint satisfies
$n\widehat\lambda_K\le1/8$, so the terminal empty-set indicator
has relative variance at most $e^{1/8}-1=O(1)$. The paired means and
terminal empty-set probability telescope exactly to $Z_G(\lambda)$
because $a_0=1$.
Since $B=O(\log n)$, we can choose
$P=O(n\delta^2/d+\log n)$;
power-series evaluation computes the endpoints in $O(MP)$ operations,
independently of $\eps$.

\paragraph{Observations and precision.}
For each step, precompute $r_i^{\pm k}$, $0\le k\le n$, where
$r_i=a_i/a_{i+1}$, using one common denominator per table in $O(n)$
operations. Maintaining $|X|$ and accumulating the corresponding
numerators then costs $O(1)$ operations per observation; the exact
sums use $O(nP+\log L_j)$ bits. Round only products of completed means,
to $p_{\rm out}=\lceil\log_2(CM/\eps)\rceil$ significant bits.
Each rounding has relative error at most $\eps/(CM)$, so all products
and the final division add at most $\eps/4$. Combined with statistical
error $\eps/2$, the final relative error is less than $\eps$.

Since each paired mean lies between $r_i^{-n}$ and $r_i^n$, and
$\sum_i\log r_i=O(\log n)$, partial products have logarithmic magnitude
$O(n\log n)$. A positive terminal empirical mean is at least $1/L_B$,
where $L_B$ is its trajectory length; if it is zero, return $1$ before division.
Total setup and rounding work, including the computation of run lengths,
is $O(M(n+P+p_{\rm out})+\log L_B)$. It is absorbed by $R_j\ge n$,
$\log(1/\eps)=O(\eps^{-2})$, and an increase of the absolute coefficient
of $n\delta^2/d$ in the exponential, without an extra accuracy logarithm.
All operand lengths are polynomial in $n,d$, numerical input lengths,
$n\delta^2/d$, and $\log(1/\eps)$: the threshold uses $O(d\log d)$ bits,
and precision depends on $\log M=O(n\delta^2/d+\log n)$.

\paragraph{Worst-case work.}
At fugacity $a$, use the coin-first Glauber update of
\cite[Section~5.1.1]{AnariEtAl2022EntropicII}: choose a uniform vertex,
propose occupation with probability $a/(1+a)$, and inspect its neighbors
only if the proposal succeeds. Set it occupied precisely when the proposal
succeeds and all neighbors are vacant. Since $a\le\lambda$ and
$d\lambda\le13$, expected update
work is $O(1+da)=O(1)$.
Let $U$ be the total number of burn-in and observation updates
over all trajectories. Chernoff's bound gives total neighbor work
$O(U+\Delta(G)\log(1/\zeta))$ except with probability $\zeta$.
Uniform-vertex rejection sampling and bitwise comparison for rational
Bernoulli coins have geometric running-time tails, giving
$O(U+\log(1/\zeta))$ random words with the same failure allowance.
For Bernoulli coins, generate the probability's binary digits by updating
a remainder modulo its denominator, keeping operand lengths bounded.
Cap both totals over all trajectories and return $1$ on exhaustion.
With failure allowance
$1/32$ per cap and statistical failure $3/32$, total failure is
$5/32<1/4$. Input preparation costs $O(n+|E|)$ and initialization costs
$O(n)$ per trajectory, absorbed by $R_j\ge n$. Charging individual
random bits adds a factor $O(\log n)$; arithmetic has polynomial binary
overhead by the operand bounds above. Together with
\cref{count:sec-improved-counting}, this proves the theorem.
\end{proof}

\paragraph{Sampling.}
The same implementation gives a sampler for the graphs and fugacities
of \cref{thm:counting-upper}. For $0<\eta<1/4$, it returns a sample
within TV distance $\eta$ of $\mu_{G,\lambda}$ using at most
\[
 O(n+|E|)+Cn\left(1+\frac nd\right)^{1+\frac1{d-2}}
       \exp\!\left\{\frac{Cn\delta^2}{d}\right\}
       \bigl[\log^2 n+\log(1/\eta)\bigr]
\]
randomized arithmetic operations in the worst case, with absolute
constants and polynomial binary overhead.
Using the upper bounds computed above, run Glauber dynamics from the
empty set for an integer number $U\ge CR(0)[\log^2 n+\log(1/\eta)]$
of updates within the displayed work bound. By
\cref{count:lem-temperature-costs}, its TV error is at most $\eta/2$.
Allocate failure probability $\eta/4$ to each work cap, returning the
empty set on exhaustion; this adds at most $\eta/2$ to the TV error.
The cap costs are absorbed by $U\ge cn\log(1/\eta)$ and
$\Delta(G)\le n-1$. Operand lengths are polynomial in $n,d$, numerical
input lengths, and $\log(1/\eta)$.

\section{Sources and implementation for the reductions}
\label{app:hardness-support}

\subsection{A constructive source for additive-gap cubic MAX-CUT}
\label{app:cubic-source}

We compose the quasilinear PCP theorem with the linear cubic
construction below.

\begin{theorem}[From SAT to additive-gap cubic MAX-CUT]
\label{thm:sat-cubic-source}
There are absolute integers $K_{\rm src},C_{\rm src}\ge1$ and a
positive rational $\nu_{\rm src}$ with the following property.
Given a SAT formula of binary encoding length $s\ge2$, a deterministic
polynomial-time algorithm produces an instance $(H,\tau)$ of
\cref{prob:cubic-gap}, with $H$ simple and cubic, having exactly
\[
 m=4K_{\rm src}s\lceil\log_2s\rceil^{C_{\rm src}}+4
\]
vertices. If the formula is satisfiable then $\MaxCut(H)=\tau$;
otherwise $\MaxCut(H)\le\tau-\nu_{\rm src}m$.
\end{theorem}

We use the bounded-occurrence SAT reduction and the MAX-CUT gadgets
as black boxes. What must still be checked is their quantitative
composition: linear output size, a computable YES threshold, and a
constant loss in the additive gap.

\subsubsection{A linear-size cubic reduction}

Write $\operatorname{unsat}(\varphi)$ for the minimum number of clauses
violated by an assignment to a CNF formula $\varphi$.

\begin{proposition}[Linear-size additive-gap cubic construction]
\label{prop:common-cubic-construction}
There are an absolute constant $C\ge1$ and a deterministic
polynomial-time procedure that transforms a CNF formula $\varphi$ with
$M\ge1$ nonempty clauses of size at most three into a pair
$(H_\varphi,\tau_\varphi)$, where $H_\varphi$ is a simple cubic graph and
$\tau_\varphi$ is an explicitly computed integer. The output satisfies
\[
 |V(H_\varphi)|\le CM,
 \qquad
 \MaxCut(H_\varphi)=\tau_\varphi-2\operatorname{unsat}(\varphi).
\]
\end{proposition}

\begin{proof}
Remove repeated literals and tautological clauses. Replace each unit clause
$(\ell)$ by $(\ell\vee y)\wedge(\ell\vee\neg y)$, using a fresh variable
$y$ for each replacement. For each assignment to the original variables,
the minimum number of violations over $y$ is unchanged; the number of
clauses is at most $2M$.

The deterministic amplifier construction
in~\cite[proof of Theorem~13.10 and
Lemma~13.2$'$]{Papadimitriou1994} produces $O(M)$ variables and clauses,
with at most three occurrences per variable, and preserves the minimum
number of violated clauses. Its added constraints are binary implications,
so every clause still has size two or three. Repeatedly satisfy pure
literals and delete their incident clauses, then remove unused variables.
Every assignment to the residual formula $\psi$ extends to satisfy the
deleted clauses. Hence each remaining variable occurs in both polarities and
\[
 \operatorname{unsat}(\psi)=\operatorname{unsat}(\varphi).
\]
If $\psi$ has no clauses, then $\varphi$ is satisfiable; output $(K_4,4)$,
since $\MaxCut(K_4)=4$.

Otherwise apply~\cite[Theorem~4.1]{AlimontiKann2000},
whose clause gadgets cover both binary and ternary clauses.
The resulting multigraph has maximum degree at most six, and its
optimum equals a computable integer offset
plus twice the maximum number of satisfied clauses. The degree reduction
in~\cite[Theorem~4.2]{AlimontiKann2000} gives a simple graph of maximum
degree three and adds a computable offset: the rounding argument aligns
the split vertices, after which all added internal edges are cut.

To make the final regularity step of~\cite[Fig.~8]{AlimontiKann2000}
explicit, fill each missing degree
with a fresh copy of $K_4$ with one edge subdivided, joining its new
degree-two vertex to the deficient vertex. This five-vertex graph has
seven edges, a triangle, and a cut of size six; its maximum cut is
therefore six, attained with either color at its attachment
vertex. Each attachment adds exactly seven to the maximum cut,
independently of the old cut, and fills one unit of the old vertex's
degree deficit.
Repeat once per missing degree. All graphs remain simple and the total
size increases only by a constant factor.
Define $\tau_\varphi$ as the sum of these computable offsets plus
twice the number of clauses of $\psi$. Then
\[
 \MaxCut(H_\varphi)
   =\tau_\varphi-2\operatorname{unsat}(\psi)
   =\tau_\varphi-2\operatorname{unsat}(\varphi).
\]
Every step has linear output size and polynomial running time.
Choose a computable absolute $C\ge4$ to cover also the empty residual case.
\end{proof}

\subsubsection{Quasilinear PCP and source-size normalization}

\begin{proof}[Proof of \Cref{thm:sat-cubic-source}]
Dinur's quasilinear PCP theorem~\cite[Definition~8.1 and
Theorem~8.1]{Dinur2007PCP}, building on~\cite{BenSassonSudan2008},
is a deterministic polynomial-time reduction from SAT of encoding
length $s$ to a Boolean constraint system with
$O(s(\log s)^C)$ constraints of constant arity, perfect completeness,
and a fixed positive fraction of violated constraints on every
unsatisfiable input.

Add a tautological predicate if the constraint list is empty, and an
unused argument to any zero-arity predicate.
For completeness, convert each constant-arity predicate to a
constant-size CNF by excluding its rejecting assignments, then split
clauses longer than three using fresh auxiliary variables private to
that predicate. This produces a constant number of clauses per
predicate. A satisfied predicate has a satisfying extension, while
an unsatisfied predicate forces at least one violated clause under
every extension. Padding a predicate's clause list by tautologies if
necessary makes the number of clauses per predicate a fixed positive
constant. Thus the resulting $3$-CNF has
$M=O(s(\log s)^C)$ clauses and a fixed positive violation fraction
on NO inputs. Apply \cref{prop:common-cubic-construction}. We obtain
a simple cubic graph $H_0$ of even order $m_0$, with a computable
threshold $\tau_0$, and an absolute rational $\nu_0>0$ such that
\[
 m_0\le B:=K_{\rm src}s\lceil\log_2s\rceil^{C_{\rm src}},
 \qquad
 \MaxCut(H_0)\le\tau_0-\nu_0m_0
\]
on NO inputs, and equality with $\tau_0$ on YES inputs. Enlarge the
absolute integers $K_{\rm src},C_{\rm src}$ to cover every $s\ge2$.

To make the number of vertices depend only on $s$, take
$r=\lceil B/m_0\rceil$ disjoint copies of $H_0$.
Their order $m'=rm_0$ lies in $[B,2B)$ and their cut deficit is at
least $\nu_0m'$. Set $m=4B+4$. The even difference $m-m'$ is at least
four, so it is $4u+6v$ for some $u\ge0$ and $v\in\{0,1\}$.
Add $u$ copies of $K_4$ and $v$ copies of $K_{3,3}$, and put
\[
 \tau=r\tau_0+4u+9v.
\]
The added maximum cuts are known exactly, so the YES optimum remains
$\tau$ and the NO deficit is at least $\nu_0B\ge(\nu_0/8)m$.
Choose a positive rational $\nu_{\rm src}\le\nu_0/8$.
All steps have polynomial running time and the claimed output size.
Replication before padding ensures that the constant source gap is
preserved even when $m_0$ is much smaller than its upper bound.
\end{proof}

\subsection{Implementation of the hardness reductions}
\label{sec:square-generator}

We construct $\ell$ independent trials from the law $\mathbb Q_1$
in \eqref{eq:reference-square-law}, then truncate the algorithm with
total TV error at most $\zeta$, as required by
\cref{prop:reduction-implementation}.

For each source vertex $v$, we must reweight the matching tuple by
$Z_{G_v}^2$. The identity
\[
 Z_{G_v}^2=\sum_{I_1,I_2\in\mathcal I(G_v)}\lambda^{|I_1|+|I_2|}
\]
allows us to generate the matchings together with two auxiliary
independent sets, giving each joint outcome weight
$\lambda^{|I_1|+|I_2|}$, and then discard the sets. We first construct
this law without conditioning on simplicity.

For one base label, the membership of its two copies in $I_1,I_2$
is recorded by four bits. Thus the state set is
$\mathcal S=\{0,1\}^4$, with a state written as
$s=(s_+^{(1)},s_-^{(1)},s_+^{(2)},s_-^{(2)})$.
Put $C_{st}=1$ if
\[
 s_+^{(r)}t_-^{(r)}=s_-^{(r)}t_+^{(r)}=0
 \qquad(r=1,2),
\]
and $C_{st}=0$ otherwise. These conditions say that the two lifts
of an edge are compatible with both independent sets.
Identifying $\mathcal S$ with $\Omega^2$, this is the matrix
$C=B\otimes B$ of \cref{sec:marked-cycle-conditioning}. The weight of
state $s$ is $w_s=\lambda^{\sum_{r=1}^2(s_+^{(r)}+s_-^{(r)})}$.

For a nonnegative count vector $\mathbf c=(c_s)_{s\in\mathcal S}$, let
$M(\mathbf c)$ count compatible perfect matchings on any fixed labeled
state assignment with these counts. It depends only on $\mathbf c$.
Set $M(\mathbf0)=1$. For a nonzero vector of even total, choose the
first state $s$ with $c_s>0$ in a fixed order and distinguish one label
of that state. Its partner has state $t$ in
$c_t-\one\{s=t\}$ possible ways, so
\[
 M(\mathbf c)=\sum_{t\in\mathcal S} C_{st}
   (c_t-\one\{s=t\})M(\mathbf c-\mathbf e_s-\mathbf e_t),
\]
where $\mathbf e_s$ is the unit vector for state $s$.
Terms with negative counts or zero coefficients contribute zero.
Dynamic programming in increasing even total, up to $N$, computes
the counts exactly using at most $(N+1)^{16}$ vectors.

With these counts available, generate one matching tuple as follows.
\begin{enumerate}[label=(\arabic*),leftmargin=*]
\item Choose a count vector $\mathbf c$ of total $N$ with probability
proportional to the computable weight
\[
 \frac{N!}{\prod_s c_s!}\prod_s w_s^{c_s}\,M(\mathbf c)^d.
\]
\item Choose uniformly an assignment of these states to the $N$ labels.
\item Generate $d$ independent compatible perfect matchings.
For each matching, use the recurrence above: distinguish a label in
the first nonempty state $s$, choose $t$ proportionally to its summand,
choose a uniform available partner of state $t$, remove the pair, and
recurse until no labels remain.
\end{enumerate}
The recurrence samples each compatible matching uniformly.
Hence steps~(2)--(3) cancel the factors
$N!/\prod_s c_s!$ and $M(\mathbf c)^d$ in step~(1), leaving joint
weight $\prod_s w_s^{c_s}=\lambda^{|I_1|+|I_2|}$. Summing over the
auxiliary sets therefore gives matching weight $Z_{G_v}^2$. Since
$\mathbb P_0$ is uniform on indexed matching tuples, discarding the
sets gives precisely its required reweighting.

Repeat these three steps until the union $B_v$ is simple. This conditions the weighted
law on simplicity, and hence gives $\mathbb P_1$ reweighted by $Z_{G_v}^2$.
After acceptance, order the edges of $M_{v,1}$ uniformly at random and
orient each independently and uniformly. Perform this procedure
independently for every gadget and trial, then connect the gadgets
as in \cref{constr:graph-reduction}. The resulting law is
$\mathbb Q_1^{\otimes\ell}$.

The following bound controls the number of proposals before acceptance.
\begin{lemma}[Acceptance probability for simple gadgets]
\label{lem:square-generator-acceptance}
Fix $d\ge3$ and a rational upper bound on $N\delta^2$.
There are a computable rational $p_{\rm acc}>0$ and a computable
integer $N_0$ such that, for every even $N\ge N_0$ and positive
rational $\delta$ satisfying the fixed bound on $N\delta^2$, the matching
tuple generated at fugacity $(1+\delta)\lambdaC(d)$ before the simplicity
test has a simple union with probability at least $p_{\rm acc}$.
\end{lemma}
\begin{proof}
The normalized count vector $\mathbf c/N$ is the two-replica type.
The critical two-replica type $\pi_2$ in
\cref{prop:global-critical-anchor} is an interior probability vector
on the sixteen states. Choose a closed neighborhood $U$ of $\pi_2$
with rational coordinate bounds on which all state proportions and
compatible coupling entries are positive, as required by
\cref{lem:square-fixed-cycles} for two replicas.
At $\lambdaC(d)$, \cref{prop:global-critical-anchor} gives a fixed
rate gap $c>0$ outside $U$.
Stirling's formula and the global upper and local lower bounds
in \cref{lem:finite-polynomial-coefficients} give an upper bound
$N^{O_d(1)}e^{-cN}$ on the fraction of the joint weighted mass outside $U$.
There are polynomially many types; the upper bound includes boundary
types, and feasible types near $\pi_2$ give the maximal exponent up to
a polynomial factor.

Increasing the fugacity to $\lambdaC(d)(1+\delta)$ multiplies each
joint weight by a number between $1$ and $(1+\delta)^{4N}$.
Hence the outside fraction is at most
\[
 N^{O_d(1)}\exp\{-cN+4N\delta\}.
\]
The fixed upper bound on $N\delta^2$ implies $N\delta=O(\sqrt N)$,
so this fraction is at most $1/2$ above a computable threshold.
On $U$, apply \cref{lem:square-fixed-cycles} with two replicas and no
exposed edges. It supplies a computable uniform lower bound $p_*>0$
for simplicity. The proposal acceptance probability is therefore at
least $p_*/2$. Fix a positive rational $p_{\rm acc}\le p_*/2$ and
choose $N_0$ to include both thresholds.
\end{proof}

\paragraph{Truncation and bit complexity.}
Report an abort if either of the following caps is exhausted.
Matching counts and fugacity powers have
$O_d(N\log N+N\bits(\lambda))$ bits. Clearing denominators reduces
each categorical choice to a uniform integer draw with polynomial bit
cost per attempt; binary rejection succeeds with probability at least
$1/2$ per attempt.
Using $p_{\rm acc}$ from \cref{lem:square-generator-acceptance}, cap
each of the $\ell m$ gadget requests at
\[
 \left\lceil p_{\rm acc}^{-1}\right\rceil\left\lceil\log_2(2\ell m/\zeta)\right\rceil
\]
proposals.
The probability that any request exceeds its cap is at most $\zeta/2$.
The total number of integer draws now has a computable deterministic upper bound
$D_{\rm draw}=\ell\poly_d(m,N,\log(\ell/\zeta))$.
Capping each draw at $\lceil\log_2(2D_{\rm draw}/\zeta)\rceil$ attempts
adds at most $\zeta/2$ failure probability by the union bound.
Except on these two failure events, the capped algorithm agrees with
the uncapped algorithm using the same random bits. Its output law is
therefore within TV distance $\zeta$ of $\mathbb Q_1^{\otimes\ell}$.
Since $\bits(\lambda)=O_d(\bits(\delta))$, the dynamic program and these
caps give the worst-case bit bound in
\cref{prop:reduction-implementation}.

The constants in
\cref{prop:uniform-annealed-soft-spin,thm:sqplant-full-moments} and
the parameter choices in the proof of \cref{thm:cut-weight-realization} give the asserted
rational choices of $C_d,\delta_0(d)$, rounding $C_d$ upward
if needed. Take $N_{\rm gen}\ge N_0$ large enough to include the
thresholds in these estimates and \cref{lem:square-generator-acceptance}.

\section{Finite-state tools for gadget moments}
\label{app:gadget-tools}

\subsection{Uniform coefficient estimates}\label{app:coefficient-tools}

The lattice local limit theorem and its coefficient form in
\cref{lem:finite-polynomial-coefficients}(i) are classical; see
\cite[Theorem~2]{Richter1958} and \cite{BhattacharyaRao1976}.
We verify the compact-uniform relative error and eventual positivity
used for residual targets after exposing connection blocks.

\begin{lemma}[Uniform moving-tilt local limit theorem]\label{lem:moving-lclt}\label{lem:uniform-centered-lclt}\label{lem:uniform-central-lclt-complete}

Let $X$ have finite support in $\mathbb Z^h$ whose difference lattice
(the lattice generated by differences of its points) is $\mathbb Z^h$.
For $\theta$ in a fixed compact set $\mathcal K\subset\mathbb R^h$, define
\[
 \Pp_\theta(X=x)=
 \frac{e^{\langle\theta,x\rangle}\Pp(X=x)}
      {\E e^{\langle\theta,X\rangle}}.
\]
Under $\Pp_\theta$, let $X_1,\ldots,X_\ell$ be independent copies of $X$,
with covariance $\Sigma(\theta)$, positive definite by the lattice assumption.

For every fixed $C>0$, uniformly for $\theta\in\mathcal K$ and integer targets
$s_\ell\in\mathbb Z^h$ with $s_\ell=\ell\E_\theta X+u$ and $\|u\|\le C\sqrt\ell$,
\begin{equation}\label{eq:uniform-centered-lclt}
 \Pp_\theta\Bigl(\sum_{i=1}^\ell X_i=s_\ell\Bigr)
 =\frac{\exp\{-u^{\mathsf T}\Sigma(\theta)^{-1}u/(2\ell)\}}
 {(2\pi\ell)^{h/2}\sqrt{\det\Sigma(\theta)}}
 (1+O_{\mathcal K,C}(\ell^{-1/2})).
\end{equation}
Every such integer target has positive probability for sufficiently large $\ell$;
its feasibility need not be assumed.
\end{lemma}

\begin{proof}[Proof of \Cref{lem:moving-lclt}]
Compactness bounds the covariance eigenvalues above and away from zero.
The centered characteristic function
$\chi_\theta(z):=\E_\theta e^{i\langle z,X-\E_\theta X\rangle}$
satisfies, uniformly for $\theta\in\mathcal K$ and $\|z\|\le\varrho$
with some fixed $0<\varrho<\pi$,
\[
 \log\chi_\theta(z)=-\tfrac12z^{\mathsf T}\Sigma(\theta)z+O(\|z\|^3),
 \qquad |\chi_\theta(z)|\le e^{-c\|z\|^2}.
\]
Fourier inversion gives
\[
 \Pp_\theta\Bigl(\sum_{i=1}^\ell X_i=s_\ell\Bigr)
 =(2\pi)^{-h}\int_{[-\pi,\pi]^h}
   \chi_\theta(z)^\ell e^{-i\langle z,u\rangle}\,\mathrm dz.
\]
Set $z=v/\sqrt\ell$. On $\|v\|\le \ell^{1/12}$,
the difference from the Gaussian integrand is bounded by
$C\ell^{-1/2}\|v\|^3e^{-c\|v\|^2}$; the phase from the target has modulus
one. Integrating this bound gives $O(\ell^{-1/2})$. The intermediate region
$\ell^{1/12}<\|v\|\le\varrho\sqrt\ell$ contributes at most $e^{-c\ell^{1/6}}$ by
the quadratic bound. The full difference lattice gives
$|\chi_\theta(z)|\le1-c$ uniformly for $\theta\in\mathcal K$ and
$z\in[-\pi,\pi]^h$ with $\|z\|\ge\varrho$, for some $c>0$.
The integral over this region is therefore exponentially small.
Thus the absolute probability error is $O(\ell^{-(h+1)/2})$.
For $\|u\|\le C\sqrt\ell$, the Gaussian main term is bounded below by
$c_{\mathcal K,C}\ell^{-h/2}$. Division gives the stated relative error and,
eventually, positivity for every integer target in this range.
With computable atom probabilities and given rational
bounds on the tilts and $C$, the covariance bounds, characteristic-function
gap and Taylor bounds can be computed, and hence so can the error constant
and size threshold.
\end{proof}

\begin{lemma}[Finite-polynomial coefficient estimates]
\label{lem:finite-polynomial-coefficients}\label{lem:coefficient-entropy-modulus}
Let $P(z)=\sum_{x\in\mathcal S}c_xz^x$ have fixed finite support
$\mathcal S\subset\mathbb Z_{\ge0}^h$ and $c_x>0$. For
$u\in\operatorname{conv}(\mathcal S)$, define
\[
 F_P(u)=\max_{p\ge0,\,\sum_xp_x=1,\,\sum_xp_xx=u}
       \left\{-\sum_xp_x\log p_x+\sum_xp_x\log c_x\right\}.
\]
The following estimates hold.
\begin{enumerate}[label=(\roman*),leftmargin=*]
\item \emph{Interior saddle estimate.}
Suppose the difference lattice of $\mathcal S$ is $\mathbb Z^h$.
For $\theta$ in a fixed compact set, define the tilted exponent law
$p_\theta(x)=c_xe^{\langle\theta,x\rangle}/P(e^\theta)$, $x\in\mathcal S$.
Let $\mu_\theta=\nabla_\theta\log P(e^\theta)$ and $\Sigma_\theta$ be its mean
and covariance. Uniformly for integer targets
$s=M\mu_\theta$,
\[
 [z^s]P(z)^M
 =\frac{e^{MF_P(s/M)}}{(2\pi M)^{h/2}\sqrt{\det\Sigma_\theta}}
       (1+O(M^{-1/2})).
\]
In particular, all these coefficients are positive for large $M$.
The determinant factor is positive and analytic in the mean on the
corresponding interior neighborhood.
\item \emph{Global rate and shifted targets.}
For every integer target $s$ with $s/M\in\operatorname{conv}(\mathcal S)$,
\[
 [z^s]P(z)^M\le (M+1)^{|\mathcal S|}e^{MF_P(s/M)}.
\]
For $u,u'\in\operatorname{conv}(\mathcal S)$ and
$0<\varrho=\|u-u'\|\le1$,
\[
 |F_P(u)-F_P(u')|\le C\varrho\log(e/\varrho).
\]
Thus a target change of size $O(\sqrt M)$ changes the upper exponent
by $O(\sqrt M\log M)$, including at boundary faces.
Targets outside the convex hull have zero coefficient.
\end{enumerate}
The constants may be chosen jointly for any fixed finite family of
such polynomials and compact tilt sets.
\end{lemma}

\begin{proof}
For any law $p$ on $\mathcal S$ of mean $u$,
\[
 \mathsf H(p)+\sum_xp_x\log c_x
 =\log P(e^\theta)-\langle\theta,u\rangle
       -\KL(p\|p_\theta).
\]
At $u=\mu_\theta$, this identifies the entropy optimizer and its value.
For $s=M\mu_\theta$ and independent $X_1,\ldots,X_M$ with law
$p_\theta$, the exact tilting identity gives
\[
 [z^s]P^M=e^{MF_P(s/M)}
          \Pp_\theta(X_1+\cdots+X_M=s).
\]
Apply \cref{lem:moving-lclt} at the mean to prove (i).
The full difference lattice makes the covariance positive definite;
the analytic inverse-function theorem gives the analytic mean coordinates.

For (ii), group the $M$ polynomial factors by their exponent counts.
There are at most $(M+1)^{|\mathcal S|}$ such types, and each
multinomial term is at most its entropy exponential. This proves the
coefficient upper bound even when some state counts vanish.
For the modulus, take an optimizer $p$ at $u$. Hoffman's error
bound~\cite{Hoffman1952} for the fixed constraints
$p'\ge0$, $\sum_xp'_x=1$, $\sum_xp'_xx=u'$ supplies a law $p'$
with $\|p-p'\|_1\le C\|u-u'\|$.
Finite-alphabet entropy continuity and the bounded values $\log c_x$
give $F_P(u)\le F_P(u')+C\varrho\log(e/\varrho)$.
Interchange $u,u'$ for the other inequality.
\end{proof}

\paragraph{Compatible matching probabilities.}\label{sec:unmarked-matching}

We also need matching probabilities for fixed finite state spaces. Let $\mathcal S$ be a fixed
finite state space and let $\mathcal B$ be a symmetric $0$--$1$ compatibility matrix.
For an interior probability law $\gamma$ on $\mathcal S$, let
$\mathcal M_N^{\mathcal B}(\gamma)$ be the number of perfect matchings of $N$ labeled
vertices, with state counts $N\gamma$, whose every edge is $\mathcal B$-compatible.
Suppose that the allowed-support graph is connected and contains an odd cycle or
an allowed loop. Assume also that the given marginal admits an edge law positive on every allowed edge. Let $P_\gamma$ be the unique entropy-maximizing symmetric edge
law with
\[
  P_\gamma\one=\gamma,
  \qquad
  \supp P_\gamma\subseteq\supp\mathcal B.
\]
It has the multiplier form
\[
  P_\gamma(s,t)=r_sr_t\mathcal B(s,t)
\]
for positive numbers $r_s$.
On the stated region of marginals, $P_\gamma$ is analytic in
$\gamma$, by the inverse-function theorem for the tilted endpoint-count mean. Put
\[
  T_\gamma=\diag(\gamma)^{-1}P_\gamma,
  \qquad
  \phi_{\mathcal B}(\gamma)=\sum_s\gamma_s\log\frac{\gamma_s}{r_s}=\tfrac12\mathsf H(P_\gamma)-\mathsf H(\gamma).
\]
Write $T_{\gamma,\perp}$ for its restriction to
$\{f:\sum_s\gamma_sf(s)=0\}$.

\begin{lemma}[Matching probabilities]\label{lem:one-color-matching-asymptotic}
For $\gamma$ as above and even $N$ for which $N\gamma$ is an integer
state-count vector admitting a compatible perfect matching,
\begin{equation}\label{eq:one-color-matching-asymptotic-new}
  \frac{\mathcal M_N^{\mathcal B}(\gamma)}{(N-1)!!}
  =e^{N\phi_{\mathcal B}(\gamma)}
   \det(I+T_{\gamma,\perp})^{-1/2}
   \left(1+O(N^{-1/2})\right).
\end{equation}
The estimate is uniform when $\gamma$ ranges in a compact interior set on which
$P_\gamma$ has all allowed entries bounded uniformly away from zero.
\end{lemma}

\begin{proof}
Let $S_P$ be the sum of $N/2$ independent endpoint-count
vectors $e_X+e_Y$, where $(X,Y)$ has law $P_\gamma$, and let $S_0$ be
the corresponding sum for independent endpoints with law $\gamma$.
On every compatible ordered-pair word with counts $N\gamma$, the
likelihood ratio of these two laws is
$\prod_s(r_s/\gamma_s)^{N\gamma_s}=e^{-N\phi_{\mathcal B}(\gamma)}$.
Conditioned on its counts, the independent-endpoint word is uniform;
pairing its consecutive positions therefore has compatibility probability
$\mathcal M_N^{\mathcal B}(\gamma)/(N-1)!!$. Consequently, exactly,
\[
 \frac{\mathcal M_N^{\mathcal B}(\gamma)}{(N-1)!!}
 =e^{N\phi_{\mathcal B}(\gamma)}
   \frac{\Pp(S_P=N\gamma)}{\Pp(S_0=N\gamma)}.
\]
Both sums have mean $N\gamma$ and the same full difference lattice
$\{z\in\mathbb Z^{\mathcal S}:\sum_s z_s=0\}$.
The connected, nonbipartite support graph has an even-length walk
between any two states. Along each two-step segment $s,v,t$,
the allowed pair vectors satisfy
$(e_s+e_v)-(e_v+e_t)=e_s-e_t$.
Summing along such walks gives every difference $e_s-e_t$;
these differences generate this lattice. As quadratic
forms on test functions in $L^2_0(\gamma)$, their covariances divided by $N$ are
$I+T_{\gamma,\perp}$ and $I$, respectively. Indeed, for a centered $f$,
\[
 \tfrac12\E_{P_\gamma}(f(X)+f(Y))^2
 =\|f\|_{2,\gamma}^2+\langle f,T_\gamma f\rangle_\gamma.
\]
The moving-tilt local limit lemma (\cref{lem:moving-lclt}), applied in
any fixed integer basis of this lattice, gives the probability ratio
$\det(I+T_{\gamma,\perp})^{-1/2}(1+O(N^{-1/2}))$.
The lattice factors cancel. Compactness of the positive tilted edge laws
makes the error uniform, proving the claim.
\end{proof}

\begin{lemma}[Edge probabilities after exposure]
\label{lem:finite-incidence-transfer}
Let $\mathcal B$ be a fixed symmetric compatibility matrix whose allowed-support graph is
connected and nonbipartite (an allowed loop suffices). Let $\gamma$ range
in a compact family of positive state laws for which $P_\gamma$ is
positive on every allowed edge, uniformly bounded away from zero there.
Take even sizes $N$ with integer, feasible state counts $N\gamma$.
Fix $q_N=O(\sqrt N)$ disjoint labeled pairs of one perfect matching,
recording their labels and states as $\mathcal R$, and let $M$ be their
union with a uniform compatible matching of the remaining labels.
The fixed pairs need not be $\mathcal B$-compatible; assume
that the residual state counts admit a compatible perfect matching and
the residual state law remains in the same compact family. Let
$F=\{\{x_i,y_i\}:1\le i\le \ell\}$, with $\ell\le N^{1/10}$, be any family of additional disjoint
compatible edges on unexposed vertices, with endpoint states $(a_i,b_i)$.  Then
\begin{equation}\label{eq:finite-incidence-transfer}
  \Pp(F\subseteq M\mid\mathcal R)
  =N^{-\ell}\prod_{i=1}^{\ell}
    \frac{P_\gamma(a_i,b_i)}{\gamma(a_i)\gamma(b_i)}
    \left[1+O\left(\frac{\ell}{\sqrt N}\right)\right],
\end{equation}
uniformly in the exposure and the edge family. The same estimate holds
with $N,\gamma$ replaced by the residual size and empirical law.
\end{lemma}

\begin{proof}
The remaining $N'=N-2q_N=N-O(\sqrt N)$ vertices have state law
$\gamma'=\gamma+O(N^{-1/2})$. For one compatible edge of states $a,b$,
its probability is exactly
\begin{equation}\label{eq:one-edge-count-ratio}
 \frac{\mathcal M_{N'-2}^{\mathcal B}(\gamma'_{a,b})}
      {\mathcal M_{N'}^{\mathcal B}(\gamma')},
 \qquad (N'-2)\gamma'_{a,b}=N'\gamma'-e_a-e_b.
\end{equation}
The envelope theorem for $P_\gamma(s,t)=r_sr_t\mathcal B(s,t)$ gives
$\partial_{\gamma(a)}\phi_{\mathcal B}=\log(\gamma(a)/r_a)+c(\gamma)$
on the simplex, where $c(\gamma)$ is independent of $a$ and
cancels for zero-sum perturbations. Taylor expansion at $\gamma'$, using
$\mathcal B(a,b)=1$, gives
\[
 (N'-2)\phi_{\mathcal B}(\gamma'_{a,b})-N'\phi_{\mathcal B}(\gamma')
 =\log\frac{P_{\gamma'}(a,b)}
                    {\gamma'(a)\gamma'(b)}+O(N'^{-1}).
\]
By \cref{lem:moving-lclt} and the lattice argument in the proof of
\cref{lem:one-color-matching-asymptotic}, all intermediate residual targets
are feasible for sufficiently large $N$.
Applying \cref{lem:one-color-matching-asymptotic} and using
$(N'-3)!!/(N'-1)!!=(N'-1)^{-1}$ shows that
\eqref{eq:one-edge-count-ratio} equals
$P_{\gamma'}(a,b)/(N'\gamma'(a)\gamma'(b))$
times $1+O(N'^{-1/2})$. Reveal the $\ell$ edges successively.
The cumulative local-limit error is $O(\ell/\sqrt N)$; bounded derivatives
of the allowed edge factors and the $O(\ell/N)$ drift of the empirical law
cost $O(\ell^2/N)$. Finally, replacing the initial residual parameters by
$N,\gamma$ costs $O(\ell/\sqrt N)$.
Since $\ell\le N^{1/10}$, the total error is
$O(\ell/\sqrt N)=o(1)$ uniformly, proving the assertion.
\end{proof}

\subsection{Critical entropy maximizers}\label{sec:unmarked-entropy}

\begin{proposition}[One- and two-replica critical maximizers]
\label{prop:global-critical-anchor}
\label{prop:unrestricted-critical-pair-anchor}

Fix $d\ge3$. Recall $\Omega=\{00,10,01,11\}$, $B$ from
\eqref{eq:base-compatibility}, and $\alpha_c$ from \eqref{eq:critical-type}.
A $j$-replica state is a tuple in $\Omega^j$.
For $j=1,2$, let $B_j=B^{\otimes j}$ and
$w_j=(1,\lambdaC,\lambdaC,\lambdaC^2)^{\otimes j}$.
For a probability law $\Gamma$ on $\Omega^j$, define its unrestricted
annealed rate by
\[
 \Upsilon^{(j)}(\Gamma)=(1-d)\mathsf H(\Gamma)+\langle\Gamma,\log w_j\rangle
 +\frac d2\max_Y\mathsf H(Y),
\]
where the maximum is over symmetric probability laws $Y$ with
$Y\one=\Gamma$ and $\supp Y\subseteq\supp B_j$.
Here $\mathsf H$ is Shannon entropy, with $0\log0=0$.

Write
$\Upsilon^{(j)}(\Gamma)=-\infty$ if no such edge law exists.
Write $\Upsilon=\Upsilon^{(1)}$, and put $\pi_j=\alpha_c^{\otimes j}$. Then
\[
 \Upsilon^{(j)}(\Gamma)\le j\Upsilon(\alpha_c),
 \qquad\text{with equality if and only if }\Gamma=\pi_j.
\]
Outside every neighborhood of $\pi_j$, the rate is smaller by a fixed
positive amount. This includes all feasible boundary types.
\end{proposition}

For $j=1$, the rate in \cref{sec:gadget-moments} satisfies
$\psi_0(\rho,b)=2\Upsilon^{(1)}(\alpha)$, where $\alpha$ is the type
specified by $(\rho,b)$. The unique maximizer $\alpha_c$ corresponds to
$(\rho,b)=(0,b_c)$.
On the zero-imbalance slice $\alpha_{10}=\alpha_{01}$, this
one-replica uniqueness was proved in~\cite[Lemma~B.5]{CCYZ2026}.

The relative-entropy identity \eqref{eq:unrestricted-critical-pair-rate}
reduces \cref{prop:global-critical-anchor} to the entropy inequality
\eqref{eq:critical-entropy-new}. We derive that inequality from a binary
matrix-norm bound by tensorization and finite-space entropy duality.
Put $D=d-1$ and recall the critical transition matrix $T_c$ and its
binary factor $T_{\rm bin}$ from \eqref{eq:critical-binary-channel}.

\begin{lemma}[Critical two-point hypercontractivity]\label{lem:critical-HC-new}
For every real-valued function $f$ on $\{0,1\}$,
\begin{equation}\label{eq:critical-HC-new}
  \|T_{\rm bin}f\|_{D+1,\bar\pi}
  \le \|f\|_{(D+1)/D,\bar\pi},
  \qquad
  \bar\pi=\left(\frac{D}{D+1},\frac1{D+1}\right).
\end{equation}
Thus $\alpha_c=\bar\pi\otimes\bar\pi$.
Equality holds only for constant functions.
\end{lemma}

\begin{proof}
Put $p=d/D$, $\ell=\lambdaC^{1/d}$, and
\[
 B_{\lambdaC}=\begin{pmatrix}1&\ell\\\ell&0\end{pmatrix},
 \qquad \hat x=\frac1{D-1},\qquad c^*=(1,\hat x/\ell).
\]
Let a nonzero $c\ge0$ maximize $\|B_{\lambdaC}c\|_d/\|c\|_p$, and put
$r=(B_{\lambdaC}c)^D$, with powers taken entrywise.
By~\cite[Equation~(10) and Theorem~4.1]{GSV2016}, both vectors
are positive and satisfy the tree recursions:
$B_{\lambdaC}$ is irreducible and has a positive diagonal entry,
which excludes boundary maxima.
In the coordinates $x=\ell c_1/c_0$, $y=\ell r_1/r_0$, the recursions are
\[
 x=\frac{\lambdaC}{(1+y)^D},\qquad
 y=\frac{\lambdaC}{(1+x)^D}.
\]
Tree uniqueness holds at $\lambda=\lambdaC$
\cite[Section~2]{Kelly1985}, \cite[Proposition~2.5]{Weitz2006}; hence its only solution is
$x=y=\hat x$. Thus $c^*$ is the unique nonnegative maximizing ray.

Write $h=B_{\lambdaC}c^*$. Direct substitution gives
\[
 \diag(h)^{-1}B_{\lambdaC}\diag(c^*)=T_{\rm bin},
 \qquad
 \frac{(c_i^*)^p}{\sum_j(c_j^*)^p}
 =\frac{h_i^d}{\sum_jh_j^d}=\bar\pi(i).
\]
Indeed, both ratios of the coordinate-$1$ weight to the coordinate-$0$
weight equal $1/D$. Therefore, for $f\ge0$,
\[
 \|T_{\rm bin}f\|_{d,\bar\pi}
 =\frac{\|B_{\lambdaC}\diag(c^*)f\|_d}{\|B_{\lambdaC}c^*\|_d}
 \le\frac{\|\diag(c^*)f\|_p}{\|c^*\|_p}
 =\|f\|_{p,\bar\pi}.
\]
Equality for nonzero $f\ge0$ makes $\diag(c^*)f$ a multiple of $c^*$,
so $f$ is constant. For real $f$, use $|T_{\rm bin}f|\le T_{\rm bin}|f|$.
Equality first forces $|f|$ to be constant and then, because the first
row of $T_{\rm bin}$ is strictly positive, forces the two values to have the
same sign. The zero function also gives equality.
\end{proof}

We use the following finite-space form of the functional/entropy
duality and equality correspondence
in~\cite[Theorems~2.1--2.2]{CarlenCorderoErausquin2009}.

\begin{lemma}[Finite-space functional-to-entropy inequality]
\label{lem:finite-functional-entropy}
Let $\Pi$ be a law on a product of two finite sets, with positive marginals $\Pi_1,\Pi_2$,
and let $p>1$. Suppose that for all nonnegative functions $f,g$,
\[
 \E_\Pi[f(X)g(Y)]\le\|f\|_{L^p(\Pi_1)}\|g\|_{L^p(\Pi_2)}.
\]
Then every joint law $\nu$ on these sets satisfies
\[
 \KL(\nu\|\Pi)\ge
 \frac{\KL(\nu_1\|\Pi_1)+\KL(\nu_2\|\Pi_2)}p.
\]
If equality in the functional inequality for nonzero $f,g$ forces both
to be constant, finite-entropy equality holds if and only if $\nu=\Pi$.
\end{lemma}

\begin{proof}
If $\nu\not\ll\Pi$, the left side is infinite. Otherwise, set
$f=(\mathrm d\nu_1/\mathrm d\Pi_1)^{1/p}$ and
$g=(\mathrm d\nu_2/\mathrm d\Pi_2)^{1/p}$.
Their $L^p$ norms are one, so $c=\E_\Pi fg\le1$.
Also $c>0$ and $\widehat\Pi=\Pi fg/c$ is positive on $\supp\nu$.
Direct substitution gives
\[
 \KL(\nu\|\Pi)-\frac{\KL(\nu_1\|\Pi_1)+\KL(\nu_2\|\Pi_2)}p
 =\KL(\nu\|\widehat\Pi)-\log c.
\]
The right side is nonnegative, proving the inequality.
Equality forces $c=1$; the assumed strictness of the functional inequality
then gives $f=g=1$, and zero relative entropy gives $\nu=\Pi$.
The converse is immediate.
\end{proof}

\begin{corollary}[Critical tensor entropy inequality]
\label{cor:critical-one-replica-entropy}
\label{cor:critical-entropy-new}
For $j=1,2$, let $\pi_j=\alpha_c^{\otimes j}$,
$K_j=T_c^{\otimes j}$, and $\Pi^{(j)}(x,y)=\pi_j(x)K_j(x,y)$.
Every law $Y$ on two $j$-replica states satisfies
\begin{equation}\label{eq:critical-entropy-new}
 \KL(Y\|\Pi^{(j)})\ge\frac{D}{D+1}
   [\KL(Y_1\|\pi_j)+\KL(Y_2\|\pi_j)].
\end{equation}
At finite entropy, equality holds if and only if $Y=\Pi^{(j)}$.
\end{corollary}

\begin{proof}
Put $p=(D+1)/D$ and $q=D+1$. The two-point inequality in
\cref{lem:critical-HC-new} tensorizes by the mixed-norm form of
Minkowski's inequality, which applies because $p\le q$
(as in~\cite[Section~3.4]{GSV2016}):
with each coordinate distributed according to $\bar\pi$ and
$T=T_{\rm bin}$ acting in the indicated coordinate,
\[
 \|T_1T_2f\|_{L^q(x,y)}
 \le\big\|\|T_1f(x,\cdot)\|_{L^p(y)}\big\|_{L^q(x)}
 \le\big\|\|T_1f(\cdot,y)\|_{L^q(x)}\big\|_{L^p(y)}
 \le\|f\|_{L^p(x,y)}.
\]
Equality in the last step makes every section $f(\cdot,y)$ constant;
the remaining one-coordinate inequality then makes $f$ constant.
Induction gives the same inequality and equality characterization for all
tensor powers. By \eqref{eq:critical-binary-channel}, $K_j$
is $T_{\rm bin}^{\otimes 2j}$ followed by a coordinate permutation.
Since $\pi_j=\bar\pi^{\otimes 2j}$, this permutation preserves the norms.
Taking all norms below under $\pi_j$, we obtain
$\|K_jg\|_q\le\|g\|_p$, with equality only for constants.

The edge law $\alpha_c(x)T_c(x,y)$ is the symmetric law $P_{\alpha_c}$, so
$K_j$ is $\pi_j$-reversible and both marginals of $\Pi^{(j)}$ equal $\pi_j$.
For nonnegative $f,g$, H\"older's inequality gives
\[
 \E_{\Pi^{(j)}}fg=\langle f,K_jg\rangle_{\pi_j}
 \le\|f\|_p\|K_jg\|_q\le\|f\|_p\|g\|_p.
\]
Equality for nonzero $f,g$ makes $g$ constant by the norm equality case,
and then $f$ constant by strict H\"older. Apply
\cref{lem:finite-functional-entropy} to obtain both the entropy inequality
and its equality characterization, including boundary marginals.
\end{proof}

\begin{proof}[Proof of \Cref{prop:global-critical-anchor}]
We first express the rate difference using relative entropy:
\begin{equation}\label{eq:unrestricted-critical-pair-rate}
 \Upsilon^{(j)}(\Gamma)-j\Upsilon(\alpha_c)
 =(d-1)\KL(\Gamma\|\pi_j)
  -\frac d2
    \inf_{\substack{Y=Y^{\mathsf T},\,Y\one=\Gamma\\
                     \supp Y\subseteq\supp \Pi^{(j)}}}
         \KL(Y\|\Pi^{(j)}).
\end{equation}

The critical edge law has multiplier form $\Pi^{(j)}(x,y)=r_xr_yB_j(x,y)$,
with marginal $\pi_{j,x}=r_x(B_jr)_x$, and it satisfies the critical
stationary equation $\pi_{j,x}\propto w_{j,x}(B_jr)_x^d$. Indeed, for $j=1$
and the states $00,10,01,11$, $r\propto(1,\hat x)^{\otimes2}$ with
$\hat x=1/(D-1)$ the fixed point of the tree recursion, so
$B_1r\propto(1+\hat x,1)^{\otimes2}$; since
$\lambdaC(1+\hat x)^{-d}=\hat x/(1+\hat x)=1/D$, both $r_x(B_1r)_x$ and
$w_{1,x}(B_1r)_x^d$ are proportional to $(1,1/D)^{\otimes2}\propto\alpha_c$.
For $j=2$ all of these are tensor squares. Thus
$D\log\pi_{j,x}+\log w_{j,x}-d\log r_x$ is constant in $x$. For admissible
$Y$, $\mathsf H(Y)=-\KL(Y\|\Pi^{(j)})-2\langle\Gamma,\log r\rangle$.
Substitute this into the rate and subtract its value at
$(\pi_j,\Pi^{(j)})$, which is $j\Upsilon(\alpha_c)$ since the entropy and
weight terms add over replicas and, by subadditivity, so does the maximal
edge entropy. The constant relation cancels every linear term and gives
\eqref{eq:unrestricted-critical-pair-rate}.
This calculation uses only finite sums with $0\log0=0$, and so includes
all feasible boundary faces. The entropy inequality makes the rate
difference nonpositive, and equality forces $\Gamma=\pi_j$.
The feasible vertex--edge pairs form a compact set on which the rate is
continuous. Maximizing over the closed subset with the vertex marginal
outside a fixed neighborhood gives a strict, uniform gap.
\end{proof}

\end{document}